\documentclass[12pt]{article}
\usepackage{a4wide,epsfig,psfrag,amsmath,amssymb,cite,scalefnt}
\usepackage{color}
\usepackage{slashed}

\graphicspath{ {figs/} }

\newcommand{\ep}{\epsilon}

\newcommand{\logtwo}{l_2}

\newcommand{\logone}{l_s}
\renewcommand{\logtwo}{l_s^2}
\newcommand{\logthree}{l_s^3}
\newcommand{\logfour}{l_s^4}
\newcommand{\logfive}{l_s^5}
\newcommand{\logsix}{l_s^6}

\newcommand{\ii}{\mathrm{i}}

\allowdisplaybreaks

\begin{document}

\title{\vskip-3cm{\baselineskip14pt
    \begin{flushleft}
     \normalsize TTP22-042, P3H-22-066
    \end{flushleft}} \vskip1.5cm
  Singlet and non-singlet three-loop massive form factors
  }

\author{
  Matteo Fael$^{a}$,
  Fabian Lange$^{a,b}$,
  Kay Sch\"onwald$^{a}$,
  Matthias Steinhauser$^{a}$
  \\
  {\small\it (a) Institut f{\"u}r Theoretische Teilchenphysik,
    Karlsruhe Institute of Technology (KIT),}\\
  {\small\it 76128 Karlsruhe, Germany}
  \\
  {\small\it (b) Institut f{\"u}r Astroteilchenphysik,
  Karlsruhe Institute of Technology (KIT),}\\
  {\small\it 76344 Eggenstein-Leopoldshafen, Germany}
}

\date{}

\maketitle

\thispagestyle{empty}

\begin{abstract}

\noindent
We consider Quantum Chromodynamics with external vector, axial-vector, scalar
and pseudo-scalar currents and compute three-loop corrections to the
corresponding vertex function taking into account massive quarks.  We consider
all non-singlet contributions as well as those singlet contributions where the
external current couples to a massive quark loop.  We apply a semi-numerical
method which is based on expansions around singular and regular kinematical
points. They are matched at intermediate values of the squared partonic
center-of-mass energy $s$ which allows to cover the whole kinematic range for
negative and positive values of $s$.  Our method permits a systematic increase
of the precision by varying the expansion depth and the choice of the
intermediate matching points.  In our current set-up we have at least 
seven significant digits for the finite contribution of all form
factors.  We present our results as a combination of series expansions and
interpolation functions which allows for a straightforward use in practical
applications.

\end{abstract}

\thispagestyle{empty}

%- }}}

\newpage

%- {{{ Introduction:

\section{Introduction}
\label{sec::introduction}

Form factors are important building blocks for a number of processes.  For
example, they constitute the virtual corrections for lepton pair production
via the Drell-Yan process, for Higgs boson production in gluon fusion and for
Higgs boson decay into heavy quarks.  For many processes there are precise
experimental results such that higher-order corrections have to be included in
the theory predictions.  This is particularly true for QCD corrections, the
topic of this paper.

There is a rapid increase in complexity when increasing
the number of loops for a given process and often one has to rely on approximations valid in
certain regions of phase space, in case one wants to perform analytic calculations.
Alternatively, it is possible to use numerical methods, which, however, are
less flexible in practical applications of the results.

In this paper we consider three-loop corrections to massive form factors
in QCD. We use a method which leads to compact expansions
of the bare three-loop expressions around kinematical points
with high-precision numerical coefficients. Their
numerical evaluation is fast and their combination covers the
whole kinematic range. The counterterms are known in analytic form
and allow for a flexible construction of finite expressions
in different renormalization schemes.

We define the external vector, axial-vector, scalar and pseudo-scalar currents
via
\begin{eqnarray}
  j_\mu^v &=& \bar{\psi}\gamma_\mu\psi\,,\nonumber\\
  j_\mu^a &=& \bar{\psi}\gamma_\mu\gamma_5\psi\,,\nonumber\\
  j^s &=& m \,\bar{\psi}\psi\,,\nonumber\\
  j^p &=& \ii m \,\bar{\psi}\gamma_5\psi\,.
  \label{eq::currents}
\end{eqnarray}
The heavy quark mass $m$ has been introduced in
the scalar and pseudo-scalar currents for convenience such that
all currents have vanishing anomalous dimensions~\cite{Chetyrkin:1994js}.

The three-point functions with an external quark-anti-quark pair,
which result from the currents in Eq.~(\ref{eq::currents}), can be decomposed
into six form factors given by
\begin{eqnarray}
  \Gamma_\mu^v(q_1,q_2) &=&
  F_1^v(q^2)\gamma_\mu - \frac{\ii}{2m}F_2^v(q^2) \sigma_{\mu\nu}q^\nu
  \,, \nonumber\\
  \Gamma_\mu^a(q_1,q_2) &=&
  F_1^a(q^2)\gamma_\mu\gamma_5 {- \frac{1}{2m}F_2^a(q^2) q_\mu }\gamma_5
  \,, \nonumber\\
  \Gamma^s(q_1,q_2) &=& {m} F^s(q^2)
  \,, \nonumber\\
  \Gamma^p(q_1,q_2) &=& {\ii m} F^p(q^2) {\gamma_5}
  \,.
  \label{eq::Gamma}
\end{eqnarray}
Here the momentum $q_1$ ($q_2$) is incoming (outgoing)
and $q=q_1-q_2$ is the outgoing momentum at the current $j^\delta$ with $q^2=s$.  The external
quarks are on-shell, i.e.\ $q_1^2=q_2^2=m^2$, and we have $\sigma^{\mu\nu} =
\ii[\gamma^\mu,\gamma^\nu]/2$.  We note that in all cases the colour structure
is a simple Kronecker delta in the fundamental colour indices of the external
quarks and not written out explicitly.

We distinguish singlet and non-singlet form factors. In the first case the
external current couples to a closed quark loop which is connected to the
quark line in the final state via gluons. In the non-singlet case the external
current couples directly to the final-state quarks. In this paper we consider
the complete non-singlet contributions as well as those singlet contributions
where the closed quark loop has the same mass as the external quarks.  While
we consider all four currents in the non-singlet case, we only consider the
vector, scalar and pseudo-scalar currents in the singlet case.  The proper
treatment of the axial-vector contribution requires the singlet contributions
with a massless closed quark loop, which is postponed to the future.

We define the perturbative expansion of the scalar form factors as
\begin{eqnarray}
  F = \sum_{n\ge0} F^{(n)}
  \left(\frac{\alpha_s(\mu)}{\pi}\right)^n
  \,,
  \label{eq::F_pert_exp}
\end{eqnarray}
where at lowest order we have $F_1^{v,(0)}=F_1^{a,(0)}=F^{s,(0)}=F^{p,(0)}=1$
and $F_2^{v,(0)}=F_2^{a,(0)}=0$.
The three-loop non-singlet form factors can be decomposed into the ten colour
factors
\begin{eqnarray}
  &&\{C_F^3,C_FC_A^2, C_F^2C_A,
  C_F^2 T_Fn_l, C_F C_A T_F n_l, C_F T_F^2n_l^2, C_F T_F^2 n_l n_h,
  C_F^2 T_F n_h,\nonumber\\&& C_F C_A T_F n_h, C_F T_F^2 n_h^2\}\,,
\end{eqnarray}
where $C_F=T_F (N_C^2-1)/N_C$ and $C_A=2 T_F N_C$ are the quadratic Casimir operators of the
$\mathrm{SU}(N_C)$ gauge group in the fundamental and adjoint representation,
respectively,  $n_l$ is the number of massless quark flavors, and $T_F=1/2$.
For convenience we introduce $n_h=1$ for closed quark loops which have the
same mass as the external quarks.

\begin{figure}[t]
  \begin{center}
    \begin{tabular}{ccc}
      \includegraphics[width=0.3\textwidth]{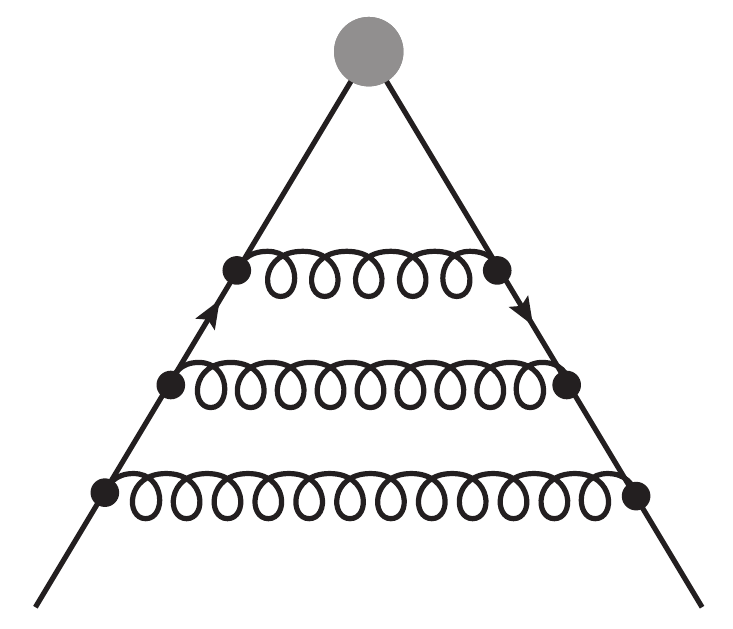}&
      \includegraphics[width=0.3\textwidth]{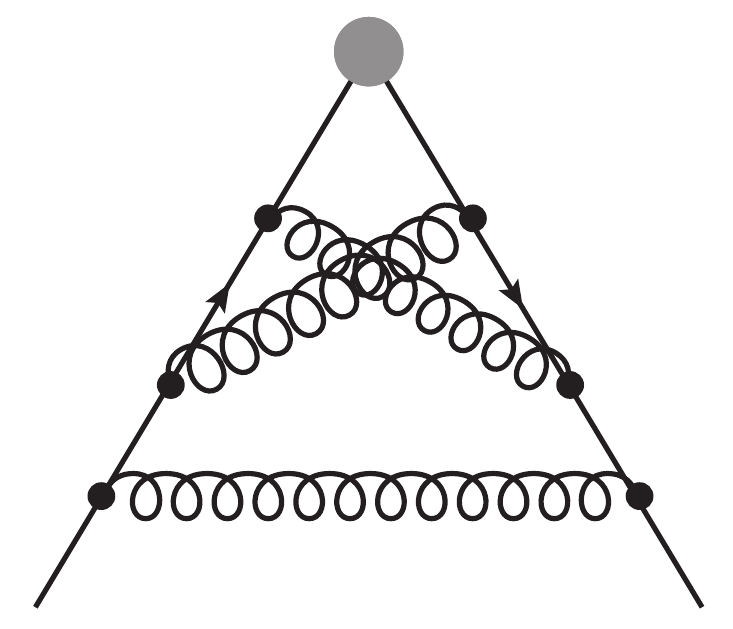}&
      \includegraphics[width=0.3\textwidth]{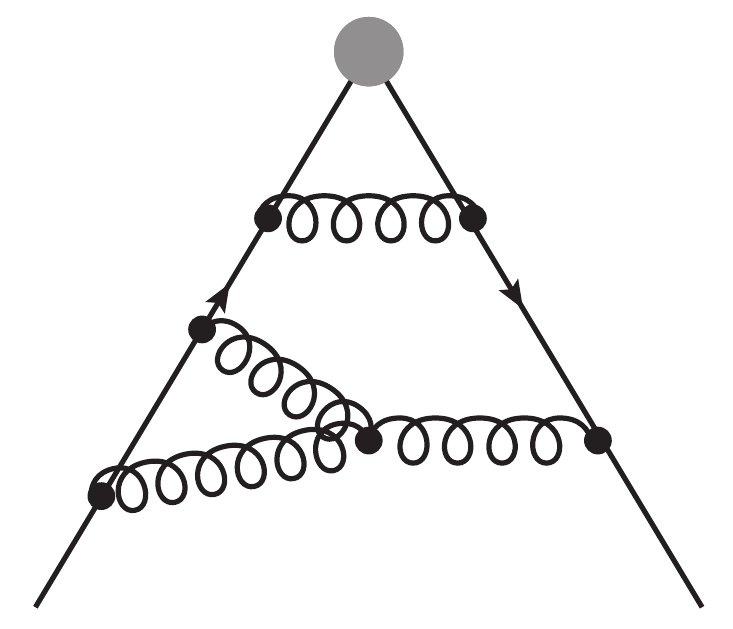}
      \\
      (a) & (b) & (c) \\
      \includegraphics[width=0.3\textwidth]{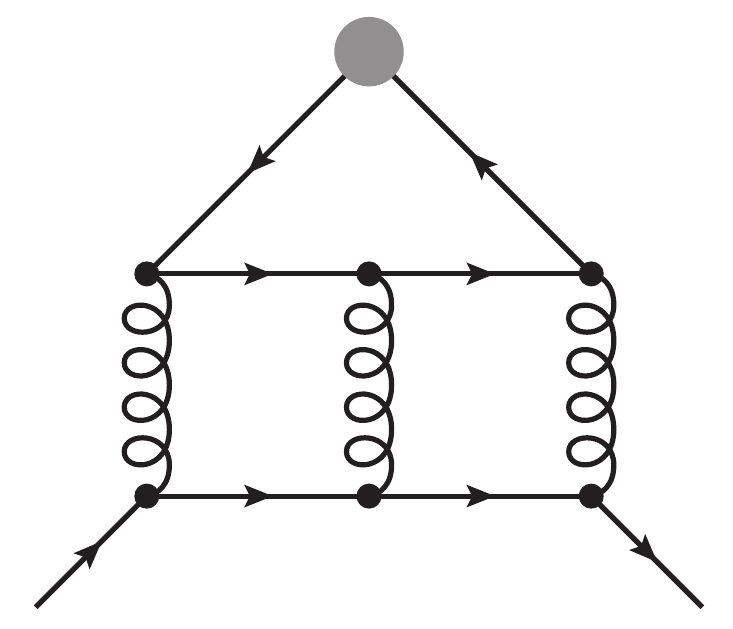}&
      \includegraphics[width=0.3\textwidth]{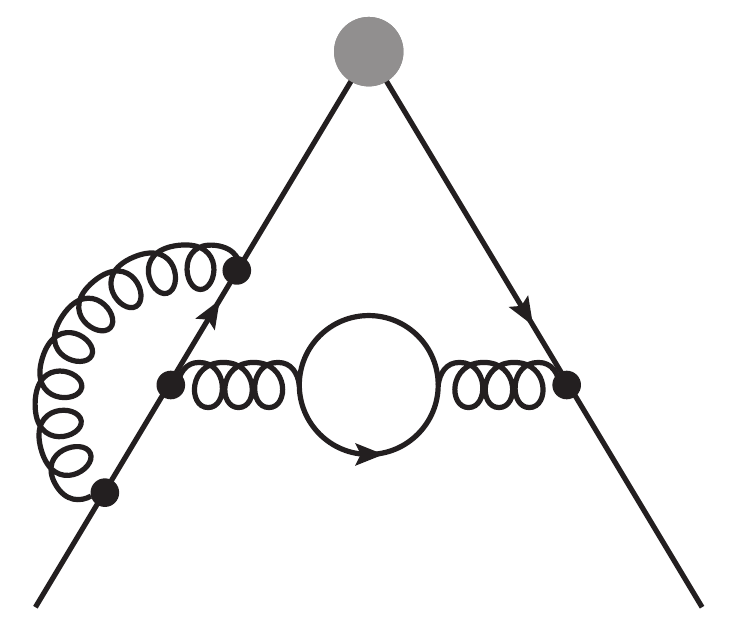}&
      \includegraphics[width=0.3\textwidth]{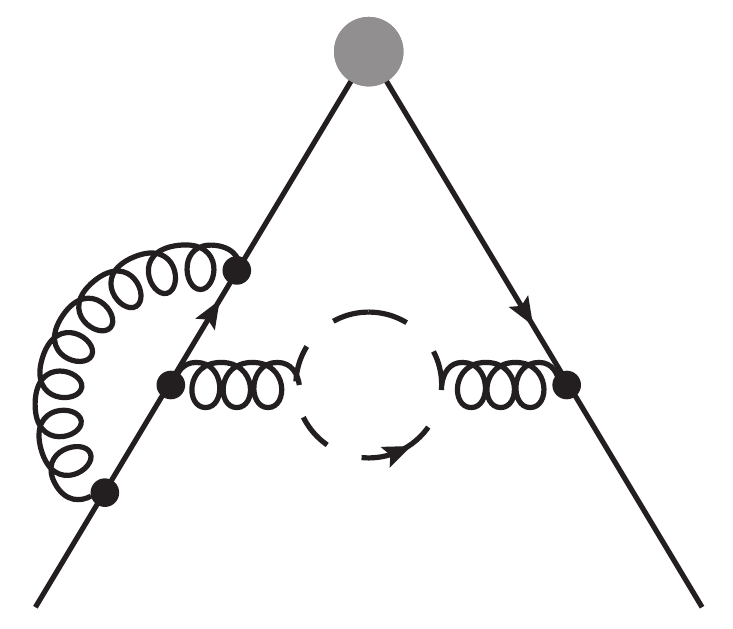}
      \\
      (d) & (e) & (f)
    \end{tabular}
    \caption{\label{fig::diags}Sample diagrams contributing to the form
      factors.  Solid, dashed and curly lines represent massive quarks,
      massless quarks and gluons,
      respectively. The blob refers to one of the external currents given
      in Eq.~(\ref{eq::currents}). Diagram~(d) is a representative
      for the singlet contribution. Here the quark in the closed loop is
      massive.}
  \end{center}
\end{figure}

The singlet form factors start at two-loop order. Due to Furry's theorem the
vector form factor is non-zero only at three-loop order, but there are
non-zero results for
$F^{s,(2)}_{\rm sing}$ and $F^{p,(2)}_{\rm sing}$.\footnote{Also the two-loop singlet axial-vector form
  factor is non-zero. However, it is not considered in this paper.}
For the three-loop singlet
form factors we have the colour factors
  \begin{eqnarray}
    \{C_F^2 T_F n_h, C_F C_A T_F n_h, C_F T_F^2 n_h^2, C_F T_F^2 n_l n_h,
    n_h (d^{abc})^2/N_C
    \}\,,
  \end{eqnarray}
  where $(d^{abc})^2 = (N_C^2-1)(N_C^2-4)/(16N_C)$ arises from Feynman diagrams in which the closed fermion
  loop is connected to the external quarks by three gluons, see Fig.~\ref{fig::diags}(d).  This is the only colour
  structure present in $F_{1,\rm sing}^{v,(3)}$ and $F_{2,\rm sing}^{v,(3)}$. On
  the other hand, the scalar and pseudo-scalar form factors do not have this
  colour structure.

Two-loop corrections to the vector form factors have been computed about 20
years ago, first in the context of QED~\cite{Mastrolia:2003yz,Bonciani:2003ai}
and later also for QCD~\cite{Bernreuther:2004ih} (see also
Ref.~\cite{Hoang:1997ca} for the fermionic contributions). Several groups have
provided cross checks and computed higher order terms in
$\epsilon$~\cite{Gluza:2009yy,Henn:2016tyf,Ahmed:2017gyt,Ablinger:2017hst,Lee:2018nxa}.
Higher order perturbative contributions in the high-energy limit of the form
factors have been predicted in
Refs.~\cite{Mitov:2006xs,Gluza:2009yy,Ahmed:2017gyt} using renormalization
group equations. Two-loop axial-vector, scalar and pseudo-scalar contributions
have been computed in
Refs.~\cite{Bernreuther:2004th,Bernreuther:2005rw,Bernreuther:2005gw,Ablinger:2017hst}.

Analytic three-loop corrections to the form factors have first been considered
in the large-$N_C$
limit~\cite{Henn:2016tyf,Lee:2018rgs,Ablinger:2018yae,Ablinger:2018zwz} where
only planar integrals have to be computed. The first non-planar contributions
appear in the light-fermion contributions which are available from
Ref.~\cite{Lee:2018nxa}.  In Ref.~\cite{Blumlein:2019oas} all contributions
with a closed heavy quark loop have been considered and around 2000 expansion
terms around $s=0$ have been computed.  Let us also mention that all-order
corrections to massive form factors in the large-$\beta_0$ limit have been
considered in Ref.~\cite{Grozin:2017aty}, where $\beta_0$ is the one-loop
correction to the QCD beta function.

In Ref.~\cite{Fael:2022rgm} we computed first complete results for the
vector form factors $F_1^v$ and $F_2^v$ taking into account all colour
factors and covering the whole $s$ range. In this work we present details
of the computational method and extend the results to the axial-vector, scalar
and pseudo-scalar currents.  We present results for the non-singlet
contributions, where the external current couples to the quarks in the final
state.  In this case it is possible to use anti-commuting $\gamma_5$.
Furthermore, we consider the singlet contributions, where the external current
couples to a closed massive quark loop.  For the treatment of $\gamma_5$ in the
pseudo-scalar singlet case we follow Ref.~\cite{Larin:1993tq}.  Sample Feynman
diagrams for the heavy-quark form factors are shown in Fig.~\ref{fig::diags}.

Form factors with massless external quarks are available to higher order
in perturbation theory. They have been computed to three-loop order in
Refs.~\cite{Baikov:2009bg,Heinrich:2009be,Lee:2010ik,Gehrmann:2010ue,Gehrmann:2010tu,vonManteuffel:2015gxa}. Recently
even four-loop corrections became
available~\cite{Henn:2016men,Lee:2016ixa,Lee:2017mip,Lee:2019zop,vonManteuffel:2020vjv,vonManteuffel:2016xki,vonManteuffel:2019wbj,Agarwal:2021zft,Lee:2021uqq,Lee:2022nhh}. In
Ref.~\cite{Chen:2021rft} three-loop singlet corrections to massless
axial-vector form factors have been considered where the external current
couples to a massive closed quark loop.

The remainder of this paper is organized as follows: In Section~\ref{sec::setup} we describe the
setup used for the computation of the amplitudes for the form
factors. In Section~\ref{sec::method} we describe our approach for the
construction of the approximations of the master integrals. 
Section~\ref{sec::singlet} is dedicated to the singlet form factors
and the renormalization and infrared
subtraction is discussed in Section~\ref{sec::ren}. 
Our results are
presented in Section~\ref{sec::res}. We conclude in Section~\ref{sec::concl}.
Supplementary material is relegated to the Appendix. In Appendix~\ref{app::initial}
we provide results for the three-loop on-shell integrals with special emphasis
on the higher order terms in the $\epsilon$ expansion, which we needed as boundary conditions to determine the master integrals.
Appendix~\ref{app::poles} contains the collection of the plots which
show the accuracy of the pole cancellation and in Appendix~\ref{app::s0}
we present semi-analytic expansions around $s=0, 4m^2$ and $\infty$.

%- }}}

%- {{{ Reduction to master integrals and choice of basis:

\section{\label{sec::setup}Reduction to master integrals and choice of basis}

In this Section we provide details for the non-singlet form factors; the
discussion for the singlet form factors is postponed to
Section~\ref{sec::singlet}.

We generate the
amplitudes with {\tt qgraf}~\cite{Nogueira:1991ex} and use {\tt q2e} and~{\tt
  exp}~\cite{Harlander:1997zb,Seidensticker:1999bb,q2eexp} to rewrite the
output to {\tt FORM}~\cite{Kuipers:2012rf} notation and map each diagram to a
predefined integral family.  In this way we can express the form factors as
linear combinations of scalar Feynman integrals with twelve indices where nine
correspond to the exponents of propagators and the remaining
three to the exponents of irreducible numerators.
In total we have $34$ different integral families.

Before performing the reduction of the Feynman integrals contributing to the
form factors, we improve the basis by getting rid of those denominators in the
coefficients of the master integrals which are multivariate polynomials.
Since the ratio $s/m^2$ is the only kinematic variable, it is possible to find
a basis such that the denominators completely factorize into univariate
polynomials of either $s/m^2$ or $d$~\cite{Smirnov:2020quc,Usovitsch:2020jrk}.
Moreover, it turns out that we can choose a basis such that all
polynomials of $s/m^2$ in the denominators are linear polynomials raised
to some power.  We call this basis \emph{good} in the following.
To construct this basis we first reduce all integrals up to the top-level
sector and with up to two dots for every integral family individually with the
help of \texttt{Kira}~\cite{Maierhofer:2017gsa,Klappert:2020nbg} employing
\texttt{Fermat}~\cite{fermat}.  As initial basis we simply take the
default master integrals found with the integral ordering~$6$ of
\texttt{Kira}, i.e.\ the sectors are primarily ordered by the number of lines
and dots are preferred over scalar products.  These reduction tables then
serve as input to search for a good basis for every family with the help of an
improved version of \texttt{ImproveMaster.m} developed in
Ref.~\cite{Smirnov:2020quc}.\footnote{We thank A.~V.~Smirnov and V.~A.~Smirnov
  for giving us access to their development version.}  The construction of a
good basis takes about three hours for the most expensive families and almost
all of that runtime is spent for the reduction.

The actual reduction of the integrals for the form factors is also performed
with \texttt{Kira} employing \texttt{Fermat}.  First, we reduce the integrals
for every family individually to the good basis of this family.  The most
expensive families run for about a week on eight cores and require about
$200$\,GiB of memory.  The resulting reduction tables, especially the
denominators, are simplified by using \texttt{Mathematica} to factorize the
expressions which is not done by \texttt{Kira} and \texttt{Fermat}.  Secondly,
we employ symmetries between the families to reduce the number of master
integrals.  To this end we sort the families by ascending number of master
integrals and use \texttt{Kira} to reduce the good master integrals for every
family to those of easier families if possible.  This reduces the number of
master integrals from $3131$ to $422$.  $136$ of these master integrals are
needed for the fermionic contributions with at least one closed massive
or massless quark loop.
This step takes about one day and can be done in parallel to the individual
reductions of all families since the non-minimal good basis is already known.

We set up the differential equations for the master integrals by
differentiating the master integrals with respect to $s/m^2$ with the help of
\texttt{LiteRed}~\cite{Lee:2012cn,Lee:2013mka} and reducing the resulting
integrals again to master integrals with \texttt{Kira}.  Also this step is
performed on a per-family basis and the already known symmetry tables are only
inserted in the very last step.  The construction of the differential
equations takes at most a few hours per family.  The only complication at this
point is that a few blocks on the diagonal of the system of differential
equations contain unfavorable pole structures that prohibit a solution of the
master integrals as series in $\epsilon$.  However, this problem can be solved
by searching the available reduction tables for integrals in the affected
sectors with a pole in front of one of the master integrals of the same
sector.  Inverting the relation, i.e.\ switching integral and master integral,
thus introduces a positive power of $\epsilon$ and improves on the pole
structure~\cite{Chetyrkin:2006dh}.  By repeating this step all sectors can be
made solvable.  For our problem this can be achieved without spoiling the good
properties of the basis.
Let us emphasize that we explicitly did not try to get rid of all poles in our reduction tables because this would most certainly spoil the good properties.

%- }}}
%- {{{ Numerical matching method:

\section{\label{sec::method}Method}

In order to (numerically) solve the master integrals we apply the method of analytical expansions
and numerical matching introduced in Ref.~\cite{Fael:2021kyg}.
Let us briefly summarize the basic idea:
\begin{itemize}
  \item[(a)] After establishing a system of differential equations for the master integrals, we
  calculate initial values at a kinematic point where the integrals simplify.
  \item[(b)] We construct an analytic series expansion around this kinematic point by inserting
  a suitable ansatz into the system of differential equations, re-expanding in $\epsilon$ and
  around the special kinematic point, establishing a system of linear equations for the expansion
  coefficients and solving it in terms of a small set of boundary coefficients.
  We use the information from step (a) to fix these coefficients analytically.
  \item[(c)] Construct an analytic expansion around a neighboring kinematic point.
  Here we cannot fix the boundary constants analytically, but we can evaluate both expansions
  at a kinematic point where the radii of convergence of both expansions overlap
  and use these values as numerical initial conditions.
  \item[(d)] Repeat this procedure until the whole kinematic interval of interest is mapped out
  with partially overlapping series expansions.
\end{itemize}
In the following we give details on how we applied this method to the calculation of the massive form factors.
We concentrate on the non-singlet form factors; details for the singlet form
factors are provided in Section~\ref{sec::singlet}.

\paragraph{(a) Calculation of initial values} \ \\

For our initial expansion we choose the value $s=0$.  At this point the master
integrals simplify to on-shell propagator integrals which are well studied in
the literature \cite{Laporta:1996mq,Melnikov:2000qh,Lee:2010ik}.  However, one
obstacle in obtaining the initial values for our system of differential
equations is the appearance of high spurious poles in the dimensional
regulator $\epsilon$ in the physical amplitude and the differential equation.
This requires the knowledge of some master integrals beyond the order given in
Ref.~\cite{Lee:2010ik}.  The same problem was also encountered in
Ref.~\cite{Blumlein:2019oas}, where a subset of integrals was already obtained
to sufficiently high order in $\epsilon$.  We extended the calculation of the
remaining integrals to the order in $\epsilon$ needed in our approach.
Details on the calculation and the results are given in
Appendix~\ref{app::initial}.  We note that due to the spurious poles of
$\mathcal{O}(1/\epsilon^8)$ we need some integrals up to transcendental weight
9, however, in the physical amplitude all contributions with weight higher
than 5 cancel.

\paragraph{(b) Construction of series expansions} \ \\

In order to construct series expansions around specific values of $s$ we insert a
suitable ansatz into the differential equation, re-expand in $\epsilon$ and around
the chosen value of $s$ and construct a system of linear equations for the expansion
coefficients of the ansatz by equating coefficients.
The suitable ansatz can be found from physical considerations.
We only have three kinematic points with non-analytical expansions.
They correspond to the two-particle threshold $s=4m^2$, the four-particle threshold
$s=16m^2$ and the high-energy expansion $s \to \pm \infty$.
For these three expansion points we have to use a power-log ansatz.
Since the threshold expansions are governed by the particle velocities, we
use the variables
$z=\sqrt{4-s}$ and $z=\sqrt{16-s}$ for the two- and four-particle thresholds respectively.
The ansatz for a master integral $I_n$ is then given by
\begin{eqnarray}
  I_n &=& \sum\limits_{i=-o}^{\infty} \sum\limits_{j=-k}^{\infty} \sum\limits_{m=0}^{i+o} c_{n,i,j,m} \epsilon^i z^l \ln^m(z)
  ~.
  \label{eq:ansatz}
\end{eqnarray}
For the present three-loop problem we have $o=3$ and we choose $k=10$ as a conservative lower bound.
In principle both lower values depend on the master integral and can be determined beforehand.
In practice, we choose them uniformly for all master integrals for simplicity.
The high-energy expansion can be performed in the variable $z=1/(-s)$, but one has to allow for
Sudakov-like double logarithms in the ansatz.
The upper summation variable of the sum over $m$ in Eq.~\eqref{eq:ansatz} has therefore to be
modified to $i+2o$.
All other expansions are given by simple Taylor expansions in the variable $z=s_0-s$, with
$s_0$ the value we want to expand around.
Thus, the sum over $m$ can be dropped and $k=0$.

The resulting system of linear equations, which does not contain any variable anymore,
is solved by \texttt{Kira} together with \texttt{Fermat} or \texttt{FireFly}~\cite{Klappert:2019emp,Klappert:2020aqs}.
For the use with \texttt{FireFly} we use a special version of \texttt{Kira} which allows
the efficient use of finite field methods and rational reconstruction even for systems without
variables.
We make sure to prefer boundary constants corresponding to low powers of $z$ and $\epsilon$
in the ansatz when solving the system.

\paragraph{(c) Numerical matching of neighboring expansions} \ \\

Once we found the symbolic expansion around $s=0$ we can use the previously calculated initial
values to fix the remaining boundary constants.
This results in an analytic expansion of the master integrals around $s=0$, where we calculated
76 terms in the expansion.
In a next step we construct a symbolic expansion around $s=1$.
For $s=1$ the boundary constants cannot be easily fixed analytically.
However, since the radii of convergence for the expansion around $s=0$ and $s=1$ overlap, we
can use the numerical evaluation of the expansion around $s=0$ at $s=1/2$ to obtain numerical
boundary conditions for the expansion around $s=1$.

In practice, we equate the expansions around $s=0$ and $s=1$ after setting
$s=1/2$ to obtain a linear system of equations for the remaining boundary
constants in the $s=1$ expansion.  For analytical values at $s=1/2$ this
system has a unique solution.  However, due to the finite numerical
precision this is not the case anymore and we have to be careful in
solving the system.  We proceed in the following way:
\begin{itemize}
  \item In order to get numerically more stable results we rationalize the numbers in the system
  with a preset accuracy.
  \item We sort the equations with ascending number of appearing boundary constants and start
  by solving the equations one by one and insert the solutions back into the remaining equations.
  After inserting we rationalize again to the preset accuracy in order to avoid
  bigger and bigger rational numbers.
  Furthermore, we set numbers with absolute values smaller than a threshold to zero.
  This is important, since sometimes boundary conditions in the system are multiplied by very
  small numbers of the order of the preset accuracy. If we accidentally solve for these constants
  we introduce large values into the system of equations which render the solution of the system unstable.
\item Sometimes after reinserting relations, which have been found before,
  back into the system, equations with no boundary constant remain.  In an
  analytic matching these numbers would be identically zero.  We use the
  absolute value of these remaining numbers to judge the accuracy of the
  matching.
\end{itemize}
To evaluate the stability of the final solution we solve the systems twice:
once with rationalized coefficients with an accuracy of 500 digits and setting all numbers smaller than $10^{-100}$ to zero
and once with rationalized coefficients with an accuracy of 100 digits and setting all numbers
smaller than $10^{-50}$ to zero.
We find agreement of both runs with the expected accuracy of $\sim 100$ digits.

\paragraph{(d) Mapping out the kinematics} \ \\

In previous calculations of the form factors the variable $x$ defined by
\begin{eqnarray}
  s = - \frac{(1-x)^2}{x}
\end{eqnarray}
has been used.
This choice of variable maps the special points of the physical amplitude
corresponding to low energy ($s=0$), production threshold of two heavy quarks ($s=4m^2$)
and the high-energy limit ($s = \pm \infty$) to the points $x=1,0,-1$ and,
up to two-loop order, allows to express the final result in terms of harmonic
polylogarithms.
However, in our approach it turns out that the variable $x$ is not a good choice.
One problem we encounter is that the production threshold of four heavy quarks, which starts
contributing at three-loop order, is mapped to $x= 4 \sqrt{3}-7 \approx -0.072 $.
As a consequence the expansion around $x=0$ has a very small radius of convergence
and large numerical coefficients appear in the expansions, making the numerical
solution less stable.
In addition, we also have to construct expansions around the value of the new threshold.
In principle this can be achieved by defining the square root as a symbol before passing the system of
equations to \texttt{Kira}.
However, this complicates the solution of the system considerably.
We therefore choose to do the expansions in the variable $s$.

To obtain the results for the non-singlet\footnote{The information for the
  singlet form factor is provided in Section~\ref{sec::singlet}.} form factors
over the whole real axis we construct expansions at the values of
\begin{eqnarray}
  \label{eq::s0}
  \frac{s_0}{m^2} &=& \Biggl\{
    - \infty, -32, -28, -24, -16, -12, -8, -4, 0, 1, 2 , 5/2 , 3 , 7/2,
    \nonumber \\ &&
    4, 9/2, 5, 6, 7, 8, 10, 12, 14, 15, 16, 17, 19, 22, 28, 40, 52
  \Biggr\}
\end{eqnarray}
starting from our initial expansion at $s=0$.\footnote{Compared to
  Ref.~\cite{Fael:2022rgm} we have added an expansion for $s/m^2=52$.}
The numerical matching between two neighboring expansions is always done close to
the center of the interval.
Some of these values are chosen to correspond to additional, unphysical poles in the
differential equation which we observe at
\begin{eqnarray}
  \frac{s}{m^2} = \Biggl\{
    -4,-2,-1,-1/2,1/2,1,2,3,16/3
  \Biggr\}
  ~.
\end{eqnarray}
However, it turns out that these additional poles are all spurious and do not
spoil the convergence of the series expansions.  Thus, expansions around
$s/m^2=4,16,\infty$ are in principle sufficient to construct the form factors
at arbitrary values of $s$.  However, we find that this approach suffers from
a slow convergence of the individual series expansions close to the singular
points, which means that very deep expansions are required to get decent
accuracy at the matching points.  Since very deep expansions are expensive in
our approach we trade the expansion depth with more expansions at intermediate
points with a default expansion depth of $51$ terms.  These intermediate
points are all Taylor expansions and therefore quite inexpensive to calculate.

For series expansions close to a singular point we employ M{\"o}bius
transformations, which have already been discussed in Ref.~\cite{Lee:2017qql}.
Assume, we want to expand around the point $x_k$ and there are singular points
of the differential equations at $x_{k-1}$ and $x_{k+1}$ with
$x_{k-1} < x_k < x_{k+1}$. Naively the radius of convergence is limited by the
distance to the closer singular point. However, the variable transformation
\begin{eqnarray}
  y_k = \frac{(x - x_k)(x_{k+1} - x_{k-1})}{(x - x_{k+1})(x_{k-1} - x_k) + (x - x_{k-1})(x_{k+1} - x_k)}
\end{eqnarray}
maps the points $x_{k-1}$, $x_k$, $x_{k+1}$ to $-1$, 0, 1.
The radius of convergence of the series expansion is therefore extended into the direction of the farther singularity
although the convergence at the boundaries can be quite slow.
For example we construct the expansion around $s/m^2=7/2$ in the difference $x=7/2 m^2 - s$, but
afterwards reexpress the result in
\begin{eqnarray}
  y &=& \frac{7 m^2 - 2 s}{2 s - 9 m^2}
  ~.
\end{eqnarray}
While the expansion in the variable $x$ converges for $s \in (3,4)$, the convergence for the
expansion in $y$ is increased to $s \in (-\infty,4)$, making the evaluation at the matching point
$s=15/4$ much more precise.

Let us comment on the resources needed to construct the series expansions in our approach.
To set up the system of linear equations for the construction of the series expansions we
use \texttt{Mathematica} and can trivially parallelize this step over all master integrals.
For the most complicated master integrals this step takes around $1\,$h on a single core.
The solution of the resulting system for a simple Taylor expansion over one prime field takes
around $6\,$h and requires about $50\,$GiB of memory.
In total around 50 evaluations are needed which can all run in parallel.
The power-log expansions are more expensive to compute.
Due to the Sudakov-like double logarithms the high-energy expansion is the most involved.
Here the solution of the linear system of equations over one prime field takes around $10\,$d
with a memory requirement of $250\,$GiB, but again the evaluation over 50 prime fields is
enough to reconstruct the full result.

The idea to utilize the associated system of differential equations to obtain
expansions or numerical evaluations of the master integrals has received a lot
of attention especially in the recent
past~\cite{Laporta:2001dd,Boughezal:2007ny,Blumlein:2017dxp,Liu:2017jxz,Lee:2017qql,Moriello:2019yhu,Dubovyk:2022frj},
and there are even public packages available which implement different
algorithms~\cite{Lee:2018ojn,Hidding:2020ytt,Liu:2022chg,Armadillo:2022ugh}.
However, most of them have restrictions on the structure of the system of
differential equations, e.g.\ it has to be triangular
for~\texttt{SeaSyde}~\cite{Armadillo:2022ugh} or Fuchsian without resonances
for \texttt{DESS}~\cite{Lee:2018ojn}, or on the solution, e.g.\ only formal
power series~\cite{Blumlein:2017dxp}.  Some methods also lack the proof that
they can handle complicated physical problems with a few hundred master
integrals.  Here, we show that the method laid out above can be used to
compute a nontrivial quantity, namely the massive form factors at three-loop order
in QCD, and obtain precise numerical results over the whole parameter space.

%- }}}
%- {{{ Singlet contributions to the form factor:

\section{\label{sec::singlet}Singlet contributions to the form factors}

In this Section we discuss the calculation of the singlet contribution to the
three-loop massive QCD form factors, i.e.\  contributions of the type shown in
Fig.~\ref{fig::diags}(d) where the current couples to a closed heavy-quark
loop.  We present results for the vector, scalar and pseudo-scalar case.  The
axial-vector contribution is anomalous and needs the inclusion of singlet
diagrams where the current couples to a closed light-quark loop.    Since many
of the technical details of the calculation closely follow the non-singlet
case, we only discuss them briefly. We put special emphasis
on the computation of the boundary conditions in the limit $s \to 0$
since the calculation is different from the one in the non-singlet case.
Results for the singlet contribution are shown in Section~\ref{sec::res}
together with the non-singlet case.

To generate the amplitude in the singlet case we need 17 different integral
families.  The resulting list of scalar integrals is again reduced with
\texttt{Kira} and \texttt{Fermat} where we find 316 master integrals after
symmetrization over all families.
In this case we not only make sure to reduce to a good
basis, but in addition reduce the number of spurious poles in $\ep$.
This is achieved with an improved version of \texttt{ImproveMasters.m}.
Afterwards we again establish a
closed system of differential equations with the help of
\texttt{LiteRed} with subsequent reductions using \texttt{Kira}.  Overall the
reduction of the singlet diagrams is significantly simpler than the one in the
non-singlet case.

In principle the whole method discussed in Sec.~\ref{sec::method} can be
directly applied to the singlet diagrams.  However, the calculation of the
boundary conditions at $s=0$ is more involved, since the singlet diagrams can
possess massless cuts.  Hence, a simple Taylor expansion around the limit
$s=0$ is not sufficient, but we have to perform an asymptotic expansion.  We
realize the asymptotic expansion by the method of
regions~\cite{Beneke:1997zp,Smirnov:2002pj}, where the relevant regions can be
found with the program \texttt{asy.m}~\cite{Jantzen:2012mw}.

Applying \texttt{asy.m} to the singlet master integrals we find that there
are regions with three different scalings:
\begin{itemize}
        \item region 1: $I \sim y^{-0 \, \epsilon}$
        \item region 2: $I \sim y^{-2 \, \epsilon}$
        \item region 3: $I \sim y^{-4 \, \epsilon}$
\end{itemize}
with $y = \sqrt{-s/m^2}$.
Region~1 corresponds to the hard region, where we can naively expand
the integrands around $s=0$ and apply the same procedure as for the non-singlet
diagrams to obtain the boundary conditions.

\begin{figure}
    \centering
    \includegraphics[width=0.4\textwidth]{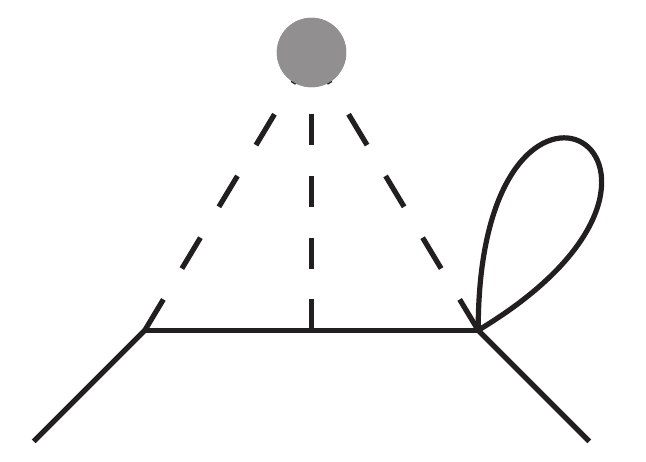}
    \caption{The graph corresponding to master integral $J$. Full (dashed)
      lines correspond to massive (massless) scalar propagators. The dot
      represents the influx of the external momentum $q$.}
    \label{fig:sing_example}
\end{figure}

The expansions in the two soft regions 2 and 3 do not permit a
straightforward diagrammatic expansion.  In order to calculate the boundary
conditions in these regions we perform the expansion on the level of the
integral representation in terms of $\alpha$ parameters and use direct
integration methods to solve the occurring integrals.  Let us exemplify the
calculation of the boundary conditions for the two soft regions for the most
involved master integral $J$ we encounter.  The corresponding graph is
shown in Fig.~\ref{fig:sing_example}.

The Symanzik polynomials $\mathcal{U}$ and $\mathcal{F}$ for this diagram are given by
\begin{eqnarray}
        \mathcal{U} &=& \alpha _5 \left(\left(\alpha _3+\alpha _4\right) \alpha _6+\alpha _1 \left(\alpha _3+\alpha _4+\alpha _6\right)+\alpha _2 \left(\alpha _3+\alpha _4+\alpha _6\right)\right)
        ~, \\
        \mathcal{F} &=& \alpha _2 \alpha _4 \alpha _5 \alpha _6 {(-s)} + \alpha _5 \biggl(\left(\alpha _3+\alpha _4+\alpha _6\right) \alpha _1^2
        +\left(\alpha _3^2+\left(\alpha _5+2 \alpha _6\right) \alpha _3+\alpha _5 \left(\alpha _4+\alpha _6\right)\right) \alpha _1
        \nonumber \\ &&
        +\left(\alpha _3^2+\alpha _5 \alpha _3+\alpha _4 \alpha _5\right) \alpha _6+\alpha _2 \left(\alpha _3^2+\alpha _5 \alpha _3+\alpha _5 \left(\alpha _4+\alpha _6\right)\right)\biggr) m^2
        ~,
\end{eqnarray}
and the master integral can be written as
\begin{eqnarray}
    J &=& \int\limits_{0}^{\infty} \left( \prod\limits_{i=1}^{6} {\rm d} \alpha_i \right) \ \mathcal{U}^{-d/2} e^{-\mathcal{F}/\mathcal{U}} .
    \label{eq:alpharep}
\end{eqnarray}
With the help of \texttt{asy.m} we find the possible scalings of the
$\alpha$-parameters to be $\{0, 0, 0, 0, 0, 0\}$ or
$\{-1, -2, -1, -2, 0, -2\}$.  The first scaling corresponds to the hard
region, which we can calculate with the same method as in the non-singlet case.
We obtain
\begin{eqnarray}
    J_1 &=& e^{-3\gamma_E \epsilon} \Biggl\{
        \frac{\pi ^2}{6 \epsilon ^2}
        +\frac{1}{\epsilon }
        \biggl(
            2 \zeta _3+\frac{\pi ^2}{2}
        \biggr)
        +6 \zeta _3 +\frac{79 \pi ^4}{360}+\frac{7 \pi ^2}{6}
        +\epsilon
        \biggl(
            \frac{5 \pi ^2 \zeta _3}{3}+14 \zeta _3+36 \zeta _5
            \nonumber \\ &&
            +\frac{79 \pi ^4}{120}+\frac{5 \pi ^2}{2}
        \biggr)
        +\epsilon ^2
        \biggl(
            6 \zeta _3^2+5 \pi ^2 \zeta _3+30 \zeta _3+108 \zeta _5+\frac{17743 \pi ^6}{60480}+\frac{553 \pi ^4}{360}+\frac{31 \pi ^2}{6}
        \biggr)
        \nonumber \\ &&
        + \mathcal{O}(\epsilon^3)
    \Biggr\}
    ~.
\end{eqnarray}
The second scaling corresponds to region~3, which can be seen by applying the
scaling relations to the integrand in Eq.~\eqref{eq:alpharep}.  In this region
the $\mathcal{U}$ and $\mathcal{F}$ polynomials reduce to
\begin{eqnarray}
        \mathcal{F}_{3} &=& \alpha _5 \left(\alpha _4 \alpha _6+\alpha _2 \left(\alpha _4+\alpha _6\right)\right)
        ~, \\
        \mathcal{U}_{3} &=& \alpha_5 \biggl(\alpha _2 \alpha _4 \alpha _6 {(-s)}
        + \Bigl(
            \left(\alpha _4+\alpha _6\right) \alpha _1^2
            +2 \alpha _3 \alpha _6 \alpha _1
            +\left(\alpha _3^2+\alpha _4 \alpha _5\right) \alpha _6
            \nonumber \\ &&
            +\alpha _2 \left(\alpha _3^2+\alpha _5 \left(\alpha _4+\alpha _6\right)\right)
        \Bigr) m^2
        \biggr)
        ~.
\end{eqnarray}
To obtain our boundary condition in region 3 we therefore have to solve the integral
\begin{eqnarray}
        J_{3} &=& \int\limits_{0}^{\infty} \left( \prod\limits_{i=1}^{6} {\rm d}\alpha_i \right)
        \left( \mathcal{U}_{3} \right)^{2\epsilon-2} \exp\left( - \frac{\mathcal{F}_{3}}{\mathcal{U}_{3}} \right)
        ~.
\end{eqnarray}
The $\alpha$ parameters $\alpha_1$, $\alpha_2$ and $\alpha_5$ can be
integrated using the formulae
\begin{eqnarray}
        \int\limits_{0}^{\infty} {\rm d}\alpha \, \alpha^a \exp\left(- A \, \alpha \right) &=& \Gamma(1+a) A^{-1-a}
        ~,
        \label{eq:directint1}
        \\
        \int\limits_{0}^{\infty} {\rm d}\alpha \, \alpha^a (1+\alpha)^b &=& \frac{\Gamma(1+a) \Gamma(-1-a-b)}{\Gamma(-b)}
        \label{eq:directint2}
\end{eqnarray}
together with suitable rescalings of the integration variables.
In the end we are left with
\begin{eqnarray}
        J_{3} &=& y^{-4 \epsilon} \frac{\Gamma(\epsilon-1) {\Gamma(2 \epsilon)}}{2}
        \int\limits_{0}^{\infty} {\rm d}\alpha_3
        \int\limits_{0}^{\infty} {\rm d}\alpha_4
        \int\limits_{0}^{\infty} {\rm d}\alpha_6
        \frac{ \alpha_4^{-2\epsilon} \, \alpha_6^{-2\epsilon} \left(\alpha_6 + \alpha_4(1+\alpha_6)\right)^{3\epsilon-1}}{\alpha_4 + \alpha_6 + 2 \alpha_3 \alpha_6 + \alpha_3^2(1+\alpha_6)}
        ~.
        \label{eq:int1}
\end{eqnarray}
The remaining integrals cannot be further reduced by applying the formulae
in Eqs.~\eqref{eq:directint1} and~\eqref{eq:directint2}.  We find it convenient
to integrate Eq.~\eqref{eq:int1} with the help of the program
\texttt{HyperInt}~\cite{Panzer:2014caa}, since the divergent pieces for
$\epsilon \to 0$ are already factored out into the $\Gamma$ functions and the
integrand can be simply expanded in $\epsilon$.  After performing the
integrals the result is given by
\begin{eqnarray}
        J_{3} &=& y^{-4 \epsilon} \pi^2 e^{-3\gamma_E \epsilon} \Biggl\{
                -\frac{1}{6 \epsilon^2}
                -\frac{7}{6 \epsilon }
                +\frac{13 \pi ^2}{72}-\frac{43}{6}
                + \epsilon \left(
                        \frac{59 \zeta_3}{6}-\frac{259}{6}+\frac{91 \pi ^2}{72}
                \right)
                \nonumber \\ &&
                + \epsilon ^2 \left(
                        \frac{413 \zeta_3}{6}-\frac{1555}{6}+\frac{559 \pi ^2}{72}+\frac{3767 \pi^4}{8640}
                \right)
                + \mathcal{O}(\epsilon^3)
        \Biggr\}
        ~.
\end{eqnarray}
Region 2 does not contribute to the master integral $J$, so we have $J_2=0$.
The complete boundary condition can be obtained by adding up the contributions
of all regions:
\begin{eqnarray}
    J &=& J_1+J_2+J_3 + \mathcal{O}(y) ~.
\end{eqnarray}
Note that for regions~2 and~3 for all other boundary conditions the
formulae in Eqs.~\eqref{eq:directint1} and~\eqref{eq:directint2} are sufficient to
obtain the boundary in terms of $\Gamma$ functions and therefore exact in
$\epsilon$.

After all boundary conditions for all three regions are found we again obtain
analytic series expansions around $s=0$.  In the singlet case we calculate
symbolic expansions at the values of
\begin{eqnarray}
    \frac{s_0}{m^2} &=& \{ -\infty , -32 , -28 , -24 , -16 , -12 , -8 , -4 , -3 , -2 , -1 , 0 ,
     1 , 2 , 3 , 7/2 , 4 ,
     \nonumber \\ &&
     9/2, 5 , 6 , 7 , 8 , 10 , 12 , 14 , 15 , 16 , 17 , 19 , 22 , 28 , 40 , 52 \}
\end{eqnarray}
and match numerically again starting from our initial expansion at $s=0$.  The
singular points at $s/m^2=4,16,\infty$ are shared among the singlet and
non-singlet case.  For the singlet diagrams also the point $s=0$ is a singular
point and we need to perform a power-log expansion. Let us mention that
in the singlet case the differential equation has the
additional unphysical poles
\begin{eqnarray}
  \frac{s}{m^2} &=& \{ -4,-2,-1/2,0,1,2,4,4 \pm 8 \ii,16/3,12,16 \}\,.
  \end{eqnarray}

%- }}}
%- {{{ UV renormalization and IR subtraction:

\section{\label{sec::ren}Ultraviolet renormalization and infrared subtraction}

The form factors develop both ultraviolet (UV) and infrared (IR) divergences.
To account for the UV divergences we have to renormalize the
quark mass, the strong coupling constant and the wave function of the external
quarks.
We denote the UV-renormalized form factors by $F^{\rm UV ren}$
which still have poles in $\epsilon$ of IR nature.
They are taken care of with the help of an IR renormalization factor
in minimal subtraction, $Z$, which is constructed from the QCD beta function
and the cusp anomalous dimension (see, e.g., Ref.~\cite{Henn:2016tyf}).
It is convenient to consider $\log Z$ which is given by
\begin{eqnarray}
  \log Z &=& -\frac{\alpha_s(\mu)}{\pi} \frac{ \Gamma_{\rm cusp}^{(1)} }{2\epsilon}
             +\left( \frac{\alpha_s(\mu)}{\pi} \right)^2
             \left( \frac{\beta_0 \Gamma_{\rm cusp}^{(1)}}{ 16\epsilon^2 }
             - \frac{ \Gamma_{\rm cusp}^{(2)} }{ 4\epsilon }
             \right)
             \nonumber\\&&\mbox{}
             +\left( \frac{\alpha_s(\mu)}{\pi} \right)^3
             \left(
             - \frac{\beta_0^2 \Gamma_{\rm cusp}^{(1)}}{ 96\epsilon^3 }
             + \frac{\beta_1 \Gamma_{\rm cusp}^{(1)} + 4\beta_0 \Gamma_{\rm cusp}^{(2)}}{ 96\epsilon^2 }
             - \frac{ \Gamma_{\rm cusp}^{(3)} }{ 6\epsilon }
             \right)
             \,,
                           \label{eq::logZ}
\end{eqnarray}
with
\begin{eqnarray}
  \beta_0 &=& \frac{11}{3}C_A - \frac{4}{3}T_Fn_l\,,\nonumber\\
  \beta_1 &=& \frac{34}{3}C_A^2 - 4 C_F T_F n_l - \frac{20}{3} C_A T_F n_l\,.
\end{eqnarray}
$\Gamma_{\rm cusp}^{(i)}$ are the expansion coefficients of the cusp anomalous
dimension defined through
\begin{eqnarray}\label{z_minimal}
  \Gamma_{\rm cusp} &=& \sum_{i \ge 1} \Gamma_{\rm cusp}^{(i)}
  \left(\frac{{\alpha}_s}{ \pi}\right)^i \,.
\end{eqnarray}
The cusp anomalous dimension in QCD is available to three loops
from Refs.~\cite{Polyakov:1980ca,Korchemsky:1987wg,Grozin:2014hna,Grozin:2015kna}.

Using $Z$, we can define a finite form factor via
\begin{eqnarray}
  \label{eq::def-Z}
  F^{\rm UV ren} &=& Z F^{f}
                     \,.
\end{eqnarray}
Spelling out this equations leads to
\begin{eqnarray}
  F^{f} &=& Z^{-1} F^{\rm UV ren}
                        \nonumber\\
  &=& Z^{-1} \left( F^{(1+2+\rm CT)} + F^{(3)} \right)
                     \,,
      \label{eq::Ffin}
\end{eqnarray}
where $F^{(1+2+\rm CT)}$ contains the exact one- and two-loop expressions
(including higher orders in $\epsilon$) and the three-loop counterterm
contribution.    The quantities $Z$ and $F^{(1+2+\rm CT)}$ are
expressed in terms of HPLs, which depend on the quantity $x$ defined
through $s/m^2 = -(1-x)^2/x$.
$ F^{(3)}$ is the bare three-loop contribution, the
main result of this paper. We construct numerical approximations
valid in the whole $s/m^2$ range.

The one- and two-loop results for the form factor are available to
order $\epsilon^2$ and $\epsilon^1$, respectively, see, e.g.,
Ref.~\cite{Lee:2018rgs}, which is necessary to obtain the
finite contributions at order $\alpha_s^3$. One has to take into account the
following counterterms, which are well established in the literature, see, e.g.,
Refs.~\cite{Melnikov:2000zc,Chetyrkin:2004mf,Marquard:2016dcn,Marquard:2018rwx}:
\begin{itemize}
\item $\alpha_s$ in the $\overline{\rm MS}$ scheme up to two loops.
  This is a multiplicative renormalization. Since the one- and two-loop
  form factors and the counterterms are $\xi$ independent also the
  induced three-loop terms are $\xi$ independent.
  We express our final results in terms of $\alpha_s^{(n_l)}$, the strong
  coupling constant with $n_l$ active quark flavours.
\item Wave-function renormalization in the on-shell scheme to three loops.
  This is a multiplicative renormalization.
  Note that $Z_2^{\rm OS}$ has a $\xi$ dependence at three loops
  in the colour factors $C_F C_A^2$ and $C_FC_A T_F n_h$.
\item Quark-mass renormalization in the on-shell scheme to two loops.  Since
  we have for the external (anti)quark momenta $q_1^2=q_2^2=m^2$, a
  multiplicative renormalization in the bare one- and two-loop expressions is
  not possible. Instead one has to generate and compute one- and two-loop
  counterterm diagrams with explicit mass insertions. While the mass
  counterterms themselves are $\xi$ independent, linear and quadratic $\xi$
  terms are generated from the mass insertions nonetheless.
\item The anomalous dimensions of the vector and axial-vector currents vanish
  and thus no renormalization is needed.  The anomalous dimensions of the
  scalar and pseudo-scalar currents only vanish because of the additional
  factor $m$ in Eq.~(\ref{eq::currents}), which has to be renormalized.
  We choose to renormalize this factor in the $\overline{\rm MS}$ scheme.
\end{itemize}

The $\overline{\rm MS}$ renormalization
constants only contain pole parts whereas for the on-shell
quantities also higher order $\epsilon$ coefficients are needed since the one-
and two-loop form factors develop $1/\epsilon$ and $1/\epsilon^2$ poles,
respectively.

We use the results from Ref.~\cite{Lee:2018rgs} and obtain
exact results for the three-loop counterterm contributions.
We expand the expression for $s\to0$ and check the cancellation in this
limit analytically.

In principle we have to expand the counterterm contributions also around all
other values of $s$ for which we perform expansions. However, this can be
quite tedious since we have to expand the iterated integrals around quite
involved arguments. Instead we check the pole cancellation numerically using
the approximated bare three-loop form factors, the exact results for the
counterterm contributions and the exact result for the cusp anomalous
dimension. We observe that the poles cancel with a relative precision of
at least nine digits in the whole $s$ range.
This is discussed in more detail in Subsection~\ref{sub::acc}.

Note that the singlet contributions start at two loops. Apart from that
the renormalization proceeds in the same way as for the non-singlet
case.  Since the singlet contributions to the vector form factors vanish at
two loops due to Furry's theorem, its three-loop amplitude is already finite
and does not need to be renormalized.

%- }}}
%- {{{ Results:

\section{\label{sec::res}Results}

In this section we present our results for the three-loop corrections of the
massive form factors.  First, we discuss how we estimate the uncertainty of
our numerical results and which cross checks we have performed to validate
them.  We then discuss the static and high-energy limits as well as the
two-particle threshold and present the leading terms of these expansions.
Furthermore, we comment on the behaviour of the four-particle threshold.
Finally, we show plots for the form factors over the whole kinematic range.
This is accompanied by a numeric package to evaluate the form factors as the
main result of this paper.  Unless stated otherwise we choose for the
renormalization scale $\mu^2=m^2$.

%- {{{ Estimation of the accuracy:

\subsection{\label{sub::acc}Estimation of the accuracy and cross checks}

We estimate the precision of our result from the numerical pole cancellations
of the renormalized and infrared-subtracted form factors: At each random sample
point, for every colour factor, and for every order in $\ep$, we add the numerical bare results and the numerical evaluations of the
counterterms as well as $Z$ defined in Eq.~(\ref{eq::def-Z}) and divide by the
absolute value of the counterterms and $Z$:
\begin{equation}
  \label{eq::delta-def}
  \delta (F^{f,(3)}\big|_{\ep^i}) = \frac{F^{(3)}\big|_{\ep^i}+F^{(\text{CT}+{Z})}\big|_{\ep^i}}{F^{(\text{CT}+{Z})}\big|_{\ep^i}} .
\end{equation}
This corresponds to the precision of the pole terms.
Real and imaginary parts are checked separately.

\begin{figure}[t]
  \begin{center}
    \begin{tabular}{cc}
      \includegraphics[width=0.47\textwidth]{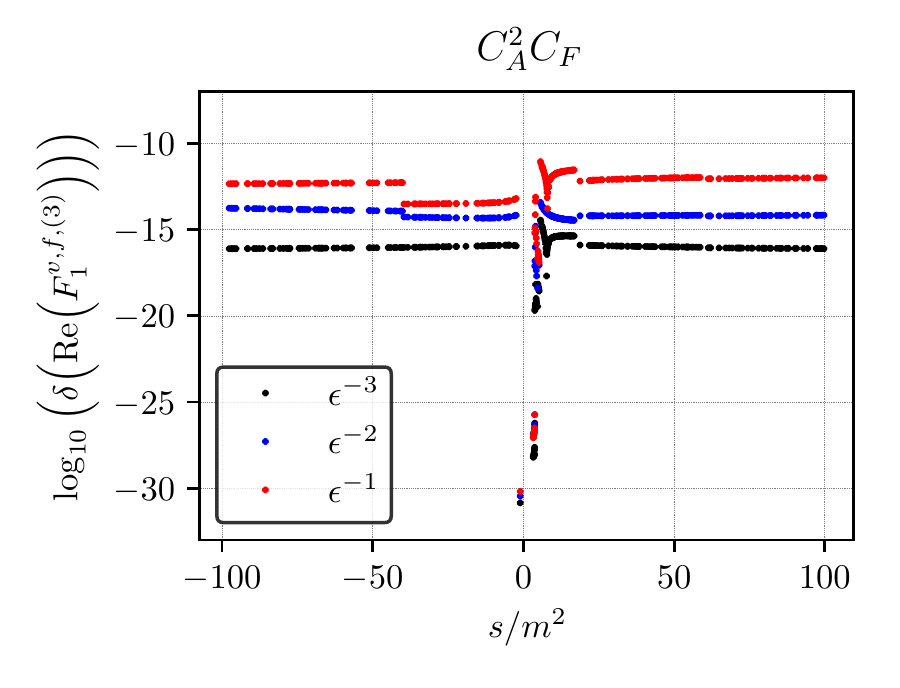}
      &
      \includegraphics[width=0.47\textwidth]{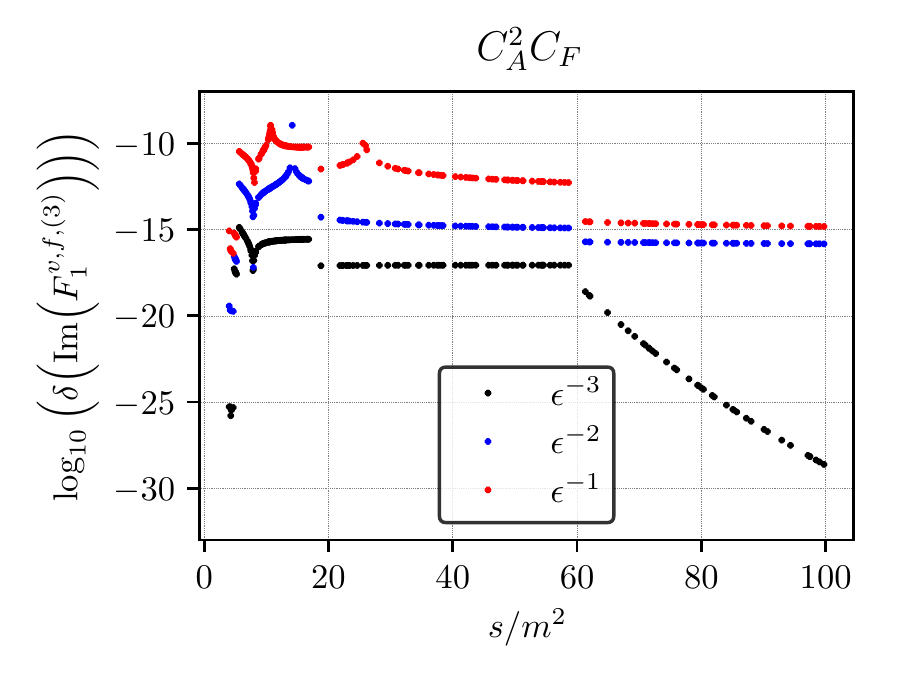}
      \\
      (a) & (b) \\
      \includegraphics[width=0.47\textwidth]{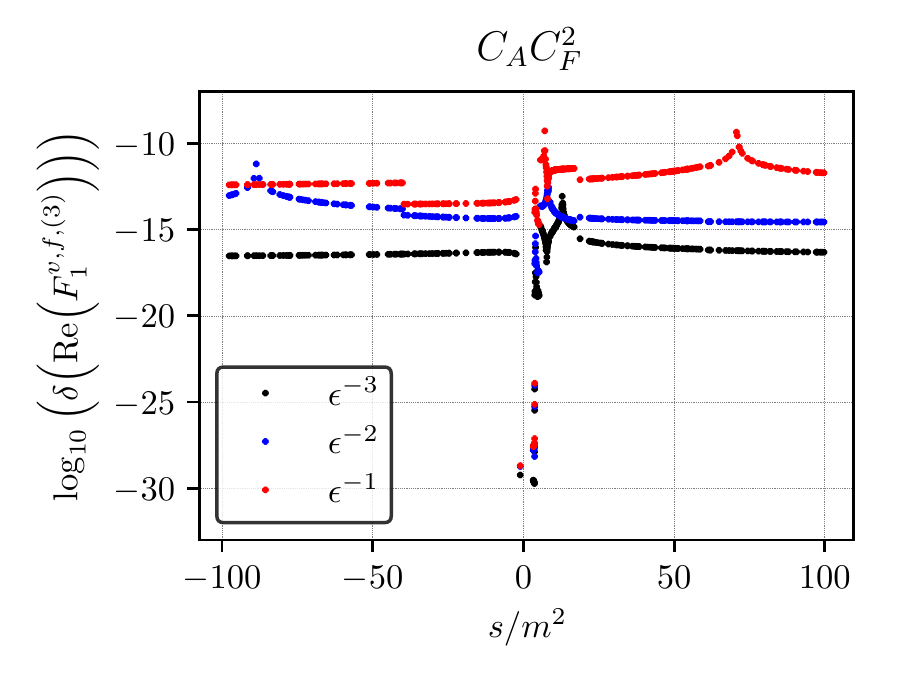}
      &
      \includegraphics[width=0.47\textwidth]{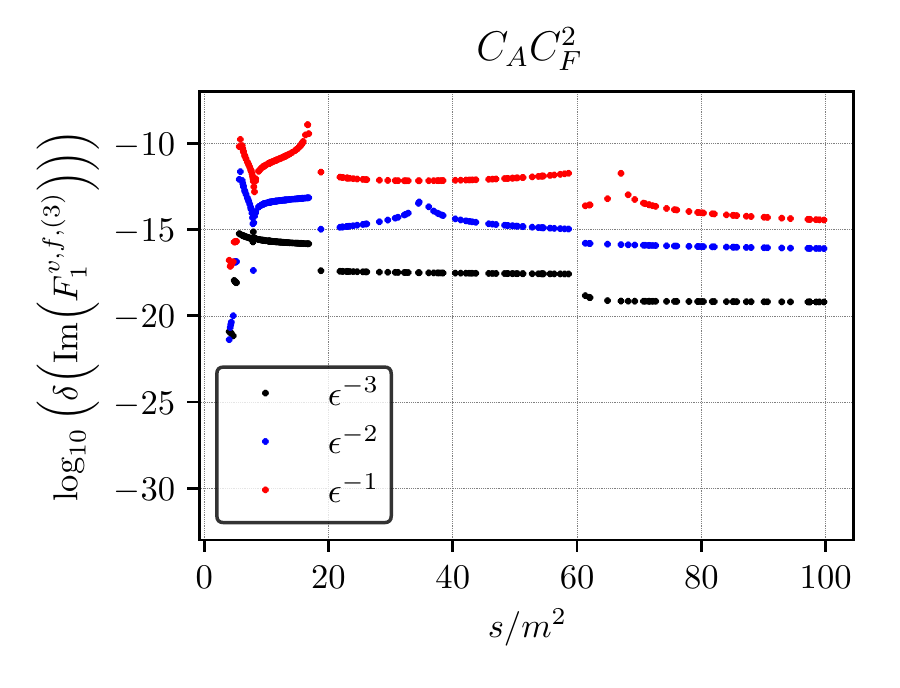}
      \\
      (c) & (d) \\
      \includegraphics[width=0.47\textwidth]{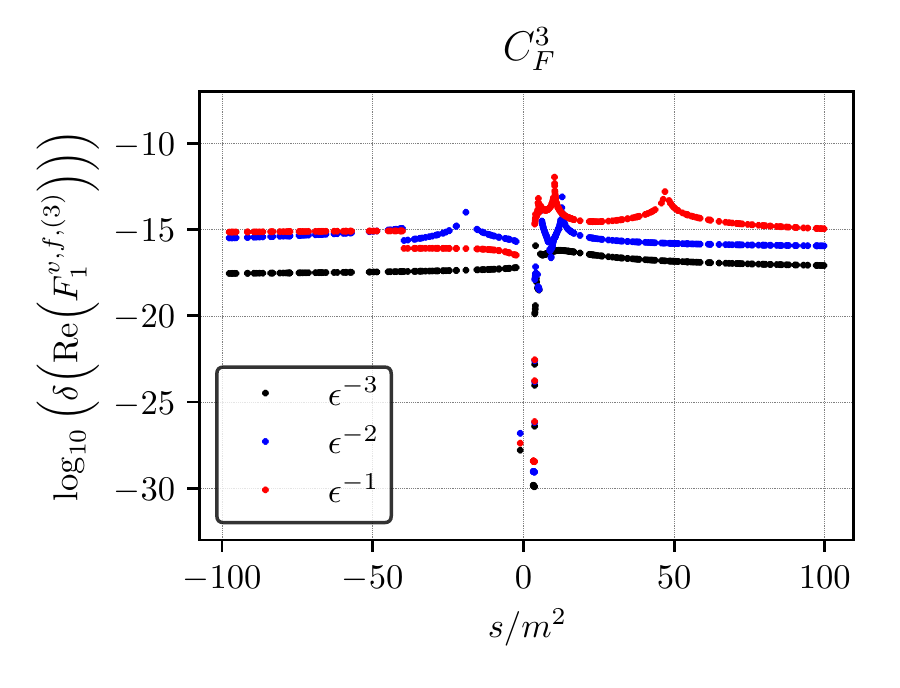}
      &
      \includegraphics[width=0.47\textwidth]{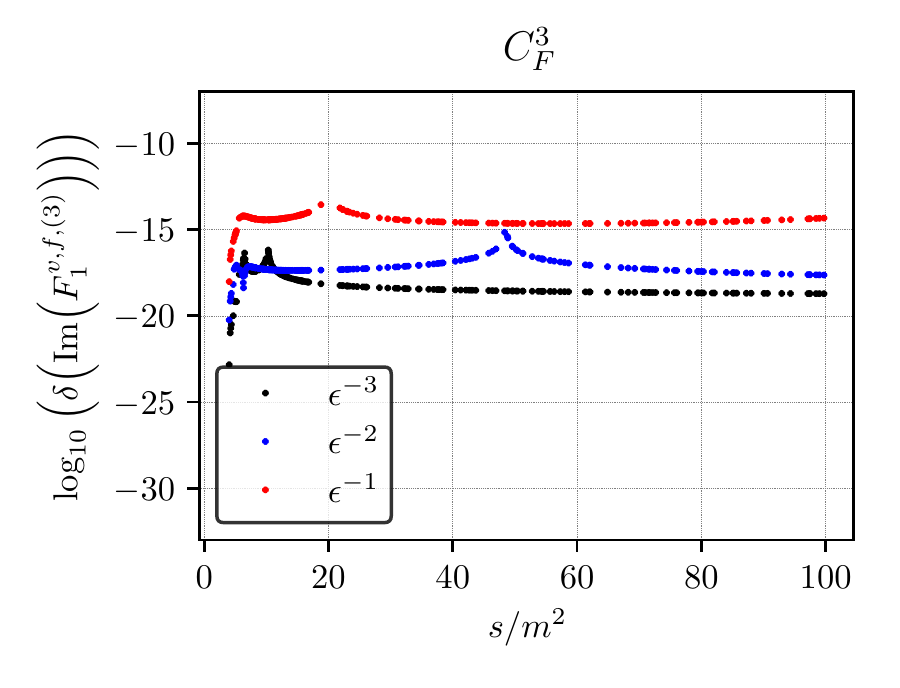}
      \\
      (e) & (f) \\
    \end{tabular}
    \caption{\label{fig::pole-cancellation-veF1}
      Relative cancellation of the real, (a), (c), (e), and imaginary parts, (b), (d), (f), of the poles for the non-fermionic colour structures of $F^{v,f,(3)}_1$, c.f.\ Eq.~(\ref{eq::delta-def}).
    }
  \end{center}
\end{figure}

For illustration we show the cancellations for the three non-fermionic colour
structures $C_F^3$, $C_F^2 C_A$, and $C_FC_A^2$ of $F_1^{v,(3)}$ in
Fig.~\ref{fig::pole-cancellation-veF1}.
Figures for the non-fermionic colour structures of the five remaining form
factors can be found in Appendix~\ref{app::poles}.
Since the precision of the fermionic colour factors is much better, we refrain from showing the associated plots.
In general we observe a
progression in the orders of $\ep$, i.e.\ the $1/\ep^{-3}$ poles cancel with
the highest precision and we lose some digits with every higher order.
Sometimes, this general progression is violated, usually if the value of the
colour factor changes sign and crosses zero.  These zero crossing are visible
in the plots, because the precision slowly decreases and then slowly increases
again, see for example the region around $s \approx -20 m^2$ for the real part of the colour
factor $C_F^3$ of $F_1^{v,f,(3)}$ in Fig.~\ref{fig::pole-cancellation-veF1}(e).

Let us now analyze the four physical regions $s < 0$, $0 \leq s < 4 m^2$,
$4 m^2 \leq s < 16 m^2$, and $16 m^2 \leq s$ separately in more detail.  In
each region we provide the minimal precision over all colour factors and form
factors.  Since the form factors are very small close to zero crossings, we
also provide the minimal precision when removing all points for which the size
of the coefficient is smaller than $2.5\,\%$ of the average in the
region.  In addition, we discard all points close to the Coulomb singularity,
i.e.\ $3.95 m^2 \leq s \leq 4.05 m^2$.
\begin{enumerate}
\item In the region $s < 0$ we sample over $250$ randomly chosen points for
  $-100 m^2 \leq s \leq 0$.  The poles generally cancel with at least $15$
  digits for the $1/\ep^{-3}$ poles, at least $10$ digits for the $1/\ep^{-2}$
  poles, and at least $10$ digits for the $1/\ep$ poles.  Removing the points
  close to the zero crossings this improves to $15$, $12$, and $11$ digits,
  respectively.

\item The region $0 \leq s < 4 m^2$ is the most precise one: For the $1000$
  random sample points, the poles cancel with at least $17$, $14$, $12$ digits
  for the $1/\ep^{-3}$, $1/\ep^{-2}$, $1/\ep$ poles, respectively.  Removing
  the points close to zero crossings with the threshold chosen above does not
  improve the precision.  However, these reported worst pole cancellations all
  belong to the power-log expansion around the two-particle threshold at
  $s = 4 m^2$.  For $0 \leq s < 3.75 m^2$ our precision is well beyond $20$
  digits as can be seen in the figures.

\item The region $4 m^2 \leq s < 16 m^2$ between the thresholds at $s = 4 m^2$
  and $s = 16 m^2$ is least precise: For the $1000$ random sample points, the
  poles of the real part only cancel with at least $11$, $11$, $8$ digits for
  the $1/\ep^{-3}$, $1/\ep^{-2}$, $1/\ep$ poles, respectively, and the poles
  for the imaginary parts with $12$, $6$, $6$ digits.  Removing the points
  close to zero crossings mildly improves the precision of the real part to
  $13$, $12$, $9$ digits.  The imaginary part on the other hand significantly
  improves to $14$, $11$, $9$ digits.

\item The region $16 m^2 \leq s$ becomes more precise again, since it is
  matched from $+ \infty$. For the $250$ random sample points between
  $16 m^2 \leq s \leq 100 m^2$, the poles cancel with at least $14$, $13$, $8$
  digits for real parts of the $1/\ep^{-3}$, $1/\ep^{-2}$, $1/\ep$ poles,
  respectively, and with $15$, $11$, $8$ digits for the imaginary parts.  This
  improves to $14$, $13$, $10$ digits for the real part and to $15$, $11$, $9$
  digits for the imaginary part when removing the points close to zero
  crossings.
\end{enumerate}
Extrapolating these numbers to the finite terms, we expect that our result is
correct up to at least $7$ digits away from the zero crossings,
with a much better precision for most colour factors and form factors over most parts of the real axis.

We have performed the calculation of the form factors for general QCD gauge
parameter $\xi$ and have checked that $\xi$ cancels in the renormalized form
factors.  Note that the mass counterterm contributions depend on $\xi$
which cancels against the bare three-loop expressions.
We have checked the cancellation numerically and observe that
the coefficient in front of $\xi$ is
of order $10^{-18}$ or smaller in most of the phase space.

After specifying to the large-$N_C$ limit via $C_F\to N_C/2$ and
$C_A\to N_C$ we can compare the $N_C^3$ terms
against the exact results from Ref.~\cite{Henn:2016tyf,Lee:2018nxa}. In
this limit only about 90 planar master integrals contribute and we observe a
significantly increased precision of our result. In fact, in the whole $s/m^2$
region we can reproduce the exact result with at least 14 digits.
with the exception very close to the singularity at $s=4$. For example,
for $s=3.9m^2$ and $s=4.1m^2$ we have an agreement of about 12 digits.

We observe similar results for the light-fermion colour factors $C_F^2n_l$,
$C_AC_Fn_l$, $C_Fn_hn_l$ and $C_Fn_l^2$, which we compare against the exact
results from Ref.~\cite{Lee:2018nxa}, and the $n_h^2$ terms where the exact
expressions can be found in Ref.~\cite{Blumlein:2019oas}.

\begin{figure}[t]
  \begin{center}
    \begin{tabular}{cc}
      \includegraphics[width=0.47\textwidth]{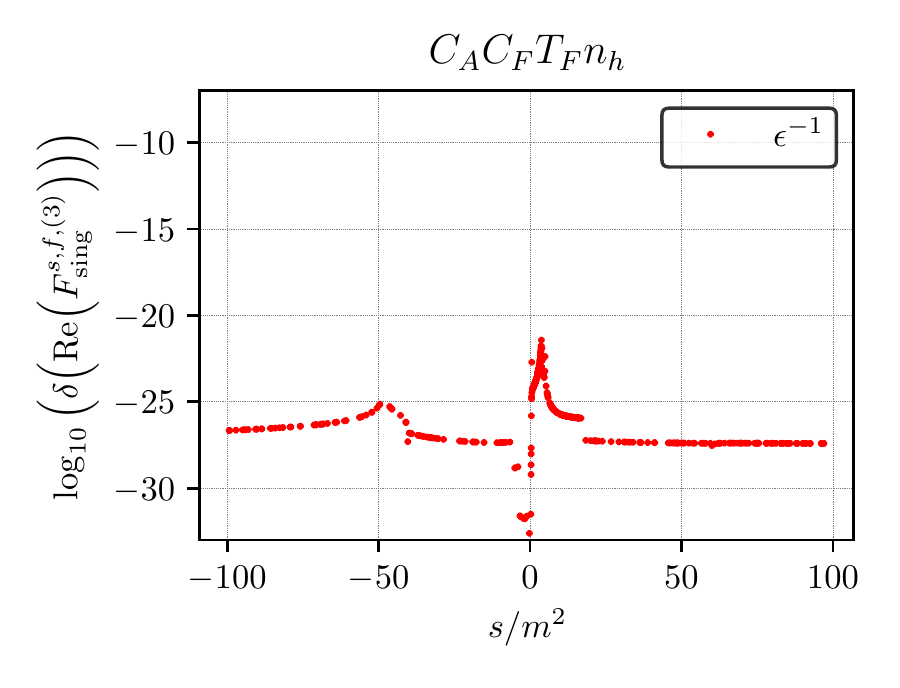}
      &
      \includegraphics[width=0.47\textwidth]{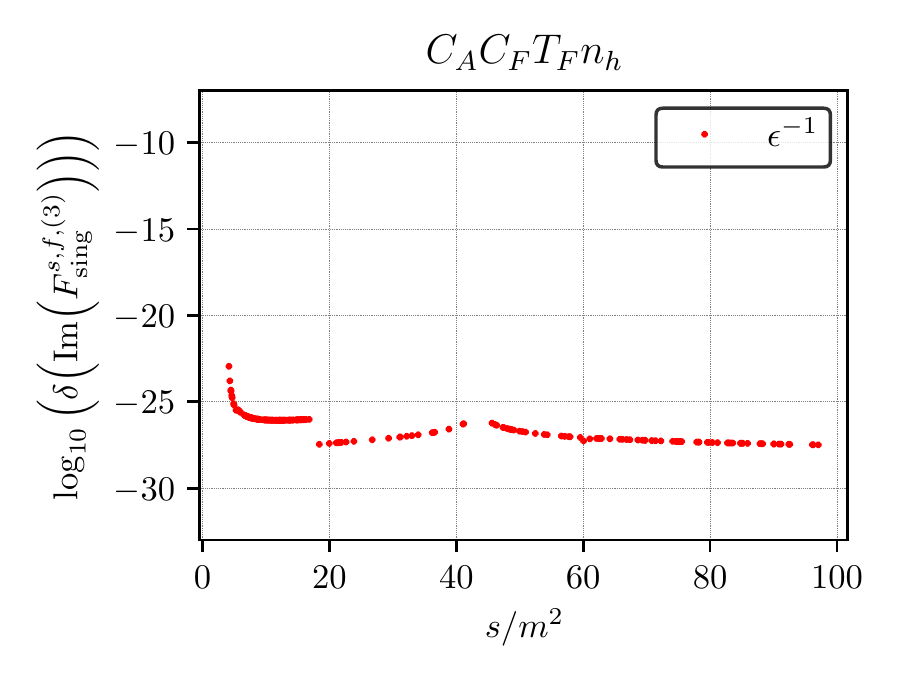}
      \\
      (a) & (b) \\
      \includegraphics[width=0.47\textwidth]{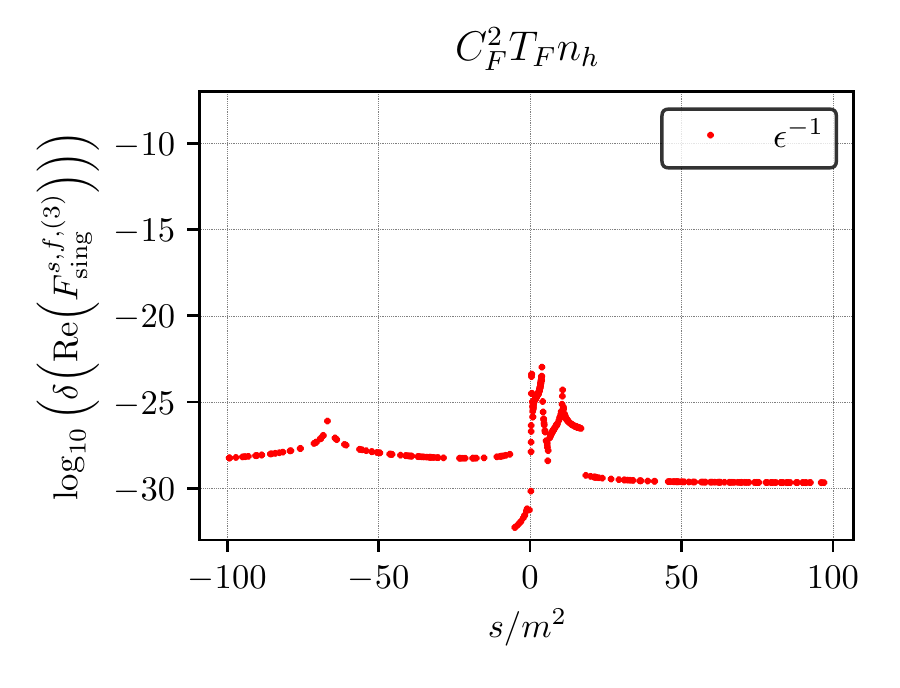}
      &
      \includegraphics[width=0.47\textwidth]{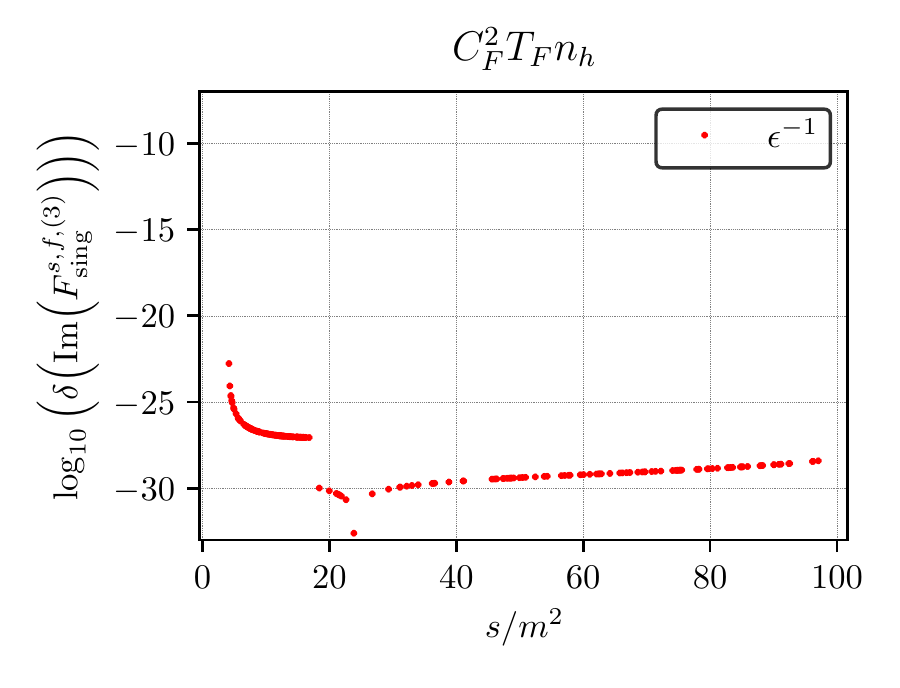}
      \\
      (c) & (d) \\
    \end{tabular}
    \caption{\label{fig::pole-cancellation-sing-nh-sc}
      Relative cancellation of the real, (a), (c), and imaginary parts, (b), (d), of the poles for the colour structures $C_A C_F T_F n_h$ and $C_F^2 T_F n_h$ of $F^{s,f,(3)}_{\text{sing}}$, c.f.\ Eq.~(\ref{eq::delta-def}).
    }
  \end{center}
\end{figure}

For the singlet contribution 
there are no $1/\epsilon^2$ and $1/\epsilon^3$ poles from the counterterms.
For the $1/\epsilon$ pole we have a much higher precision as can be seen in
Fig.~\ref{fig::pole-cancellation-sing-nh-sc} for the colour structures
$C_A C_F T_F n_h$ and $C_F^2 T_F n_h$ of $F^{s,f,(3)}_\text{sing}$.

We observe that the poles cancel with at least 20 digits, and with several
more digits for most regions of the phase space.  The cancellation for the
other two colour factors with a second fermion loop is even more precise.  Due
to the high precision we refrain from showing plots for these two colour
factors and $F^{p,f,(3)}_\text{sing}$.  As a conservative estimate we claim a
precision of at least 10 digits for the $\epsilon^0$ terms of the
singlet form factors.  Since contributions to $F^{v}_\text{sing}$ only start
at three-loop order, we cannot check the pole cancellation in this case.
However, we do not observe any difference in the behavior of the master
integrals which only contribute in this case and, thus, assume the same
precision.

%- }}}
%- {{{ Analytic and numeric expansions:

\subsection{Analytic and numeric expansions}

Next we present expansions around the special kinematic points $s=0$ and
$s=\pm\infty$.  For
$s=4m^2$ we construct expansions for the cross sections and decay rates,
respectively. Such expansions around $s=0, 4m^2$ and $\infty$ may
serve as input for approximation procedures as those based on Pad\'e
approximations, see, e.g.,
Refs.~\cite{Chetyrkin:1996cf,Maier:2009fz,Hoang:2008qy}.

%- {{{ Static limit: $s\to0$:

\subsubsection{\label{subsub::s0}Static limit: \boldmath{$s\to0$}}

In the static limit we construct an analytic expansion including\footnote{We have
  a deeper expansion of the master integrals. However, there are spurious
  poles (in the non-singlet case up to $1/s^9$) in the amplitude which reduce
  the expansion depths for the form factors.}  $s^{66}$ from the
boundary values at $s=0$.  We restrict ourselves again to the five colour
structures which are not known analytically.  All colour factors as well as
higher orders in the expansion are available in the ancillary file
accompanying this paper~\cite{progdata}.  For illustration we show the results for $F^v_1$
and $F^s$ in the main part of the paper and relegate the remaining four form
factors to Appendix~\ref{app::s0}. For the non-singlet form factors the 
expansions up to $s/m^2$ are given by
\begin{align}
  &F_1^{v,f,(3)}\Big|_{s\to0} =
  \frac{s}{m^2} \Bigg\{
  C_A C_F^2 \Bigg[
    \frac{19 a_4}{2}+\frac{17725 \zeta_3}{3456}-\frac{\pi ^2 \zeta_3}{9}-\frac{55 \zeta_5}{32}+\frac{707}{288}-\frac{4829 \pi ^2}{10368}-\frac{347 \pi ^4}{17280} \nonumber\\
    &\quad +\frac{19 l_2^4}{48}-\frac{97}{720} \pi ^2 l_2^2+\frac{29}{240} \pi ^2 l_2
  \Bigg]
  + C_A^2 C_F \Bigg[
    -a_4+\frac{4045 \zeta_3}{5184}+\frac{7 \pi ^2 \zeta_3}{96}-\frac{5 \zeta_5}{64}-\frac{7876}{2187} \nonumber\\
    &\quad +\frac{172285 \pi ^2}{186624}+\frac{67 \pi ^4}{8640}-\frac{l_2^4}{24}+\frac{67}{360} \pi ^2 l_2^2-\frac{5131 \pi ^2 l_2}{2880}
  \Bigg]
  + C_F^3 \Bigg[
    -15 a_4-\frac{18367 \zeta_3}{1728}-\frac{17 \pi ^2 \zeta_3}{24} \nonumber\\
    &\quad +\frac{25 \zeta_5}{8}+\frac{13135}{20736}-\frac{24463 \pi ^2}{7776}+\frac{3037 \pi ^4}{25920}-\frac{5 l_2^4}{8}-\frac{19}{40} \pi ^2 l_2^2+\frac{4957}{720} \pi ^2 l_2
  \Bigg]
  + C_F^2 T_F n_h \Bigg[
    -\frac{32 a_4}{9} \nonumber\\
    &\quad +\frac{1441 \zeta_3}{1728}-\frac{2273}{1296}+\frac{1057 \pi ^2}{2430}-\frac{13 \pi ^4}{1620}-\frac{4 l_2^4}{27}+\frac{4}{27} \pi ^2 l_2^2-\frac{2}{9} \pi ^2 l_2
  \Bigg]
  + C_F C_A T_F n_h \Bigg[
    \frac{17 a_4}{6} \nonumber\\
    &\quad +\frac{1775 \zeta_3}{864}-\frac{\pi ^2 \zeta_3}{18}+\frac{5 \zeta_5}{12}-\frac{23089}{5184}+\frac{4813 \pi ^2}{5184}+\frac{803 \pi ^4}{51840}+\frac{17 l_2^4}{144}-\frac{17}{144} \pi ^2 l_2^2-\frac{149}{108} \pi ^2 l_2
  \Bigg]
  \Bigg\} \nonumber\\
  &\quad + {\cal O}\left(\frac{s^2}{m^4}\right) + \mbox{$n_l$, $n_l^2$ and $n_h^2$ terms} ,
\end{align}

\begin{align}
  &F^{s,f,(3)}\Big|_{s\to0} =
  C_A C_F^2 \Bigg[
    \frac{4 a_4}{3}+\frac{491 \zeta_3}{96}+\frac{19 \pi ^2 \zeta_3}{16}-\frac{45 \zeta_5}{16}+\frac{26117}{4608}-\frac{1193 \pi^2}{576}-\frac{65 \pi ^4}{432}+\frac{l_2^4}{18} \nonumber\\
    &\quad + \frac{31}{36} \pi ^2 l_2^2+\frac{43}{18} \pi ^2 l_2
  \Bigg]
  + C_A^2 C_F \Bigg[
    -\frac{11 a_4}{3}-\frac{947 \zeta_3}{288}-\frac{51 \pi ^2 \zeta_3}{64}+\frac{65 \zeta_5}{32}-\frac{584447}{124416}+ \frac{3011\pi ^2}{3456} \nonumber\\
    &\quad+\frac{179 \pi ^4}{3456}-\frac{11 l_2^4}{72}-\frac{11}{36} \pi ^2 l_2^2+\frac{49}{72} \pi ^2 l_2
  \Bigg]
  + C_F^3 \Bigg[
    12 a_4+\frac{87 \zeta_3}{16}+\frac{\pi ^2 \zeta_3}{16}- \frac{5 \zeta_5}{8}+\frac{55}{96}+\frac{643 \pi ^2}{192} \nonumber\\
    &\quad +\frac{\pi^4}{48}+\frac{l_2^4}{2}-\frac{1}{2} \pi ^2 l_2^2-\frac{15}{2} \pi ^2 l_2
  \Bigg]
  + C_F^2 T_F n_h \Bigg[
    8 a_4+\frac{17 \zeta_3}{18}- \frac{2083}{432}-\frac{52 \pi ^2}{81}-\frac{\pi ^4}{720}+\frac{l_2^4}{3} \nonumber\\
    &\quad -\frac{1}{3} \pi ^2 l_2^2+\frac{8}{9} \pi ^2 l_2
  \Bigg]
  + C_F C_A T_F n_h \Bigg[
    -6 a_4-\frac{199 \zeta_3}{144}+\frac{\pi ^2 \zeta_3}{8} - \frac{5 \zeta_5}{8}+\frac{209857}{15552}-\frac{4351 \pi^2}{1296} \nonumber\\
    &\quad -\frac{\pi ^4}{288}-\frac{l_2^4}{4}+\frac{1}{4} \pi ^2 l_2^2+\frac{32}{9} \pi ^2 l_2
  \Bigg] \nonumber\\
  &\quad + \frac{s}{m^2} \Bigg\{
  C_A C_F^2 \Bigg[
    -\frac{8 a_4}{9}+\frac{2515 \zeta_3}{2304}-\frac{29 \pi ^2 \zeta_3}{144}-\frac{95 \zeta_5}{48}+\frac{11191}{41472}-\frac{15101\pi ^2}{62208}+ \frac{37 \pi ^4}{4320}-\frac{l_2^4}{27} \nonumber\\
    &\quad +\frac{1259 \pi ^2 l_2^2}{2160}+\frac{1409 \pi ^2 l_2}{1440}
  \Bigg]
  + C_A^2 C_F \Bigg[
    \frac{5 a_4}{4}+\frac{8675 \zeta_3}{10368}-\frac{73 \pi ^2 \zeta_3}{1152}+ \frac{125 \zeta_5}{384}-\frac{851465}{279936} \nonumber\\
    &\quad +\frac{130417 \pi ^2}{186624}+\frac{689 \pi ^4}{103680}+\frac{5 l_2^4}{96}-\frac{1}{20} \pi ^2 l_2^2-\frac{12253 \pi ^2 l_2}{8640}
  \Bigg]
  + C_F^3 \Bigg[
    -\frac{29 a_4}{9}-\frac{12401 \zeta_3}{3456} \nonumber\\
    &\quad -\frac{67 \pi ^2 \zeta_3}{288}+\frac{85 \zeta_5}{32}+\frac{22613}{41472}-\frac{69355 \pi ^2}{31104}+\frac{1727 \pi ^4}{25920}- \frac{29 l_2^4}{216}-\frac{1043 \pi ^2 l_2^2}{1080}+\frac{4013 \pi ^2 l_2}{1080}
  \Bigg] \nonumber\\
  &\quad + C_F^2 T_F n_h \Bigg[
    \frac{8 a_4}{9}+\frac{9889 \zeta_3}{6912}-\frac{8059}{5184}-\frac{4261 \pi ^2}{25920} +\frac{7 \pi ^4}{1620}+\frac{\log^4(2)}{27}-\frac{1}{27} \pi ^2 l_2^2+\frac{4}{27} \pi ^2 l_2
  \Bigg] \nonumber\\
  &\quad + C_F C_A T_F n_h \Bigg[
    -\frac{5 a_4}{18}+ \frac{1657 \zeta_3}{1728}+\frac{\pi ^2 \zeta_3}{96}+\frac{5 \zeta_5}{96}+\frac{2257}{3888}-\frac{13663 \pi^2}{25920}-\frac{121 \pi ^4}{51840}-\frac{5 l_2^4}{432} \nonumber\\
    &\quad +\frac{5}{432} \pi ^2 l_2^2+ \frac{55}{108} \pi ^2 l_2
  \Bigg]
  \Bigg\}
  + {\cal O}\left(\frac{s^2}{m^4}\right) + \mbox{$n_l$, $n_l^2$ and $n_h^2$ terms} ,
\end{align}
where $l_2=\log(2)$, $a_4=\mbox{Li}_4(1/2)$ and $\zeta_n$ is Riemann's zeta
function evaluated at $n$.  For $s=0$ our results for $F_2^v(0)$ and
$F_1^a(0)$ agree with Refs.~\cite{Grozin:2007fh}
and~\cite{Archambault:2004zs}, respectively.
Note that for $s=0$ the cusp anomalous dimension in Eq.~(\ref{eq::logZ})
vanishes and we have $Z=1$.

The $n_h$-singlet contribution to the scalar form factor reads
\begin{align}
  &F_\text{sing}^{s,f,(3)}\Big|_{s\to0} =
  C_A C_F T_F n_h \Bigg[
    \frac{22 a_4}{3}+\frac{113 \zeta_3}{36}-\frac{\pi ^2 \zeta_3}{4}+\frac{5 \zeta_5}{4}-\frac{643}{54}+\frac{466 \pi ^2}{81}+\frac{187 \pi ^4}{4320}+\frac{11 l_2^4}{36} \nonumber\\
    &\quad -\frac{11}{36} \pi ^2 l_2^2-\frac{61}{9} \pi ^2 l_2
  \Bigg]
  + C_F^2 T_F n_h \Bigg[
    -\frac{32 a_4}{3}+\frac{55 \zeta_3}{72}+\frac{445}{108}+\frac{517 \pi ^2}{324}-\frac{11 \pi ^4}{270}-\frac{4 l_2^4}{9}+\frac{4}{9} \pi ^2 l_2^2 \nonumber\\
    &\quad -\frac{22}{9} \pi ^2 l_2
  \Bigg]
  + C_F T_F^2 n_h^2 \Bigg[
    -\frac{8 \zeta_3}{3}+\frac{16}{9}+\frac{26 \pi ^2}{135}
  \Bigg]
  + C_F T_F^2 n_h n_l \Bigg[
    \frac{20}{9}-\frac{10 \pi ^2}{27}
  \Bigg] \nonumber\\
  &\quad + \frac{\sqrt{-s}}{m} \Bigg\{
  C_A C_F T_F n_h \Bigg[
    \frac{11}{36} \pi ^2 l_{\sqrt{-s}/m}+\frac{\pi ^4}{72}-\frac{263 \pi ^2}{432}
  \Bigg]
  + C_F^2 T_F n_h \Bigg[
    \frac{\pi ^2}{16}
  \Bigg]+ C_F T_F^2 n_h n_l \Bigg[
    \frac{4 \pi ^2}{27} \nonumber\\
    &\quad -\frac{1}{9} \pi ^2 l_{\sqrt{-s}/m}
  \Bigg]
  \Bigg\} \nonumber\\
  &\quad {- \frac{s}{m^2}} \Bigg\{
  C_A C_F T_F n_h \Bigg[
    -\frac{13 a_4}{10}+\frac{11 l_{\sqrt{-s}/m}^2}{36}+\frac{1}{18} \pi ^2 l_{\sqrt{-s}/m}-\frac{115 l_{\sqrt{-s}/m}}{108}-\frac{5 \zeta_5}{12}+\frac{\pi ^2 \zeta_3}{12} \nonumber\\
    &\quad -\frac{429 \zeta_3}{800}-\frac{3661 \pi ^4}{259200}-\frac{12708229 \pi ^2}{6804000}+\frac{2494529}{518400}-\frac{13 l_2^4}{240}+\frac{13}{240} \pi ^2 l_2^2+\frac{2929 \pi ^2 l_2}{1350}
  \Bigg] \nonumber\\
  &\quad + C_F^2 T_F n_h \Bigg[
    \frac{34 a_4}{15}+\frac{l_{\sqrt{-s}/m}}{8}+\frac{15083 \zeta_3}{172800}+\frac{29 \pi ^4}{2400}-\frac{15689483 \pi ^2}{27216000}-\frac{3190951}{1036800} \nonumber\\
    &\quad +\frac{17 l_2^4}{180}-\frac{17}{180} \pi ^2 l_2^2+\frac{1417 \pi ^2 l_2}{1350}
  \Bigg]
  + C_F T_F^2 n_h^2 \Bigg[
    \frac{1531 \zeta_3}{1728}-\frac{1661}{2592}-\frac{397 \pi ^2}{8505}
  \Bigg] \nonumber\\
  &\quad + C_F T_F^2 n_h n_l \Bigg[
    -\frac{1}{9} l_{\sqrt{-s}/m}^2+\frac{13 l_{\sqrt{-s}/m}}{54}+\frac{89 \pi ^2}{972}-\frac{485}{648}
  \Bigg]
  \Bigg\} \nonumber\\
  &\quad + {\cal O}\left(\frac{\sqrt{-s}^3}{m^3}\right)\,,
  \label{eq::Fssing_s0}
\end{align}
with $l_{\sqrt{-s}/m} = \log (\sqrt{-s}/m)$.  This logarithm as well as the
expansion in powers of $\sqrt{-s}/m$, instead of $s/m^2$ as for the
non-singlet contributions show above, originate from the massless cuts through
singlet diagrams discussed in Section~\ref{sec::singlet}.  The results for the
vector and pseudo-scalar form factors can be found in Appendix~\ref{app::s0}.

%- }}}
%- {{{ High-energy expansion:

\subsubsection{High-energy expansion: \boldmath{$s\to-\infty$}}
\label{subsub:high-energy}

Also for the high-energy expansion we focus on the colour factors $C_F^3$,
$C_F^2 C_A$, $C_FC_A^2$, $C_F^2 T_F n_h$, and $C_F C_A T_F n_h$ and refer to
the literature~\cite{Lee:2018nxa,Lee:2018rgs} for the remaining fermionic
contributions.  The high-energy expansions of the non-singlet form factors up
to $m^2/s$ read
\begin{align}
  & F_1^{v,f,(3)}\Big|_{s\to-\infty} =
    8.3501 C_A^2 C_F-20.762 C_A C_F^2+10.425 C_A C_F T_F n_h+4.7318 C_F^3 \nonumber\\
    &\quad -3.2872 C_F^2 T_F n_h
  + \Big[
    -6.3561 C_A^2 C_F-4.0082 C_A C_F^2+7.6917 C_A C_F T_F n_h+3.4586 C_F^3 \nonumber\\
    &\quad -2.8785 C_F^2 T_F n_h
  \Big] \logone
  + \Big[
    -2.2488 C_A^2 C_F+0.51078 C_A C_F^2+2.2962 C_A C_F T_F n_h \nonumber\\
    &\quad +1.4025 C_F^3-1.8900 C_F^2 T_F n_h
  \Big] \logtwo
  + \Big[
    -0.42778 C_A^2 C_F+0.90267 C_A C_F^2 \nonumber\\
    &\quad +0.33008 C_A C_F T_F n_h+0.062184 C_F^3-0.55727 C_F^2 T_F n_h
  \Big] \logthree
  + \Big[
    -0.035012 C_A^2 C_F \nonumber\\
    &\quad +0.20814 C_A C_F^2+0.025463 C_A C_F T_F n_h-0.075860 C_F^3-0.086806 C_F^2 T_F n_h
  \Big] \logfour \nonumber\\
  &\quad + \Big[
    0.019097 C_A C_F^2-0.023438 C_F^3-0.0069444 C_F^2 T_F n_h
  \Big] \logfive
  + \Big[
    -0.0026042 C_F^3
  \Big] \logsix \nonumber\\
  &\quad + \frac{m^2}{-s} \Big\{
    -47.821 C_A^2 C_F+123.65 C_A C_F^2-52.115 C_A C_F T_F n_h-92.918 C_F^3 \nonumber\\
    &\quad -5.2612 C_F^2 T_F n_h
  + \Big[
    17.305 C_A^2 C_F+2.3223 C_A C_F^2-25.912 C_A C_F T_F n_h-10.381 C_F^3 \nonumber\\
    &\quad +3.3633 C_F^2 T_F n_h
  \Big] \logone
  + \Big[
    8.0183 C_A^2 C_F-19.097 C_A C_F^2-7.8739 C_A C_F T_F n_h+4.9856 C_F^3 \nonumber\\
    &\quad +8.4570 C_F^2 T_F n_h
  \Big] \logtwo
  + \Big[
    1.9149 C_A^2 C_F-6.8519 C_A C_F^2-1.4464 C_A C_F T_F n_h+3.0499 C_F^3 \nonumber\\
    &\quad +2.3758 C_F^2 T_F n_h
  \Big] \logthree
  + \Big[
    0.24069 C_A^2 C_F-0.91213 C_A C_F^2-0.067130 C_A C_F T_F n_h \nonumber\\
    &\quad +0.67172 C_F^3+0.48843 C_F^2 T_F n_h
  \Big] \logfour
  + \Big[
    0.0043403 C_A^2 C_F-0.051389 C_A C_F^2 \nonumber\\
    &\quad -0.0034722 C_A C_F T_F n_h+0.13229 C_F^3+0.0069444 C_F^2 T_F n_h
  \Big] \logfive
  + \Big[
    -0.00052083 C_A^2 C_F \nonumber\\
    &\quad -0.0010417 C_A C_F^2+0.0041667 C_F^3
  \Big] \logsix
  \Big\}
  + {\cal O}\left(\frac{m^4}{(-s)^2}\right) + \mbox{$n_l$, $n_l^2$ and $n_h^2$ terms} ,
  \label{eq::F_v1_sminf}
\end{align}

\begin{align}
  & F_2^{v,f,(3)}\Big|_{s\to-\infty} =
  \frac{m^2}{-s} \Big\{
    38.118 C_A^2 C_F-45.924 C_A C_F^2+32.493 C_A C_F T_F n_h+22.519 C_F^3 \nonumber\\
    &\quad -2.2135 C_F^2 T_F n_h
  + \Big[
    -6.6042 C_A^2 C_F-29.110 C_A C_F^2+11.444 C_A C_F T_F n_h+13.611 C_F^3 \nonumber\\
    &\quad +0.099988 C_F^2 T_F n_h
  \Big] \logone
  + \Big[
    0.95557 C_A^2 C_F-7.0847 C_A C_F^2+1.4122 C_A C_F T_F n_h \nonumber\\
    &\quad +5.1299 C_F^3-1.2305 C_F^2 T_F n_h
  \Big] \logtwo
  + \Big[
    0.31532 C_A^2 C_F+0.16787 C_A C_F^2 \nonumber\\
    &\quad +0.037037 C_A C_F T_F n_h+0.90641 C_F^3-0.44444 C_F^2 T_F n_h
  \Big] \logthree+ \Big[
    0.065972 C_A C_F^2 \nonumber\\
    &\quad +0.031250 C_F^3-0.069444 C_F^2 T_F n_h
  \Big] \logfour
  + \Big[
    -0.031250 C_F^3
  \Big] \logfive
  \Big\} \nonumber\\
  &\quad + {\cal O}\left(\frac{m^4}{(-s)^2}\right) + \mbox{$n_l$, $n_l^2$ and $n_h^2$ terms} ,
\end{align}

\begin{align}
  & F_1^{a,f,(3)}\Big|_{s\to-\infty} =
    8.3501 C_A^2 C_F-20.762 C_A C_F^2+10.425 C_A C_F T_F n_h+4.7318 C_F^3 \nonumber\\
    &\quad -3.2872 C_F^2 T_F n_h
  + \Big[
    -6.3561 C_A^2 C_F-4.0082 C_A C_F^2+7.6917 C_A C_F T_F n_h+3.4586 C_F^3 \nonumber\\
    &\quad -2.8785 C_F^2 T_F n_h
  \Big] \logone
  + \Big[
    -2.2488 C_A^2 C_F+0.51078 C_A C_F^2+2.2962 C_A C_F T_F n_h \nonumber\\
    &\quad +1.4025 C_F^3-1.8900 C_F^2 T_F n_h
  \Big] \logtwo
  + \Big[
    -0.42778 C_A^2 C_F+0.90267 C_A C_F^2 \nonumber\\
    &\quad +0.33008 C_A C_F T_F n_h+0.062184 C_F^3-0.55727 C_F^2 T_F n_h
  \Big] \logthree
  + \Big[
    -0.035012 C_A^2 C_F \nonumber\\
    &\quad +0.20814 C_A C_F^2+0.025463 C_A C_F T_F n_h-0.075860 C_F^3-0.086806 C_F^2 T_F n_h
  \Big] \logfour \nonumber\\
  &\quad + \Big[
    0.019097 C_A C_F^2-0.023438 C_F^3-0.0069444 C_F^2 T_F n_h
  \Big] \logfive
  + \Big[
    -0.0026042 C_F^3
  \Big] \logsix \nonumber\\
  &\quad + \frac{m^2}{-s} \Big\{
    -57.129 C_A^2 C_F+113.95 C_A C_F^2-59.667 C_A C_F T_F n_h-59.120 C_F^3 \nonumber\\
    &\quad +8.8792 C_F^2 T_F n_h
  + \Big[
    10.011 C_A^2 C_F+41.904 C_A C_F^2-35.338 C_A C_F T_F n_h-39.202 C_F^3 \nonumber\\
    &\quad +13.451 C_F^2 T_F n_h
  \Big] \logone
  + \Big[
    2.2965 C_A^2 C_F+8.6681 C_A C_F^2-9.8312 C_A C_F T_F n_h-13.759 C_F^3 \nonumber\\
    &\quad +9.8343 C_F^2 T_F n_h
  \Big] \logtwo
  + \Big[
    -0.24620 C_A^2 C_F-0.66843 C_A C_F^2-1.3242 C_A C_F T_F n_h \nonumber\\
    &\quad -0.50775 C_F^3+1.8351 C_F^2 T_F n_h
  \Big] \logthree
  + \Big[
    -0.064891 C_A^2 C_F-0.17408 C_A C_F^2 \nonumber\\
    &\quad +0.011574 C_A C_F T_F n_h-0.057746 C_F^3+0.40046 C_F^2 T_F n_h
  \Big] \logfour
  + \Big[
    -0.0085069 C_A^2 C_F \nonumber\\
    &\quad +0.022222 C_A C_F^2+0.0034722 C_A C_F T_F n_h+0.067708 C_F^3-0.0069444 C_F^2 T_F n_h
  \Big] \logfive \nonumber\\
  &\quad + \Big[
    0.00052083 C_A^2 C_F+0.0010417 C_A C_F^2-0.0041667 C_F^3
  \Big] \logsix
  \Big\} \nonumber\\
  &\quad + {\cal O}\left(\frac{m^4}{(-s)^2}\right) + \mbox{$n_l$, $n_l^2$ and $n_h^2$ terms} ,
\end{align}

\begin{align}
  & F_2^{a,f,(3)}\Big|_{s\to-\infty} =
  \frac{m^2}{-s} \Big\{
    -5.2743 C_A^2 C_F+34.122 C_A C_F^2+60.378 C_A C_F T_F n_h+0.51545 C_F^3 \nonumber\\
    &\quad -14.591 C_F^2 T_F n_h
  + \Big[
    -44.280 C_A^2 C_F+6.7642 C_A C_F^2+32.524 C_A C_F T_F n_h+8.6355 C_F^3 \nonumber\\
    &\quad -16.414 C_F^2 T_F n_h
  \Big] \logone
  + \Big[
    -9.8090 C_A^2 C_F+6.8860 C_A C_F^2+6.4028 C_A C_F T_F n_h \nonumber\\
    &\quad -0.26335 C_F^3-6.2035 C_F^2 T_F n_h
  \Big] \logtwo
  + \Big[
    -0.84028 C_A^2 C_F+4.3484 C_A C_F^2 \nonumber\\
    &\quad +0.61111 C_A C_F T_F n_h-0.32065 C_F^3-1.5556 C_F^2 T_F n_h
  \Big] \logthree
  + \Big[
    0.57292 C_A C_F^2-0.34375 C_F^3 \nonumber\\
    &\quad -0.20833 C_F^2 T_F n_h
  \Big] \logfour
  + \Big[
    -0.093750 C_F^3
  \Big] \logfive
  \Big\}
  + {\cal O}\left(\frac{m^4}{(-s)^2}\right) + \mbox{$n_l$, $n_l^2$ and $n_h^2$ terms} ,
\end{align}

\begin{align}
  & F^{s,f,(3)}\Big|_{s\to-\infty} =
    25.519 C_A^2 C_F-16.796 C_A C_F^2-3.3697 C_A C_F T_F n_h-0.82112 C_F^3 \nonumber\\
    &\quad +0.46640 C_F^2 T_F n_h
  + \Big[
    4.7139 C_A^2 C_F-5.0600 C_A C_F^2-0.43937 C_A C_F T_F n_h+0.73034 C_F^3 \nonumber\\
    &\quad +2.0176 C_F^2 T_F n_h
  \Big] \logone
  + \Big[
    0.20351 C_A^2 C_F-2.3029 C_A C_F^2+0.69548 C_A C_F T_F n_h+1.0101 C_F^3 \nonumber\\
    &\quad -0.23908 C_F^2 T_F n_h
  \Big] \logtwo
  + \Big[
    -0.21771 C_A^2 C_F-0.26083 C_A C_F^2+0.17730 C_A C_F T_F n_h \nonumber\\
    &\quad +0.14235 C_F^3-0.14060 C_F^2 T_F n_h
  \Big] \logthree
  + \Big[
    -0.035012 C_A^2 C_F+0.064915 C_A C_F^2 \nonumber\\
    &\quad +0.025463 C_A C_F T_F n_h+0.041327 C_F^3-0.034722 C_F^2 T_F n_h
  \Big] \logfour
  + \Big[
    0.019097 C_A C_F^2 \nonumber\\
    &\quad -0.0069444 C_F^2 T_F n_h
  \Big] \logfive
  + \Big[
    -0.0026042 C_F^3
  \Big] \logsix \nonumber\\
  &\quad + \frac{m^2}{-s} \Big\{
    -40.495 C_A^2 C_F-37.175 C_A C_F^2-33.918 C_A C_F T_F n_h+32.272 C_F^3 \nonumber\\
    &\quad +1.5879 C_F^2 T_F n_h
  + \Big[
    31.911 C_A^2 C_F+5.1716 C_A C_F^2-28.235 C_A C_F T_F n_h-14.201 C_F^3 \nonumber\\
    &\quad +0.60352 C_F^2 T_F n_h
  \Big] \logone
  + \Big[
    6.9504 C_A^2 C_F+20.896 C_A C_F^2-3.5948 C_A C_F T_F n_h-6.8796 C_F^3 \nonumber\\
    &\quad +0.96287 C_F^2 T_F n_h
  \Big] \logtwo
  + \Big[
    0.20466 C_A^2 C_F-0.065462 C_A C_F^2-0.69163 C_A C_F T_F n_h \nonumber\\
    &\quad +1.1505 C_F^3-0.24638 C_F^2 T_F n_h
  \Big] \logthree
  + \Big[
    0.052838 C_A^2 C_F-0.76691 C_A C_F^2-0.69218 C_F^3 \nonumber\\
    &\quad +0.27778 C_F^2 T_F n_h
  \Big] \logfour
  + \Big[
    0.12500 C_F^3
  \Big] \logfive
  \Big\}
  + {\cal O}\left(\frac{m^4}{(-s)^2}\right) + \mbox{$n_l$, $n_l^2$ and $n_h^2$ terms} ,
\end{align}

\begin{align}
  & F^{p,f,(3)}\Big|_{s\to-\infty} =
    25.519 C_A^2 C_F-16.796 C_A C_F^2-3.3697 C_A C_F T_F n_h-0.82112 C_F^3 \nonumber\\
    &\quad +0.46640 C_F^2 T_F n_h
  + \Big[
    4.7139 C_A^2 C_F-5.0600 C_A C_F^2-0.43937 C_A C_F T_F n_h+0.73034 C_F^3 \nonumber\\
    &\quad +2.0176 C_F^2 T_F n_h
  \Big] \logone
  + \Big[
    0.20351 C_A^2 C_F-2.3029 C_A C_F^2+0.69548 C_A C_F T_F n_h+1.0101 C_F^3 \nonumber\\
    &\quad -0.23908 C_F^2 T_F n_h
  \Big] \logtwo
  + \Big[
    -0.21771 C_A^2 C_F-0.26083 C_A C_F^2+0.17730 C_A C_F T_F n_h \nonumber\\
    &\quad +0.14235 C_F^3-0.14060 C_F^2 T_F n_h
  \Big] \logthree
  + \Big[
    -0.035012 C_A^2 C_F+0.064915 C_A C_F^2 \nonumber\\
    &\quad +0.025463 C_A C_F T_F n_h+0.041327 C_F^3-0.034722 C_F^2 T_F n_h
  \Big] \logfour
  + \Big[
    0.019097 C_A C_F^2 \nonumber\\
    &\quad -0.0069444 C_F^2 T_F n_h
  \Big] \logfive
  + \Big[
    -0.0026042 C_F^3
  \Big] \logsix \nonumber\\
  &\quad + \frac{m^2}{-s} \Big\{
    -23.466 C_A^2 C_F+0.55954 C_A C_F^2+0.60203 C_A C_F T_F n_h+50.838 C_F^3 \nonumber\\
    &\quad -0.95432 C_F^2 T_F n_h
  + \Big[
    10.183 C_A^2 C_F-15.480 C_A C_F^2-11.531 C_A C_F T_F n_h+7.4435 C_F^3 \nonumber\\
    &\quad +4.5632 C_F^2 T_F n_h
  \Big] \logone
  + \Big[
    4.6124 C_A^2 C_F+4.9688 C_A C_F^2-0.78449 C_A C_F T_F n_h-9.1395 C_F^3 \nonumber\\
    &\quad +2.5514 C_F^2 T_F n_h
  \Big] \logtwo
  + \Big[
    0.17134 C_A^2 C_F-0.029534 C_A C_F^2-0.45089 C_A C_F T_F n_h \nonumber\\
    &\quad +1.3184 C_F^3-0.30193 C_F^2 T_F n_h
  \Big] \logthree
  + \Big[
    0.052838 C_A^2 C_F-0.50996 C_A C_F^2-0.19218 C_F^3 \nonumber\\
    &\quad +0.13889 C_F^2 T_F n_h
  \Big] \logfour
  + \Big[
    0.062500 C_F^3
  \Big] \logfive
  \Big\}
  + {\cal O}\left(\frac{m^4}{(-s)^2}\right) + \mbox{$n_l$, $n_l^2$ and $n_h^2$ terms} ,
  \label{eq::F_p_sminf}
\end{align}
with $l_s = \log(m^2/(-s-i\delta))$. The leading logarithmic contributions of order
$\alpha_s^n \log^{2n}(m^2/s)$ are given by the Sudakov
exponent~\cite{Sudakov:1954sw,Frenkel:1976bj}
$\mbox{exp}[-C_F\alpha_s/(4\pi) \times \log^2 (m^2/s)]$ for $F_1^{v,f,(3)}$, $F_1^{a,f,(3)}$, $F^{s,f,(3)}$, and $F^{p,f,(3)}$.
Our numerical results (shown above with lower precision) are sufficiently precise to reconstruct the analytic coefficient
\begin{equation}
  F_1^{v,f,(3)} \Big|_{m^0/(-s)^0,\,\logsix} = F_1^{a,f,(3)} \Big|_{m^0/(-s)^0,\,\logsix} = F^{s,f,(3)} \Big|_{m^0/(-s)^0,\,\logsix} = F^{p,f,(3)} \Big|_{m^0/(-s)^0,\,\logsix} = -\frac{C_F^3}{384} .
  \label{eq::F1sminf_Sudakov}
\end{equation}
Similarly, we can reconstruct the analytic coefficients for the leading logarithms of the first mass corrections and find
\begin{align}
  F_1^{v,f,(3)} \Big|_{m^2/(-s)^1,\,\logsix} = - F_1^{a,f,(3)} \Big|_{m^2/(-s)^1,\,\logsix} = \frac{C_F^3}{240} - \frac{C_F^2C_A}{960} - \frac{C_FC_A^2}{1920} .
  \label{eq::F1sminf_lead}
\end{align}
The latter agree with
Refs.~\cite{Liu:2017axv,Liu:2017vkm,Liu:2018czl} where the results in
Eq.~(\ref{eq::F1sminf_lead}) have been obtained using an involved asymptotic
expansion of the three-loop vertex diagrams.  Moreover, we confirm that there
are only subleading contributions from the non-singlet diagrams to the
remaining form factors.  While $F_2^{v,f,(3)} |_{m^2/(-s)^1,\,\logsix}$ and
$F_2^{a,f,(3)} |_{m^2/(-s)^1,\,\logsix}$ vanish completely,
$F^{s,f,(3)} |_{m^2/(-s)^1,\,\logsix}$ and
$F^{p,f,(3)} |_{m^2/(-s)^1,\,\logsix}$ should receive contributions only through
the singlet diagrams which is discussed below.

Our numerical results also allow the reconstruction of the analytic result for
the quartic mass corrections of $F_2^{v,f,(3)}$ which is given by (not shown
in numerical form above)
\begin{equation}
  F_2^{v,f,(3)} \Big|_{m^4/(-s)^2,\,\logsix} = -\frac{C_A^2 C_F}{180}+\frac{C_A C_F^2}{160}+\frac{7 C_F^3}{720} .
\end{equation}
This result disagrees with Ref.~\cite{Liu:2021chn}.  However, we can make both
results agree by modifying Eq.~(2.14) in Ref.~\cite{Liu:2021chn} to\footnote{
  Our method does not provide the squared and cubic terms in this
  equation and we cannot make any statement about them.  }
\begin{equation}
  f(z) = 1 - \frac{8}{3} \cdot \frac{z}{5} + \dots .
\end{equation}
The correctness of our result has been confirmed by the authors of Ref.~\cite{Liu:2021chn}.

Finally, we show the reconstructed analytic coefficients for the remaining leading and first subleading logarithms for the first two terms in the high-energy expansion for all currents:
\begin{align}
  F_1^{v,f,(3)} \Big|_{m^0/(-s)^0,\,\logfive} &= \frac{11 C_A C_F^2}{576}-\frac{3 C_F^3}{128}-\frac{C_F^2 T_F n_h}{144}-\frac{C_F^2 T_F n_l}{144} \,,\nonumber\\
  F_1^{v,f,(3)} \Big|_{m^2/(-s)^1,\,\logfive} &= \frac{5 C_A^2 C_F}{1152}-\frac{37 C_A C_F^2}{720}-\frac{C_A C_F T_F n_h}{288}-\frac{C_A C_F T_F n_l}{288}+\frac{127 C_F^3}{960}+\frac{C_F^2 T_F n_h}{144} \nonumber\\
  &\quad +\frac{C_F^2 T_F n_l}{144} \,,\nonumber\\
  F_2^{v,f,(3)} \Big|_{m^2/(-s)^1,\,\logfive} &= -\frac{C_F^3}{32} \,,\nonumber\\
  F_2^{v,f,(3)} \Big|_{m^4/(-s)^2,\,\logfive} &= \frac{13 C_A^2 C_F}{1440}-\frac{71 C_A C_F^2}{360}-\frac{C_A C_F T_F n_h}{72}-\frac{C_A C_F T_F n_l}{72}+\frac{101 C_F^3}{240} \nonumber\\
  &\quad +\frac{C_F^2 T_F n_h}{36}+\frac{C_F^2 T_F n_l}{36} \,,\nonumber\\
  F_1^{a,f,(3)} \Big|_{m^0/(-s)^0,\,\logfive} &= \frac{11 C_A C_F^2}{576}-\frac{3 C_F^3}{128}-\frac{C_F^2 T_F n_h}{144}-\frac{C_F^2 T_F n_l}{144} \,,\nonumber\\
  F_1^{a,f,(3)} \Big|_{m^2/(-s)^1,\,\logfive} &= -\frac{49 C_A^2 C_F}{5760}+\frac{C_A C_F^2}{45}+\frac{C_A C_F T_F n_h}{288}+\frac{C_A C_F T_F n_l}{288}+\frac{13 C_F^3}{192}-\frac{C_F^2 T_F n_h}{144} \nonumber\\
  &\quad -\frac{C_F^2 T_F n_l}{144} \,,\nonumber\\
  F_2^{a,f,(3)} \Big|_{m^2/(-s)^1,\,\logfive} &= -\frac{3 C_F^3}{32} \,,\nonumber\\
  F_2^{a,f,(3)} \Big|_{m^4/(-s)^2,\,\logsix} &= \frac{C_A^2 C_F}{480}+\frac{C_A C_F^2}{240}-\frac{C_F^3}{60} \,,\nonumber\\
  F_2^{a,f,(3)} \Big|_{m^4/(-s)^2,\,\logfive} &= -\frac{49 C_A^2 C_F}{1440}+\frac{4 C_A C_F^2}{45}+\frac{C_A C_F T_F n_h}{72}+\frac{C_A C_F T_F n_l}{72}+\frac{C_F^3}{48}-\frac{C_F^2 T_F n_h}{36} \nonumber\\
  &\quad -\frac{C_F^2 T_F n_l}{36} \,,\nonumber\\
  F^{s,f,(3)} \Big|_{m^0/(-s)^0,\,\logfive} &= \frac{11 C_A C_F^2}{576}-\frac{C_F^2 T_F n_h}{144}-\frac{C_F^2 T_F n_l}{144} \,,\nonumber\\
  F^{s,f,(3)} \Big|_{m^2/(-s)^1,\,\logfive} &= \frac{C_F^3}{8} \,,\nonumber\\
  F^{p,f,(3)} \Big|_{m^0/(-s)^0,\,\logfive} &= \frac{11 C_A C_F^2}{576}-\frac{C_F^2 T_F n_h}{144}-\frac{C_F^2 T_F n_l}{144} \,,\nonumber\\
  F^{p,f,(3)} \Big|_{m^2/(-s)^1,\,\logfive} &= \frac{C_F^3}{16} \,.
\end{align}

Apart form the leading and subleading logarithms discussed above, our approach provides the whole tower of logarithms and also higher order contributions in $m^2/(-s)$.
We estimate the accuracy of the non-logarithmic terms in Eqs.~(\ref{eq::F_v1_sminf})-(\ref{eq::F_p_sminf}) to ten digits.
For the subleading terms the accuracy decreases.
Note, however, that we use the $s\to\infty$ expansion only for $|s/m^2| \gtrsim 45$ and that $1/45^3 \approx {\cal O}(10^{-5})$.

For the scalar and pseudo-scalar singlet contributions we obtain the
following results for the leading logarithmic contributions of the
power-suppressed term
\begin{eqnarray}
  F_{\rm sing}^{s,f,(3)} \Big|_{m^2/(-s),\,\logsix} =
  F_{\rm sing}^{p,f,(3)} \Big|_{m^2/(-s),\,\logsix} =
  \frac{C_A C_F T_F}{960} + \frac{C_F^2 T_F}{240}
  \,,
\end{eqnarray}
which is in agreement with Ref.~\cite{Liu:2018czl}.

%- }}}
%- {{{ Threshold expansions: $s\to4m^2$ and $s\to16m^2$:

\subsubsection{\label{sub::thr}Threshold expansions: \boldmath{$s\to4m^2$} and \boldmath{$s\to16m^2$}}

Let us next discuss the two- and four-particle thresholds at $s=4m^2$ and $s=16m^2$.
Close to the two-particle threshold $F_1$ develops the famous Coulomb
singularity with negative powers in the velocity of the produced quarks,
$\beta = \sqrt{1 - 4 m^2/s}$, up to
third order multiplied by $\log(\beta)$ terms.
In this limit real radiation is suppressed by three powers of $\beta$ and
it is thus possible to construct physical quantities from the square of
the form factors. For the four currents under consideration we follow
Ref.~\cite{Lee:2018rgs} and define
\begin{eqnarray}
R^v &=& \beta \left( |F_1^v+F_2^v|^2 + \frac{|(1-\beta^2){F_1^v+F_2^v}|^2}{2(1-\beta^2)}\right)
\,,\nonumber\\
R^a &=& \beta^3 |F_1^a|^2
\,,\nonumber\\
R^s &=& \beta^3 |F^s|^2
\,,\nonumber\\
R^p &=& \beta |F^p|^2
\label{eq::R_F}
\,,
\end{eqnarray}
These quantities form building blocks for, e.g., cross sections of heavy quark
production in electron-positron annihilation or decay rates for scalar or
pseudo-scalar Higgs bosons (see also Ref.~\cite{Lee:2018rgs}).  For reference
we provide the (exact) leading order results which are given by
\begin{eqnarray}
  R^{v,(0)} &=& \frac{3\beta}{2}\left(1-\frac{\beta^2}{3}\right)\,,\nonumber\\
  R^{a,(0)} &=& \beta^3\,,\nonumber\\
  R^{s,(0)} &=& \beta^3\,,\nonumber\\
  R^{p,(0)} &=& \beta\,,
\end{eqnarray}
where we adapt the notation from Eq.~(\ref{eq::F_pert_exp}).  We parametrize
the QCD corrections to $R^\delta$ with the quantities $\Delta^{\delta,(i)}$
which we introduce as
\begin{eqnarray}
  R^\delta &=& R^{\delta,(0)}
  + K_\delta \beta^{n_\delta}
  \sum_{i\ge1} \left(\frac{\alpha_s(m)}{\pi}\right)^i \Delta^{\delta,(i)}
  \,,
  \label{eq::Rdelta}
\end{eqnarray}
with $K_v=3/2$, $K_a=K_s=K_p=1$, $n_v=n_p=1$ and $n_a=n_s=3$.
For convenience we set $\mu=m$.
In contrast to Ref.~\cite{Lee:2018rgs} we present
results parametrized in terms of $\alpha_s^{(n_l)}$ (and not
$\alpha_s^{(n_l+1)}$). Furthermore, for the scalar and
pseudo-scalar current we keep the factor $m$ in the definition of the currents
(see Eq.~(\ref{eq::currents})) in the $\overline{\rm MS}$ scheme and refrain
from transformation to the on-shell scheme.
Note that this is the natural choice for Higgs decays where the factor
$m$ takes over the role of the Yukawa couplings.

The three leading terms in $\beta$ for the four currents read
\begin{align}
  & \Delta^{v,(3)} = \Bigg\{
    C_F^3 \bigg[ -\frac{32.470}{\beta ^2} + \frac{1}{\beta} ( 14.998-32.470 l_{2\beta}) \bigg]
    + C_A C_F^2 \bigg[ \frac{1}{\beta^2} ( -29.764 l_{2\beta}-7.7703) \nonumber \\
    & \quad + \frac{1}{\beta} (-12.516 l_{2 \beta}-11.435) \bigg]
    + C_A^2 C_F \bigg[ \frac{1}{\beta} (16.586 l^2_{2 \beta}-22.572 l_{2\beta}+42.936) \bigg]
  + \mathcal{O}(\beta^0) \Bigg\} \nonumber \\
  & \quad + \Bigg\{
    C_F^2 T_F n_l \bigg[ \frac{1}{\beta^2} (-1.1101 + 10.823 l_{2\beta}) + \frac{1}{\beta} (17.275-10.692 l_{2\beta}) +10.357+59.345 l_{2\beta} \nonumber \\
    & \quad -4.3865 l_{2\beta}^2 \bigg]
    + C_A C_F T_F n_l \bigg[ \frac{1}{\beta} (-33.609+19.831 l_{2\beta}-12.063 l_{2\beta}^2) +52.985+26.593 l_{2\beta} \nonumber \\
    & \quad -6.5797 l_{2\beta}^2 \bigg]
    + C_F^2 T_F n_h \bigg[ \frac{2.4792}{\beta} -2.0339-1.3159 l_{2\beta} \bigg]
    + C_A C_F T_F n_h \bigg[ -0.20495 \bigg] \nonumber \\
    & \quad + C_F T_F^2 n_l^2 \bigg[ \frac{1}{\beta} (5.1308-3.6554 l_{2\beta}+2.1932 l_{2\beta}^2) -4.9367 \bigg]
    + C_F T_F^2 n_l n_h \bigg[ -0.54050 \bigg] \nonumber \\
    & \quad + C_F T_F^2 n_h^2 \bigg[ 0.10248 \bigg]
  + \mathcal{O}(\beta^1) \Bigg\}\,,
\label{eq::Delta3v}
\end{align}
\begin{align}
  & \Delta^{a,(3)} = \Bigg\{
    C_F^3 \bigg[ \frac{12.176}{\beta^3} - \frac{21.170}{\beta^2} + \frac{1}{\beta} (64.060-40.587 l_{2\beta}) \bigg]
    + C_A C_F^2 \bigg[ \frac{1}{\beta^2} (48.861-38.811 l_{2\beta}) \nonumber \\
    & \quad + \frac{1}{\beta} (-32.967+1.8594 l_{2\beta}) \bigg]
    + C_A^2 C_F \bigg[ \frac{1}{\beta} (82.095-55.745 l_{2\beta}+16.586 l_{2\beta}^2) \bigg]
  + \mathcal{O}(\beta^0) \Bigg\} \nonumber \\
  & \quad + \Bigg\{
    C_F^2 T_F n_l \bigg[ \frac{1}{\beta^2} (-22.900+14.113 l_{2\beta}) + \frac{1}{\beta} (13.711-4.1123 l_{2\beta}) -22.081+59.937 l_{2\beta} \nonumber \\
    & \quad -5.4831 l_{2\beta}^2 \bigg]
    + C_A C_F T_F n_l \bigg[ \frac{1}{\beta} (-65.503+43.956 l_{2\beta}-12.063 l_{2\beta}^2) + 24.409+14.713 l_{2\beta} \nonumber \\
    & \quad -2.1932 l_{2\beta}^2 \bigg]
    + C_F^2 T_F n_h \bigg[ \frac{0.28599}{\beta} + 1.7363 \bigg]
    + C_A C_F T_F n_h \bigg[ -0.43382 \bigg] \nonumber \\
    & \quad + C_F T_F^2 n_l^2 \bigg[ \frac{1}{\beta} (10.979-8.0419 l_{2\beta}+2.1932 l_{2\beta}^2) -2.2646 \bigg]
    + C_F T_F^2 n_l n_h \bigg[ -0.24958 \bigg] \nonumber \\
    & \quad + C_F T_F^2 n_h^2 \bigg[ -0.022729 \bigg]
  + \mathcal{O}(\beta^1) \Bigg\}\,,
\label{eq::Delta3a}
\end{align}
\begin{align}
  & \Delta^{s,(3)} = \Bigg\{
    C_F^3 \bigg[ \frac{12.176}{\beta^3} + \frac{10.585}{\beta^2} + \frac{1}{\beta} (90.512-64.939 l_{2\beta}) \bigg]
    + C_A C_F^2 \bigg[ \frac{1}{\beta^2} (48.861-38.811 l_{2\beta}) \nonumber \\
    & \quad + \frac{1}{\beta} (39.492-25.282 l_{2\beta}) \bigg]
    + C_A^2 C_F \bigg[ \frac{1}{\beta} (82.095-55.745 l_{2\beta}+16.586 l_{2\beta}^2) \bigg]
  + \mathcal{O}(\beta^0) \Bigg\} \nonumber \\
  & \quad + \Bigg\{
    C_F^2 T_F n_l \bigg[ \frac{1}{\beta^2} (-22.900+14.113 l_{2\beta}) + \frac{1}{\beta} (-25.009+5.7573 l_{2\beta}) -4.5086+69.532 l_{2\beta} \nonumber \\
    & \quad -8.7730 l_{2\beta}^2 \bigg]
    + C_A C_F T_F n_l \bigg[ \frac{1}{\beta} (-65.503+43.956 l_{2\beta}-12.063 l_{2\beta}^2) -5.3053+14.713 l_{2\beta} \nonumber \\
    & \quad -2.1932 l_{2\beta}^2 \bigg]
    + C_F^2 T_F n_h \bigg[ \frac{2.2364}{\beta} + 1.4814 \bigg]
    + C_A C_F T_F n_h \bigg[ 1.8511 \bigg] \nonumber \\
    & \quad + C_F T_F^2 n_l^2 \bigg[ \frac{1}{\beta} (10.979-8.0419 l_{2\beta}+2.1932 l_{2\beta}^2) + 3.6357 \bigg]
    + C_F T_F^2 n_l n_h \bigg[ -0.087673 \bigg] \nonumber \\
    & \quad + C_F T_F^2 n_h^2 \bigg[ 0.39896 \bigg]
  + \mathcal{O}(\beta^1) \Bigg\}\,,
\label{eq::Delta3s}
\end{align}
\begin{align}
  & \Delta^{p,(3)} = \Bigg\{
    C_F^3 \bigg[ -\frac{8.1174}{\beta^2} + \frac{1}{\beta} (19.153-97.409 l_{2\beta}) \bigg]
    + C_A C_F^2 \bigg[ \frac{1}{\beta^2} (-7.7703-29.764 l_{2\beta}) \nonumber \\
    & \quad + \frac{1}{\beta} (17.393-39.657 l_{2\beta}) \bigg]
    + C_A^2 C_F \bigg[ \frac{1}{\beta} (42.936-22.572 l_{2\beta}+16.586 l_{2\beta}^2) \bigg]
  + \mathcal{O}(\beta^0) \Bigg\} \nonumber \\
  & \quad + \Bigg\{
    C_F^2 T_F n_l \bigg[ \frac{1}{\beta^2} (-1.1101+10.823 l_{2\beta}) + \frac{1}{\beta} (-11.575-0.82247 l_{2\beta}) +45.228+73.188 l_{2\beta} \nonumber \\
    & \quad -13.159 l_{2\beta}^2 \bigg]
    + C_A C_F T_F n_l \bigg[ \frac{1}{\beta} (-33.609+19.831 l_{2\beta}-12.063 l_{2\beta}^2) +27.382+26.593 l_{2\beta} \nonumber \\
    & \quad -6.5797 l_{2\beta}^2 \bigg]
    + C_F^2 T_F n_h \bigg[ \frac{4.4296}{\beta} + 4.0027-1.3159 l_{2\beta} \bigg]
    + C_A C_F T_F n_h \bigg[ 0.80851 \bigg] \nonumber \\
    & \quad + C_F T_F^2 n_l^2 \bigg[ \frac{1}{\beta} (5.1308-3.6554 l_{2\beta}+2.1932 l_{2\beta}^2) + 0.96364 \bigg]
    + C_F T_F^2 n_l n_h \bigg[ -0.37860 \bigg] \nonumber \\
    & \quad + C_F T_F^2 n_h^2 \bigg[ 0.52417 \bigg]
  + \mathcal{O}(\beta^1) \Bigg\}\,,
\label{eq::Delta3p}
\end{align}
with $l_{2\beta}=\log(2\beta)$.  Note that the fermionic contributions are
suppressed by one additional power of $\beta$ and, thus, we can show the term
$\beta^0$ for them in contrast to the non-fermionic contributions.  The
non-fermionic part of $\Delta^{v,(3)}$ has already been shown in
Ref.~\cite{Fael:2022rgm}.  In Eqs.~(\ref{eq::Delta3v}) to~(\ref{eq::Delta3p})
we only show five digits for each coefficient, however, our results for
$\Delta^{\delta,(3)}$ contain more significant digits. For example, in the
vector and pseudo-scalar case our numerical results reproduce the analytic
expressions from Ref.~\cite{Kiyo:2009gb} (see also
Refs.~\cite{Pineda:2006ri,Hoang:2008qy}) with at least 13 digits accuracy.
The light-fermion contributions can be compared with the analytic results of
Ref.~\cite{Lee:2018rgs} and agreement is found for 19 digits. Similarly,
after specifying to the large-$N_C$ limit we can reproduce the first 14
digits of Ref.~\cite{Lee:2018rgs} for all four currents.

The four-particle thresholds are much less pronounced in the results for the
form factors. For the individual master integrals we observe behaviours of the
form $(\beta_4)^k \times \log^l(\beta_4)$ with $k=3$ and $l=5$
where $\beta_4 = \sqrt{1 - 16 m^2/s}$.  However, all form factors have a
smooth behaviour for $s\to 16 m^2$. In fact, in all cases there are no
$\log(\beta_4)$ terms in our expansions for $\beta_4\to0$.  Furthermore, we
observe the first non-analytic terms, i.e.\ terms where $\beta_4$ is raised to
an odd power, at order $(\beta_4)^7$ for the axial-vector and scalar form
factor and $(\beta_4)^9$ in the vector and pseudo-scalar case. This
statement is true both for the non-singlet and singlet form factors. Such a high
suppression can partly be explained by the fact that the massive four-particle
phase-space, which is one of our master integrals, already provides a factor
$(\beta_4)^7$.  Due to the more divergent behaviour of the master integrals it
is nevertheless necessary to perform a careful matching to $s=16 m^2$, both from
above and below.  This means we have to go quite close to $s=16 m^2$ using
Taylor expansions around regular points (in our case we chose $s=15 m^2$ and
$s=17m^2$). Furthermore, we constructed 100 terms for the expansion in
$\beta_4$.

%- }}}

%- }}}
%- {{{ Numerical results for finite form factor:

\subsection{Numerical results for finite three-loop form factors}

\begin{figure}[h]
  \begin{center}
    \begin{tabular}{cc}
      \includegraphics[width=0.47\textwidth]{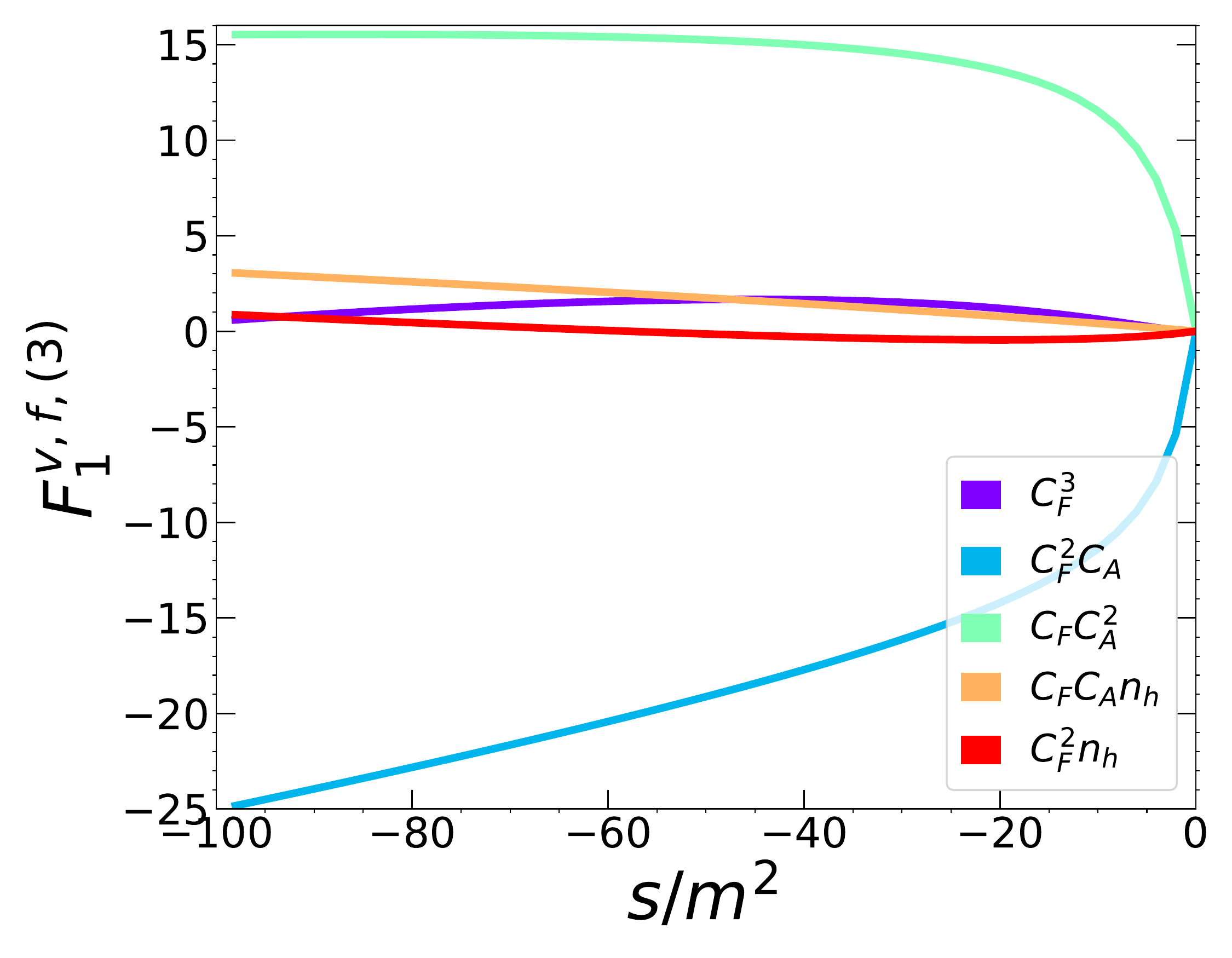}
      &
      \includegraphics[width=0.47\textwidth]{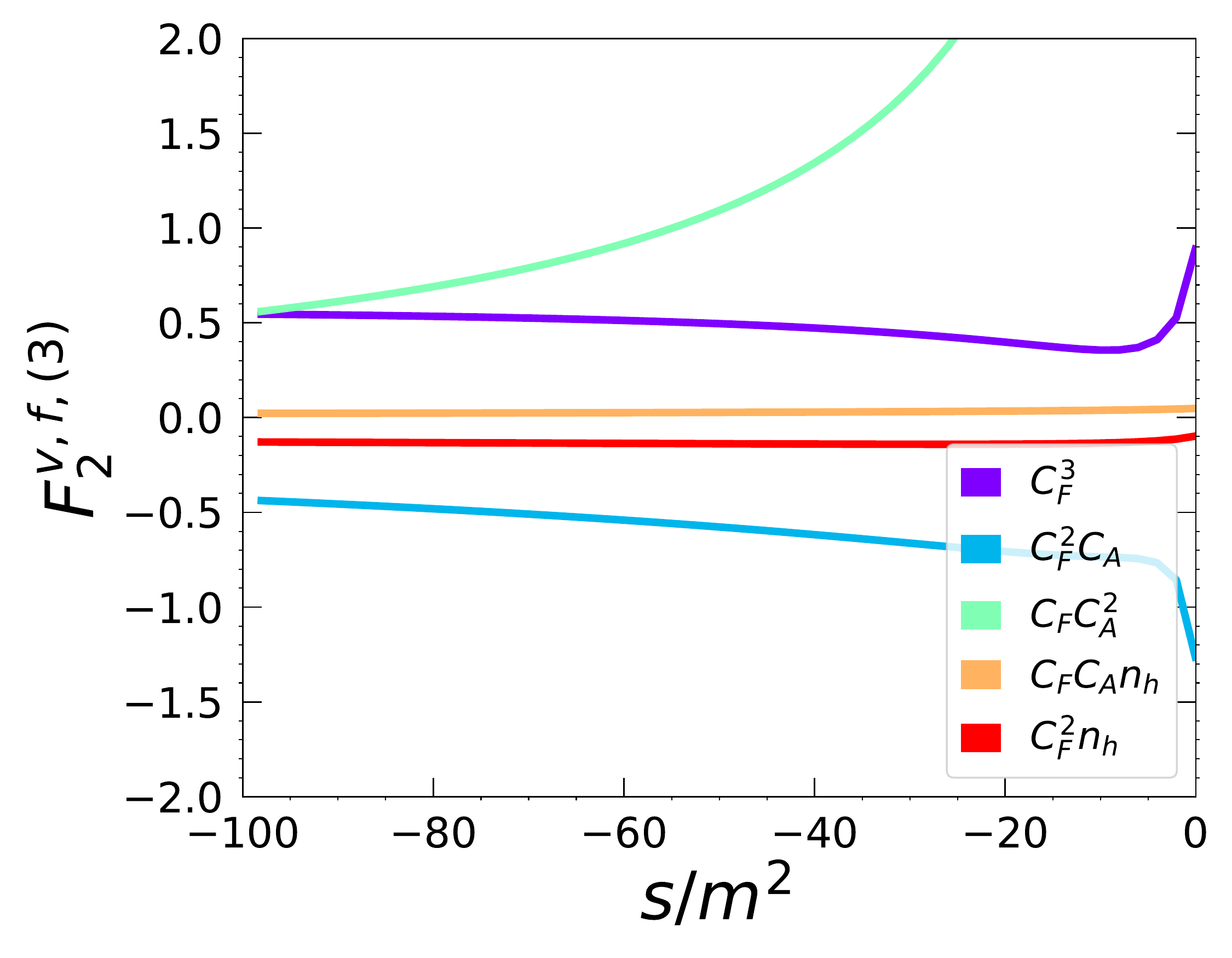}
      \\
      (a) & (b) \\
      \includegraphics[width=0.47\textwidth]{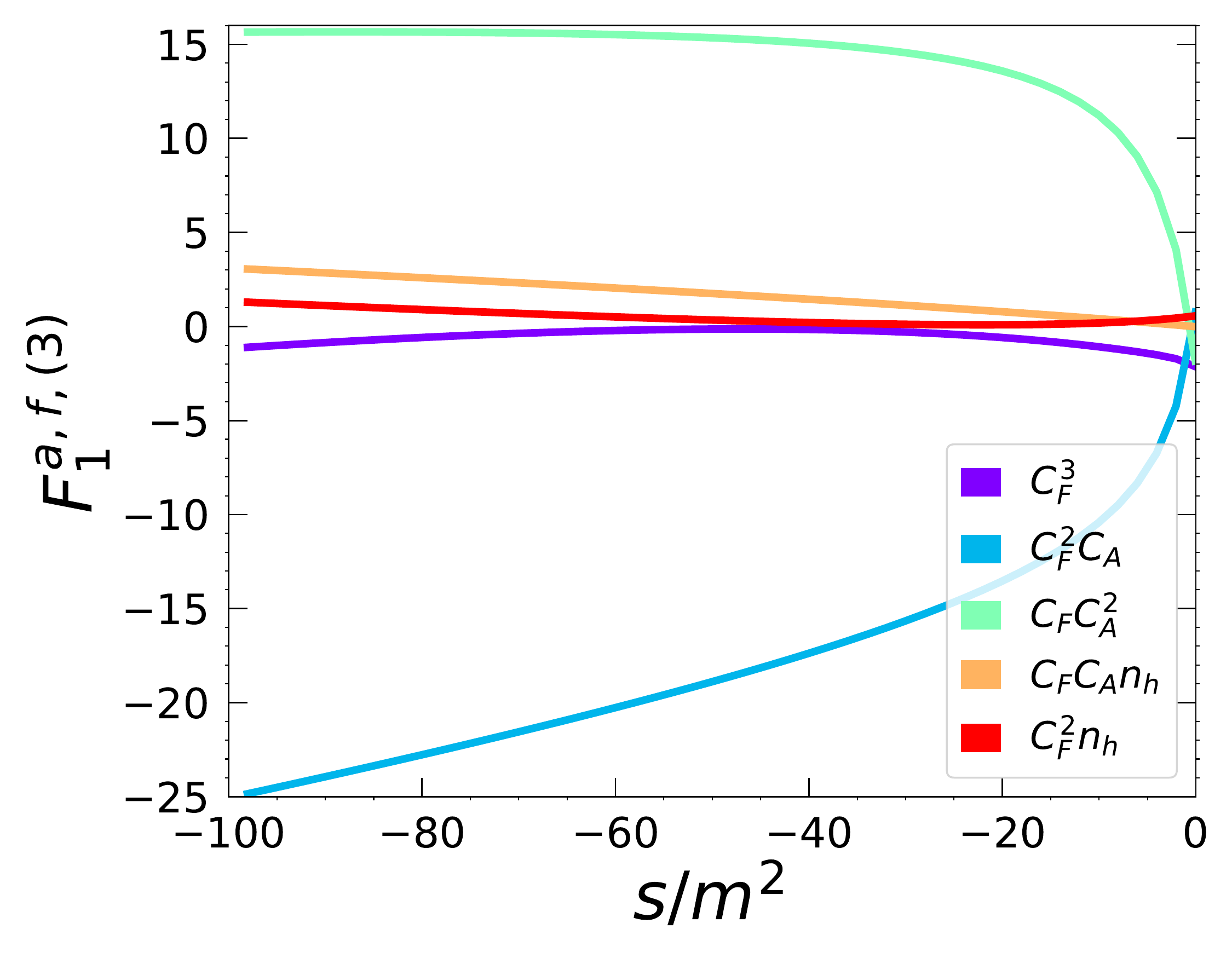}
      &
      \includegraphics[width=0.47\textwidth]{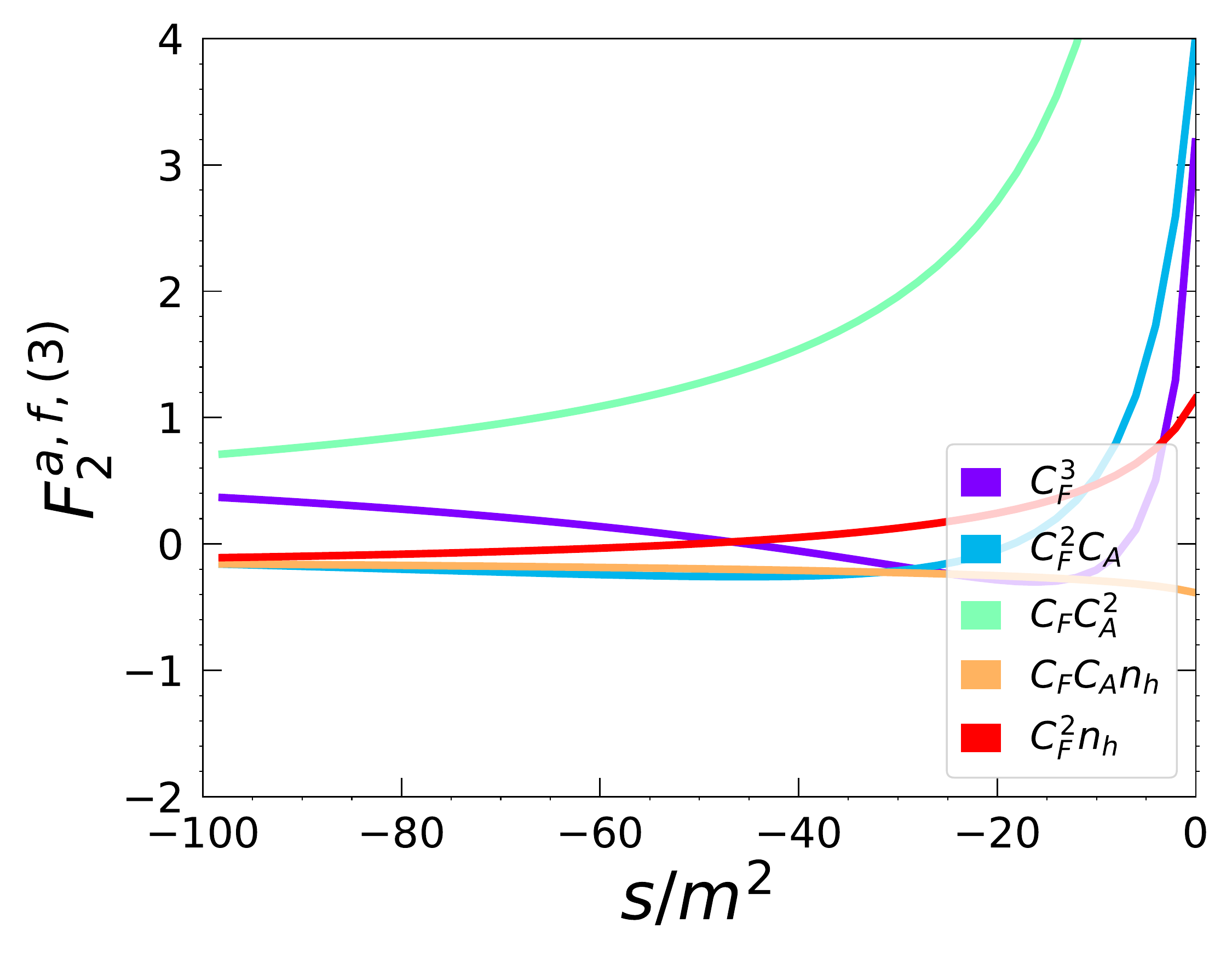}
      \\
      (c) & (d) \\
      \includegraphics[width=0.47\textwidth]{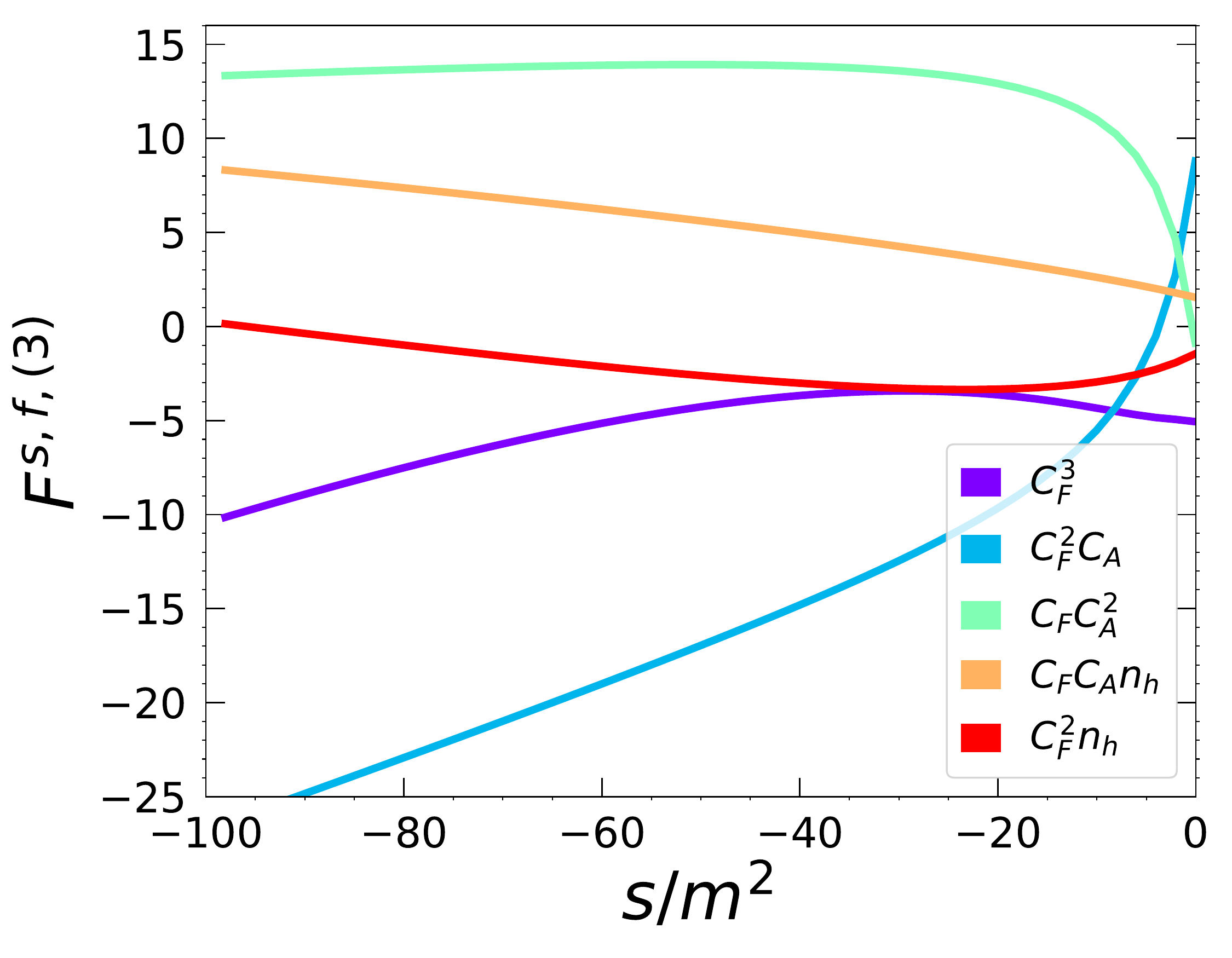}
      &
      \includegraphics[width=0.47\textwidth]{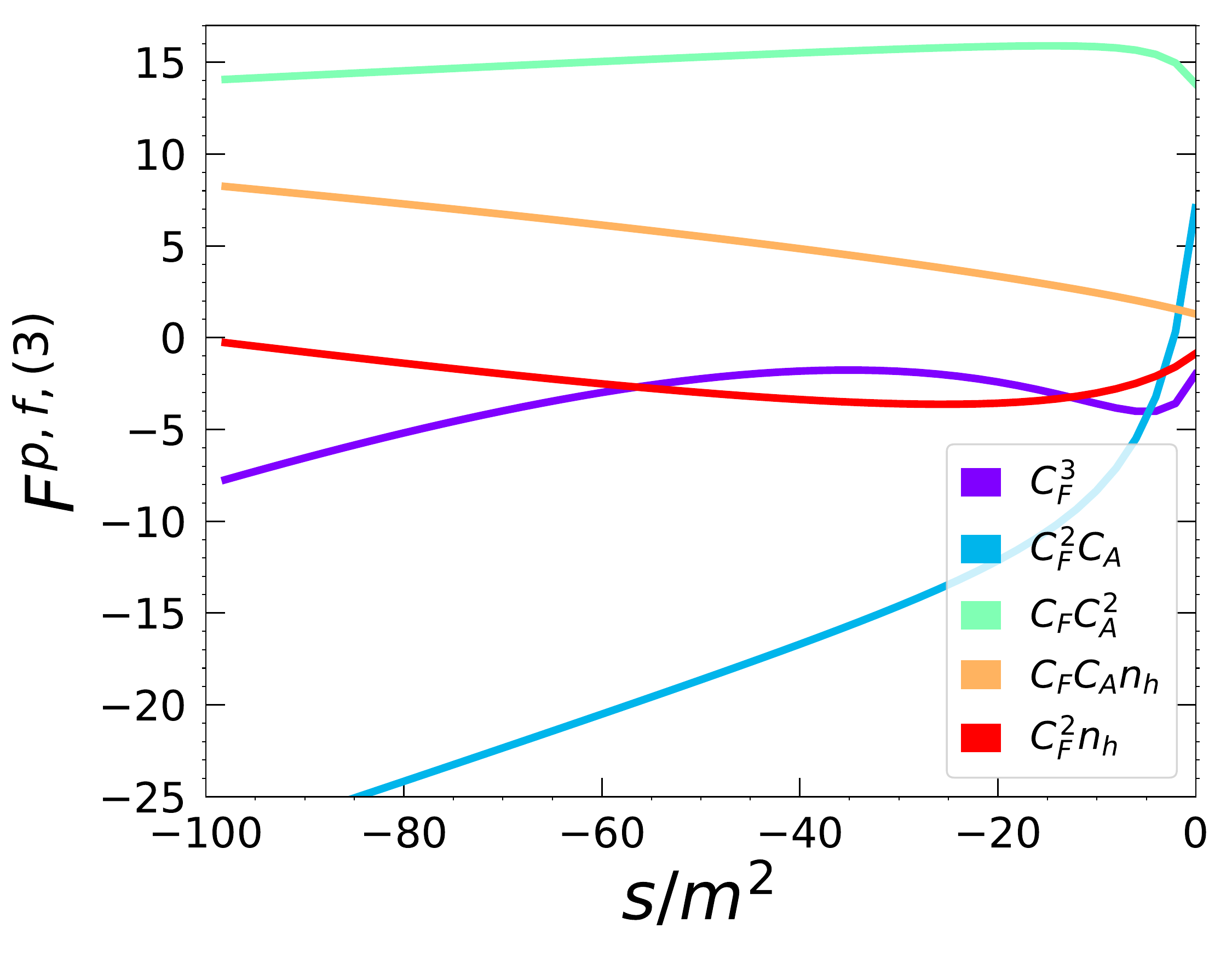}
      \\
      (e) & (f) \\
    \end{tabular}
    \caption{\label{fig::FF1}Non-singlet form factors as a function of $s$ for $s<0$.}
  \end{center}
\end{figure}

\begin{figure}[h]
  \begin{center}
    \begin{tabular}{cc}
      \includegraphics[width=0.47\textwidth]{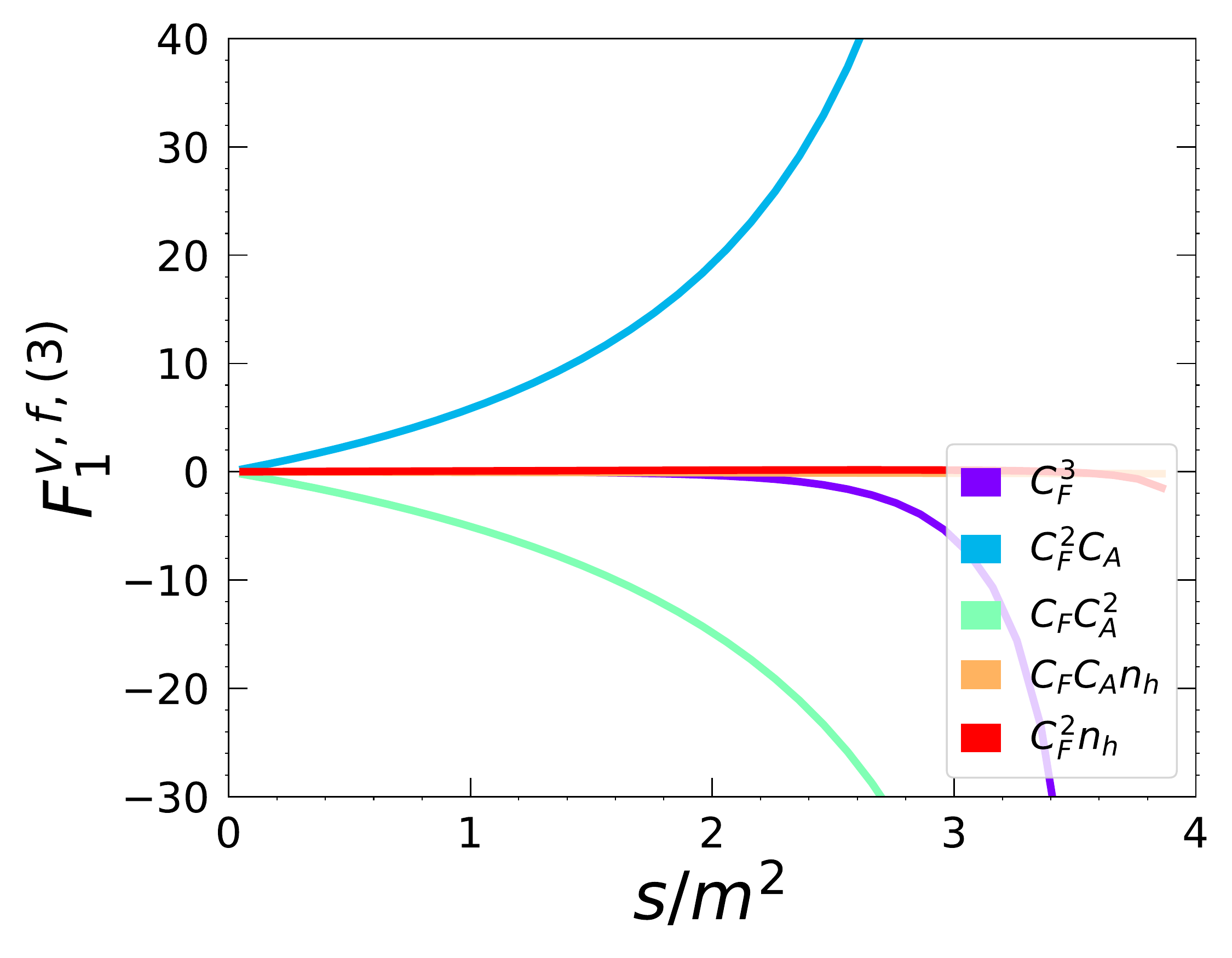}
      &
      \includegraphics[width=0.47\textwidth]{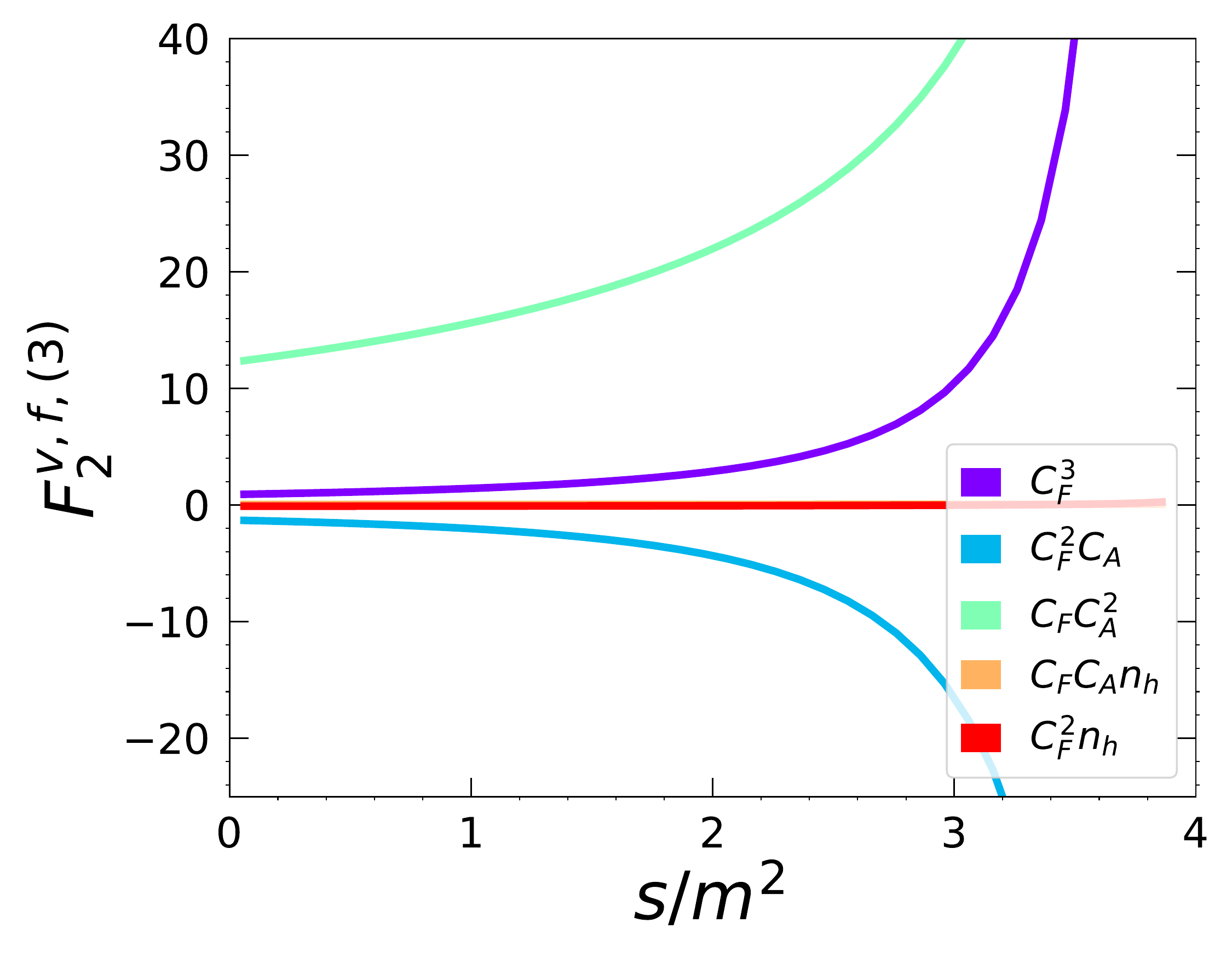}
      \\
      (a) & (b) \\
      \includegraphics[width=0.47\textwidth]{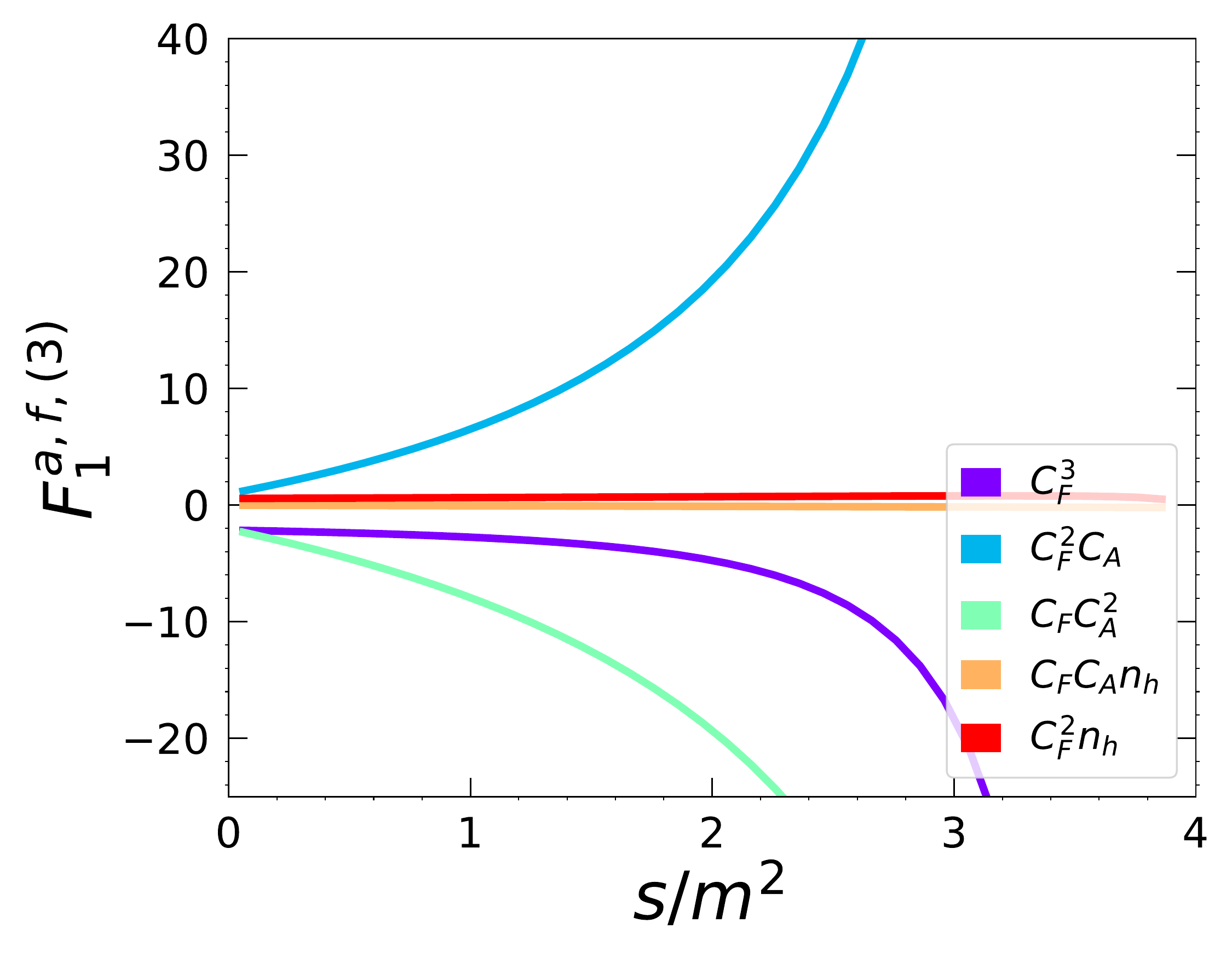}
      &
      \includegraphics[width=0.47\textwidth]{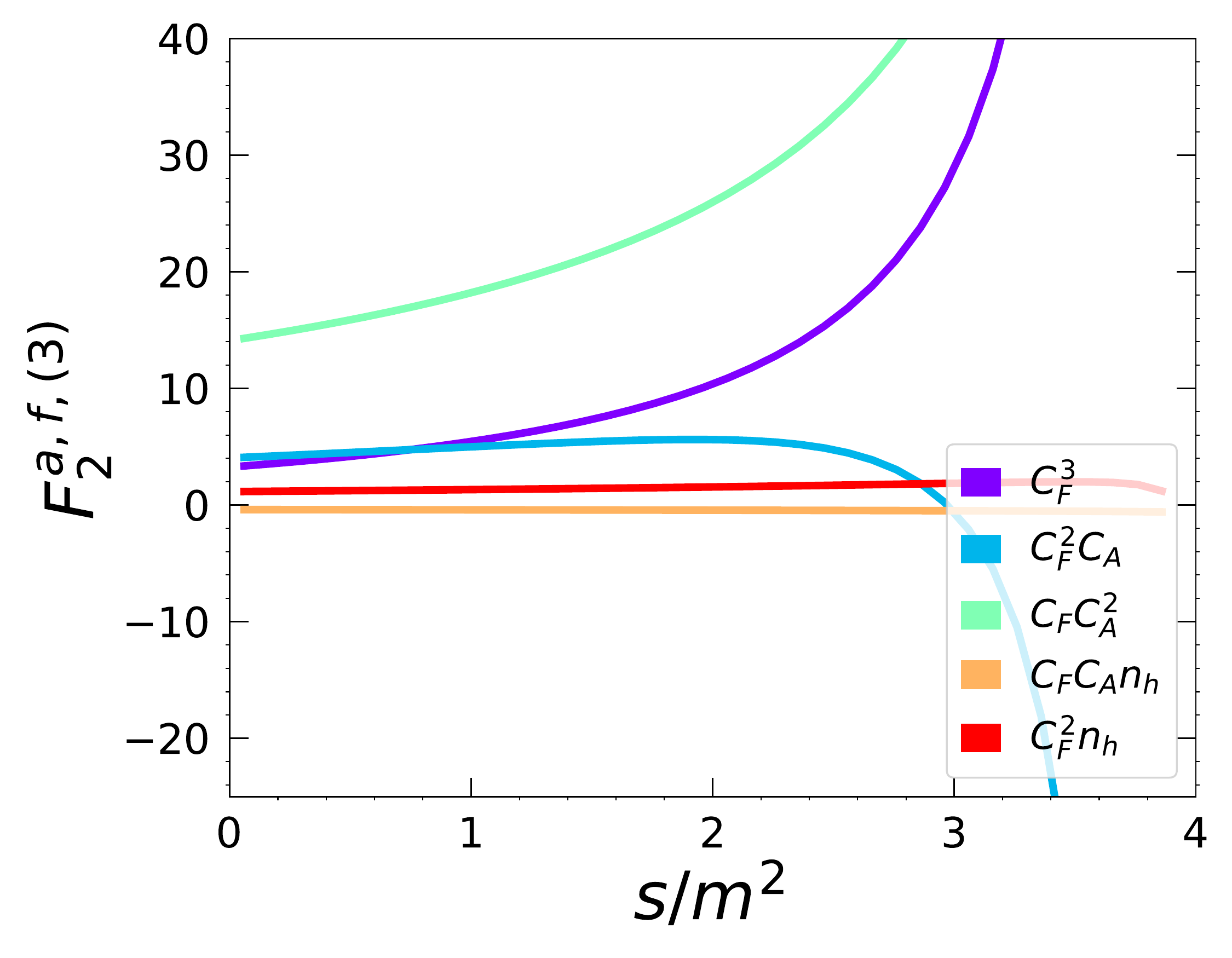}
      \\
      (c) & (d) \\
      \includegraphics[width=0.47\textwidth]{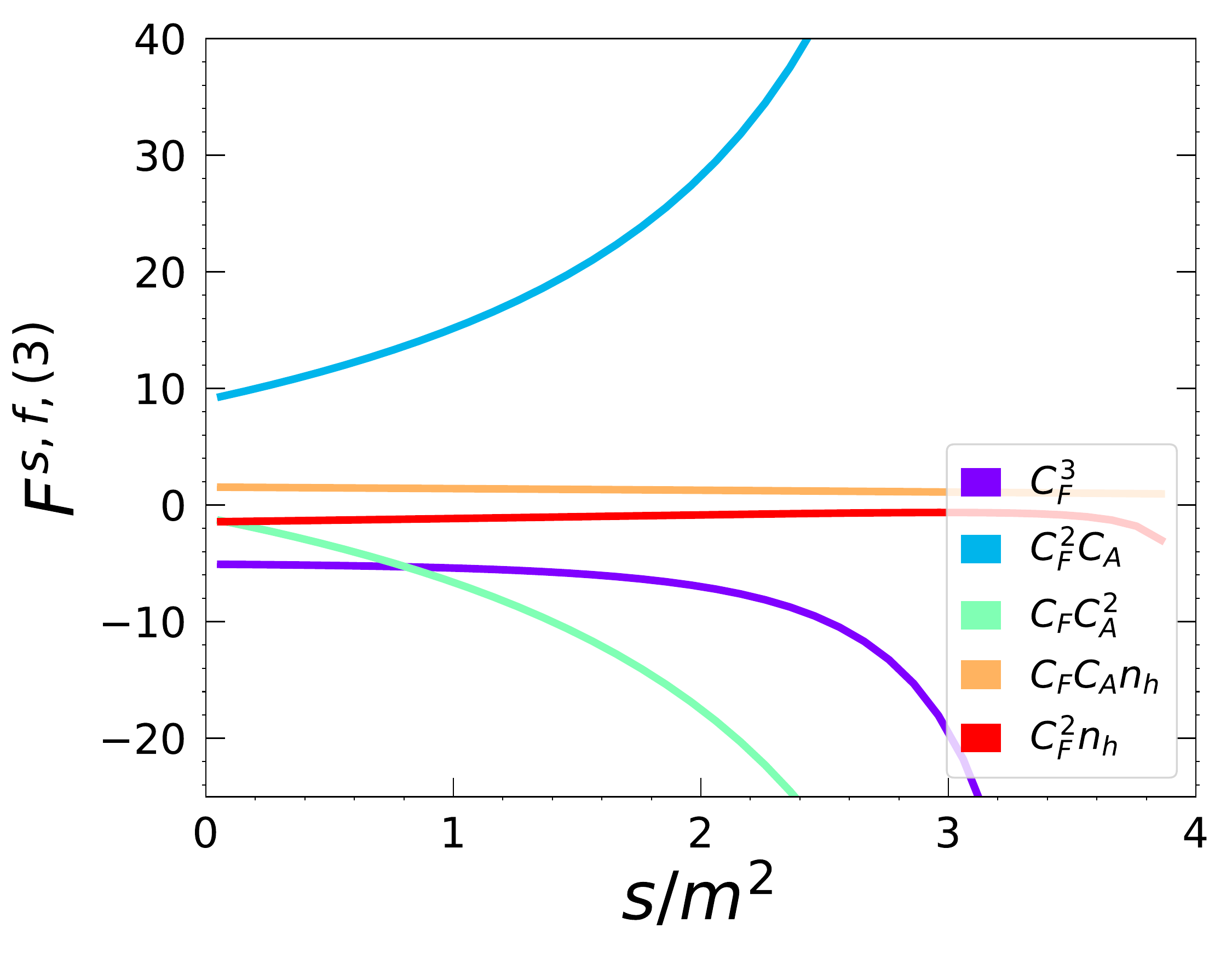}
      &
      \includegraphics[width=0.47\textwidth]{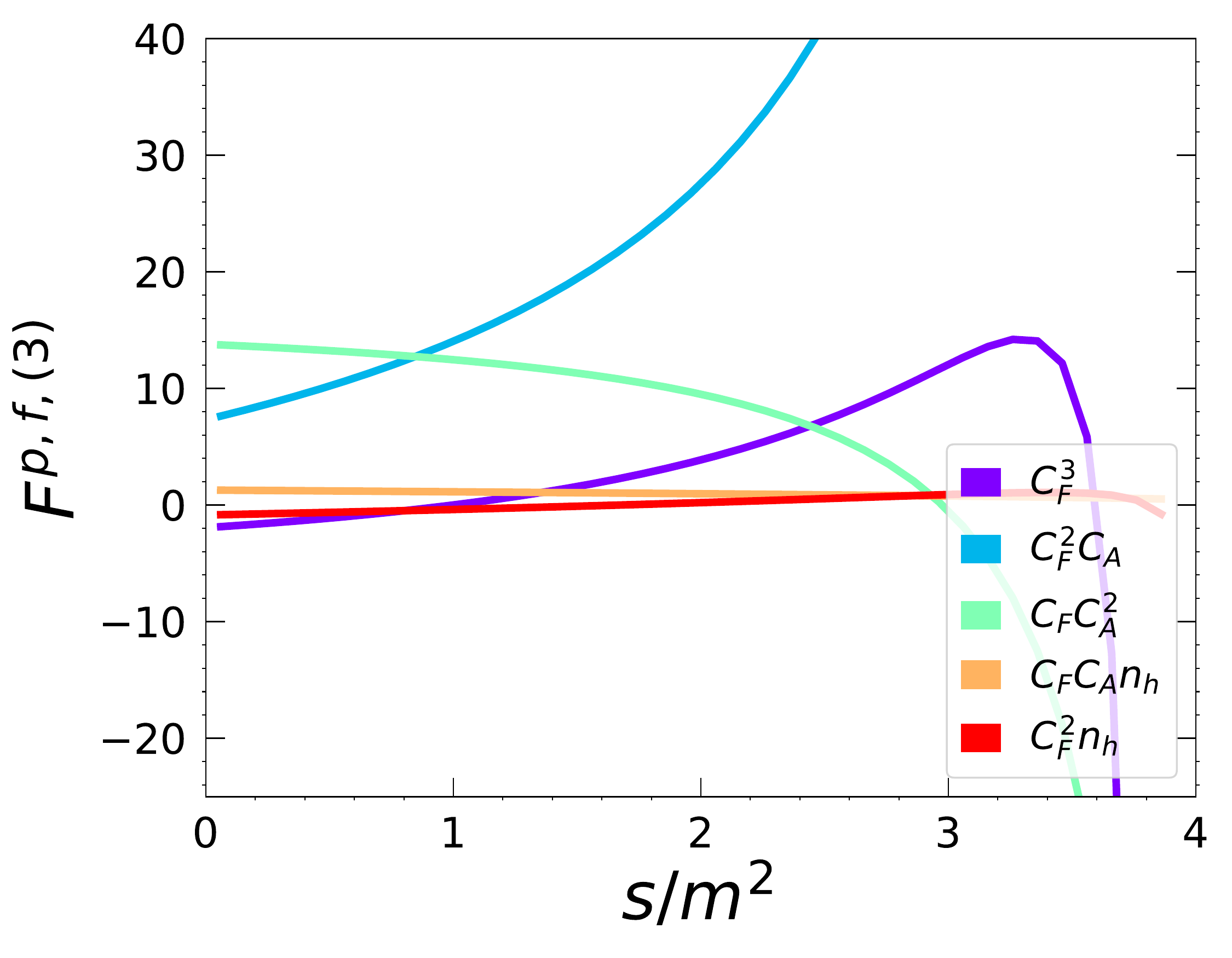}
      \\
      (e) & (f) \\
    \end{tabular}
    \caption{\label{fig::FF2}Non-singlet form factors as a function of $s$ for $0<s<4m^2$.}
  \end{center}
\end{figure}

\begin{figure}[h]
  \begin{center}
    \begin{tabular}{cc}
      \includegraphics[width=0.46\textwidth]{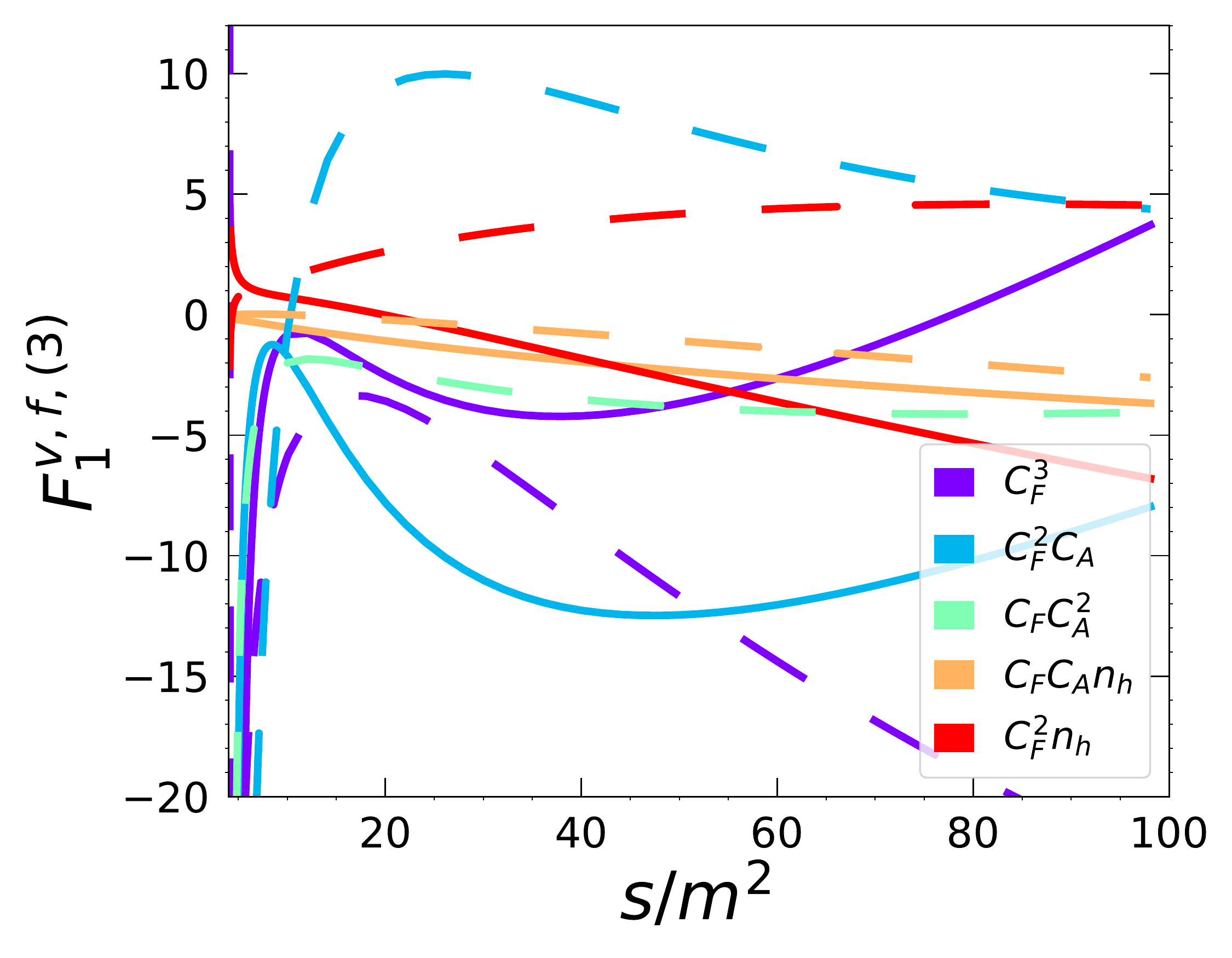}
      &
      \includegraphics[width=0.46\textwidth]{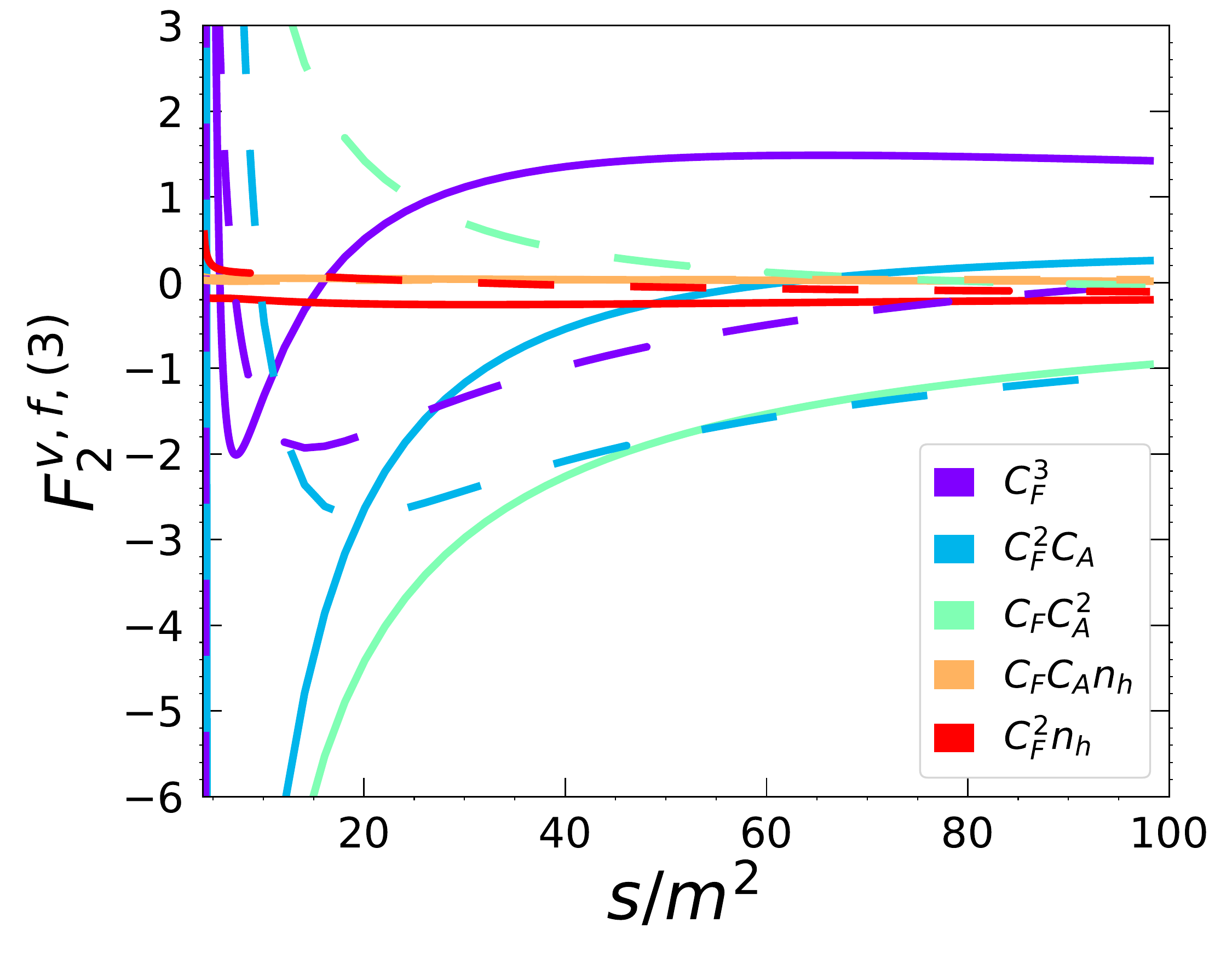}
      \\
      (a) & (b) \\
      \includegraphics[width=0.46\textwidth]{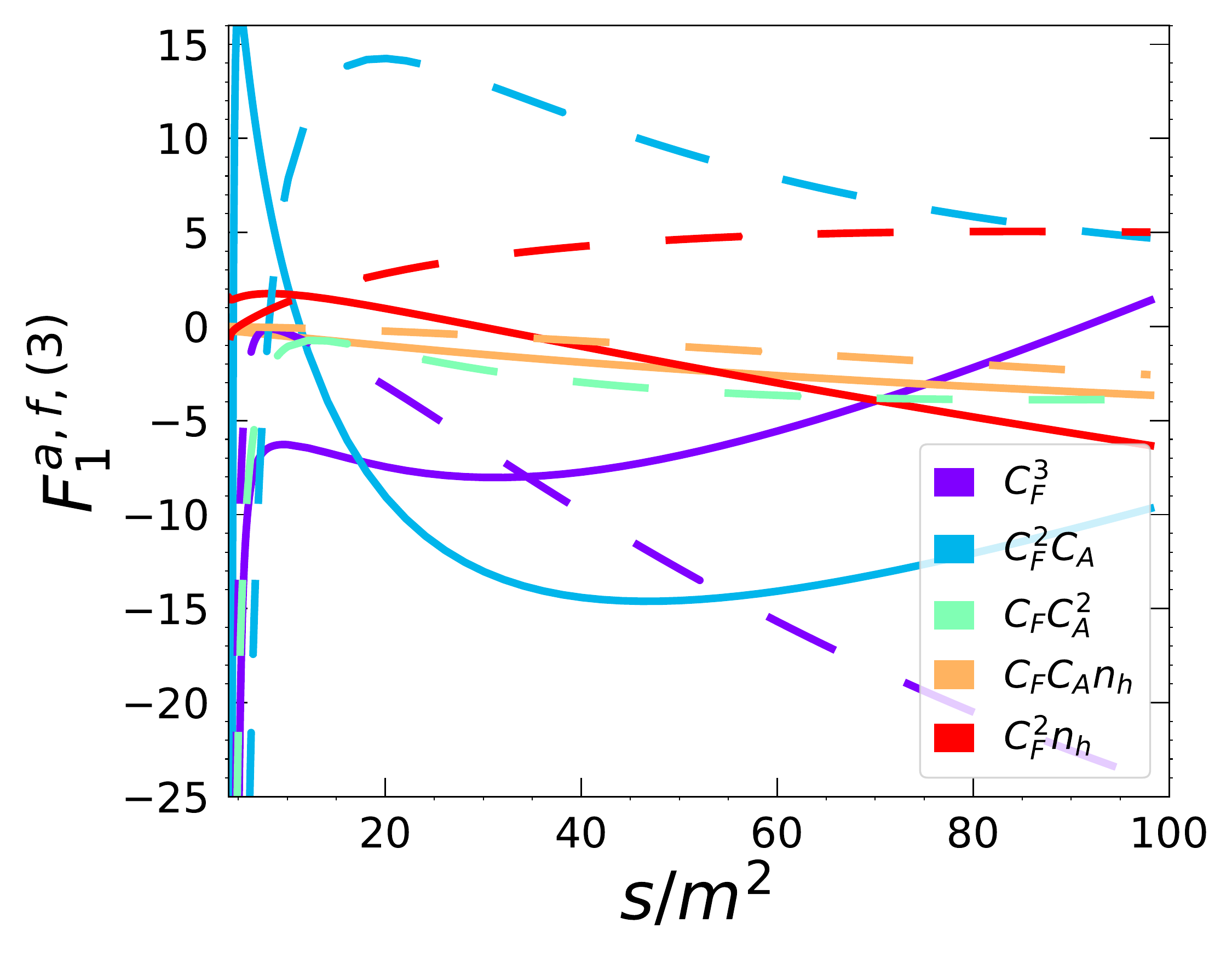}
      &
      \includegraphics[width=0.46\textwidth]{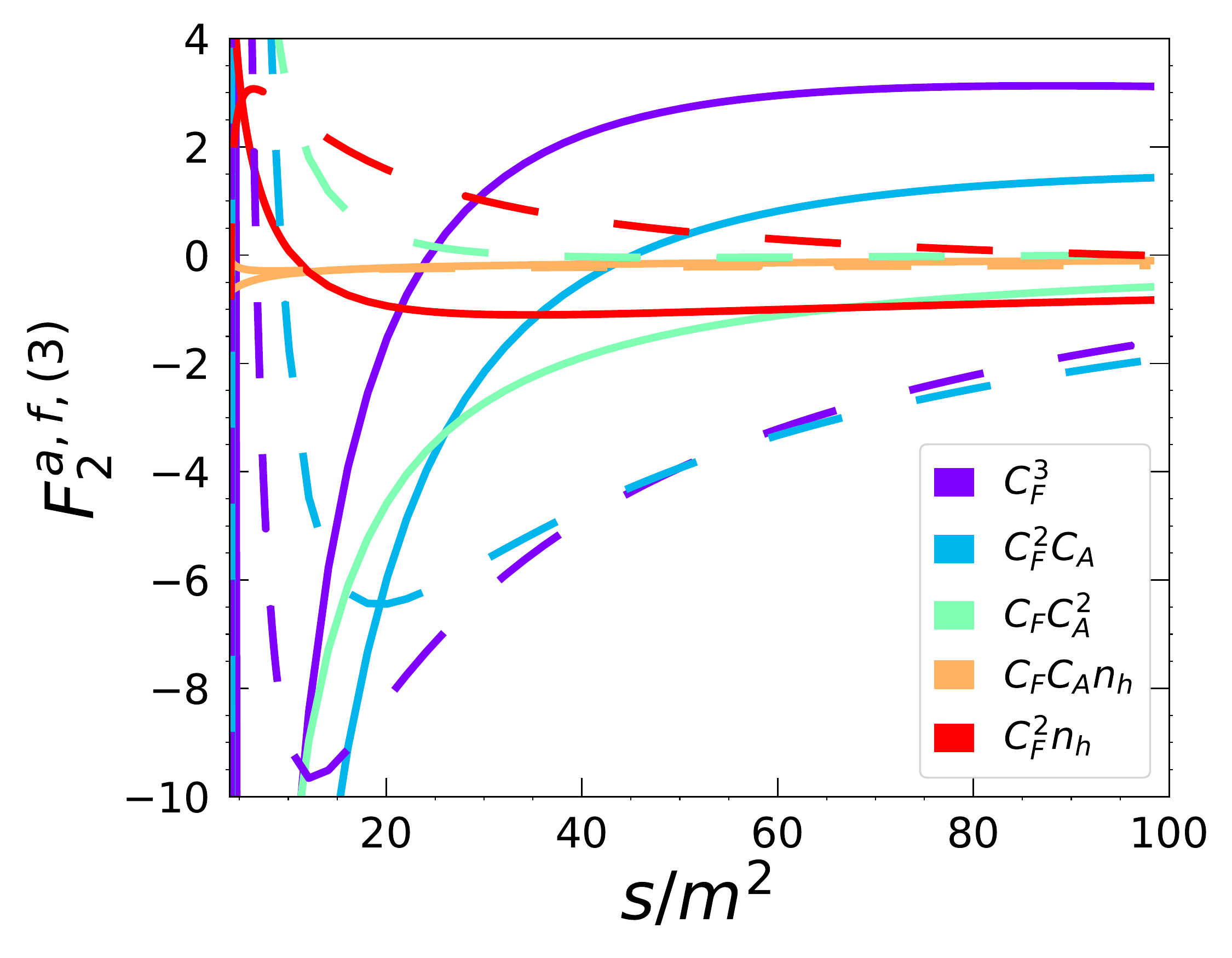}
      \\
      (c) & (d) \\
      \includegraphics[width=0.46\textwidth]{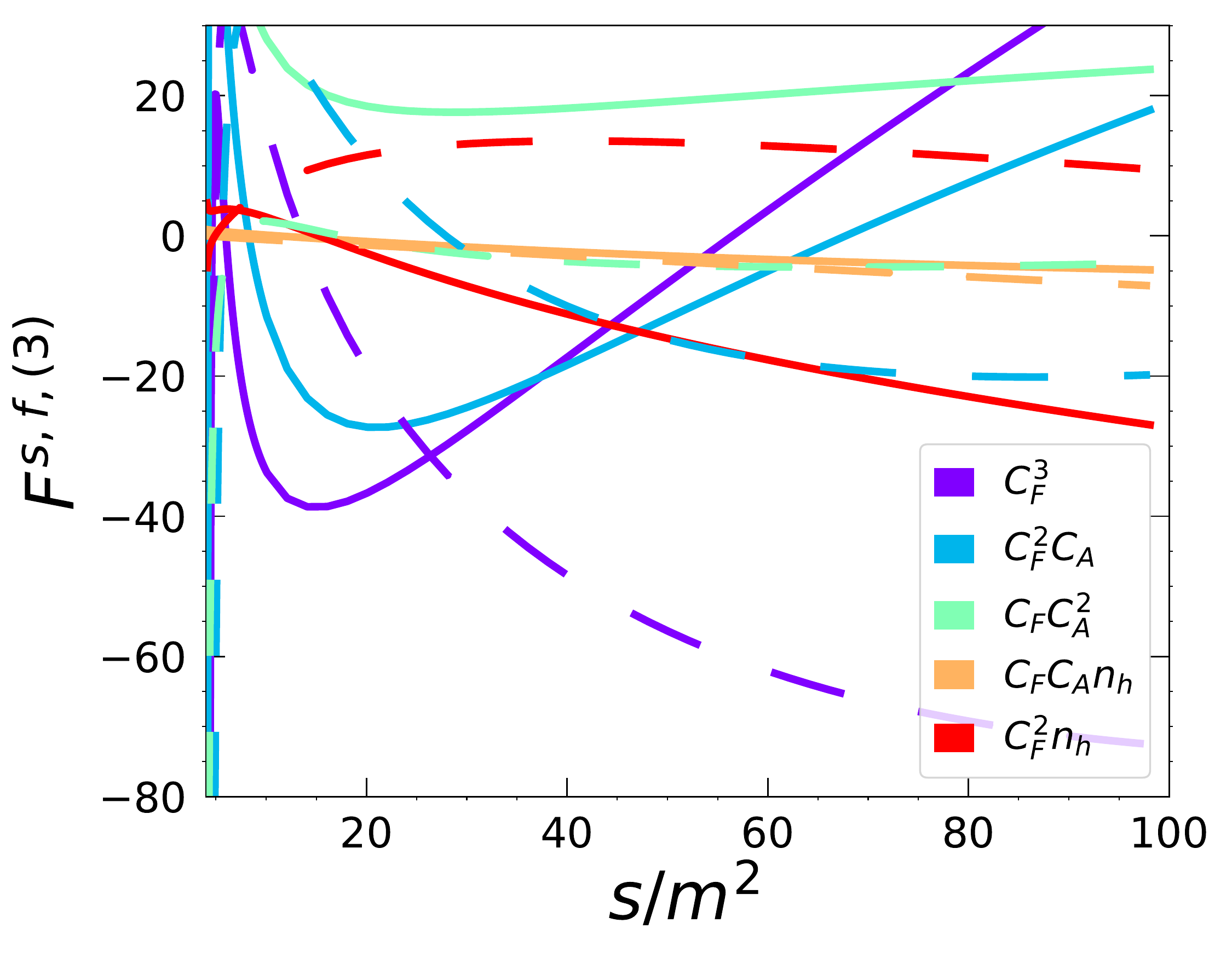}
      &
      \includegraphics[width=0.46\textwidth]{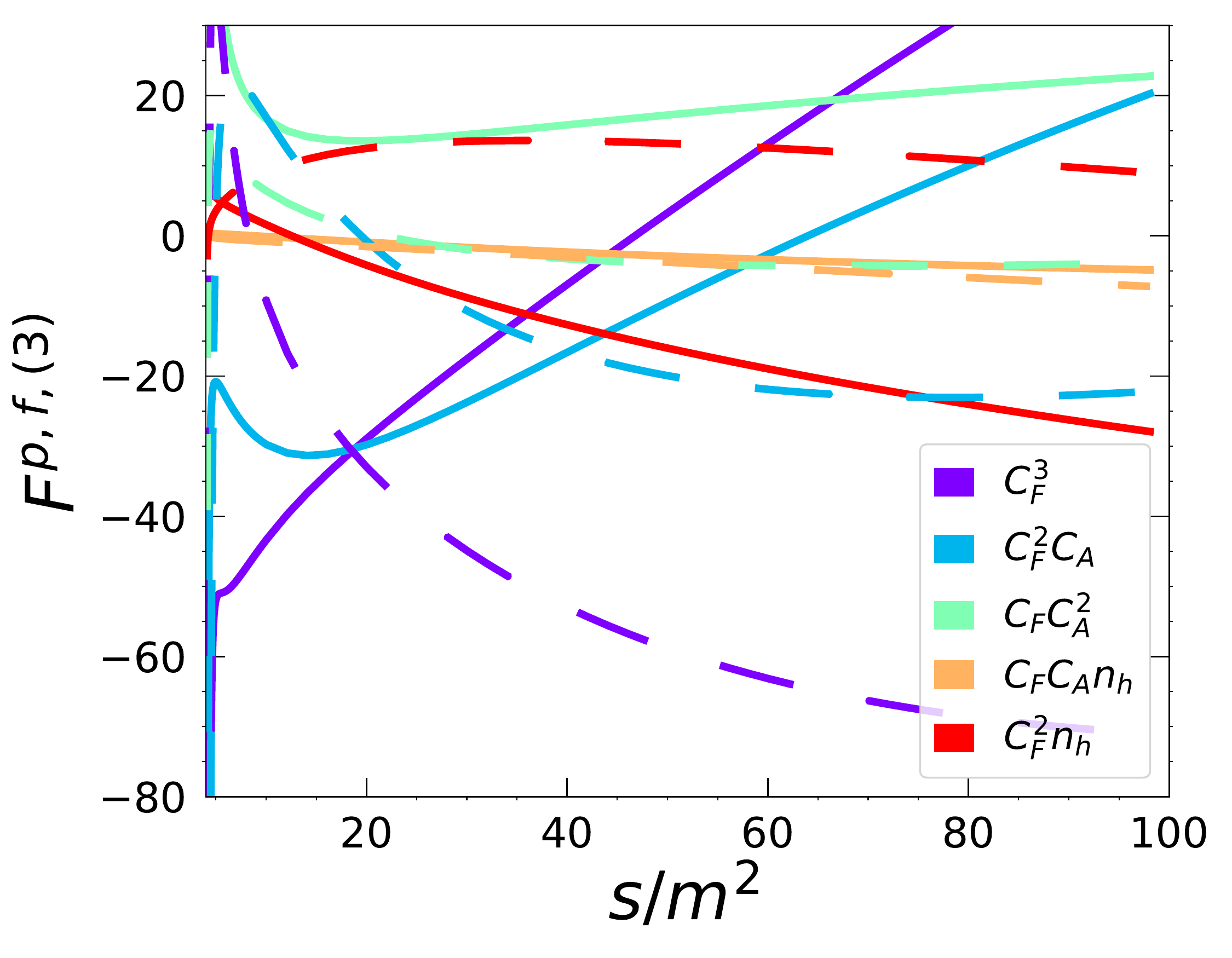}
      \\
      (e) & (f) \\
    \end{tabular}
    \caption{\label{fig::FF3}Non-singlet form factors as a function of $s$ for $s>4m^2$.
      Real and imaginary parts are shown as solid and dashed lines, respectively.}
  \end{center}
\end{figure}

For illustration we show in Figs.~\ref{fig::FF1},~\ref{fig::FF2}
and~\ref{fig::FF3} the finite non-singlet form factors (see
Eq.~(\ref{eq::Ffin})) for negative $s$, for $0<s<4m^2$ and above the
two-particle threshold, respectively. Only in the latter case the imaginary
parts are different from zero.  We restrict ourselves to the non-fermionic
colour factors and the contributions containing a closed heavy quark loop.  In
total we present results for the colour factors $C_F^3$, $C_F^2 C_A$,
$C_FC_A^2$, $C_F^2 T_F n_h$, and $C_F C_A T_F n_h$.  The remaining fermionic
contributions are available in the
literature~\cite{Lee:2018nxa,Lee:2018rgs,Blumlein:2019oas}.  Our 
calculations have been performed for general renormalization scale $\mu$;
in the plots we choose $\mu^2=m^2$.

%- {{{ FF plots, non-singlet:

%- }}}

For $s=0$ we have $F_1^{v,f,(3)}=0$ as can be seen in Figs.~\ref{fig::FF1}(a)
and~\ref{fig::FF2}(a), however, the other form factors have in general a
finite non-zero value in this limit.  For negative $s$ one observes that in
general the non-abelian colour structures $C_F^2C_A$ and $C_FC_A^2$ have large
coefficients. For the vector and axial-vector contribution the $n_h$ terms are
numerically smaller whereas for the scalar and pseudo-scalar case they have a
similar order of magnitude as the other colour structures.

In Fig.~\ref{fig::FF2} one can clearly see the Coulomb singularities for the
non-fermionic contributions close to $s=4m^2$. 
In the $n_h$ contributions
the closed heavy-quark loop  regularizes the $1/\beta$ behaviour and leads to a
finite limit for $s\to 4m^2$, see also Section~\ref{sub::thr}.

Fig.~\ref{fig::FF3} shows the results for $s>4m^2$
where the form factors develop imaginary parts, see the dashed curves.
One again notices the Coulomb singularity on the left part
of the plot and the logarithmic behaviour for large values of $s$.

\begin{figure}[h]
  \begin{center}
    \begin{tabular}{cc}
      \includegraphics[width=0.47\textwidth]{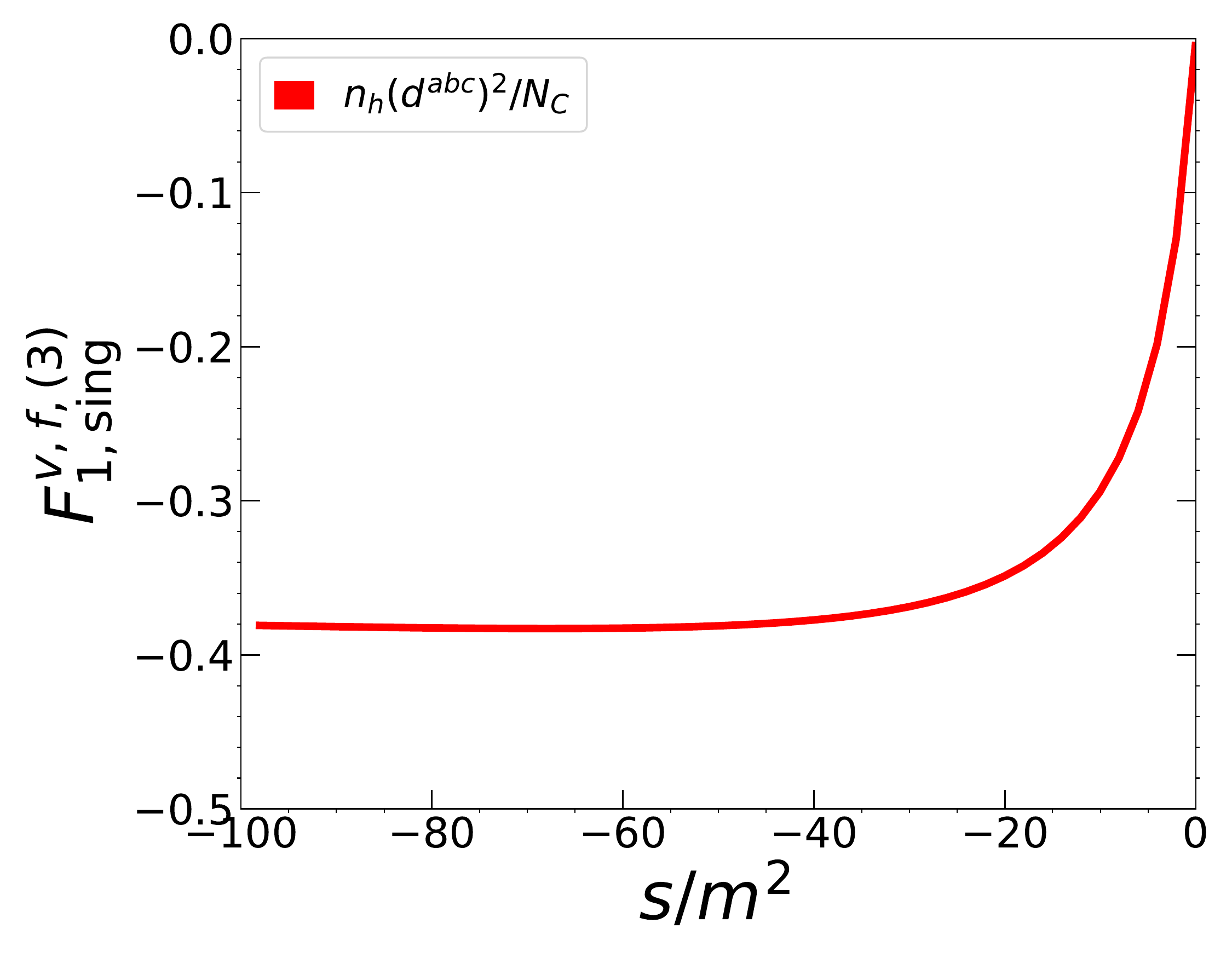}
      &
      \includegraphics[width=0.47\textwidth]{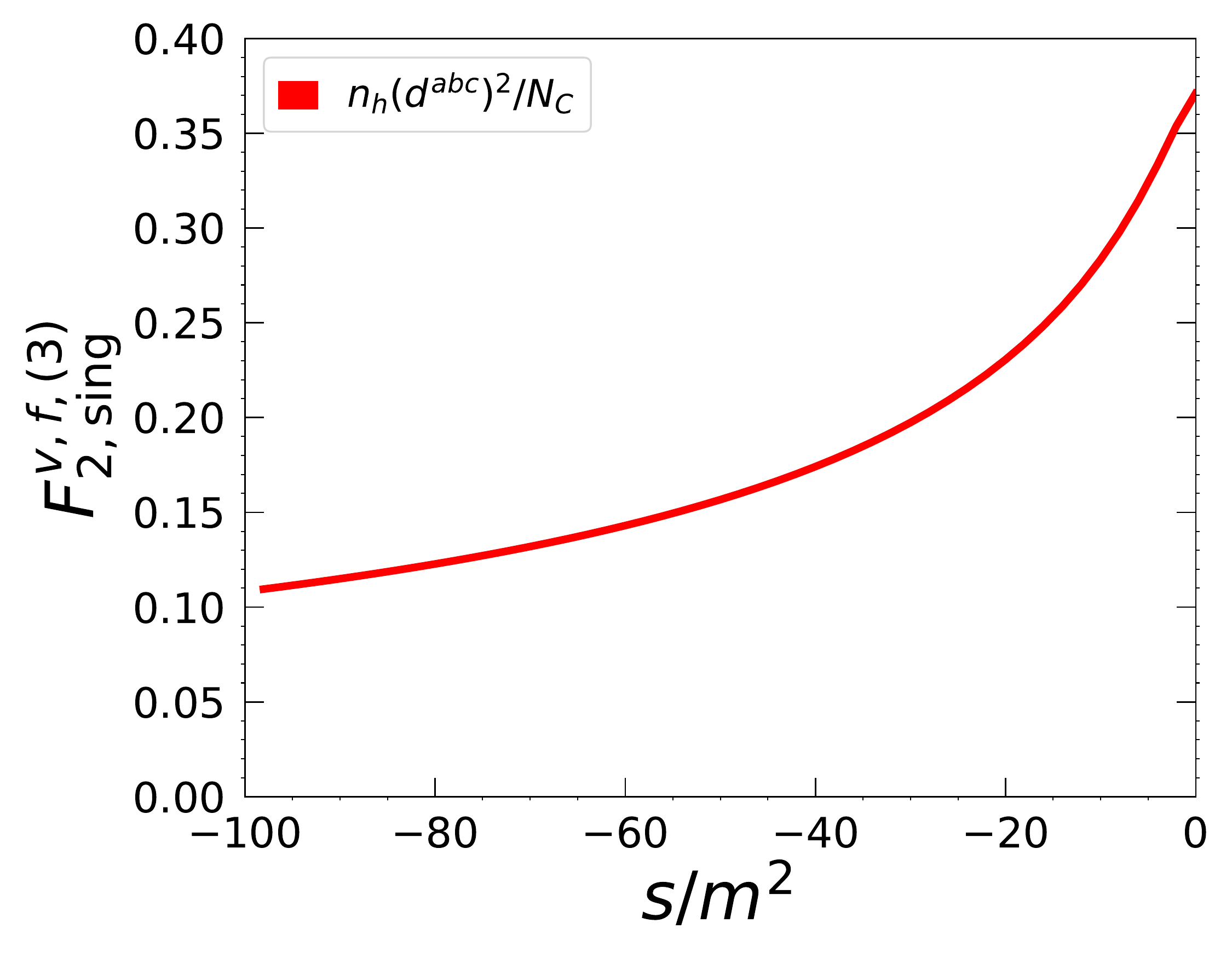}
      \\
      (a) & (b) \\
      \includegraphics[width=0.47\textwidth]{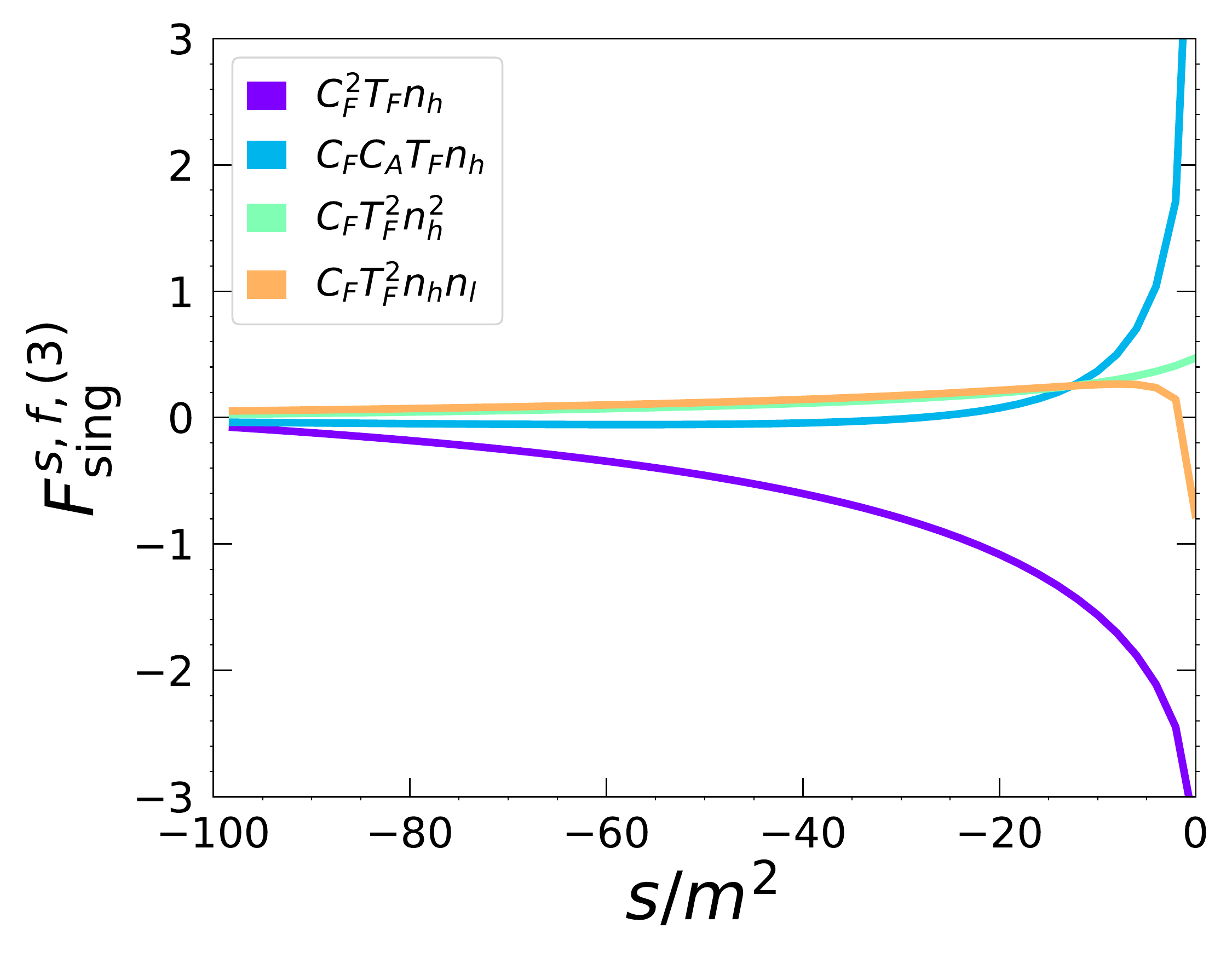}
      &
      \includegraphics[width=0.47\textwidth]{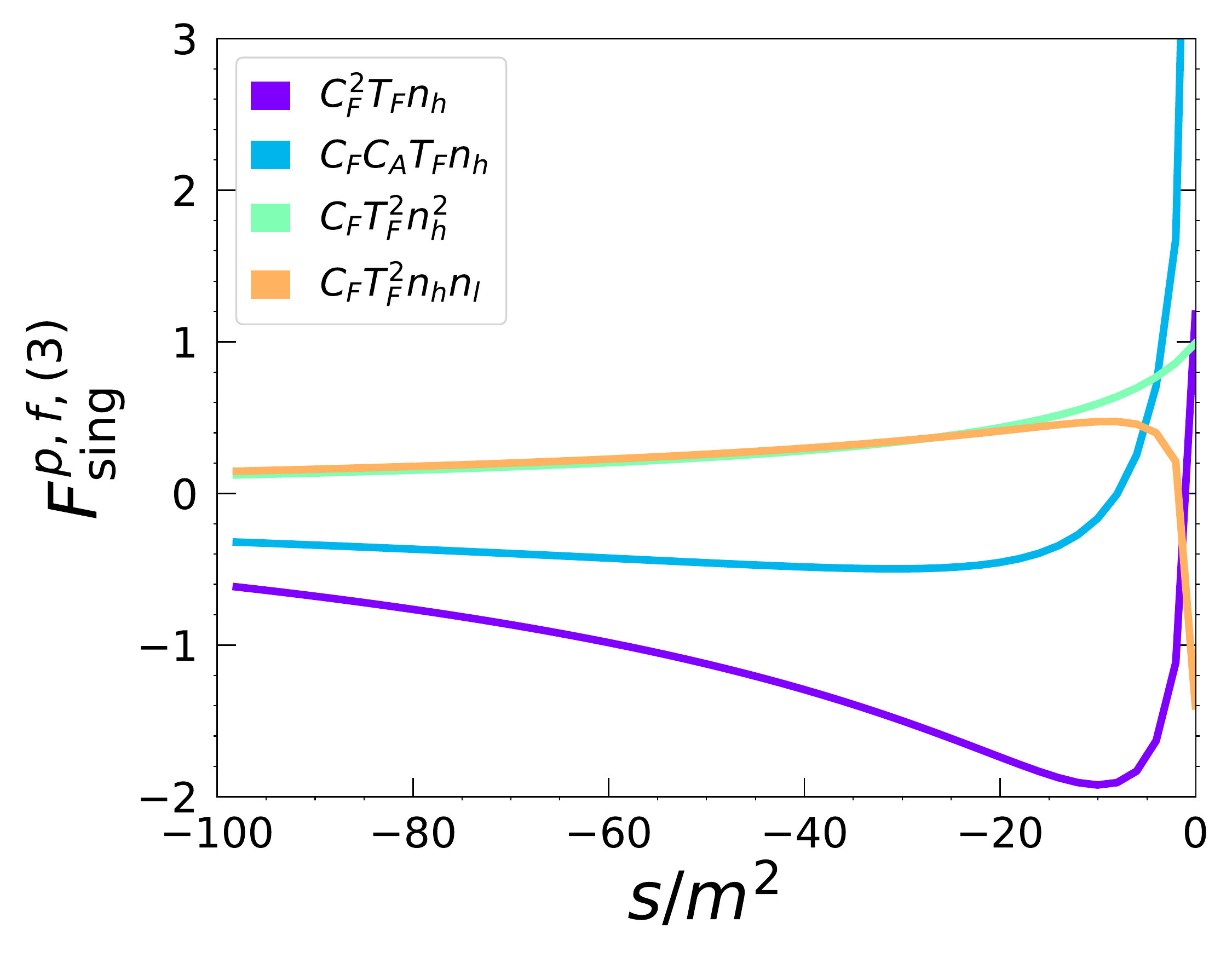}
      \\
      (c) & (d) \\
    \end{tabular}
    \caption{\label{fig::FF1_nhsing}Singlet form factors as a function of $s$ for $s<0$.}
  \end{center}
\end{figure}

\begin{figure}[h]
  \begin{center}
    \begin{tabular}{cc}
      \includegraphics[width=0.47\textwidth]{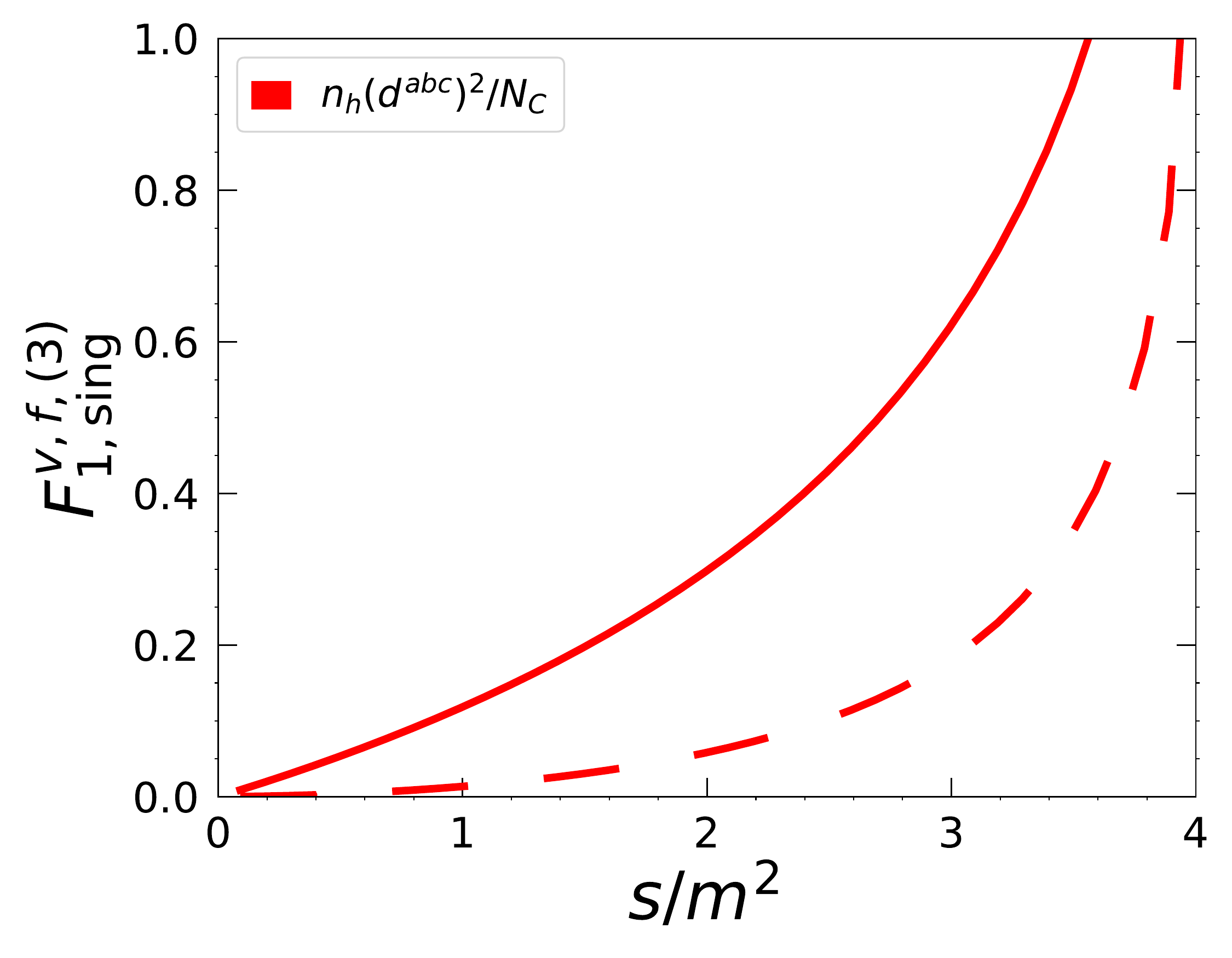}
      &
      \includegraphics[width=0.47\textwidth]{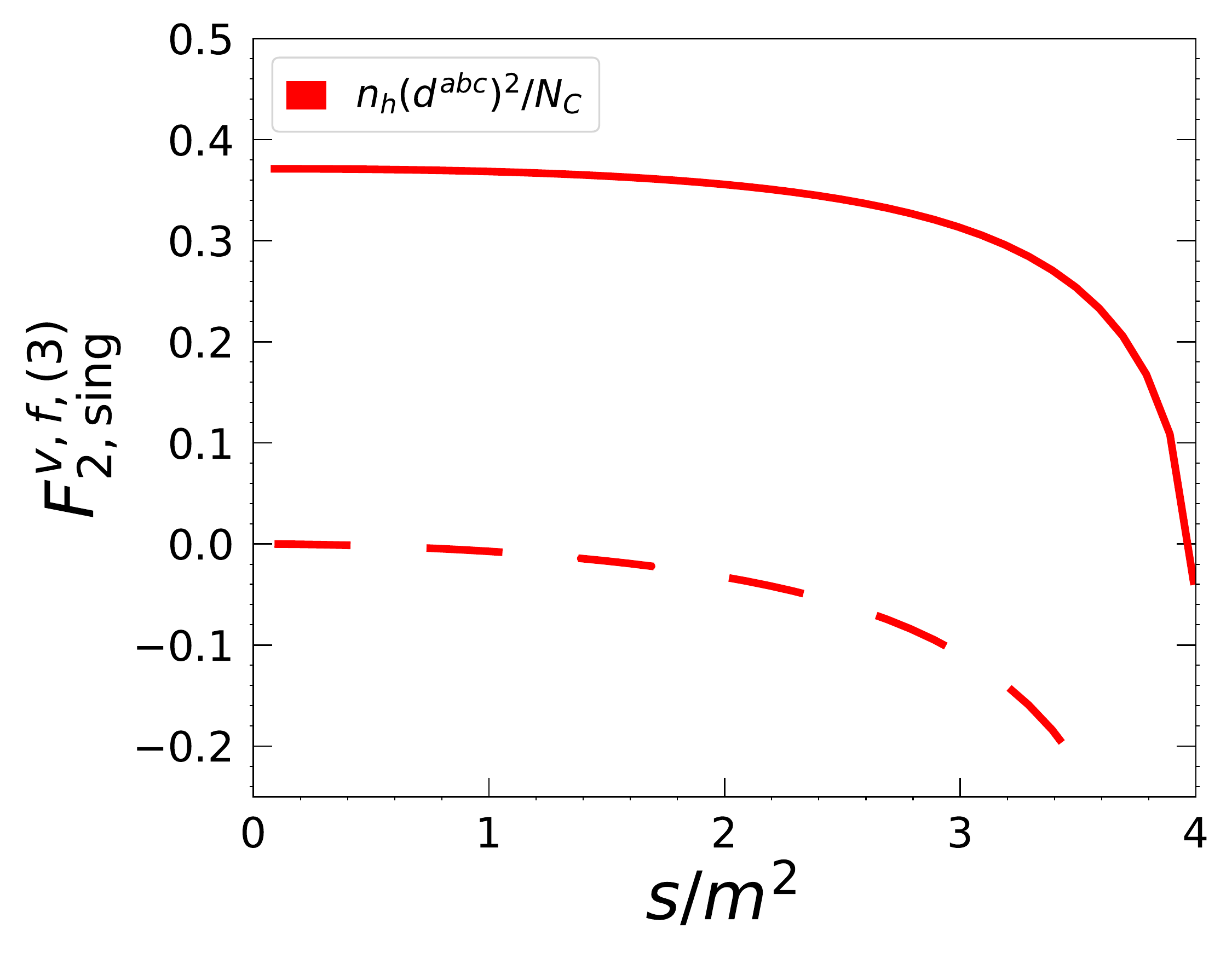}
      \\
      (a) & (b) \\
      \includegraphics[width=0.47\textwidth]{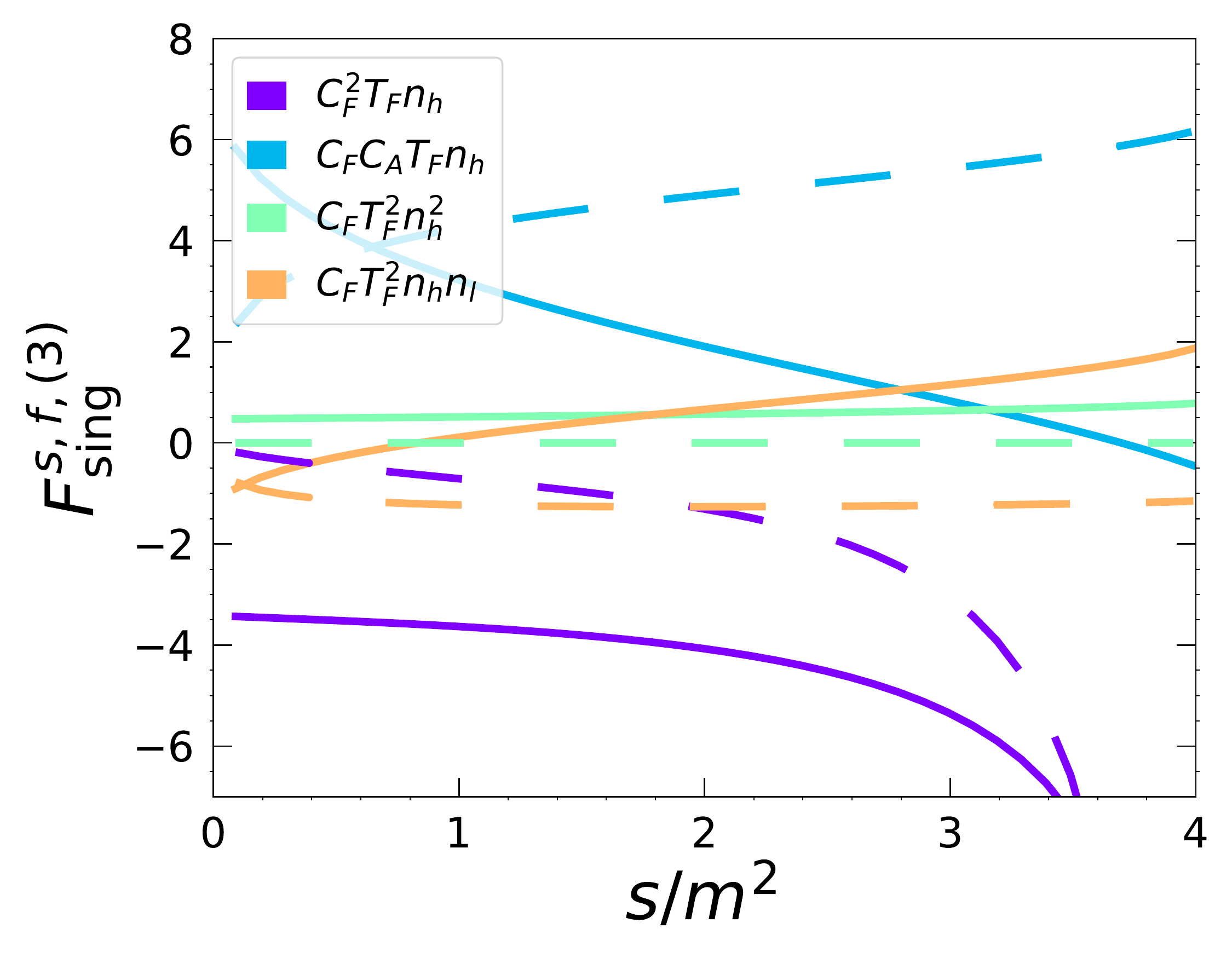}
      &
      \includegraphics[width=0.47\textwidth]{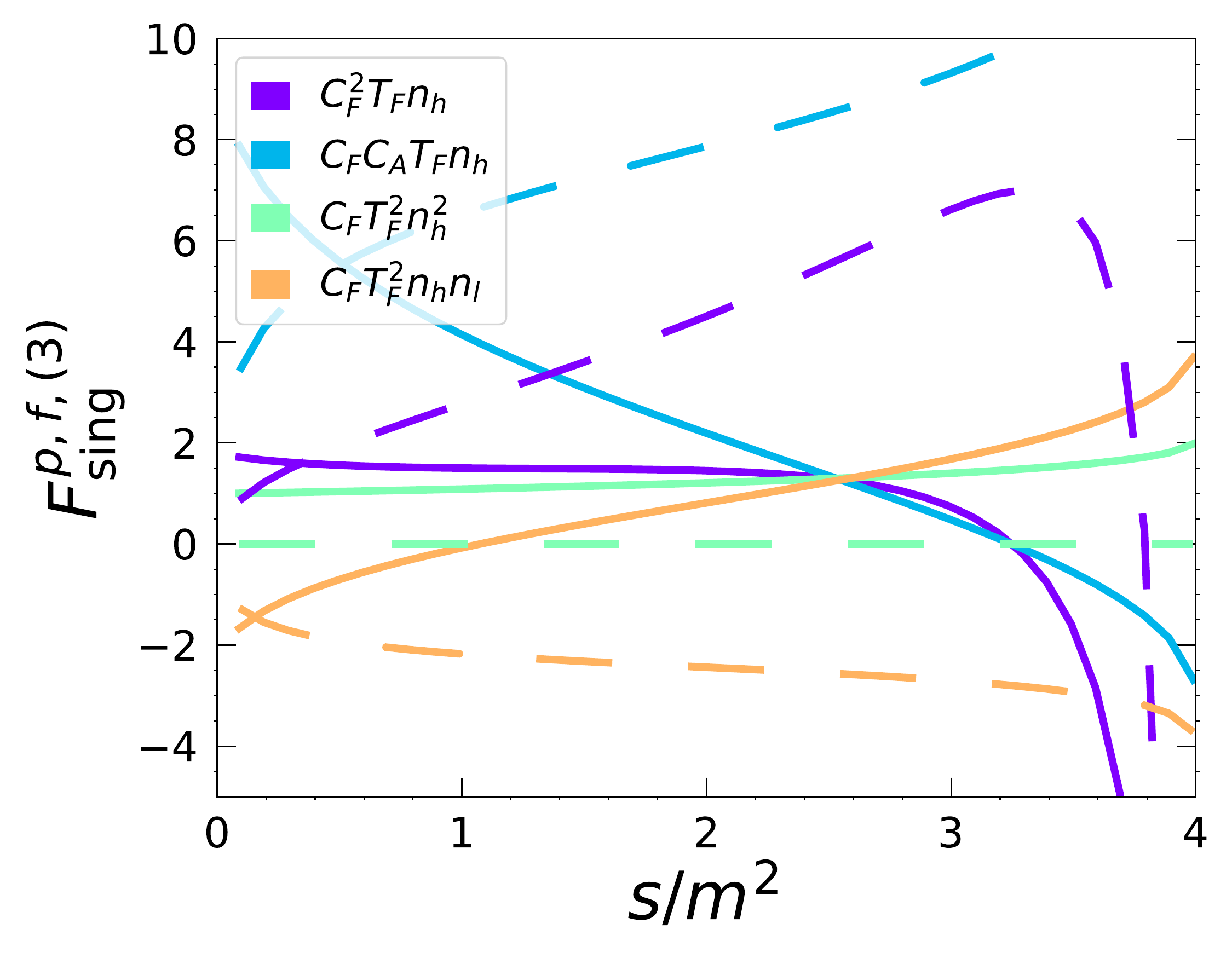}
      \\
      (c) & (d) \\
    \end{tabular}
    \caption{\label{fig::FF2_nhsing}Singlet form factors as a function of $s$
      for $0<s<4m^2$.
      Real and imaginary parts are shown as solid and dashed lines, respectively.}
  \end{center}
\end{figure}

\begin{figure}[h]
  \begin{center}
    \begin{tabular}{cc}
      \includegraphics[width=0.47\textwidth]{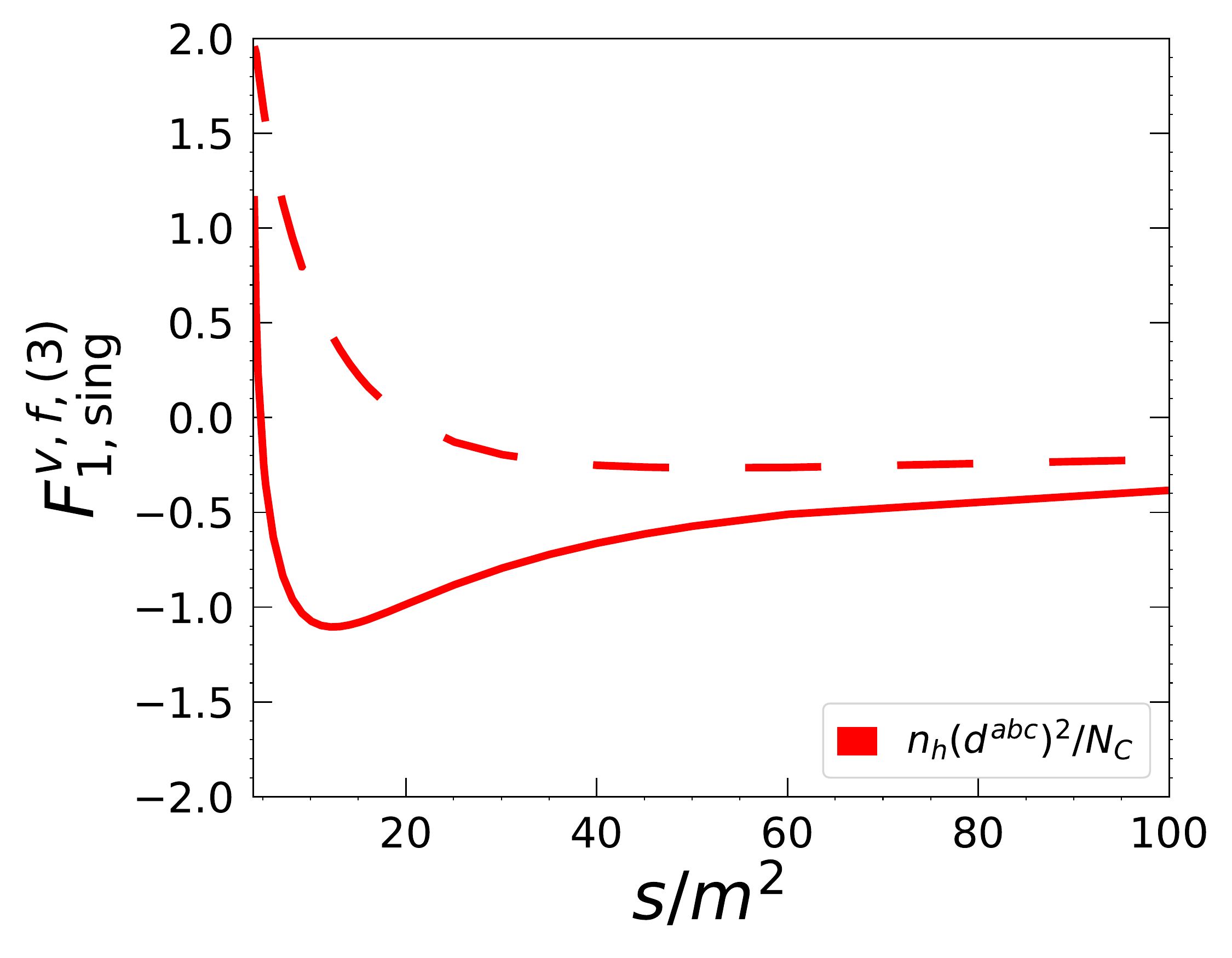}
      &
      \includegraphics[width=0.47\textwidth]{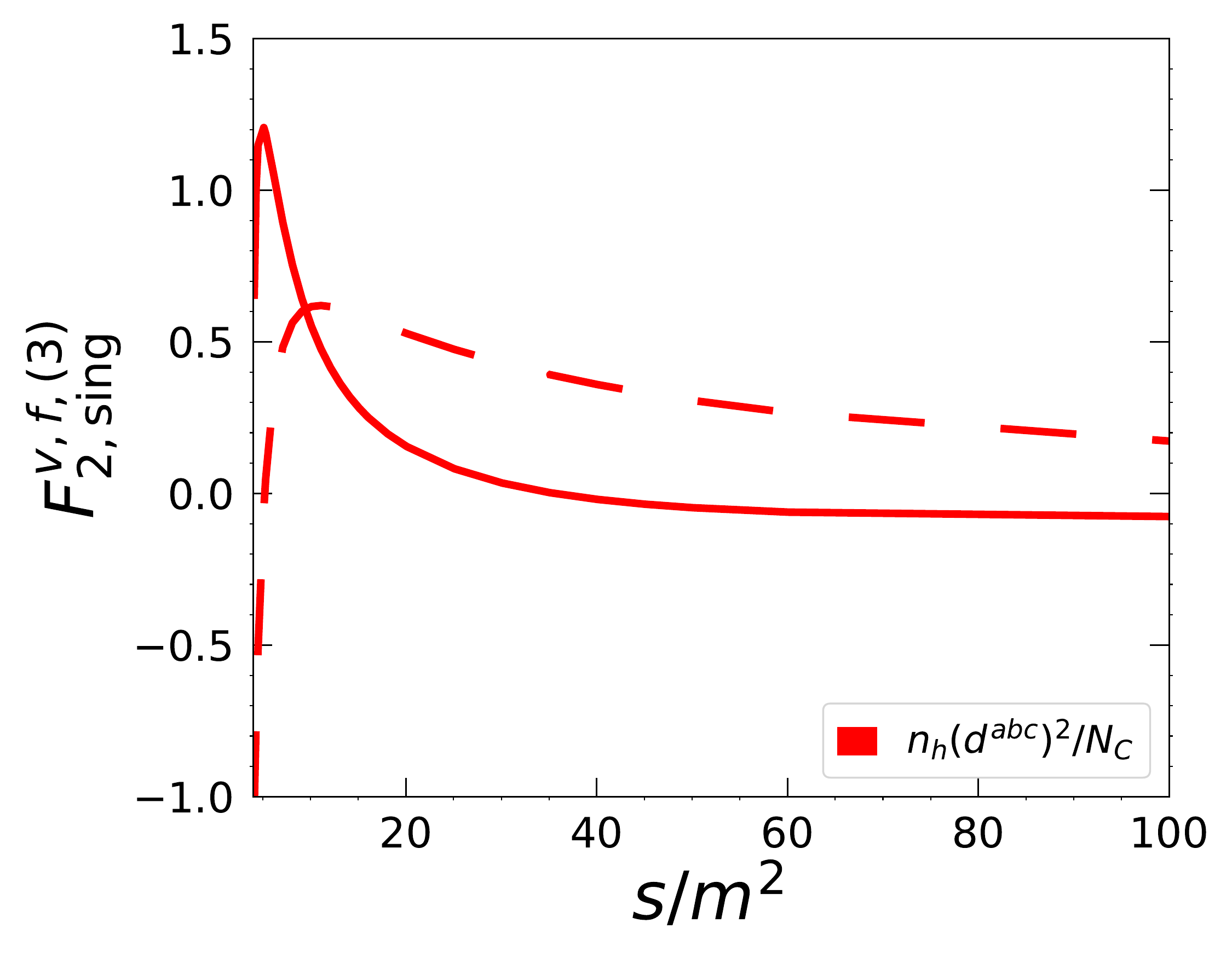}
      \\
      (a) & (b) \\
      \includegraphics[width=0.47\textwidth]{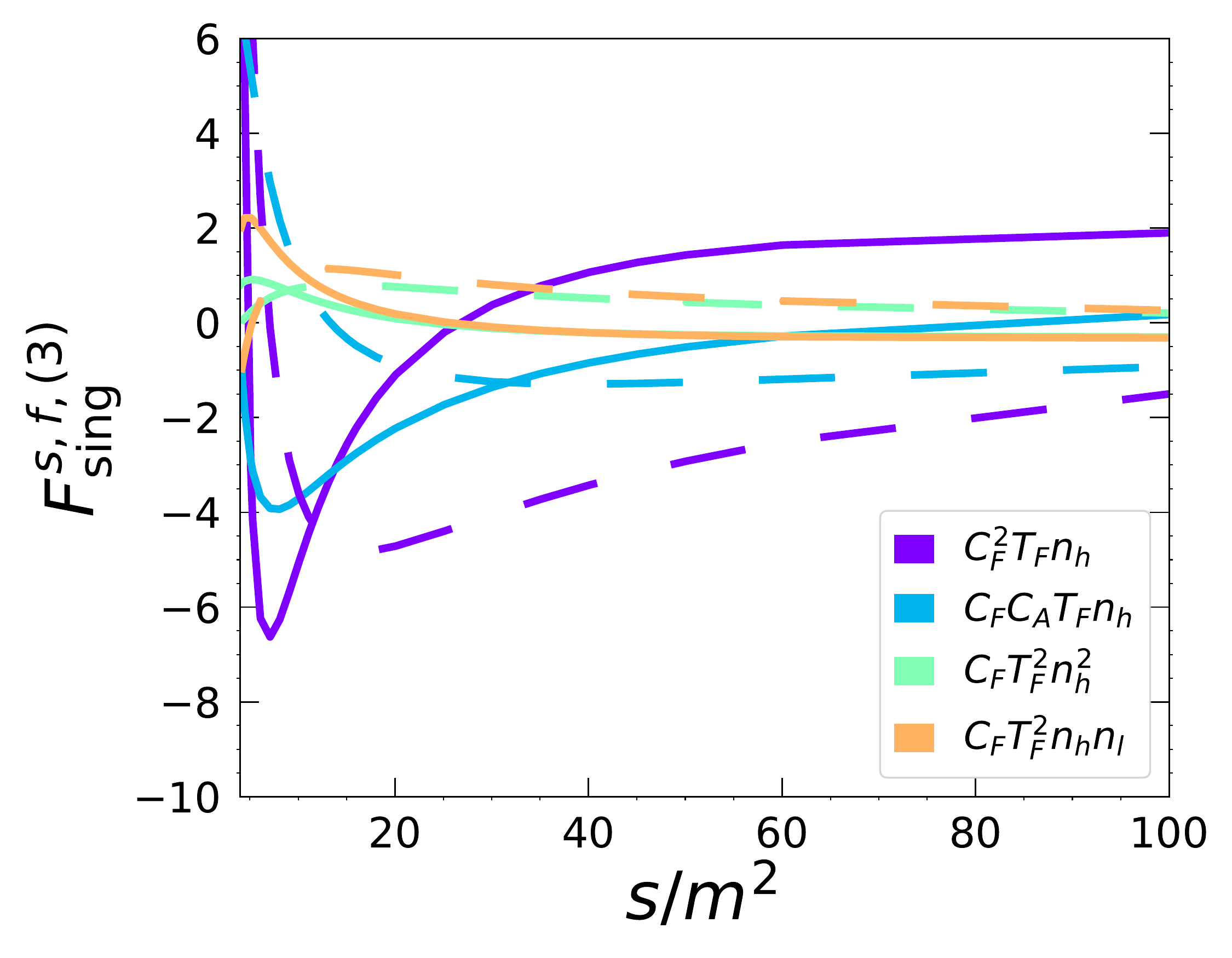}
      &
      \includegraphics[width=0.47\textwidth]{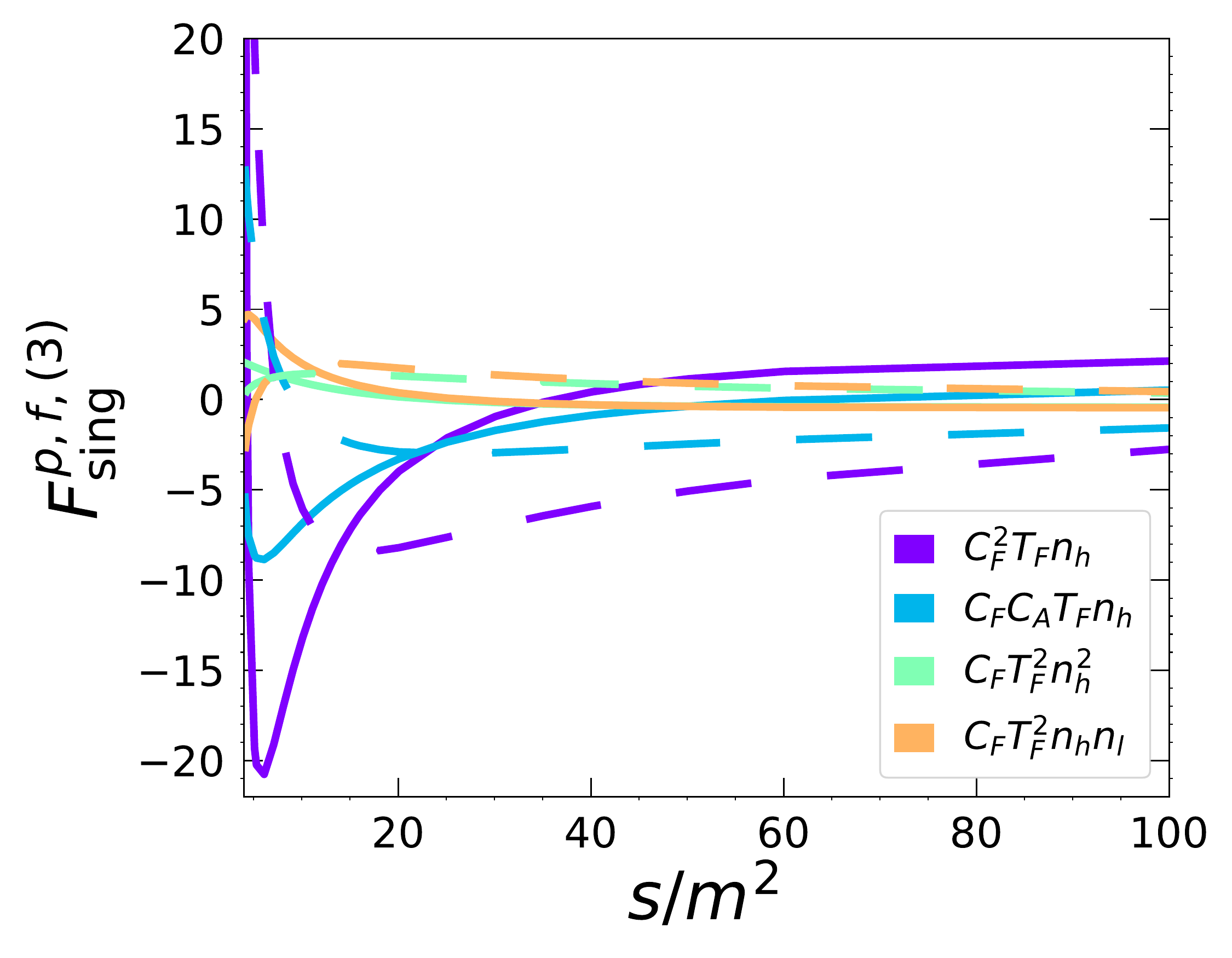}
      \\
      (c) & (d) \\
    \end{tabular}
    \caption{\label{fig::FF3_nhsing}Singlet form factors as a function of $s$ for $s>4m^2$.
      Real and imaginary parts are shown as solid and dashed lines, respectively.}
  \end{center}
\end{figure}

Results for the singlet form factors are shown in
Figs.~\ref{fig::FF1_nhsing},~\ref{fig::FF2_nhsing}
and~\ref{fig::FF3_nhsing} for the three regions $s<0$, $0<s<4m^2$ and
$s>4m^2$, respectively.  In each figure we show plots for the two
vector-current form
factors and for the scalar and pseudo-scalar
currents. Note that non-zero results for the vector form factor are
only obtained from the colour factor proportional to $(d^{abc})^2$
whereas in the scalar and pseudo-scalar case the other four colour
factors have non-zero coefficients.

%- {{{ FF plots, singlet:

%- }}}

It is interesting to note that for $s\to0$ we have $F_{1,\rm
sing}^{v,f,(3)}(0)=0$ and $F_{1,\rm sing}^{v,f,(3)}(0)=\mbox{const}$
whereas the scalar and pseudo-scalar currents behave as
$\log^2(-s/m^2)$. The logarithms, which
are present in higher expansion terms for all form factors and colour factors, are the reason
for the imaginary parts for $s>0$.

The behaviour around $s=4$ is smoother than for the non-singlet form
factors. In particular, there are no $1/\beta^3$ or $1/\beta^2$
singularities.  The two vector form factors have a smooth limit and
only the colour factor $C_F^2 T_Fn_h$ of the scalar and pseudo-scalar
form factors develop $1/\beta$ and $\log(\beta)/\beta$ singularities,
both for the real and imaginary parts. The other three colour factors
have a finite limit for $s\to4$.

In the high-energy region all form factors vanish except $F_{1,\rm
sing}^{v,f,(3)}$ which approaches a constant. At subleading order there
are logarithmic contributions, see also the discussion in
Section~\ref{subsub:high-energy}.

As the central result of this paper we provide a package which allows for the
evaluation of both the bare and finite ($F^{f}$) form factors on the publicly
accessible webpage~\cite{git}.  The package is based on expansions combined
with interpolations. For the latter we evaluate all six form factors at about
4500 $s/m^2$ values.

%- }}}

%- }}}
%- {{{ Conclusions:

\section{\label{sec::concl}Conclusions}

In this paper we compute three-loop corrections to massive quark form
factors with external vector, axial-vector, scalar and pseudo-scalar
currents and provide precise numerical results in the whole $s/m^2$
range.  We apply the method developed in Ref.~\cite{Fael:2021kyg} to
obtain expansions around regular and singular points. It is based on
differential equations which are used to construct the expansions.
Two neighboring expansions are numerically matched at an intermediate
value of $s$.

We consider both the non-singlet and the singlet contributions, with the restriction that
the external currents couple to a closed massive quark
loop in the latter case .  The main difference between the two contributions is the
computation of the boundary conditions at $s=0$. While we have simple Taylor expansions in the
non-singlet case, it is necessary to
perform an asymptotic expansion for the singlet contributions.  Our
expansions around $s=0$ are analytic; the expansions around the other
$s$ values have precise numeric coefficients.  In some cases the
precision is sufficient to reconstruct coefficients
analytically as, e.g., for leading and sub-leading logarithmic
contributions in the high-energy limit.  We also provide expansions
close to threshold and make the Coulomb singularity explicit.

Our results can be downloaded from the website~\cite{git} where an
easy-to-use routine is provided which provides numerical results for
all six colour factors, both for the singlet and non-singlet
contributions. We provide results for each individual colour factor
which makes it straightforward to specify to QED.

%- }}}

%- {{{ Acknowledgments:

\section*{\label{sec::ack}Acknowledgments}

We thank Roman Lee for discussions about the M\"obius transformations and
Alexander Smirnov and Vladimir Smirnov for discussions about the basis change
for the master integrals and providing an improved version of the {\tt
  Mathematica} code from Ref.~\cite{Smirnov:2020quc}.  We thank Andreas
Maier for communications concerning the threshold behaviour. This research
was supported by the Deutsche Forschungsgemeinschaft (DFG, German Research
Foundation) under grant 396021762 --- TRR 257 ``Particle Physics Phenomenology
after the Higgs Discovery''.  The Feynman diagrams were drawn with the help of
Axodraw~\cite{Vermaseren:1994je} and JaxoDraw~\cite{Binosi:2003yf}.

%- }}}

\begin{appendix}

%- {{{ Three-loop on-shell integrals to higher orders in $\epsilon$:

\section{\label{app::initial} Three-loop on-shell integrals to higher \mbox{orders in \boldmath{$\epsilon$}}}

For the expansion around $s=0$ we can analytically fix the boundary conditions,
since in this case all integrals reduce to massive three-loop on-shell propagators.
The most recent calculation of these integrals is given in Ref.~\cite{Lee:2010ik} and extends
to weight 7, formally enough for four-loop calculations.
However, since we encounter spurious poles in our reductions of the full system,
this is not enough to fix all boundary constants.
Some integrals have to be calculated up to weight 9.

A subset of the integrals can be found in the appendix of
Ref.~\cite{Blumlein:2019oas}.  Translating to the notation of
Ref.~\cite{Lee:2010ik} we can fix $I_{12} = G_{45}$, $I_{11} = G_{46}$, and
$I_{9} = G_{52}$.  But we still need the expansions of the integrals $G_{43}$,
$G_{51}$, $G_{53}^{(\star)}$, $G_{61}$, $G_{63}^{(\star)}$, $G_{64}$, $G_{66}$
up to weight 9 (integrals with a $(\star)$ actually only to weight 8).  In the
following we will describe the steps of the calculation and provide results
for all master integrals which cannot be expressed in terms of
$\Gamma$-functions up to weight 9.

In a first step we use the {\tt Mathematica} package {\tt Summertime} \cite{Lee:2015eva}
to calculate the needed integrals up to the necessary order in $\epsilon$
with $5000$ digits accuracy,
for the integral $G_8$ we only obtain with $3000$ digits accuracy.
This is the input we use to apply the {\tt PSLQ} algorithm~\cite{PSLQ} to
fix the analytic form of the expansions.
In order to apply this algorithm we need a basis of constants.
In all the orders obtained by now, the basis of constants given by harmonic
polylogarithms evaluated at argument $1$ was sufficient.
We use the basis given in Ref.~\cite{Blumlein:2009cf}.
This basis is convenient since the harmonic polylogarithms can be calculated to
(in principle) arbitrary precision with {\tt ginac}.
For the usage in the {\tt PSLQ} algorithm we have calculated them with $7500$ digits
accuracy.
Since we are not dealing with integrals of uniform transcendentality one has
to use all products of constants up to the desired weight in the ansatz.
Therefore, the number of unknown constants from weight 1 to weight 9 is
given by $2$, $4$, $7$, $12$, $20$, $33$, $54$, $88$, $143$.
After the successful reconstruction we change to the notation of {\tt SUMMER} \cite{Vermaseren:1998uu}.
This has the advantage that most of the transcendental constants are
known to {\tt Mathematica}.
The additional constants are given by:
\begin{eqnarray}
  s_6 &=& S_{0,0,0,0,-1,-1}(\infty) = 0.9874414264...
  ~, \nonumber\\
  s_{7a} &=& S_{0,0,0,0,-1,1,1}(\infty) = -0.9529600758...
  ~, \nonumber\\
  s_{7b} &=& S_{0,0,0,0,1,-1,-1}(\infty) = 1.029121263...
  ~, \nonumber\\
  s_{8a} &=& S_{0,0,0,0,1,0,0,1}(\infty) = 1.041785029...
  ~, \nonumber\\
  s_{8b} &=& S_{0,0,0,0,0,0,-1,-1}(\infty) = 0.9964477484...
  ~, \nonumber\\
  s_{8c} &=& S_{0,0,0,0,-1,-1,-1,-1}(\infty) = 0.9839666738...
  ~, \nonumber\\
  s_{8d} &=& S_{0,0,0,0,-1,-1,1,1}(\infty) = 0.9999626135...
  ~, \nonumber\\
  s_{9a} &=& S_{0,0,0,0,0,0,1,-1,-1}(\infty) = 1.006401963...
  ~, \nonumber\\
  s_{9b} &=& S_{0,0,0,0,0,0,-1,-1,1}(\infty) = 0.9984295251...
  ~, \nonumber\\
  s_{9c} &=& S_{0,0,0,0,0,-1,0,-1,-1}(\infty) = -0.9874751576...
  ~, \nonumber\\
  s_{9d} &=& S_{0,0,0,0,-1,-1,1,1,1}(\infty) = 1.002198174...
  ~, \nonumber\\
  s_{9e} &=& S_{0,0,0,0,-1,-1,-1,-1,1}(\infty) = 0.9859117196...
  ~, \nonumber\\
  s_{9f} &=& S_{0,0,0,0,-1,-1,-1,1,-1}(\infty) = 0.9784811713...
  ~.
\end{eqnarray}
Finally, the new results read:\footnote{The analytic expressions can be
  obtained from the website~\cite{progdata}.}
\begin{align}
  G_{43} &= \Gamma(1+\epsilon)^3
  \Biggl\{
         \frac{3}{2 \epsilon ^3}
        +\frac{23}{4 \epsilon ^2}
        +\frac{105}{8 \epsilon }
        +\frac{275}{16}
        +\frac{4 \pi ^2}{3}
        +\epsilon
        \biggl(
                -\frac{567}{32}
                +10 \pi ^2
                -8 \pi ^2 l_ 2
                +28 \zeta_3
        \biggr)
        \nonumber\\ &
+\epsilon ^2
\biggl(
        -\frac{14917}{64}
        +\frac{145 \pi ^2}{3}
        -\frac{62 \pi ^4}{45}
        +192 a_ 4
        -60 \pi ^2 l_ 2
        +16 \pi ^2 l_ 2^2
        +8 l_ 2^4
        +210 \zeta_3
\biggr)
\nonumber\\ &
+\epsilon ^3
\biggl(
        -\frac{144015}{128}
        +\frac{385 \pi ^2}{2}
        -\frac{31 \pi ^4}{3}
        +1440 a_ 4
        +1152 a_ 5
        -290 \pi ^2 l_ 2
        +\frac{124 \pi ^4 l_ 2}{15}
        \nonumber\\ &
        +120 \pi ^2 l_ 2^2
        -32 \pi ^2 l_ 2^3
        +60 l_ 2^4
        -\frac{48 l_ 2^5}{5}
        +1015 \zeta_3
        -\frac{40}{3} \pi ^2 \zeta_3
        -930 \zeta_5
\biggr)
+\epsilon ^4
\biggl(
        -\frac{1108525}{256}
        \nonumber\\ &
        +\frac{8281 \pi ^2}{12}
        -\frac{899 \pi ^4}{18}
        -\frac{562 \pi ^6}{135}
        +6960 a_ 4
        +8640 a_ 5
        +6912 a_ 6
        -1155 \pi ^2 l_ 2
        +62 \pi ^4 l_ 2
        \nonumber\\ &
        +580 \pi ^2 l_ 2^2
        -\frac{124}{5} \pi ^4 l_ 2^2
        -240 \pi ^2 l_ 2^3
        +290 l_ 2^4
        +48 \pi ^2 l_ 2^4
        -72 l_ 2^5
        +\frac{48 l_ 2^6}{5}
        +2880 s_ 6
        \nonumber\\ &
        +\frac{8085 \zeta_3}{2}
        -100 \pi ^2 \zeta_3
        +80 \pi ^2 l_ 2 \zeta_3
        -1220 \zeta_3^2
        -6975 \zeta_5
\biggr)
+\epsilon ^5
\biggl(
        -\frac{7710087}{512}
        +\frac{18585 \pi ^2}{8}
        \nonumber\\ &
        -\frac{2387 \pi ^4}{12}
        -\frac{281 \pi ^6}{9}
        +27720 a_ 4
        +41760 a_ 5
        +51840 a_ 6
        +41472 a_ 7
        -\frac{8281}{2} \pi ^2 l_ 2
        \nonumber\\ &
        +\frac{899 \pi ^4 l_ 2}{3}
        +\frac{784 \pi ^6 l_ 2}{45}
        +2310 \pi ^2 l_ 2^2
        -186 \pi ^4 l_ 2^2
        -1160 \pi ^2 l_ 2^3
        +\frac{248}{5} \pi ^4 l_ 2^3
        +1155 l_ 2^4
        \nonumber\\ &
        +360 \pi ^2 l_ 2^4
        -348 l_ 2^5
        -\frac{288}{5} \pi ^2 l_ 2^5
        +72 l_ 2^6
        -\frac{288 l_ 2^7}{35}
        +21600 s_ 6
        -\frac{55680}{7} l_ 2 s_ 6
        +\frac{55680 s_{7 a}}{7}
        \nonumber\\ &
        -\frac{65280 s_{7 b}}{7}
        +\frac{57967 \zeta_3}{4}
        -\frac{1450}{3} \pi ^2 \zeta_3
        +\frac{4300 \pi ^4 \zeta_3}{63}
        +600 \pi ^2 l_ 2 \zeta_3
        -240 \pi ^2 l_ 2^2 \zeta_3
        -9150 \zeta_3^2
        \nonumber\\ &
        +\frac{69600}{7} l_ 2 \zeta_3^2
        -\frac{67425 \zeta_5}{2}
        +\frac{32086 \pi ^2 \zeta_5}{7}
        +16740 l_ 2^2 \zeta_5
        -\frac{579651 \zeta_7}{7}
\biggr)
+\epsilon ^6
\biggl(
        -\frac{50743957}{1024}
        \nonumber\\ &
        +\frac{360865 \pi ^2}{48}
        -\frac{256711 \pi ^4}{360}
        -\frac{8149 \pi ^6}{54}
        -\frac{429467 \pi ^8}{33075}
        +99372 a_ 4
        +\frac{1856 \pi ^4 a_ 4}{7}
        \nonumber\\ &
        +166320 a_ 5
        +250560 a_ 6
        +311040 a_ 7
        +248832 a_ 8
        -\frac{55755}{4} \pi ^2 l_ 2
        +\frac{2387 \pi ^4 l_ 2}{2}
        \nonumber\\ &
        +\frac{392 \pi ^6 l_ 2}{3}
        +8281 \pi ^2 l_ 2^2
        -899 \pi ^4 l_ 2^2
        +\frac{124}{35} \pi ^6 l_ 2^2
        -4620 \pi ^2 l_ 2^3
        +372 \pi ^4 l_ 2^3
        +\frac{8281 l_ 2^4}{2}
        \nonumber\\ &
        +1740 \pi ^2 l_ 2^4
        -\frac{6652}{105} \pi ^4 l_ 2^4
        -1386 l_ 2^5
        -432 \pi ^2 l_ 2^5
        +348 l_ 2^6
        +\frac{288}{5} \pi ^2 l_ 2^6
        -\frac{432 l_ 2^7}{7}
        +\frac{216 l_ 2^8}{35}
        \nonumber\\ &
        +104400 s_ 6
        -\frac{171840}{7} \pi ^2 s_ 6
        -\frac{417600}{7} l_ 2 s_ 6
        -\frac{195840}{7} l_ 2^2 s_ 6
        +\frac{417600 s_{7 a}}{7}
        +\frac{575628 s_{8 a}}{7}
        \nonumber\\ &
        -\frac{489600 s_{7 b}}{7}
        +\frac{3647232 s_{8 b}}{7}
        +\frac{391680 s_{8 c}}{7}
        +\frac{334080 s_{8 d}}{7}
        +\frac{390285 \zeta_3}{8}
        -1925 \pi ^2 \zeta_3
        \nonumber\\ &
        +\frac{10750 \pi ^4 \zeta_3}{21}
        -\frac{668160}{7} a_ 5 \zeta_3
        +2900 \pi ^2 l_ 2 \zeta_3
        -\frac{37844}{21} \pi ^4 l_ 2 \zeta_3
        -1800 \pi ^2 l_ 2^2 \zeta_3
        \nonumber\\ &
        -\frac{5920}{7} \pi ^2 l_ 2^3 \zeta_3
        +\frac{5568}{7} l_ 2^5 \zeta_3
        -44225 \zeta_3^2
        -\frac{214780}{21} \pi ^2 \zeta_3^2
        +\frac{522000}{7} l_ 2 \zeta_3^2
        -\frac{55080}{7} l_ 2^2 \zeta_3^2
        \nonumber\\ &
        -\frac{537075 \zeta_5}{4}
        +\frac{240645 \pi ^2 \zeta_5}{7}
        -\frac{192816}{7} \pi ^2 l_ 2 \zeta_5
        +125550 l_ 2^2 \zeta_5
        +\frac{477156 \zeta_3 \zeta_5}{7}
        \nonumber\\ &
        -\frac{8694765 \zeta_7}{14}
\biggr)
+\epsilon ^7
\biggl(
        -\frac{323148735}{2048}
        +\frac{758945 \pi ^2}{32}
        -\frac{38409 \pi ^4}{16}
        -\frac{21637 \pi ^6}{36}
        -\frac{429467 \pi ^8}{4410}
        \nonumber\\ &
        +334530 a_ 4
        +\frac{13920 \pi ^4 a_ 4}{7}
        +596232 a_ 5
        +\frac{43776 \pi ^4 a_ 5}{7}
        +997920 a_ 6
        +1503360 a_ 7
        \nonumber\\ &
        +1866240 a_ 8
        +1492992 a_ 9
        -\frac{360865}{8} \pi ^2 l_ 2
        +\frac{256711 \pi ^4 l_ 2}{60}
        +\frac{5684 \pi ^6 l_ 2}{9}
        \nonumber\\ &
        +\frac{12840131 \pi ^8 l_ 2}{14700}
        +\frac{55755}{2} \pi ^2 l_ 2^2
        -\frac{7161}{2} \pi ^4 l_ 2^2
        +\frac{186}{7} \pi ^6 l_ 2^2
        -16562 \pi ^2 l_ 2^3
        +1798 \pi ^4 l_ 2^3
        \nonumber\\ &
        +\frac{2812}{35} \pi ^6 l_ 2^3
        +\frac{55755 l_ 2^4}{4}
        +6930 \pi ^2 l_ 2^4
        -\frac{3326}{7} \pi ^4 l_ 2^4
        -\frac{24843 l_ 2^5}{5}
        -2088 \pi ^2 l_ 2^5
        +\frac{6504}{175} \pi ^4 l_ 2^5
        \nonumber\\ &
        +1386 l_ 2^6
        +432 \pi ^2 l_ 2^6
        -\frac{2088 l_ 2^7}{7}
        -\frac{1728}{35} \pi ^2 l_ 2^7
        +\frac{324 l_ 2^8}{7}
        -\frac{144 l_ 2^9}{35}
        +415800 s_ 6
        \nonumber\\ &
        -\frac{1288800}{7} \pi ^2 s_ 6
        -\frac{2018400}{7} l_ 2 s_ 6
        +\frac{2908800}{49} \pi ^2 l_ 2 s_ 6
        -\frac{1468800}{7} l_ 2^2 s_ 6
        +\frac{195840}{7} l_ 2^3 s_ 6
        \nonumber\\ &
        +\frac{2018400 s_{7 a}}{7}
        -\frac{2223360}{49} \pi ^2 s_{7 a}
        +\frac{4317210 s_{8 a}}{7}
        -\frac{2840598}{7} l_ 2 s_{8 a}
        +\frac{11116416 s_{9 a}}{7}
        \nonumber\\ &
        -\frac{2366400 s_{7 b}}{7}
        +\frac{3623040}{49} \pi ^2 s_{7 b}
        +\frac{27354240 s_{8 b}}{7}
        -\frac{17823360}{7} l_ 2 s_{8 b}
        -713856 s_{9 b}
        \nonumber\\ &
        +\frac{2937600 s_{8 c}}{7}
        -\frac{2970560 s_{9 c}}{7}
        +\frac{2505600 s_{8 d}}{7}
        +\frac{829440 s_{9 d}}{7}
        +\frac{1175040 s_{9 e}}{7}
        \nonumber\\ &
        +\frac{2526055 \zeta_3}{16}
        -\frac{41405}{6} \pi ^2 \zeta_3
        +\frac{155875 \pi ^4 \zeta_3}{63}
        -\frac{27807877 \pi ^6 \zeta_3}{13230}
        -\frac{691200}{7} \pi ^2 a_ 4 \zeta_3
        \nonumber\\ &
        -\frac{5011200}{7} a_ 5 \zeta_3
        -\frac{4008960}{7} a_ 6 \zeta_3
        +11550 \pi ^2 l_ 2 \zeta_3
        -\frac{94610}{7} \pi ^4 l_ 2 \zeta_3
        -8700 \pi ^2 l_ 2^2 \zeta_3
        \nonumber\\ &
        +\frac{49814}{7} \pi ^4 l_ 2^2 \zeta_3
        -\frac{44400}{7} \pi ^2 l_ 2^3 \zeta_3
        -\frac{19920}{7} \pi ^2 l_ 2^4 \zeta_3
        +\frac{41760}{7} l_ 2^5 \zeta_3
        -\frac{5568}{7} l_ 2^6 \zeta_3
        \nonumber\\ &
        -2258496 s_ 6 \zeta_3
        -\frac{352275 \zeta_3^2}{2}
        -\frac{536950}{7} \pi ^2 \zeta_3^2
        +\frac{2523000}{7} l_ 2 \zeta_3^2
        -\frac{2606020}{49} \pi ^2 l_ 2 \zeta_3^2
        \nonumber\\ &
        -\frac{413100}{7} l_ 2^2 \zeta_3^2
        +\frac{55080}{7} l_ 2^3 \zeta_3^2
        +\frac{1602131 \zeta_3^3}{3}
        -\frac{3850665 \zeta_5}{8}
        +\frac{2326235 \pi ^2 \zeta_5}{14}
        \nonumber\\ &
        -\frac{690564211 \pi ^4 \zeta_5}{5880}
        +\frac{27820944 a_ 4 \zeta_5}{7}
        -\frac{1446120}{7} \pi ^2 l_ 2 \zeta_5
        +606825 l_ 2^2 \zeta_5
        \nonumber\\ &
        -\frac{1683618}{7} \pi ^2 l_ 2^2 \zeta_5
        +\frac{1159206}{7} l_ 2^4 \zeta_5
        +\frac{3578670 \zeta_3 \zeta_5}{7}
        +\frac{22000710}{7} l_ 2 \zeta_3 \zeta_5
        -\frac{84049395 \zeta_7}{28}
        \nonumber\\ &
        -\frac{211530187}{392} \pi ^2 \zeta_7
        -\frac{11029569}{7} l_ 2^2 \zeta_7
        +\frac{648378363 \zeta_9}{56}
\biggr)
\Biggr\}
+ \mathcal{O}(\epsilon^9)~,
\end{align}

\begin{align}
  G_{51} &= \Gamma(1+\epsilon)^3
  \Biggl\{
        -\frac{1}{3 \epsilon ^3}
        -\frac{5}{3 \epsilon ^2}
        - \frac{1}{\epsilon}
        \biggl(
                4
                + \frac{2 \pi^2}{3}
        \biggr)
        +\frac{10}{3}
        -\frac{7 \pi ^2}{3}
        -\frac{26 \zeta_3}{3}
        +\epsilon
        \biggl(
                \frac{302}{3}
                -\pi ^2
                \nonumber\\ &
                -\frac{35 \pi ^4}{18}
                -\frac{94 \zeta_3}{3}
        \biggr)
+\epsilon ^2
\biggl(
        734
        +\frac{101 \pi ^2}{3}
        -\frac{551 \pi ^4}{90}
        -20 \zeta_3
        -\frac{76}{3} \pi ^2 \zeta_3
        -462 \zeta_5
\biggr)
+\epsilon ^3
\biggl(
        \frac{12254}{3}
        \nonumber\\ &
        +\frac{775 \pi ^2}{3}
        +\frac{28 \pi ^4}{15}
        -\frac{2353 \pi ^6}{378}
        +\frac{1232 \zeta_3}{3}
        -\frac{236}{3} \pi ^2 \zeta_3
        -\frac{482 \zeta_3^2}{3}
        -1482 \zeta_5
\biggr)
+\epsilon ^4
\biggl(
        \frac{60346}{3}
        \nonumber\\ &
        +1383 \pi ^2
        +\frac{5249 \pi ^4}{45}
        -\frac{36031 \pi ^6}{1890}
        +\frac{9904 \zeta_3}{3}
        +32 \pi ^2 \zeta_3
        -\frac{3571}{45} \pi ^4 \zeta_3
        -\frac{1510 \zeta_3^2}{3}
        +252 \zeta_5
        \nonumber\\ &
        -894 \pi ^2 \zeta_5
        -15307 \zeta_7
\biggr)
+\epsilon ^5
\biggl(
        92474
        +\frac{19327 \pi ^2}{3}
        +\frac{35659 \pi ^4}{45}
        +\frac{1021 \pi ^6}{105}
        -\frac{1247749 \pi ^8}{63000}
        \nonumber\\ &
        -\frac{6984 s_{8 a}}{5}
        +18064 \zeta_3
        +\frac{4648 \pi ^2 \zeta_3}{3}
        -\frac{10817}{45} \pi ^4 \zeta_3
        +172 \zeta_3^2
        -\frac{1564}{3} \pi ^2 \zeta_3^2
        +26976 \zeta_5
        \nonumber\\ &
        -2730 \pi ^2 \zeta_5
        -18636 \zeta_3 \zeta_5
        -47249 \zeta_7
\biggr)
+\epsilon ^6
\biggl(
        \frac{1221898}{3}
        +\frac{84125 \pi ^2}{3}
        +\frac{11927 \pi ^4}{3}
        \nonumber\\ &
        +\frac{366544 \pi ^6}{945}
        -\frac{11341381 \pi ^8}{189000}
        -\frac{20952 s_{8 a}}{5}
        +\frac{256432 \zeta_3}{3}
        +\frac{31160 \pi ^2 \zeta_3}{3}
        +\frac{2138 \pi ^4 \zeta_3}{15}
        \nonumber\\ &
        -\frac{244121}{945} \pi ^6 \zeta_3
        +\frac{29120 \zeta_3^2}{3}
        -\frac{4724}{3} \pi ^2 \zeta_3^2
        -\frac{19732 \zeta_3^3}{9}
        +186720 \zeta_5
        +1452 \pi ^2 \zeta_5
        -2587 \pi ^4 \zeta_5
        \nonumber\\ &
        -56676 \zeta_3 \zeta_5
        +21318 \zeta_7
        -\frac{52885}{2} \pi ^2 \zeta_7
        -\frac{4189826 \zeta_9}{9}
\biggr)
\Biggr\}
+ \mathcal{O}(\epsilon^7)~,
\end{align}

\begin{align}
  G_{53} &= \Gamma(1+\epsilon)^3
  \Biggl\{
-\frac{1}{\epsilon ^3}
-\frac{16}{3 \epsilon ^2}
-\frac{16}{\epsilon }
-20
-\frac{8 \pi ^2}{3}
+2 \zeta_3
+\epsilon
\biggl(
        \frac{364}{3}
        -28 \pi ^2
        -\frac{3 \pi ^4}{10}
        +16 \pi ^2 l_2
        \nonumber\\ &
        -\frac{200 \zeta_3}{3}
\biggr)
+\epsilon ^2
\biggl(
        1244
        -188 \pi ^2
        +\frac{46 \pi ^4}{15}
        -512 a_4
        +168 \pi ^2 l_2
        -\frac{80}{3} \pi ^2 l_2^2
        -\frac{64 l_2^4}{3}
        \nonumber\\ &
        -776 \zeta_3
        +21 \pi ^2 \zeta_3
        -126 \zeta_5
\biggr)
+\epsilon ^3
\biggl(
        7572
        -\frac{3100 \pi ^2}{3}
        +\frac{218 \pi ^4}{5}
        +\frac{22 \pi ^6}{35}
        -5376 a_4
        \nonumber\\ &
        -144 \pi ^2 a_4
        -3072 a_5
        +1128 \pi ^2 l_2
        -\frac{128}{5} \pi ^4 l_2
        -280 \pi ^2 l_2^2
        +6 \pi ^4 l_2^2
        +\frac{160}{3} \pi ^2 l_2^3
        \nonumber\\ &
        -224 l_2^4
        -6 \pi ^2 l_2^4
        +\frac{128 l_2^5}{5}
        -5360 \zeta_3
        +\frac{332 \pi ^2 \zeta_3}{3}
        -126 \pi ^2 l_2 \zeta_3
        +332 \zeta_3^2
        +1976 \zeta_5
\biggr)
\nonumber\\ &
+\epsilon ^4
\biggl(
        \frac{116084}{3}
        -5084 \pi ^2
        +\frac{1576 \pi ^4}{5}
        +\frac{13268 \pi ^6}{945}
        -36096 a_4
        -576 \pi ^2 a_4
        -32256 a_5
        \nonumber\\ &
        +864 \pi ^2 a_5
        -18432 a_6
        +6200 \pi ^2 l_2
        -\frac{1344}{5} \pi ^4 l_2
        -\frac{28}{15} \pi ^6 l_2
        +864 \pi ^2 a_4 l_2
        \nonumber\\ &
        -1880 \pi ^2 l_2^2
        +\frac{504}{5} \pi ^4 l_2^2
        +560 \pi ^2 l_2^3
        -24 \pi ^4 l_2^3
        -1504 l_2^4
        -104 \pi ^2 l_2^4
        +\frac{1344 l_2^5}{5}
        \nonumber\\ &
        +\frac{144}{5} \pi ^2 l_2^5
        -\frac{128 l_2^6}{5}
        -7680 s_6
        +2304 l_2 s_6
        -2304 s_{7 a}
        -2304 s_{7 b}
        -\frac{89392 \zeta_3}{3}
        \nonumber\\ &
        +364 \pi ^2 \zeta_3
        +\frac{287 \pi ^4 \zeta_3}{5}
        -4032 a_4 \zeta_3
        -664 \pi ^2 l_2 \zeta_3
        +546 \pi ^2 l_2^2 \zeta_3
        -168 l_2^4 \zeta_3
        +\frac{13744 \zeta_3^2}{3}
        \nonumber\\ &
        -2880 l_2 \zeta_3^2
        +25536 \zeta_5
        +942 \pi ^2 \zeta_5
        -19302 \zeta_7
\biggr)
+\epsilon ^5
\biggl(
        181108
        -23380 \pi ^2
        \nonumber\\ &
        +\frac{26744 \pi ^4}{15}
        +\frac{38914 \pi ^6}{315}
        +\frac{1019 \pi ^8}{280}
        -198400 a_4
        -576 \pi ^2 a_4
        +\frac{1152 \pi ^4 a_4}{5}
        \nonumber\\ &
        -13824 a_4^2
        -216576 a_5
        +3456 \pi ^2 a_5
        -193536 a_6
        -5184 \pi ^2 a_6
        -110592 a_7
        \nonumber\\ &
        +30504 \pi ^2 l_2
        -\frac{9024}{5} \pi ^4 l_2
        -\frac{7624}{135} \pi ^6 l_2
        +3456 \pi ^2 a_4 l_2
        -5184 \pi ^2 a_5 l_2
        -\frac{31000}{3} \pi ^2 l_2^2
        \nonumber\\ &
        +\frac{4152}{5} \pi ^4 l_2^2
        -\frac{48}{5} \pi ^6 l_2^2
        -1440 \pi ^2 a_4 l_2^2
        +3760 \pi ^2 l_2^3
        -\frac{1248}{5} \pi ^4 l_2^3
        -\frac{24800 l_2^4}{3}
        \nonumber\\ &
        -864 \pi ^2 l_2^4
        +\frac{198}{5} \pi ^4 l_2^4
        -1152 a_4 l_2^4
        +\frac{9024 l_2^5}{5}
        +\frac{1056}{5} \pi ^2 l_2^5
        -\frac{1344 l_2^6}{5}
        -24 \pi ^2 l_2^6
        \nonumber\\ &
        +\frac{768 l_2^7}{35}
        -24 l_2^8
        -80640 s_6
        -576 \pi ^2 s_6
        +\frac{212992 l_2 s_6}{7}
        -\frac{212992 s_{7 a}}{7}
        -4338 s_{8 a}
        \nonumber\\ &
        +\frac{109568 s_{7 b}}{7}
        -147344 \zeta_3
        +872 \pi ^2 \zeta_3
        +\frac{13124 \pi ^4 \zeta_3}{315}
        -16128 a_4 \zeta_3
        +24192 a_5 \zeta_3
        \nonumber\\ &
        -2184 \pi ^2 l_2 \zeta_3
        +\frac{1008}{5} \pi ^4 l_2 \zeta_3
        +2664 \pi ^2 l_2^2 \zeta_3
        -420 \pi ^2 l_2^3 \zeta_3
        -672 l_2^4 \zeta_3
        -\frac{1008}{5} l_2^5 \zeta_3
        \nonumber\\ &
        +35488 \zeta_3^2
        +510 \pi ^2 \zeta_3^2
        -\frac{266240}{7} l_2 \zeta_3^2
        +180888 \zeta_5
        -\frac{178568}{21} \pi ^2 \zeta_5
        +1116 \pi ^2 l_2 \zeta_5
        \nonumber\\ &
        -44640 l_2^2 \zeta_5
        -24318 \zeta_3 \zeta_5
        +\frac{1005280 \zeta_7}{7}
\biggr)
+\epsilon ^6
\biggl(
        805212
        -\frac{308852 \pi ^2}{3}
        +\frac{44416 \pi ^4}{5}
        \nonumber\\ &
        +\frac{246458 \pi ^6}{315}
        +\frac{1091513 \pi ^8}{22050}
        -976128 a_4
        +6912 \pi ^2 a_4
        +\frac{22528 \pi ^4 a_4}{105}
        -55296 a_4^2
        \nonumber\\ &
        -1190400 a_5
        +3456 \pi ^2 a_5
        +3456 \pi ^4 a_5
        -165888 a_4 a_5
        -1299456 a_6
        -20736 \pi ^2 a_6
        \nonumber\\ &
        -1161216 a_7
        +31104 \pi ^2 a_7
        -663552 a_8
        +140280 \pi ^2 l_2
        -9920 \pi ^4 l_2
        -\frac{23492}{45} \pi ^6 l_2
        \nonumber\\ &
        +\frac{481517 \pi ^8 l_2}{500}
        +3456 \pi ^2 a_4 l_2
        -\frac{32832}{5} \pi ^4 a_4 l_2
        -20736 \pi ^2 a_5 l_2
        +31104 \pi ^2 a_6 l_2
        \nonumber\\ &
        -50840 \pi ^2 l_2^2
        +\frac{25632}{5} \pi ^4 l_2^2
        -\frac{12664}{315} \pi ^6 l_2^2
        -5760 \pi ^2 a_4 l_2^2
        +22464 \pi ^2 a_5 l_2^2
        \nonumber\\ &
        +\frac{62000}{3} \pi ^2 l_2^3
        -\frac{8544}{5} \pi ^4 l_2^3
        +\frac{404}{5} \pi ^6 l_2^3
        +2880 \pi ^2 a_4 l_2^3
        -40672 l_2^4
        -5352 \pi ^2 l_2^4
        \nonumber\\ &
        +\frac{113192}{315} \pi ^4 l_2^4
        -4608 a_4 l_2^4
        -6912 a_5 l_2^4
        +9920 l_2^5
        +\frac{5616}{5} \pi ^2 l_2^5
        -\frac{1464}{5} \pi ^4 l_2^5
        \nonumber\\ &
        +\frac{6912}{5} a_4 l_2^5
        -\frac{9024 l_2^6}{5}
        -192 \pi ^2 l_2^6
        +\frac{1152 l_2^7}{5}
        -\frac{1056}{35} \pi ^2 l_2^7
        -\frac{3936 l_2^8}{35}
        +\frac{288 l_2^9}{5}
        \nonumber\\ &
        -541440 s_6
        +\frac{442112 \pi ^2 s_6}{7}
        +231936 l_2 s_6
        +\frac{293760}{7} \pi ^2 l_2 s_6
        +\frac{522240}{7} l_2^2 s_6
        \nonumber\\ &
        +48384 l_2^3 s_6
        -231936 s_{7 a}
        +\frac{165888}{7} \pi ^2 s_{7 a}
        -\frac{1656472 s_{8 a}}{7}
        -\frac{2470338}{5} l_2 s_{8 a}
        \nonumber\\ &
        +2272896 s_{9 a}
        +251904 s_{7 b}
        +\frac{335232}{7} \pi ^2 s_{7 b}
        +41472 l_2^2 s_{7 b}
        -\frac{9725952 s_{8 b}}{7}
        \nonumber\\ &
        -3100032 l_2 s_{8 b}
        -1190016 s_{9 b}
        -\frac{1044480 s_{8 c}}{7}
        -516672 s_{9 c}
        -\frac{890880 s_{8 d}}{7}
        \nonumber\\ &
        -41472 s_{9 d}
        +207360 s_{9 e}
        +82944 s_{9 f}
        -679088 \zeta_3
        +\frac{3784 \pi ^2 \zeta_3}{3}
        -\frac{26156}{15} \pi ^4 \zeta_3
        \nonumber\\ &
        -\frac{165286}{35} \pi ^6 \zeta_3
        -16128 a_4 \zeta_3
        +4896 \pi ^2 a_4 \zeta_3
        +\frac{2459136 a_5 \zeta_3}{7}
        +20736 a_6 \zeta_3
        \nonumber\\ &
        -5232 \pi ^2 l_2 \zeta_3
        +\frac{197664}{35} \pi ^4 l_2 \zeta_3
        +7224 \pi ^2 l_2^2 \zeta_3
        -\frac{8886}{5} \pi ^4 l_2^2 \zeta_3
        +\frac{18800}{21} \pi ^2 l_2^3 \zeta_3
        \nonumber\\ &
        -672 l_2^4 \zeta_3
        +1266 \pi ^2 l_2^4 \zeta_3
        -\frac{102464}{35} l_2^5 \zeta_3
        +\frac{144}{5} l_2^6 \zeta_3
        -2522880 s_6 \zeta_3
        +213424 \zeta_3^2
        \nonumber\\ &
        +\frac{616520}{21} \pi ^2 \zeta_3^2
        -289920 l_2 \zeta_3^2
        +\frac{231300}{7} \pi ^2 l_2 \zeta_3^2
        +\frac{146880}{7} l_2^2 \zeta_3^2
        +13608 l_2^3 \zeta_3^2
        \nonumber\\ &
        +\frac{1703845 \zeta_3^3}{3}
        +1015432 \zeta_5
        -125080 \pi ^2 \zeta_5
        -\frac{40164351}{280} \pi ^4 \zeta_5
        +5058000 a_4 \zeta_5
        \nonumber\\ &
        +\frac{547440}{7} \pi ^2 l_2 \zeta_5
        -468720 l_2^2 \zeta_5
        -226950 \pi ^2 l_2^2 \zeta_5
        +170574 l_2^4 \zeta_5
        -\frac{1953320}{7} \zeta_3 \zeta_5
        \nonumber\\ &
        +3826602 l_2 \zeta_3 \zeta_5
        +2241396 \zeta_7
        -\frac{75881325}{56} \pi ^2 \zeta_7
        -1472679 l_2^2 \zeta_7
        +\frac{585155695 \zeta_9}{24}
\biggr)
\Biggr\}
\nonumber\\ &
+ \mathcal{O}(\epsilon^7)~,
\end{align}

\begin{align}
    G_{61} &= \Gamma(1+\epsilon)^3
    \Biggl\{
   \frac{1}{6 \epsilon ^3}
  +\frac{3}{2 \epsilon ^2}
  + \frac{1}{\epsilon}
  \biggl(
           \frac{55}{6}
          -\frac{\pi ^2}{3}
  \biggr)
  +\frac{95}{2}
  -2 \pi ^2
  -\frac{8 \zeta_3}{3}
  -\frac{\pi ^4}{15}
  +\epsilon
  \biggl(
          \frac{1351}{6}
          \nonumber\\ &
          -\frac{17 \pi ^2}{3}
          -\frac{47 \pi ^4}{45}
          -14 \zeta_3
          +6 \pi ^2 \zeta_3
          -64 \zeta_5
  \biggr)
  +\epsilon ^2
  \biggl(
          \frac{2023}{2}
          +\frac{16 \pi ^2}{3}
          -\frac{457 \pi ^4}{90}
          +\frac{1471 \pi ^6}{2835}
          \nonumber\\ &
          -64 \pi ^2 a_4
          -16 \pi ^2 l_2
          +\frac{8}{3} \pi ^4 l_2^2
          -\frac{8}{3} \pi ^2 l_2^4
          +\frac{16 \zeta_3}{3}
          +\frac{26 \pi ^2 \zeta_3}{3}
          -56 \pi ^2 l_2 \zeta_3
          +62 \zeta_3^2
          -342 \zeta_5
  \biggr)
  \nonumber\\ &
  +\epsilon ^3 \biggr(
          \frac{26335}{6}
          +187 \pi ^2
          -\frac{277 \pi ^4}{15}
          -\frac{299 \pi ^6}{378}
          +512 a_4
          -192 \pi ^2 a_4
          +768 \pi ^2 a_5
          -224 \pi ^2 l_2
          \nonumber\\ &
          -\frac{112}{135} \pi ^6 l_2
          +768 \pi ^2 a_4 l_2
          +\frac{224}{3} \pi ^2 l_2^2
          +8 \pi ^4 l_2^2
          -\frac{64}{3} \pi ^4 l_2^3
          +\frac{64 l_2^4}{3}
          -8 \pi ^2 l_2^4
          +\frac{128}{5} \pi ^2 l_2^5
          \nonumber\\ &
          +1024 l_2 s_6
          -1024 s_{7 a}
          -1024 s_{7 b}
          +598 \zeta_3
          -\frac{58}{3} \pi ^2 \zeta_3
          +\frac{1651 \pi ^4 \zeta_3}{45}
          -1792 a_4 \zeta_3
          \nonumber\\ &
          -168 \pi ^2 l_2 \zeta_3
          +\frac{1232}{3} \pi ^2 l_2^2 \zeta_3
          -\frac{224}{3} l_2^4 \zeta_3
          +\frac{400 \zeta_3^2}{3}
          -1280 l_2 \zeta_3^2
          -1082 \zeta_5
          \nonumber\\ &
          -\frac{499}{3} \pi ^2 \zeta_5
          -8365 \zeta_7
  \biggl)
  +\epsilon ^4
  \biggl(
          \frac{37215}{2}
          +\frac{4598 \pi ^2}{3}
          -\frac{1195 \pi ^4}{18}
          -\frac{55943 \pi ^6}{5670}
          +\frac{1119323 \pi ^8}{243000}
          \nonumber\\ &
          +7168 a_4
          -320 \pi ^2 a_4
          +\frac{2752 \pi ^4 a_4}{15}
          -12288 a_4^2
          +6144 a_5
          +2304 \pi ^2 a_5
          -9216 \pi ^2 a_6
          \nonumber\\ &
          -1904 \pi ^2 l_2
          +\frac{688 \pi ^4 l_2}{15}
          -\frac{112}{45} \pi ^6 l_2
          +2304 \pi ^2 a_4 l_2
          -9216 \pi ^2 a_5 l_2
          +\frac{3136}{3} \pi ^2 l_2^2
          \nonumber\\ &
          +\frac{40}{3} \pi ^4 l_2^2
          -\frac{344}{45} \pi ^6 l_2^2
          -3584 \pi ^2 a_4 l_2^2
          -\frac{896}{3} \pi ^2 l_2^3
          -64 \pi ^4 l_2^3
          +\frac{896 l_2^4}{3}
          -\frac{40}{3} \pi ^2 l_2^4
          \nonumber\\ &
          +\frac{3704}{45} \pi ^4 l_2^4
          -1024 a_4 l_2^4
          -\frac{256 l_2^5}{5}
          +\frac{384}{5} \pi ^2 l_2^5
          -\frac{256}{3} \pi ^2 l_2^6
          -\frac{64 l_2^8}{3}
          -4288 \pi ^2 s_6
          \nonumber\\ &
          +3072 l_2 s_6
          -3072 s_{7 a}
          +\frac{468 s_{8 a}}{5}
          -3072 s_{7 b}
          +\frac{17356 \zeta_3}{3}
          -\frac{518}{3} \pi ^2 \zeta_3
          +\frac{3692 \pi ^4 \zeta_3}{45}
          \nonumber\\ &
          -5376 a_4 \zeta_3
          +21504 a_5 \zeta_3
          -280 \pi ^2 l_2 \zeta_3
          +\frac{2408}{15} \pi ^4 l_2 \zeta_3
          +1232 \pi ^2 l_2^2 \zeta_3
          \nonumber\\ &
          -\frac{3136}{3} \pi ^2 l_2^3 \zeta_3
          -224 l_2^4 \zeta_3
          -\frac{896}{5} l_2^5 \zeta_3
          +44 \zeta_3^2
          +1592 \pi ^2 \zeta_3^2
          -3840 l_2 \zeta_3^2
          -7090 \zeta_5
          \nonumber\\ &
          -813 \pi ^2 \zeta_5
          +8308 \pi ^2 l_2 \zeta_5
          -19116 \zeta_3 \zeta_5
          -30142 \zeta_7
  \biggr)
  +\epsilon ^5
  \biggl(
          \frac{465751}{6}
          +\frac{28331 \pi ^2}{3}
          \nonumber\\ &
          -\frac{12826 \pi ^4}{45}
          -\frac{436571 \pi ^6}{5670}
          +\frac{1941047 \pi ^8}{283500}
          +60928 a_4
          +64 \pi ^2 a_4
          +\frac{2752 \pi ^4 a_4}{5}
          \nonumber\\ &
          -36864 a_4^2
          +86016 a_5
          +3840 \pi ^2 a_5
          +6400 \pi ^4 a_5
          -294912 a_4 a_5
          +73728 a_6
          \nonumber\\ &
          -27648 \pi ^2 a_6
          +110592 \pi ^2 a_7
          -12736 \pi ^2 l_2
          +\frac{9632 \pi ^4 l_2}{15}
          -\frac{112}{27} \pi ^6 l_2
          \nonumber\\ &
          +\frac{13250626 \pi ^8 l_2}{7875}
          +3840 \pi ^2 a_4 l_2
          -\frac{57088}{5} \pi ^4 a_4 l_2
          -27648 \pi ^2 a_5 l_2
          +110592 \pi ^2 a_6 l_2
          \nonumber\\ &
          +\frac{26656}{3} \pi ^2 l_2^2
          -\frac{4168}{15} \pi ^4 l_2^2
          -\frac{344}{15} \pi ^6 l_2^2
          -10752 \pi ^2 a_4 l_2^2
          +67584 \pi ^2 a_5 l_2^2
          \nonumber\\ &
          -\frac{12544}{3} \pi ^2 l_2^3
          -\frac{320}{3} \pi ^4 l_2^3
          +\frac{2048}{15} \pi ^6 l_2^3
          +14336 \pi ^2 a_4 l_2^3
          +\frac{7616 l_2^4}{3}
          +\frac{2696}{3} \pi ^2 l_2^4
          \nonumber\\ &
          +\frac{3704}{15} \pi ^4 l_2^4
          -3072 a_4 l_2^4
          -12288 a_5 l_2^4
          -\frac{3584 l_2^5}{5}
          +128 \pi ^2 l_2^5
          -\frac{3328}{5} \pi ^4 l_2^5
          \nonumber\\ &
          +\frac{12288}{5} a_4 l_2^5
          +\frac{512 l_2^6}{5}
          -256 \pi ^2 l_2^6
          +\frac{17408}{105} \pi ^2 l_2^7
          -64 l_2^8
          +\frac{512 l_2^9}{5}
          +33792 s_6
          \nonumber\\ &
          -12864 \pi ^2 s_6
          +5120 l_2 s_6
          +\frac{723456}{7} \pi ^2 l_2 s_6
          +86016 l_2^3 s_6
          -5120 s_{7 a}
          \nonumber\\ &
          +\frac{410880}{7} \pi ^2 s_{7 a}
          -\frac{924 s_{8 a}}{5}
          -\frac{4391712}{5} l_2 s_{8 a}
          +4099072 s_{9 a}
          -5120 s_{7 b}
          \nonumber\\ &
          +\frac{394752}{7} \pi ^2 s_{7 b}
          +73728 l_2^2 s_{7 b}
          -5511168 l_2 s_{8 b}
          -2057216 s_{9 b}
          -918528 s_{9 c}
          \nonumber\\ &
          -73728 s_{9 d}
          +368640 s_{9 e}
          +147456 s_{9 f}
          +39538 \zeta_3
          -782 \pi ^2 \zeta_3
          +\frac{1762 \pi ^4 \zeta_3}{45}
          \nonumber\\ &
          -\frac{23014981 \pi ^6 \zeta_3}{2835}
          -8960 a_4 \zeta_3
          +7936 \pi ^2 a_4 \zeta_3
          +64512 a_5 \zeta_3
          +36864 a_6 \zeta_3
          \nonumber\\ &
          +504 \pi ^2 l_2 \zeta_3
          +\frac{2408}{5} \pi ^4 l_2 \zeta_3
          +\frac{6160}{3} \pi ^2 l_2^2 \zeta_3
          -\frac{45232}{15} \pi ^4 l_2^2 \zeta_3
          -3136 \pi ^2 l_2^3 \zeta_3
          \nonumber\\ &
          -\frac{1120}{3} l_2^4 \zeta_3
          +\frac{12704}{3} \pi ^2 l_2^4 \zeta_3
          -\frac{2688}{5} l_2^5 \zeta_3
          +\frac{256}{5} l_2^6 \zeta_3
          -4518144 s_6 \zeta_3
          -\frac{41540 \zeta_3^2}{3}
          \nonumber\\ &
          +\frac{13796}{3} \pi ^2 \zeta_3^2
          -6400 l_2 \zeta_3^2
          +\frac{463088}{7} \pi ^2 l_2 \zeta_3^2
          +24192 l_2^3 \zeta_3^2
          +\frac{3053372 \zeta_3^3}{3}
          \nonumber\\ &
          -69418 \zeta_5
          -\frac{7349}{3} \pi ^2 \zeta_5
          -\frac{15096776}{63} \pi ^4 \zeta_5
          +8899072 a_4 \zeta_5
          +24924 \pi ^2 l_2 \zeta_5
          \nonumber\\ &
          -\frac{1198784}{3} \pi ^2 l_2^2 \zeta_5
          +\frac{898112}{3} l_2^4 \zeta_5
          -63528 \zeta_3 \zeta_5
          +6802848 l_2 \zeta_3 \zeta_5
          -67226 \zeta_7
          \nonumber\\ &
          -\frac{110755595}{42} \pi ^2 \zeta_7
          -2676008 l_2^2 \zeta_7
          +\frac{391372085 \zeta_9}{9}
  \biggr)
  \Biggr\}
  +\mathcal{O}(\epsilon^6)~,
\end{align}

\begin{align}
  G_{62} &= \Gamma(1+\epsilon)^3
  \Biggl\{
     \frac{1}{3 \epsilon ^3}
    +\frac{7}{3 \epsilon ^2}
    +\frac{31}{3 \epsilon }
    +\frac{103}{3}
    +\frac{\pi ^2}{3}
    -\frac{4 \pi ^4}{45}
    +\frac{2}{3} \zeta_3
    +\epsilon
    \biggl(
            \frac{235}{3}
            +4 \pi ^2
            -\frac{3 \pi ^4}{10}
            \nonumber\\ &
            +\frac{20}{3} \zeta_3
            +\frac{2}{3} \pi ^2 \zeta_3
            -2 \zeta_5
    \biggr)
    +\epsilon ^2
    \biggl(
            \frac{19}{3}
            +\frac{91 \pi ^2}{3}
            +\frac{14 \pi ^4}{45}
            +\frac{1009 \pi ^6}{1890}
            +\frac{206}{3} \zeta_3
            +2 \pi ^2 \zeta_3
            \nonumber\\ &
            +32 \zeta_3^2
            +6 \zeta_5
            -64 \pi ^2 a_4
            -16 \pi ^2 l_2
            -56 \pi ^2 \zeta_3 l_2
            +\frac{8}{3} \pi ^4 l_2^2
            -\frac{8}{3} \pi ^2 l_2^4
    \biggr)
    +\epsilon ^3
    \biggl(
            -\frac{3953}{3}
            \nonumber\\ &
            +186 \pi ^2
            +\frac{307 \pi ^4}{90}
            +\frac{979 \pi ^6}{630}
            +\frac{1760}{3} \zeta_3
            +14 \pi ^2 \zeta_3
            +\frac{976}{45} \pi ^4 \zeta_3
            +\frac{296}{3} \zeta_3^2
            +222 \zeta_5
            \nonumber\\ &
            -215 \pi ^2 \zeta_5
            -6114 \zeta_7
            +512 a_4
            -192 \pi ^2 a_4
            -1792 \zeta_3 a_4
            +768 \pi ^2 a_5
            -224 \pi ^2 l_2
            \nonumber\\ &
            -\frac{112}{135} \pi ^6 l_2
            -168 \pi ^2 \zeta_3 l_2
            -1280 \zeta_3^2 l_2
            +768 \pi ^2 a_4 l_2
            +\frac{224}{3} \pi ^2 l_2^2
            +8 \pi ^4 l_2^2
            +\frac{1232}{3} \pi ^2 \zeta_3 l_2^2
            \nonumber\\ &
            -\frac{64}{3} \pi ^4 l_2^3
            +\frac{64 l_2^4}{3}
            -8 \pi ^2 l_2^4
            -\frac{224}{3} \zeta_3 l_2^4
            +\frac{128}{5} \pi ^2 l_2^5
            +1024 l_2 s_6
            -1024 s_{7 a}
            -1024 s_{7 b}
    \biggr)
    \nonumber\\ &
    +\epsilon ^4
    \biggl(
            -\frac{31889}{3}
            +\frac{3025 \pi ^2}{3}
            -\frac{211 \pi ^4}{15}
            +\frac{3167 \pi ^6}{630}
            +\frac{2567 \pi ^8}{500}
            +\frac{12386}{3} \zeta_3
            +\frac{142}{3} \pi ^2 \zeta_3
            \nonumber\\ &
            +\frac{324}{5} \pi ^4 \zeta_3
            +\frac{752}{3} \zeta_3^2
            +\frac{4444}{3} \pi ^2 \zeta_3^2
            -3994 \zeta_5
            -645 \pi ^2 \zeta_5
            -21656 \zeta_3 \zeta_5
            -18176 \zeta_7
            \nonumber\\ &
            +7168 a_4
            -320 \pi ^2 a_4
            +\frac{2752 \pi ^4 a_4}{15}
            -5376 \zeta_3 a_4
            -12288 a_4^2
            +6144 a_5
            +2304 \pi ^2 a_5
            \nonumber\\ &
            +21504 \zeta_3 a_5
            -9216 \pi ^2 a_6
            -1904 \pi ^2 l_2
            +\frac{688 \pi ^4 l_2}{15}
            -\frac{112}{45} \pi ^6 l_2
            -280 \pi ^2 \zeta_3 l_2
            \nonumber\\ &
            +\frac{2408}{15} \pi ^4 \zeta_3 l_2
            -3840 \zeta_3^2 l_2
            +8308 \pi ^2 \zeta_5 l_2
            +2304 \pi ^2 a_4 l_2
            -9216 \pi ^2 a_5 l_2
            +\frac{3136}{3} \pi ^2 l_2^2
            \nonumber\\ &
            +\frac{40}{3} \pi ^4 l_2^2
            -\frac{344}{45} \pi ^6 l_2^2
            +1232 \pi ^2 \zeta_3 l_2^2
            -3584 \pi ^2 a_4 l_2^2
            -\frac{896}{3} \pi ^2 l_2^3
            -64 \pi ^4 l_2^3
            -\frac{3136}{3} \pi ^2 \zeta_3 l_2^3
            \nonumber\\ &
            +\frac{896 l_2^4}{3}
            -\frac{40}{3} \pi ^2 l_2^4
            +\frac{3704}{45} \pi ^4 l_2^4
            -224 \zeta_3 l_2^4
            -1024 a_4 l_2^4
            -\frac{256 l_2^5}{5}
            +\frac{384}{5} \pi ^2 l_2^5
            -\frac{896}{5} \zeta_3 l_2^5
            \nonumber\\ &
            -\frac{256}{3} \pi ^2 l_2^6
            -\frac{64 l_2^8}{3}
            -4288 \pi ^2 s_6
            +3072 l_2 s_6
            -3072 s_{7 a}
            -\frac{2228 s_{8 a}}{5}
            -3072 s_{7 b}
    \biggr)
    \nonumber\\ &
    +\epsilon ^5
    \biggl(
            -\frac{188141}{3}
            +5048 \pi ^2
            -\frac{29239 \pi ^4}{90}
            -\frac{43927 \pi ^6}{1890}
            +\frac{69049 \pi ^8}{4500}
            +\frac{75500}{3} \zeta_3
            \nonumber\\ &
            -\frac{550}{3} \pi ^2 \zeta_3
            +\frac{1208}{9} \pi ^4 \zeta_3
            -\frac{856952}{105} \pi ^6 \zeta_3
            -\frac{37768}{3} \zeta_3^2
            +4444 \pi ^2 \zeta_3^2
            +\frac{3051872}{3} \zeta_3^3
            \nonumber\\ &
            -71210 \zeta_5
            -659 \pi ^2 \zeta_5
            -\frac{75528449}{315} \pi ^4 \zeta_5
            -64872 \zeta_3 \zeta_5
            -24070 \zeta_7
            -\frac{55362142}{21} \pi ^2 \zeta_7
            \nonumber\\ &
            +43543603 \zeta_9
            +60928 a_4
            +64 \pi ^2 a_4
            +\frac{2752 \pi ^4 a_4}{5}
            -8960 \zeta_3 a_4
            +7936 \pi ^2 \zeta_3 a_4
            \nonumber\\ &
            +8899072 \zeta_5 a_4
            -36864 a_4^2
            +86016 a_5
            +3840 \pi ^2 a_5
            +6400 \pi ^4 a_5
            +64512 \zeta_3 a_5
            \nonumber\\ &
            -294912 a_4 a_5
            +73728 a_6
            -27648 \pi ^2 a_6
            +36864 \zeta_3 a_6
            +110592 \pi ^2 a_7
            -12736 \pi ^2 l_2
            \nonumber\\ &
            +\frac{9632 \pi ^4 l_2}{15}
            -\frac{112}{27} \pi ^6 l_2
            +\frac{13250626 \pi ^8 l_2}{7875}
            +504 \pi ^2 \zeta_3 l_2
            +\frac{2408}{5} \pi ^4 \zeta_3 l_2
            -6400 \zeta_3^2 l_2
            \nonumber\\ &
            +\frac{463088}{7} \pi ^2 \zeta_3^2 l_2
            +24924 \pi ^2 \zeta_5 l_2
            +6802848 \zeta_3 \zeta_5 l_2
            +3840 \pi ^2 a_4 l_2
            -\frac{57088}{5} \pi ^4 a_4 l_2
            \nonumber\\ &
            -27648 \pi ^2 a_5 l_2
            +110592 \pi ^2 a_6 l_2
            +\frac{26656}{3} \pi ^2 l_2^2
            -\frac{4168}{15} \pi ^4 l_2^2
            -\frac{344}{15} \pi ^6 l_2^2
            +\frac{6160}{3} \pi ^2 \zeta_3 l_2^2
            \nonumber\\ &
            -\frac{45232}{15} \pi ^4 \zeta_3 l_2^2
            -\frac{1198784}{3} \pi ^2 \zeta_5 l_2^2
            -2676008 \zeta_7 l_2^2
            -10752 \pi ^2 a_4 l_2^2
            +67584 \pi ^2 a_5 l_2^2
            \nonumber\\ &
            -\frac{12544}{3} \pi ^2 l_2^3
            -\frac{320}{3} \pi ^4 l_2^3
            +\frac{2048}{15} \pi ^6 l_2^3
            -3136 \pi ^2 \zeta_3 l_2^3
            +24192 \zeta_3^2 l_2^3
            +14336 \pi ^2 a_4 l_2^3
            \nonumber\\ &
            +\frac{7616 l_2^4}{3}
            +\frac{2696}{3} \pi ^2 l_2^4
            +\frac{3704}{15} \pi ^4 l_2^4
            -\frac{1120}{3} \zeta_3 l_2^4
            +\frac{12704}{3} \pi ^2 \zeta_3 l_2^4
            +\frac{898112}{3} \zeta_5 l_2^4
            \nonumber\\ &
            -3072 a_4 l_2^4
            -12288 a_5 l_2^4
            -\frac{3584 l_2^5}{5}
            +128 \pi ^2 l_2^5
            -\frac{3328}{5} \pi ^4 l_2^5
            -\frac{2688}{5} \zeta_3 l_2^5
            +\frac{12288}{5} a_4 l_2^5
            \nonumber\\ &
            +\frac{512 l_2^6}{5}
            -256 \pi ^2 l_2^6
            +\frac{256}{5} \zeta_3 l_2^6
            +\frac{17408}{105} \pi ^2 l_2^7
            -64 l_2^8
            +\frac{512 l_2^9}{5}
            +33792 s_6
            -12864 \pi ^2 s_6
            \nonumber\\ &
            -4518144 \zeta_3 s_6
            +5120 l_2 s_6
            +\frac{723456}{7} \pi ^2 l_2 s_6
            +86016 l_2^3 s_6
            -5120 s_{7 a}
            +\frac{410880}{7} \pi ^2 s_{7 a}
            \nonumber\\ &
            -\frac{6684 s_{8 a}}{5}
            -\frac{4391712}{5} l_2 s_{8 a}
            +4099072 s_{9 a}
            -5120 s_{7 b}
            +\frac{394752}{7} \pi ^2 s_{7 b}
            +73728 l_2^2 s_{7 b}
            \nonumber\\ &
            -5511168 l_2 s_{8 b}
            -2057216 s_{9 b}
            -918528 s_{9 c}
            -73728 s_{9 d}
            +368640 s_{9 e}
            +147456 s_{9 f}
    \biggr)
    \nonumber\\ &
  \Biggr\}
  +\mathcal{O}(\epsilon^6)~,
\end{align}

\begin{align}
  G_{63} &= \Gamma(1+\epsilon)^3
  \Biggl\{
         \frac{1}{6 \epsilon ^3}
        +\frac{3}{2 \epsilon ^2}
        + \frac{1}{\epsilon}
        \biggl(
                \frac{55}{6}
                - \frac{\pi^2}{3}
        \biggr)
        +\frac{95}{2}
        -\frac{7 \pi ^2}{3}
        -\frac{14 \zeta_3}{3}
        -\frac{4 \pi ^4}{45}
        \nonumber\\ &
        +\epsilon
        \biggl(
                \frac{1351}{6}
                -\frac{31 \pi ^2}{3}
                -\frac{23 \pi ^4}{30}
                +4 \pi ^2 l_2
                -42 \zeta_3
                +\frac{25 \pi ^2 \zeta_3}{6}
                -\frac{49 \zeta_5}{2}
        \biggr)
+\epsilon ^2
\biggl(
        \frac{2023}{2}
        \nonumber\\ &
        -\frac{103 \pi ^2}{3}
        -\frac{148 \pi ^4}{45}
        +\frac{23 \pi ^6}{54}
        -160 a_4
        -56 \pi ^2 a_4
        +32 \pi ^2 l_2
        -\frac{40}{3} \pi ^2 l_2^2
        +\frac{7}{3} \pi ^4 l_2^2
        -\frac{20 l_2^4}{3}
        \nonumber\\ &
        -\frac{7}{3} \pi ^2 l_2^4
        -\frac{698 \zeta_3}{3}
        +\frac{109 \pi ^2 \zeta_3}{6}
        -49 \pi ^2 l_2 \zeta_3
        +\frac{191 \zeta_3^2}{4}
        -\frac{413 \zeta_5}{2}
\biggr)
+\epsilon ^3
\biggl(
        \frac{26335}{6}
        -\frac{235 \pi ^2}{3}
        \nonumber\\ &
        -\frac{371 \pi ^4}{30}
        +\frac{457 \pi ^6}{270}
        -1344 a_4
        -280 \pi ^2 a_4
        -1472 a_5
        +656 \pi ^2 a_5
        +140 \pi ^2 l_2
        -\frac{182}{45} \pi ^4 l_2
        \nonumber\\ &
        -\frac{35}{54} \pi ^6 l_2
        +656 \pi ^2 a_4 l_2
        -80 \pi ^2 l_2^2
        +\frac{35}{3} \pi ^4 l_2^2
        +\frac{368}{9} \pi ^2 l_2^3
        -\frac{164}{9} \pi ^4 l_2^3
        -56 l_2^4
        -\frac{35}{3} \pi ^2 l_2^4
        \nonumber\\ &
        +\frac{184 l_2^5}{15}
        +\frac{328}{15} \pi ^2 l_2^5
        +800 l_2 s_6
        -800 s_{7 a}
        -800 s_{7 b}
        -994 \zeta_3
        +\frac{129 \pi ^2 \zeta_3}{2}
        +\frac{3803 \pi ^4 \zeta_3}{180}
        \nonumber\\ &
        -1400 a_4 \zeta_3
        -245 \pi ^2 l_2 \zeta_3
        +\frac{1036}{3} \pi ^2 l_2^2 \zeta_3
        -\frac{175}{3} l_2^4 \zeta_3
        +\frac{2641 \zeta_3^2}{12}
        -1000 l_2 \zeta_3^2
        +\frac{107 \zeta_5}{2}
        \nonumber\\ &
        -\frac{2057}{12} \pi ^2 \zeta_5
        -\frac{44413 \zeta_7}{8}
\biggr)
+\epsilon ^4
\biggl(
        \frac{37215}{2}
        -\frac{19 \pi ^2}{3}
        -\frac{2467 \pi ^4}{45}
        +\frac{25507 \pi ^6}{1890}
        +\frac{2351431 \pi ^8}{567000}
        \nonumber\\ &
        -6496 a_4
        -1192 \pi ^2 a_4
        +\frac{1580 \pi ^4 a_4}{9}
        -9920 a_4^2
        -10624 a_5
        +3280 \pi ^2 a_5
        -12928 a_6
        \nonumber\\ &
        -8032 \pi ^2 a_6
        +328 \pi ^2 l_2
        -\frac{124}{45} \pi ^4 l_2
        -\frac{175}{54} \pi ^6 l_2
        +3280 \pi ^2 a_4 l_2
        -8032 \pi ^2 a_5 l_2
        -\frac{280}{3} \pi ^2 l_2^2
        \nonumber\\ &
        +\frac{3149}{45} \pi ^4 l_2^2
        -\frac{395}{54} \pi ^6 l_2^2
        -\frac{9568}{3} \pi ^2 a_4 l_2^2
        +\frac{1504}{9} \pi ^2 l_2^3
        -\frac{820}{9} \pi ^4 l_2^3
        -\frac{812 l_2^4}{3}
        -\frac{1255}{9} \pi ^2 l_2^4
        \nonumber\\ &
        +\frac{3983}{54} \pi ^4 l_2^4
        -\frac{2480}{3} a_4 l_2^4
        +\frac{1328 l_2^5}{15}
        +\frac{328}{3} \pi ^2 l_2^5
        -\frac{808 l_2^6}{45}
        -\frac{694}{9} \pi ^2 l_2^6
        -\frac{155 l_2^8}{9}
        -4704 s_6
        \nonumber\\ &
        -3592 \pi ^2 s_6
        +4000 l_2 s_6
        -4000 s_{7 a}
        -\frac{1591 s_{8 a}}{5}
        -4000 s_{7 b}
        -\frac{10346 \zeta_3}{3}
        +\frac{1177 \pi ^2 \zeta_3}{6}
        \nonumber\\ &
        +\frac{18551 \pi ^4 \zeta_3}{180}
        -7000 a_4 \zeta_3
        +17360 a_5 \zeta_3
        -1067 \pi ^2 l_2 \zeta_3
        +\frac{2765}{18} \pi ^4 l_2 \zeta_3
        +\frac{5180}{3} \pi ^2 l_2^2 \zeta_3
        \nonumber\\ &
        -\frac{8372}{9} \pi ^2 l_2^3 \zeta_3
        -\frac{875}{3} l_2^4 \zeta_3
        -\frac{434}{3} l_2^5 \zeta_3
        +\frac{10741 \zeta_3^2}{4}
        +\frac{7757}{6} \pi ^2 \zeta_3^2
        -5000 l_2 \zeta_3^2
        +\frac{5667 \zeta_5}{2}
        \nonumber\\ &
        -\frac{10861}{12} \pi ^2 \zeta_5
        +\frac{13919}{2} \pi ^2 l_2 \zeta_5
        -\frac{66869}{4} \zeta_3 \zeta_5
        -\frac{231361 \zeta_7}{8}
\biggr)
+\epsilon ^5
\biggl(
        \frac{465751}{6}
        +\frac{3953 \pi ^2}{3}
        \nonumber\\ &
        -\frac{9017 \pi ^4}{30}
        +\frac{4567 \pi ^6}{70}
        +\frac{455579 \pi ^8}{22680}
        -20736 a_4
        -5048 \pi ^2 a_4
        +\frac{7900 \pi ^4 a_4}{9}
        -49600 a_4^2
        \nonumber\\ &
        -35392 a_5
        +11440 \pi ^2 a_5
        +\frac{46856 \pi ^4 a_5}{9}
        -250112 a_4 a_5
        -71936 a_6
        \nonumber\\ &
        -40160 \pi ^2 a_6
        -110336 a_7
        +99392 \pi ^2 a_7
        -636 \pi ^2 l_2
        +\frac{12278 \pi ^4 l_2}{45}
        -\frac{13331}{210} \pi ^6 l_2
        \nonumber\\ &
        +\frac{119841319 \pi ^8 l_2}{84000}
        +12464 \pi ^2 a_4 l_2
        -\frac{445712}{45} \pi ^4 a_4 l_2
        -40160 \pi ^2 a_5 l_2
        +99392 \pi ^2 a_6 l_2
        \nonumber\\ &
        +2080 \pi ^2 l_2^2
        +\frac{7693}{45} \pi ^4 l_2^2
        -\frac{1975}{54} \pi ^6 l_2^2
        -\frac{47840}{3} \pi ^2 a_4 l_2^2
        +\frac{180352}{3} \pi ^2 a_5 l_2^2
        -\frac{7280}{9} \pi ^2 l_2^3
        \nonumber\\ &
        -\frac{57088}{135} \pi ^4 l_2^3
        +\frac{98777}{810} \pi ^6 l_2^3
        +\frac{117824}{9} \pi ^2 a_4 l_2^3
        -864 l_2^4
        -\frac{2933}{9} \pi ^2 l_2^4
        +\frac{19915}{54} \pi ^4 l_2^4
        \nonumber\\ &
        -\frac{12400}{3} a_4 l_2^4
        -\frac{31264}{3} a_5 l_2^4
        +\frac{4424 l_2^5}{15}
        +\frac{25976}{45} \pi ^2 l_2^5
        -\frac{79303}{135} \pi ^4 l_2^5
        +\frac{31264}{15} a_4 l_2^5
        \nonumber\\ &
        -\frac{4496 l_2^6}{45}
        -\frac{3470}{9} \pi ^2 l_2^6
        +\frac{6896 l_2^7}{315}
        +\frac{153872}{945} \pi ^2 l_2^7
        -\frac{775 l_2^8}{9}
        +\frac{3908 l_2^9}{45}
        -19392 s_6
        \nonumber\\ &
        -17960 \pi ^2 s_6
        +\frac{241376 l_2 s_6}{7}
        +\frac{1847792}{21} \pi ^2 l_2 s_6
        +\frac{218848}{3} l_2^3 s_6
        -\frac{241376 s_{7 a}}{7}
        \nonumber\\ &
        +\frac{1058272}{21} \pi ^2 s_{7 a}
        -1591 s_{8 a}
        -\frac{14898273}{20} l_2 s_{8 a}
        +3476912 s_{9 a}
        +\frac{42208 s_{7 b}}{7}
        \nonumber\\ &
        +\frac{997232}{21} \pi ^2 s_{7 b}
        +62528 l_2^2 s_{7 b}
        -4673968 l_2 s_{8 b}
        -1744176 s_{9 b}
        -\frac{2336984 s_{9 c}}{3}
        \nonumber\\ &
        -62528 s_{9 d}
        +312640 s_{9 e}
        +125056 s_{9 f}
        -9218 \zeta_3
        +\frac{817 \pi ^2 \zeta_3}{2}
        +\frac{325943 \pi ^4 \zeta_3}{1260}
        \nonumber\\ &
        -\frac{104344579 \pi ^6 \zeta_3}{15120}
        -28648 a_4 \zeta_3
        +\frac{20960}{3} \pi ^2 a_4 \zeta_3
        +86800 a_5 \zeta_3
        +31264 a_6 \zeta_3
        \nonumber\\ &
        -4273 \pi ^2 l_2 \zeta_3
        +\frac{13825}{18} \pi ^4 l_2 \zeta_3
        +\frac{21692}{3} \pi ^2 l_2^2 \zeta_3
        -\frac{479699}{180} \pi ^4 l_2^2 \zeta_3
        -\frac{41860}{9} \pi ^2 l_2^3 \zeta_3
        \nonumber\\ &
        -\frac{3581}{3} l_2^4 \zeta_3
        +\frac{34256}{9} \pi ^2 l_2^4 \zeta_3
        -\frac{2170}{3} l_2^5 \zeta_3
        +\frac{1954}{45} l_2^6 \zeta_3
        -3828168 s_6 \zeta_3
        +\frac{130813 \zeta_3^2}{12}
        \nonumber\\ &
        +\frac{12907}{2} \pi ^2 \zeta_3^2
        -\frac{301720}{7} l_2 \zeta_3^2
        +\frac{1184933}{21} \pi ^2 l_2 \zeta_3^2
        +20517 l_2^3 \zeta_3^2
        +\frac{6894063 \zeta_3^3}{8}
        -\frac{1281 \zeta_5}{2}
        \nonumber\\ &
        -\frac{1111457}{84} \pi ^2 \zeta_5
        -\frac{1364278829 \pi ^4 \zeta_5}{6720}
        +7555066 a_4 \zeta_5
        +\frac{69595}{2} \pi ^2 l_2 \zeta_5
        -39246 l_2^2 \zeta_5
        \nonumber\\ &
        -\frac{4070633}{12} \pi ^2 l_2^2 \zeta_5
        +\frac{3050645}{12} l_2^4 \zeta_5
        -\frac{337033}{4} \zeta_3 \zeta_5
        +\frac{23077717}{4} l_2 \zeta_3 \zeta_5
        +\frac{4815827 \zeta_7}{56}
        \nonumber\\ &
        -\frac{3014004103 \pi ^2 \zeta_7}{1344}
        -\frac{18160169}{8} l_2^2 \zeta_7
        +\frac{21258293207 \zeta_9}{576}
\biggr)
\Biggr\}
+ \mathcal{O}(\epsilon^6)~,
\end{align}

\begin{align}
  G_{64} &= \Gamma(1+\epsilon)^3
  \Biggl\{
         \frac{2 \zeta_3}{\epsilon }
        +2 \zeta_3
        +\frac{\pi ^2}{3}
        -\frac{7 \pi ^4}{90}
  +\epsilon
  \biggl(
        \frac{14 \pi ^2}{3}
        -\frac{41 \pi ^4}{90}
        -12 \zeta_3
        -\frac{2}{3} \pi ^2 \zeta_3
        +44 \zeta_5
  \biggr)
  \nonumber\\ &
  +\epsilon ^2
  \biggl(
          \frac{119 \pi ^2}{3}
          -\frac{14 \pi ^4}{45}
          +\frac{1447 \pi ^6}{2835}
          -64 \pi ^2 a_4
          -16 \pi ^2 l_2
          +\frac{8}{3} \pi ^4 l_2^2
          -\frac{8}{3} \pi ^2 l_2^4
          -38 \zeta_3
          \nonumber\\ &
          -\frac{26}{3} \pi ^2 \zeta_3
          -56 \pi ^2 l_2 \zeta_3
          +54 \zeta_3^2
          -42 \zeta_5
  \biggr)
  +\epsilon ^3
  \biggl(
          \frac{796 \pi ^2}{3}
          +\frac{51 \pi ^4}{10}
          +\frac{31 \pi ^6}{630}
          +512 a_4
          \nonumber\\ &
          -192 \pi ^2 a_4
          +768 \pi ^2 a_5
          -224 \pi ^2 l_2
          -\frac{112}{135} \pi ^6 l_2
          +768 \pi ^2 a_4 l_2
          +\frac{224}{3} \pi ^2 l_2^2
          +8 \pi ^4 l_2^2
          \nonumber\\ &
          -\frac{64}{3} \pi ^4 l_2^3
          +\frac{64 l_2^4}{3}
          -8 \pi ^2 l_2^4
          +\frac{128}{5} \pi ^2 l_2^5
          +1024 l_2 s_6
          -1024 s_{7 a}
          -1024 s_{7 b}
          +256 \zeta_3
          \nonumber\\ &
          -26 \pi ^2 \zeta_3
          +\frac{931 \pi ^4 \zeta_3}{45}
          -1792 a_4 \zeta_3
          -168 \pi ^2 l_2 \zeta_3
          +\frac{1232}{3} \pi ^2 l_2^2 \zeta_3
          -\frac{224}{3} l_2^4 \zeta_3
          +104 \zeta_3^2
          \nonumber\\ &
          -1280 l_2 \zeta_3^2
          -574 \zeta_5
          -217 \pi ^2 \zeta_5
          -5655 \zeta_7
  \biggr)
  +\epsilon ^4
  \biggl(
          1539 \pi ^2
          +\frac{331 \pi ^4}{15}
          -\frac{10667 \pi ^6}{5670}
          \nonumber\\ &
          +\frac{8360501 \pi ^8}{1701000}
          +7168 a_4
          -320 \pi ^2 a_4
          +\frac{2752 \pi ^4 a_4}{15}
          -12288 a_4^2
          +6144 a_5
          +2304 \pi ^2 a_5
          \nonumber\\ &
          -9216 \pi ^2 a_6
          -1904 \pi ^2 l_2
          +\frac{688 \pi ^4 l_2}{15}
          -\frac{112}{45} \pi ^6 l_2
          +2304 \pi ^2 a_4 l_2
          -9216 \pi ^2 a_5 l_2
          \nonumber\\ &
          +\frac{3136}{3} \pi ^2 l_2^2
          +\frac{40}{3} \pi ^4 l_2^2
          -\frac{344}{45} \pi ^6 l_2^2
          -3584 \pi ^2 a_4 l_2^2
          -\frac{896}{3} \pi ^2 l_2^3
          -64 \pi ^4 l_2^3
          +\frac{896 l_2^4}{3}
          \nonumber\\ &
          -\frac{40}{3} \pi ^2 l_2^4
          +\frac{3704}{45} \pi ^4 l_2^4
          -1024 a_4 l_2^4
          -\frac{256 l_2^5}{5}
          +\frac{384}{5} \pi ^2 l_2^5
          -\frac{256}{3} \pi ^2 l_2^6
          -\frac{64 l_2^8}{3}
          \nonumber\\ &
          -4288 \pi ^2 s_6
          +3072 l_2 s_6
          -3072 s_{7 a}
          -\frac{1452 s_{8 a}}{5}
          -3072 s_{7 b}
          +3750 \zeta_3
          -\frac{58}{3} \pi ^2 \zeta_3
          \nonumber\\ &
          +\frac{1756 \pi ^4 \zeta_3}{45}
          -5376 a_4 \zeta_3
          +21504 a_5 \zeta_3
          -280 \pi ^2 l_2 \zeta_3
          +\frac{2408}{15} \pi ^4 l_2 \zeta_3
          +1232 \pi ^2 l_2^2 \zeta_3
          \nonumber\\ &
          -\frac{3136}{3} \pi ^2 l_2^3 \zeta_3
          -224 l_2^4 \zeta_3
          -\frac{896}{5} l_2^5 \zeta_3
          +36 \zeta_3^2
          +\frac{4472}{3} \pi ^2 \zeta_3^2
          -3840 l_2 \zeta_3^2
          -7046 \zeta_5
          \nonumber\\ &
          -917 \pi ^2 \zeta_5
          +8308 \pi ^2 l_2 \zeta_5
          -20172 \zeta_3 \zeta_5
          -22344 \zeta_7
  \biggr)
  +\epsilon ^5
  \biggl(
          8126 \pi ^2
          -\frac{611 \pi ^4}{18}
          \nonumber\\ &
          -\frac{38155 \pi ^6}{1134}
          +\frac{2534927 \pi ^8}{283500}
          +60928 a_4
          +64 \pi ^2 a_4
          +\frac{2752 \pi ^4 a_4}{5}
          -36864 a_4^2
          +86016 a_5
          \nonumber\\ &
          +3840 \pi ^2 a_5
          +6400 \pi ^4 a_5
          -294912 a_4 a_5
          +73728 a_6
          -27648 \pi ^2 a_6
          +110592 \pi ^2 a_7
          \nonumber\\ &
          -12736 \pi ^2 l_2
          +\frac{9632 \pi ^4 l_2}{15}
          -\frac{112}{27} \pi ^6 l_2
          +\frac{13250626 \pi ^8 l_2}{7875}
          +3840 \pi ^2 a_4 l_2
          -\frac{57088}{5} \pi ^4 a_4 l_2
          \nonumber\\ &
          -27648 \pi ^2 a_5 l_2
          +110592 \pi ^2 a_6 l_2
          +\frac{26656}{3} \pi ^2 l_2^2
          -\frac{4168}{15} \pi ^4 l_2^2
          -\frac{344}{15} \pi ^6 l_2^2
          \nonumber\\ &
          -10752 \pi ^2 a_4 l_2^2
          +67584 \pi ^2 a_5 l_2^2
          -\frac{12544}{3} \pi ^2 l_2^3
          -\frac{320}{3} \pi ^4 l_2^3
          +\frac{2048}{15} \pi ^6 l_2^3
          \nonumber\\ &
          +14336 \pi ^2 a_4 l_2^3
          +\frac{7616 l_2^4}{3}
          +\frac{2696}{3} \pi ^2 l_2^4
          +\frac{3704}{15} \pi ^4 l_2^4
          -3072 a_4 l_2^4
          \nonumber\\ &
          -12288 a_5 l_2^4
          -\frac{3584 l_2^5}{5}
          +128 \pi ^2 l_2^5
          -\frac{3328}{5} \pi ^4 l_2^5
          +\frac{12288}{5} a_4 l_2^5
          +\frac{512 l_2^6}{5}
          -256 \pi ^2 l_2^6
          \nonumber\\ &
          +\frac{17408}{105} \pi ^2 l_2^7
          -64 l_2^8
          +\frac{512 l_2^9}{5}
          +33792 s_6
          -12864 \pi ^2 s_6
          +5120 l_2 s_6
          +\frac{723456}{7} \pi ^2 l_2 s_6
          \nonumber\\ &
          +86016 l_2^3 s_6
          -5120 s_{7 a}
          +\frac{410880}{7} \pi ^2 s_{7 a}
          -\frac{6684 s_{8 a}}{5}
          -\frac{4391712}{5} l_2 s_{8 a}
          +4099072 s_{9 a}
          \nonumber\\ &
          -5120 s_{7 b}
          +\frac{394752}{7} \pi ^2 s_{7 b}
          +73728 l_2^2 s_{7 b}
          -5511168 l_2 s_{8 b}
          -2057216 s_{9 b}
          -918528 s_{9 c}
          \nonumber\\ &
          -73728 s_{9 d}
          +368640 s_{9 e}
          +147456 s_{9 f}
          +28484 \zeta_3
          +\frac{338 \pi ^2 \zeta_3}{3}
          +\frac{262 \pi ^4 \zeta_3}{15}
          \nonumber\\ &
          -\frac{23122573 \pi ^6 \zeta_3}{2835}
          -8960 a_4 \zeta_3
          +7936 \pi ^2 a_4 \zeta_3
          +64512 a_5 \zeta_3
          +36864 a_6 \zeta_3
          +504 \pi ^2 l_2 \zeta_3
          \nonumber\\ &
          +\frac{2408}{5} \pi ^4 l_2 \zeta_3
          +\frac{6160}{3} \pi ^2 l_2^2 \zeta_3
          -\frac{45232}{15} \pi ^4 l_2^2 \zeta_3
          -3136 \pi ^2 l_2^3 \zeta_3
          -\frac{1120}{3} l_2^4 \zeta_3
          \nonumber\\ &
          +\frac{12704}{3} \pi ^2 l_2^4 \zeta_3
          -\frac{2688}{5} l_2^5 \zeta_3
          +\frac{256}{5} l_2^6 \zeta_3
          -4518144 s_6 \zeta_3
          -13492 \zeta_3^2
          +\frac{12916}{3} \pi ^2 \zeta_3^2
          \nonumber\\ &
          -6400 l_2 \zeta_3^2
          +\frac{463088}{7} \pi ^2 l_2 \zeta_3^2
          +24192 l_2^3 \zeta_3^2
          +\frac{3052412 \zeta_3^3}{3}
          -75326 \zeta_5
          -1999 \pi ^2 \zeta_5
          \nonumber\\ &
          -\frac{25177384}{105} \pi ^4 \zeta_5
          +8899072 a_4 \zeta_5
          +24924 \pi ^2 l_2 \zeta_5
          -\frac{1198784}{3} \pi ^2 l_2^2 \zeta_5
          +\frac{898112}{3} l_2^4 \zeta_5
          \nonumber\\ &
          -66888 \zeta_3 \zeta_5
          +6802848 l_2 \zeta_3 \zeta_5
          -50828 \zeta_7
          -\frac{36911121}{14} \pi ^2 \zeta_7
          -2676008 l_2^2 \zeta_7
          \nonumber\\ &
          +\frac{391891853 \zeta_9}{9}
  \biggr)
  \Biggr\}
  + \mathcal{O}(\epsilon^6)~,
\end{align}

\begin{align}
  G_{65} &= \Gamma(1+\epsilon)^3
  \Biggl\{
     \frac{1}{3 \epsilon ^3}
    +\frac{7}{3 \epsilon ^2}
    +\frac{31}{3 \epsilon }
    +\frac{103}{3}
    -\frac{2 \pi ^4}{15}
    -\frac{4}{3} \zeta_3
    +\epsilon
    \biggl(
            \frac{235}{3}
            +\frac{8 \pi ^2}{3}
            -\frac{3 \pi ^4}{5}
            +\frac{32}{3} \zeta_3
            \nonumber\\ &
            +\frac{28}{3} \pi ^2 \zeta_3
            -78 \zeta_5
    \biggr)
    +\epsilon ^2
    \biggl(
            \frac{19}{3}
            +\frac{16 l_2^4}{3}
            +\frac{112 \pi ^2}{3}
            -32 l_2 \pi ^2
            -\frac{16}{3} l_2^2 \pi ^2
            -\frac{16}{3} l_2^4 \pi ^2
            \nonumber\\ &
            -\frac{164 \pi ^4}{45}
            +\frac{16 l_2^2 \pi ^4}{3}
            +\frac{928 \pi ^6}{945}
            +\frac{692}{3} \zeta_3
            +\frac{140}{3} \pi ^2 \zeta_3
            -112 l_2 \pi ^2 \zeta_3
            +169 \zeta_3^2
            -414 \zeta_5
            \nonumber\\ &
            +128 a_4
            -128 \pi ^2 a_4
    \biggr)
    +\epsilon ^3
    \biggl(
            -\frac{3953}{3}
            +96 l_2^4
            -\frac{32 l_2^5}{5}
            +\frac{952 \pi ^2}{3}
            -448 l_2 \pi ^2
            +96 l_2^2 \pi ^2
            \nonumber\\ &
            +\frac{32 l_2^3 \pi ^2}{3}
            -\frac{80}{3} l_2^4 \pi ^2
            +\frac{256 l_2^5 \pi ^2}{5}
            -\frac{269 \pi ^4}{9}
            +\frac{136 l_2 \pi ^4}{15}
            +\frac{80 l_2^2 \pi ^4}{3}
            -\frac{128}{3} l_2^3 \pi ^4
            +\frac{946 \pi ^6}{189}
            \nonumber\\ &
            -\frac{266 l_2 \pi ^6}{135}
            +\frac{6392}{3} \zeta_3
            -\frac{532}{3} l_2^4 \zeta_3
            +\frac{532}{3} \pi ^2 \zeta_3
            -560 l_2 \pi ^2 \zeta_3
            +\frac{2548}{3} l_2^2 \pi ^2 \zeta_3
            +\frac{367}{5} \pi ^4 \zeta_3
            \nonumber\\ &
            +\frac{2519}{3} \zeta_3^2
            -3040 l_2 \zeta_3^2
            -2246 \zeta_5
            -132 \pi ^2 \zeta_5
            -\frac{35591}{2} \zeta_7
            +2304 a_4
            -640 \pi ^2 a_4
            \nonumber\\ &
            +1536 l_2 \pi ^2 a_4
            -4256 \zeta_3 a_4
            +768 a_5
            +1536 \pi ^2 a_5
            +2432 l_2 s_6
            -2432 s_{7 a}
            -2432 s_{7 b}
    \biggr)
    \nonumber\\ &
    +\epsilon ^4
    \biggl(
            -\frac{31889}{3}
            +\frac{2800 l_2^4}{3}
            -\frac{832 l_2^5}{5}
            +\frac{32 l_2^6}{5}
            -\frac{140 l_2^8}{3}
            +\frac{6368 \pi ^2}{3}
            -3808 l_2 \pi ^2
            \nonumber\\ &
            +\frac{5264 l_2^2 \pi ^2}{3}
            -\frac{1472}{3} l_2^3 \pi ^2
            -\frac{352}{3} l_2^4 \pi ^2
            +256 l_2^5 \pi ^2
            -\frac{488}{3}l_2^6 \pi ^2
            -\frac{10162 \pi ^4}{45}
            +\frac{912 l_2 \pi ^4}{5}
            \nonumber\\ &
            +\frac{1112 l_2^2 \pi ^4}{15}
            -\frac{640}{3} l_2^3 \pi ^4
            +\frac{1466 l_2^4 \pi ^4}{9}
            +\frac{2164 \pi ^6}{135}
            -\frac{266 l_2 \pi ^6}{27}
            -\frac{158}{9} l_2^2 \pi ^6
            +\frac{128117 \pi ^8}{13500}
            \nonumber\\ &
            +\frac{43964}{3} \zeta_3
            -\frac{2660}{3} l_2^4 \zeta_3
            -392 l_2^5 \zeta_3
            +532 \pi ^2 \zeta_3
            -2128 l_2 \pi ^2 \zeta_3
            +\frac{12740}{3} l_2^2 \pi ^2 \zeta_3
            \nonumber\\ &
            -\frac{6104}{3} l_2^3 \pi ^2 \zeta_3
            +\frac{5513}{15} \pi ^4 \zeta_3
            +\frac{1106}{3} l_2 \pi ^4 \zeta_3
            +\frac{7097}{3} \zeta_3^2
            -15200 l_2 \zeta_3^2
            +\frac{8936}{3} \pi ^2 \zeta_3^2
            \nonumber\\ &
            -22734 \zeta_5
            -660 \pi ^2 \zeta_5
            +15872 l_2 \pi ^2 \zeta_5
            -42455 \zeta_3 \zeta_5
            -\frac{178619}{2} \zeta_7
            +22400 a_4
            \nonumber\\ &
            -2240 l_2^4 a_4
            -2432 \pi ^2 a_4
            +7680 l_2 \pi ^2 a_4
            -6976 l_2^2 \pi ^2 a_4
            +\frac{1264 \pi ^4 a_4}{3}
            -21280 \zeta_3 a_4
            \nonumber\\ &
            -26880 a_4^2
            +19968 a_5
            +7680 \pi ^2 a_5
            -18432 l_2 \pi ^2 a_5
            +47040 \zeta_3 a_5
            +4608 a_6
            \nonumber\\ &
            -18432 \pi ^2 a_6
            +1920 s_6
            +12160 l_2 s_6
            -8192 \pi ^2 s_6
            -12160 s_{7 a}
            -\frac{2884 s_{8 a}}{5}
            \nonumber\\ &
            -12160 s_{7 b}
    \biggr)
    +\epsilon ^5
    \biggl(
            -\frac{188141}{3}
            +6784 l_2^4
            -\frac{9184 l_2^5}{5}
            +\frac{1344 l_2^6}{5}
            -\frac{192 l_2^7}{35}
            -\frac{700 l_2^8}{3}
            \nonumber\\ &
            +\frac{1072 l_2^9}{5}
            +12312 \pi ^2
            -25472 l_2 \pi ^2
            +16064 l_2^2 \pi ^2
            -\frac{23072}{3} l_2^3 \pi ^2
            +\frac{3856 l_2^4 \pi ^2}{3}
            +992 l_2^5 \pi ^2
            \nonumber\\ &
            -\frac{2440}{3} l_2^6 \pi ^2
            +\frac{32128 l_2^7 \pi ^2}{105}
            -\frac{66227 \pi ^4}{45}
            +\frac{27832 l_2 \pi ^4}{15}
            -\frac{7136}{15} l_2^2 \pi ^4
            -\frac{11344}{15} l_2^3 \pi ^4
            \nonumber\\ &
            +\frac{7330 l_2^4 \pi ^4}{9}
            -\frac{6844}{5} l_2^5 \pi ^4
            -\frac{16462 \pi ^6}{315}
            -\frac{3142 l_2 \pi ^6}{135}
            -\frac{790}{9} l_2^2 \pi ^6
            +\frac{1446 l_2^3 \pi ^6}{5}
            +\frac{128429 \pi ^8}{2700}
            \nonumber\\ &
            +\frac{222279223 l_2 \pi ^8}{63000}
            +\frac{257840}{3} \zeta_3
            -\frac{10108}{3} l_2^4 \zeta_3
            -1960 l_2^5 \zeta_3
            +\frac{536}{5} l_2^6 \zeta_3
            +924 \pi ^2 \zeta_3
            \nonumber\\ &
            -6384 l_2 \pi ^2 \zeta_3
            +\frac{48412}{3} l_2^2 \pi ^2 \zeta_3
            -\frac{30520}{3} l_2^3 \pi ^2 \zeta_3
            +\frac{25444}{3} l_2^4 \pi ^2 \zeta_3
            +\frac{50723}{35} \pi ^4 \zeta_3
            \nonumber\\ &
            +\frac{5530}{3} l_2 \pi ^4 \zeta_3
            -\frac{95387}{15} l_2^2 \pi ^4 \zeta_3
            -\frac{7166381}{420} \pi ^6 \zeta_3
            -\frac{70981}{3} \zeta_3^2
            -\frac{357920}{7} l_2 \zeta_3^2
            +50652 l_2^3 \zeta_3^2
            \nonumber\\ &
            +\frac{44680}{3} \pi ^2 \zeta_3^2
            +\frac{958072}{7} l_2 \pi ^2 \zeta_3^2
            +\frac{12770399}{6} \zeta_3^3
            -209918 \zeta_5
            +11160 l_2^2 \zeta_5
            \nonumber\\ &
            +\frac{1882375}{3} l_2^4 \zeta_5
            +\frac{12512}{21} \pi ^2 \zeta_5
            +79360 l_2 \pi ^2 \zeta_5
            -\frac{2511907}{3} l_2^2 \pi ^2 \zeta_5
            -\frac{508541849 \pi ^4 \zeta_5}{1008}
            \nonumber\\ &
            -212467 \zeta_3 \zeta_5
            +14243463 l_2 \zeta_3 \zeta_5
            -\frac{5534359}{14} \zeta_7
            -\frac{11188067}{2} l_2^2 \zeta_7
            -\frac{1842855461}{336} \pi ^2 \zeta_7
            \nonumber\\ &
            +\frac{4372301119}{48} \zeta_9
            +162816 a_4
            -11200 l_2^4 a_4
            +\frac{25728 l_2^5 a_4}{5}
            -8320 \pi ^2 a_4
            +29184 l_2 \pi ^2 a_4
            \nonumber\\ &
            -34880 l_2^2 \pi ^2 a_4
            +28288 l_2^3 \pi ^2 a_4
            +\frac{6320 \pi ^4 a_4}{3}
            -\frac{120128}{5} l_2 \pi ^4 a_4
            -80864 \zeta_3 a_4
            \nonumber\\ &
            +16448 \pi ^2 \zeta_3 a_4
            +18648056 \zeta_5 a_4
            -134400 a_4^2
            +220416 a_5
            -25728 l_2^4 a_5
            +29184 \pi ^2 a_5
            \nonumber\\ &
            -92160 l_2 \pi ^2 a_5
            +136320 l_2^2 \pi ^2 a_5
            +13280 \pi ^4 a_5
            +235200 \zeta_3 a_5
            -617472 a_4 a_5
            \nonumber\\ &
            +193536 a_6
            -92160 \pi ^2 a_6
            +221184 l_2 \pi ^2 a_6
            +77184 \zeta_3 a_6
            +27648 a_7
            +221184 \pi ^2 a_7
            \nonumber\\ &
            +86784 s_6
            +\frac{286336 l_2 s_6}{7}
            +180096 l_2^3 s_6
            -40960 \pi ^2 s_6
            +\frac{1477440}{7} l_2 \pi ^2 s_6
            \nonumber\\ &
            -9454176 \zeta_3 s_6
            -\frac{286336 s_{7 a}}{7}
            +\frac{831552}{7} \pi ^2 s_{7 a}
            -2884 s_{8 a}
            -\frac{9195147}{5} l_2 s_{8 a}
            \nonumber\\ &
            +8573504 s_{9 a}
            -\frac{366976 s_{7 b}}{7}
            +154368 l_2^2 s_{7 b}
            +\frac{863808}{7} \pi ^2 s_{7 b}
            -11539008 l_2 s_{8 b}
            \nonumber\\ &
            -4316224 s_{9 b}
            -1923168 s_{9 c}
            -154368 s_{9 d}
            +771840 s_{9 e}
            +308736 s_{9 f}
    \biggr)
  \Biggr\}
  + \mathcal{O}(\epsilon^6)~,
\end{align}

\begin{align}
  G_{66} &= \Gamma(1+\epsilon)^3
  \Biggl\{
         \frac{2 \zeta_3}{\epsilon }
        +10 \zeta_3
        -\frac{\pi ^2}{3}
        -\frac{13 \pi ^4}{90}
  +\epsilon
  \biggl(
        -\frac{4 \pi ^2}{3}
        -\frac{13 \pi ^4}{18}
        +4 \pi ^2 l_2
        +24 \zeta_3
        \nonumber\\ &
        +\frac{49 \pi ^2 \zeta_3}{6}
        -\frac{85 \zeta_5}{2}
  \biggr)
  +\epsilon ^2 \biggl(
          7 \pi ^2
          -\frac{217 \pi ^4}{90}
          +\frac{1751 \pi ^6}{1890}
          -160 a_4
          -120 \pi ^2 a_4
          +16 \pi ^2 l_2
          \nonumber\\ &
          -\frac{40}{3} \pi ^2 l_2^2
          +5 \pi ^4 l_2^2
          -\frac{20 l_2^4}{3}
          -5 \pi ^2 l_2^4
          +42 \zeta_3
          +\frac{245 \pi ^2 \zeta_3}{6}
          -105 \pi ^2 l_2 \zeta_3
          +\frac{495 \zeta_3^2}{4}
          -\frac{425 \zeta_5}{2}
  \biggr)
  \nonumber\\ &
  +\epsilon ^3 \biggl(
        \frac{394 \pi ^2}{3}
        -\frac{223 \pi ^4}{18}
        +\frac{1751 \pi ^6}{378}
        -832 a_4
        -600 \pi ^2 a_4
        -1472 a_5
        +1424 \pi ^2 a_5
        -84 \pi ^2 l_2
        \nonumber\\ &
        -\frac{182}{45} \pi ^4 l_2
        -\frac{133}{90} \pi ^6 l_2
        +1424 \pi ^2 a_4 l_2
        -\frac{16}{3} \pi ^2 l_2^2
        +25 \pi ^4 l_2^2
        +\frac{368}{9} \pi ^2 l_2^3
        -\frac{356}{9} \pi ^4 l_2^3
        \nonumber\\ &
        -\frac{104 l_2^4}{3}
        -25 \pi ^2 l_2^4
        +\frac{184 l_2^5}{15}
        +\frac{712}{15} \pi ^2 l_2^5
        +1824 l_2 s_6
        -1824 s_{7 a}
        -1824 s_{7 b}
        +268 \zeta_3
        \nonumber\\ &
        +\frac{313 \pi ^2 \zeta_3}{2}
        +\frac{9259 \pi ^4 \zeta_3}{180}
        -3192 a_4 \zeta_3
        -525 \pi ^2 l_2 \zeta_3
        +756 \pi ^2 l_2^2 \zeta_3
        -133 l_2^4 \zeta_3
        +\frac{2475 \zeta_3^2}{4}
        \nonumber\\ &
        -2280 l_2 \zeta_3^2
        +\frac{431 \zeta_5}{2}
        -\frac{3841}{12} \pi ^2 \zeta_5
        -\frac{105133 \zeta_7}{8}
\biggr)
  +\epsilon ^4 \biggl(
          \frac{3343 \pi ^2}{3}
          -\frac{8833 \pi ^4}{90}
          +\frac{48437 \pi ^6}{1890}
          \nonumber\\ &
          +\frac{1024391 \pi ^8}{113400}
          +672 a_4
          -2408 \pi ^2 a_4
          +\frac{16156 \pi ^4 a_4}{45}
          -22208 a_4^2
          -4480 a_5
          +7120 \pi ^2 a_5
          \nonumber\\ &
          -12928 a_6
          -17248 \pi ^2 a_6
          -1576 \pi ^2 l_2
          +\frac{388 \pi ^4 l_2}{9}
          -\frac{133}{18} \pi ^6 l_2
          +7120 \pi ^2 a_4 l_2
          \nonumber\\ &
          -17248 \pi ^2 a_5 l_2
          +952 \pi ^2 l_2^2
          +\frac{5429}{45} \pi ^4 l_2^2
          -\frac{4039}{270} \pi ^6 l_2^2
          -\frac{20320}{3} \pi ^2 a_4 l_2^2
          -\frac{1184}{9} \pi ^2 l_2^3
          \nonumber\\ &
          -\frac{1780}{9} \pi ^4 l_2^3
          +28 l_2^4
          -\frac{1711}{9} \pi ^2 l_2^4
          +\frac{42139}{270} \pi ^4 l_2^4
          -\frac{5552}{3} a_4 l_2^4
          +\frac{112 l_2^5}{3}
          +\frac{712}{3} \pi ^2 l_2^5
          \nonumber\\ &
          -\frac{808 l_2^6}{45}
          -\frac{1462}{9} \pi ^2 l_2^6
          -\frac{347 l_2^8}{9}
          -4704 s_6
          -7688 \pi ^2 s_6
          +9120 l_2 s_6
          -9120 s_{7 a}
          \nonumber\\ &
          -547 s_{8 a}
          -9120 s_{7 b}
          +3038 \zeta_3
          +\frac{2881 \pi ^2 \zeta_3}{6}
          +\frac{9259 \pi ^4 \zeta_3}{36}
          -15960 a_4 \zeta_3
          +38864 a_5 \zeta_3
          \nonumber\\ &
          -2131 \pi ^2 l_2 \zeta_3
          +\frac{28273}{90} \pi ^4 l_2 \zeta_3
          +3780 \pi ^2 l_2^2 \zeta_3
          -\frac{17780}{9} \pi ^2 l_2^3 \zeta_3
          -665 l_2^4 \zeta_3
          -\frac{4858}{15} l_2^5 \zeta_3
          \nonumber\\ &
          +\frac{16965 \zeta_3^2}{4}
          +\frac{16661}{6} \pi ^2 \zeta_3^2
          -11400 l_2 \zeta_3^2
          -\frac{3721 \zeta_5}{2}
          -\frac{19205}{12} \pi ^2 \zeta_5
          +\frac{29791}{2} \pi ^2 l_2 \zeta_5
          \nonumber\\ &
          -\frac{143445}{4} \zeta_3 \zeta_5
          -\frac{525665 \zeta_7}{8}
  \biggr)
  +\epsilon ^5 \biggl(
          7264 \pi ^2
          -\frac{13375 \pi ^4}{18}
          +\frac{122261 \pi ^6}{1890}
          +\frac{1024391 \pi ^8}{22680}
          \nonumber\\ &
          +40192 a_4
          -9208 \pi ^2 a_4
          +\frac{16156 \pi ^4 a_4}{9}
          -111040 a_4^2
          +50624 a_5
          +26032 \pi ^2 a_5
          \nonumber\\ &
          +\frac{104456 \pi ^4 a_5}{9}
          -545024 a_4 a_5
          +1792 a_6
          -86240 \pi ^2 a_6
          -110336 a_7
          +209984 \pi ^2 a_7
          \nonumber\\ &
          -13372 \pi ^2 l_2
          +\frac{41174 \pi ^4 l_2}{45}
          -\frac{149771 \pi ^6 l_2}{1890}
          +\frac{783894389 \pi ^8 l_2}{252000}
          +27056 \pi ^2 a_4 l_2
          \nonumber\\ &
          -\frac{959504}{45} \pi ^4 a_4 l_2
          -86240 \pi ^2 a_5 l_2
          +209984 \pi ^2 a_6 l_2
          +\frac{32896}{3} \pi ^2 l_2^2
          +\frac{3109}{45} \pi ^4 l_2^2
          \nonumber\\ &
          -\frac{4039}{54} \pi ^6 l_2^2
          -\frac{101600}{3} \pi ^2 a_4 l_2^2
          +\frac{383104}{3} \pi ^2 a_5 l_2^2
          -\frac{44912}{9} \pi ^2 l_2^3
          -\frac{111808}{135} \pi ^4 l_2^3
          \nonumber\\ &
          +\frac{209369}{810} \pi ^6 l_2^3
          +\frac{246848}{9} \pi ^2 a_4 l_2^3
          +\frac{5024 l_2^4}{3}
          +\frac{3571}{9} \pi ^2 l_2^4
          +\frac{42139}{54} \pi ^4 l_2^4
          -\frac{27760}{3} a_4 l_2^4
          \nonumber\\ &
          -\frac{68128}{3} a_5 l_2^4
          -\frac{6328 l_2^5}{15}
          +\frac{47864}{45} \pi ^2 l_2^5
          -\frac{169159}{135} \pi ^4 l_2^5
          +\frac{68128}{15} a_4 l_2^5
          +\frac{112 l_2^6}{45}
          \nonumber\\ &
          -\frac{7310}{9} \pi ^2 l_2^6
          +\frac{6896 l_2^7}{315}
          +\frac{310544}{945} \pi ^2 l_2^7
          -\frac{1735 l_2^8}{9}
          +\frac{8516 l_2^9}{45}
          +14400 s_6
          -38440 \pi ^2 s_6
          \nonumber\\ &
          +\frac{377568 l_2 s_6}{7}
          +\frac{3990512}{21} \pi ^2 l_2 s_6
          +\frac{476896}{3} l_2^3 s_6
          -\frac{377568 s_{7 a}}{7}
          +\frac{2270176}{21} \pi ^2 s_{7 a}
          \nonumber\\ &
          -2735 s_{8 a}
          -\frac{32465121}{20} l_2 s_{8 a}
          +7572912 s_{9 a}
          -\frac{93984 s_{7 b}}{7}
          +\frac{2209136}{21} \pi ^2 s_{7 b}
          \nonumber\\ &
          +136256 l_2^2 s_{7 b}
          -10185136 l_2 s_{8 b}
          -3804464 s_{9 b}
          -\frac{5092568 s_{9 c}}{3}
          -136256 s_{9 d}
          \nonumber\\ &
          +681280 s_{9 e}
          +272512 s_{9 f}
          +25448 \zeta_3
          +\frac{5467 \pi ^2 \zeta_3}{6}
          +\frac{1071079 \pi ^4 \zeta_3}{1260}
          \nonumber\\ &
          -\frac{227461763 \pi ^6 \zeta_3}{15120}
          -62696 a_4 \zeta_3
          +\frac{44768}{3} \pi ^2 a_4 \zeta_3
          +194320 a_5 \zeta_3
          +68128 a_6 \zeta_3
          \nonumber\\ &
          -7465 \pi ^2 l_2 \zeta_3
          +\frac{28273}{18} \pi ^4 l_2 \zeta_3
          +\frac{45100}{3} \pi ^2 l_2^2 \zeta_3
          -\frac{1022483}{180} \pi ^4 l_2^2 \zeta_3
          -\frac{88900}{9} \pi ^2 l_2^3 \zeta_3
          \nonumber\\ &
          -\frac{7837}{3} l_2^4 \zeta_3
          +\frac{72368}{9} \pi ^2 l_2^4 \zeta_3
          -\frac{4858}{3} l_2^5 \zeta_3
          +\frac{4258}{45} l_2^6 \zeta_3
          -8344008 s_6 \zeta_3
          +\frac{12303 \zeta_3^2}{4}
          \nonumber\\ &
          +\frac{83305}{6} \pi ^2 \zeta_3^2
          -\frac{471960}{7} l_2 \zeta_3^2
          +\frac{2566421}{21} \pi ^2 l_2 \zeta_3^2
          +44709 l_2^3 \zeta_3^2
          +\frac{15028463 \zeta_3^3}{8}
          \nonumber\\ &
          -\frac{150221 \zeta_5}{2}
          -\frac{1324537}{84} \pi ^2 \zeta_5
          -\frac{8938707847 \pi ^4 \zeta_5}{20160}
          +16460282 a_4 \zeta_5
          +\frac{148955}{2} \pi ^2 l_2 \zeta_5
          \nonumber\\ &
          -39246 l_2^2 \zeta_5
          -\frac{8868841}{12} \pi ^2 l_2^2 \zeta_5
          +\frac{6646165}{12} l_2^4 \zeta_5
          -\frac{717225}{4} \zeta_3 \zeta_5
          +\frac{50289109}{4} l_2 \zeta_3 \zeta_5
          \nonumber\\ &
          -\frac{2869501 \zeta_7}{56}
          -\frac{6548672647 \pi ^2 \zeta_7}{1344}
          -\frac{39543849}{8} l_2^2 \zeta_7
          +\frac{46315776599 \zeta_9}{576}
  \big)
  \Biggr\}
  + \mathcal{O}(\epsilon^6)~,
\end{align}

\begin{align}
  G_{7} &= \Gamma(1+\epsilon)^3
  \Biggl\{
     2 \pi ^2 \zeta_3
    -5 \zeta_5
    +\epsilon
    \biggl(
        \frac{16 \pi ^6}{189}
        +4 \pi ^2 \zeta_3
        +7 \zeta_3^2
        -10 \zeta_5
    \biggr)
    +\epsilon ^2
    \biggl(
        \frac{32 \pi ^6}{189}
        +8 \pi ^2 \zeta_3
        \nonumber\\ &
        -\frac{181}{30} \pi ^4 \zeta_3
        +14 \zeta_3^2
        -20 \zeta_5
        +166 \pi ^2 \zeta_5
        -212 \zeta_7
    \biggr)
    +\epsilon ^3
    \biggl(
        \frac{64 \pi ^6}{189}
        +\frac{6241 \pi ^8}{8400}
        +16 \pi ^2 \zeta_3
        \nonumber\\ &
        -\frac{181}{15} \pi ^4 \zeta_3
        +28 \zeta_3^2
        -10 \pi ^2 \zeta_3^2
        -40 \zeta_5
        +332 \pi ^2 \zeta_5
        -1398 \zeta_3 \zeta_5
        -424 \zeta_7
        -728 s_{8 a}
    \biggr)
    \nonumber\\ &
    +\epsilon ^4
    \biggl(
        \frac{128 \pi ^6}{189}
        +\frac{6241 \pi ^8}{4200}
        +32 \pi ^2 \zeta_3
        -\frac{362}{15} \pi ^4 \zeta_3
        -\frac{21449}{945} \pi ^6 \zeta_3
        +56 \zeta_3^2
        -20 \pi ^2 \zeta_3^2
        \nonumber \\ &
        -\frac{1006}{3} \zeta_3^3
        -80 \zeta_5
        +664 \pi ^2 \zeta_5
        +\frac{131}{2} \pi ^4 \zeta_5
        -2796 \zeta_3 \zeta_5
        -848 \zeta_7
        +5364 \pi ^2 \zeta_7
        \nonumber \\ &
        -\frac{13255}{3} \zeta_9
        -1456 s_{8 a}
    \biggr)
  \Biggr\}
  + \mathcal{O}(\epsilon^5)~,
\end{align}

\begin{align}
  G_{8} &= \Gamma(1+\epsilon)^3
  \Biggl\{
     4 \pi ^2 l_2^2
    -\frac{\pi ^4}{6}
    +\epsilon  \biggl(
            \frac{2 \pi ^4}{3}
            +\frac{17}{6} \pi ^2 \zeta_3
            -291 \zeta_5
            +256 a_5
            +\frac{166 \pi ^4 l_2}{45}
            -16 \pi ^2 l_2^2
            \nonumber \\ &
            -\frac{208}{9} \pi ^2 l_2^3
            -\frac{32 l_2^5}{15}
    \biggr)
    +\epsilon ^2 \biggl(
            -\frac{14 \pi ^4}{3}
            -\frac{21743 \pi ^6}{11340}
            -\frac{34}{3} \pi ^2 \zeta_3
            -953 \zeta_3^2
            +1164 \zeta_5
            -104 \pi ^2 a_4
            \nonumber \\ &
            -1024 a_5
            +5120 a_6
            -\frac{664}{45} \pi ^4 l_2
            -51 \pi ^2 \zeta_3 l_2
            +112 \pi ^2 l_2^2
            -\frac{197}{9} \pi ^4 l_2^2
            +\frac{832}{9} \pi ^2 l_2^3
            +\frac{713}{9} \pi ^2 l_2^4
            \nonumber \\ &
            +\frac{128 l_2^5}{15}
            +\frac{64 l_2^6}{9}
            +2688 s_6
    \biggr)
    +\epsilon ^3 \biggl(
            \frac{80 \pi ^4}{3}
            +\frac{21743 \pi ^6}{2835}
            +\frac{238}{3} \pi ^2 \zeta_3
            +\frac{4003}{21} \pi ^4 \zeta_3
            +3812 \zeta_3^2
            \nonumber \\ &
            -8148 \zeta_5
            +\frac{875561}{84} \pi ^2 \zeta_5
            -\frac{1325727}{7} \zeta_7
            +416 \pi ^2 a_4
            -2528 \zeta_3 a_4
            +7168 a_5
            +1520 \pi ^2 a_5
            \nonumber \\ &
            -20480 a_6
            +77824 a_7
            +\frac{4648 \pi ^4 l_2}{45}
            +\frac{4868 \pi ^6 l_2}{189}
            +204 \pi ^2 \zeta_3 l_2
            +\frac{133600}{7} \zeta_3^2 l_2
            +1776 \pi ^2 a_4 l_2
            \nonumber \\ &
            -640 \pi ^2 l_2^2
            +\frac{788}{9} \pi ^4 l_2^2
            +\frac{2167}{3} \pi ^2 \zeta_3 l_2^2
            +37200 \zeta_5 l_2^2
            -\frac{5824}{9} \pi ^2 l_2^3
            +\frac{8492}{135} \pi ^4 l_2^3
            -\frac{2852}{9} \pi ^2 l_2^4
            \nonumber \\ &
            -\frac{316}{3} \zeta_3 l_2^4
            -\frac{896 l_2^5}{15}
            -\frac{7288}{45} \pi ^2 l_2^5
            -\frac{256 l_2^6}{9}
            -\frac{4864 l_2^7}{315}
            -10752 s_6
            -\frac{106880}{7} l_2 s_6
            +\frac{106880 s_{7 a}}{7}
            \nonumber \\ &
            -\frac{161920 s_{7 b}}{7}
    \biggr)
    +\epsilon ^4 \biggl(
            -\frac{488 \pi ^4}{3}
            -\frac{21743 \pi ^6}{405}
            -\frac{393464657 \pi ^8}{11907000}
            -\frac{1360}{3} \pi ^2 \zeta_3
            -\frac{16012}{21} \pi ^4 \zeta_3
            \nonumber \\ &
            -26684 \zeta_3^2
            -\frac{664737}{14} \pi ^2 \zeta_3^2
            +46560 \zeta_5
            -\frac{875561}{21} \pi ^2 \zeta_5
            +\frac{2191601}{7} \zeta_3 \zeta_5
            +\frac{5302908}{7} \zeta_7
            \nonumber \\ &
            -2912 \pi ^2 a_4
            +\frac{197728 \pi ^4 a_4}{105}
            +10112 \zeta_3 a_4
            -28416 a_4^2
            -40960 a_5
            -6080 \pi ^2 a_5
            \nonumber \\ &
            -\frac{2764864}{7} \zeta_3 a_5
            +143360 a_6
            -31136 \pi ^2 a_6
            -311296 a_7
            +1064960 a_8
            -\frac{5312}{9} \pi ^4 l_2
            \nonumber \\ &
            -\frac{19472}{189} \pi ^6 l_2
            -1428 \pi ^2 \zeta_3 l_2
            -\frac{2650744}{315} \pi ^4 \zeta_3 l_2
            -\frac{534400}{7} \zeta_3^2 l_2
            -\frac{1705401}{14} \pi ^2 \zeta_5 l_2
            \nonumber \\ &
            -7104 \pi ^2 a_4 l_2
            -29088 \pi ^2 a_5 l_2
            +3904 \pi ^2 l_2^2
            -\frac{5516}{9} \pi ^4 l_2^2
            +\frac{49876}{945} \pi ^6 l_2^2
            -\frac{8668}{3} \pi ^2 \zeta_3 l_2^2
            \nonumber \\ &
            -\frac{285120}{7} \zeta_3^2 l_2^2
            -148800 \zeta_5 l_2^2
            -12176 \pi ^2 a_4 l_2^2
            +\frac{33280}{9} \pi ^2 l_2^3
            -\frac{33968}{135} \pi ^4 l_2^3
            -\frac{557750}{63} \pi ^2 \zeta_3 l_2^3
            \nonumber \\ &
            +\frac{19964}{9} \pi ^2 l_2^4
            -\frac{28297}{945} \pi ^4 l_2^4
            +\frac{1264}{3} \zeta_3 l_2^4
            -2368 a_4 l_2^4
            +\frac{1024 l_2^5}{3}
            +\frac{29152}{45} \pi ^2 l_2^5
            +\frac{345608}{105} \zeta_3 l_2^5
            \nonumber \\ &
            +\frac{1792 l_2^6}{9}
            +\frac{4556}{27} \pi ^2 l_2^6
            +\frac{19456 l_2^7}{315}
            -\frac{1444 l_2^8}{63}
            +75264 s_6
            -\frac{899720}{7} \pi ^2 s_6
            +\frac{427520 l_2 s_6}{7}
            \nonumber \\ &
            -\frac{1013760}{7} l_2^2 s_6
            -\frac{427520 s_{7 a}}{7}
            +\frac{14237854 s_{8 a}}{35}
            +\frac{647680 s_{7 b}}{7}
            +\frac{17512704 s_{8 b}}{7}
            +\frac{2027520 s_{8 c}}{7}
            \nonumber \\ &
            +\frac{1542144 s_{8 d}}{7}
    \biggr)
    +\epsilon ^5
    \biggl(
        \frac{2912 \pi ^4}{3}
        +\frac{173944 \pi ^6}{567}
        +\frac{393464657 \pi ^8}{2976750}
        +\frac{8296}{3} \pi ^2 \zeta_3
        +\frac{16012}{3} \pi ^4 \zeta_3
        \nonumber \\ &
        -\frac{1000756616 \pi ^6 \zeta_3}{19845}
        +152480 \zeta_3^2
        +\frac{1329474}{7} \pi ^2 \zeta_3^2
        +\frac{52173865}{6} \zeta_3^3
        -284016 \zeta_5
        \nonumber \\ &
        +\frac{875561}{3} \pi ^2 \zeta_5
        -\frac{71023973521 \pi ^4 \zeta_5}{35280}
        -\frac{8766404}{7} \zeta_3 \zeta_5
        -5302908 \zeta_7
        -\frac{32711242967 \pi ^2 \zeta_7}{2352}
        \nonumber \\ &
        +\frac{89086967387}{336} \zeta_9
        +16640 \pi ^2 a_ 4
        -\frac{790912}{105} \pi ^4 a_ 4
        -70784 \zeta_3 a_ 4
        -\frac{5581776}{7} \pi ^2 \zeta_3 a_ 4
        \nonumber \\ &
        +\frac{482384376}{7} \zeta_5 a_ 4
        +113664 a_ 4^2
        +249856 a_ 5
        +42560 \pi ^2 a_ 5
        +\frac{8900032 \pi ^4 a_ 5}{105}
        \nonumber \\ &
        +\frac{11059456}{7} \zeta_3 a_ 5
        -930816 a_ 4 a_ 5
        -819200 a_ 6
        +124544 \pi ^2 a_ 6
        -\frac{39031424}{7} \zeta_3 a_ 6
        +2179072 a_ 7
        \nonumber \\ &
        +397760 \pi ^2 a_ 7
        -4259840 a_ 8
        +13828096 a_ 9
        +\frac{162016 \pi ^4 l_ 2}{45}
        +\frac{19472 \pi ^6 l_ 2}{27}
        \nonumber \\ &
        +\frac{1264935019 \pi ^8 l_ 2}{88200}
        +8160 \pi ^2 \zeta_3 l_ 2
        +\frac{10602976}{315} \pi ^4 \zeta_3 l_ 2
        +534400 \zeta_3^2 l_ 2
        -\frac{10368279}{49} \pi ^2 \zeta_3^2 l_ 2
        \nonumber \\ &
        +\frac{3410802}{7} \pi ^2 \zeta_5 l_ 2
        +\frac{376591815}{7} \zeta_3 \zeta_5 l_ 2
        +49728 \pi ^2 a_ 4 l_ 2
        -\frac{190624}{5} \pi ^4 a_ 4 l_ 2
        +116352 \pi ^2 a_ 5 l_ 2
        \nonumber \\ &
        +414144 \pi ^2 a_ 6 l_ 2
        -23296 \pi ^2 l_ 2^2
        +\frac{31520}{9} \pi ^4 l_ 2^2
        -\frac{199504}{945} \pi ^6 l_ 2^2
        +\frac{60676}{3} \pi ^2 \zeta_3 l_ 2^2
        \nonumber \\ &
        +\frac{17272063}{315} \pi ^4 \zeta_3 l_ 2^2
        +\frac{1140480}{7} \zeta_3^2 l_ 2^2
        +1041600 \zeta_5 l_ 2^2
        -\frac{25559713}{7} \pi ^2 \zeta_5 l_ 2^2
        -\frac{338072499}{14} \zeta_7 l_ 2^2
        \nonumber \\ &
        +48704 \pi ^2 a_ 4 l_ 2^2
        +245856 \pi ^2 a_ 5 l_ 2^2
        -\frac{203008}{9} \pi ^2 l_ 2^3
        +\frac{237776}{135} \pi ^4 l_ 2^3
        +\frac{3211342 \pi ^6 l_ 2^3}{2835}
        \nonumber \\ &
        +\frac{2231000}{63} \pi ^2 \zeta_3 l_ 2^3
        +\frac{1138716}{7} \zeta_3^2 l_ 2^3
        +56096 \pi ^2 a_ 4 l_ 2^3
        -\frac{114080}{9} \pi ^2 l_ 2^4
        +\frac{113188}{945} \pi ^4 l_ 2^4
        \nonumber \\ &
        -\frac{8848}{3} \zeta_3 l_ 2^4
        -\frac{117077}{63} \pi ^2 \zeta_3 l_ 2^4
        +\frac{18521325}{7} \zeta_5 l_ 2^4
        +9472 a_ 4 l_ 2^4
        -38784 a_ 5 l_ 2^4
        -\frac{31232 l_ 2^5}{15}
        \nonumber \\ &
        -\frac{204064}{45} \pi ^2 l_ 2^5
        -\frac{9436108 \pi ^4 l_ 2^5}{4725}
        -\frac{1382432}{105} \zeta_3 l_ 2^5
        +\frac{38784}{5} a_ 4 l_ 2^5
        -\frac{10240 l_ 2^6}{9}
        -\frac{18224}{27} \pi ^2 l_ 2^6
        \nonumber \\ &
        -\frac{2439464}{315} \zeta_3 l_ 2^6
        -\frac{19456 l_ 2^7}{45}
        -\frac{9104}{135} \pi ^2 l_ 2^7
        +\frac{5776 l_ 2^8}{63}
        +\frac{161648 l_ 2^9}{567}
        -430080 s_ 6
        +\frac{3598880 \pi ^2 s_ 6}{7}
        \nonumber \\ &
        -37566624 \zeta_3 s_ 6
        -427520 l_ 2 s_ 6
        +\frac{41516192}{49} \pi ^2 l_ 2 s_ 6
        +\frac{4055040}{7} l_ 2^2 s_ 6
        +\frac{4048768}{7} l_ 2^3 s_ 6
        \nonumber \\ &
        +427520 s_{7 a}
        -\frac{9030480}{49} \pi ^2 s_{7 a}
        -\frac{56951416 s_{8 a}}{35}
        -\frac{48623247}{7} l_ 2 s_{8 a}
        +\frac{201342528 s_{9 a}}{7}
        \nonumber \\ &
        -647680 s_{7 b}
        +\frac{48783584}{49} \pi ^2 s_{7 b}
        +232704 l_ 2^2 s_{7 b}
        -\frac{70050816 s_{8 b}}{7}
        -\frac{305087040}{7} l_ 2 s_{8 b}
        \nonumber \\ &
        -14695872 s_{9 b}
        -\frac{8110080 s_{8 c}}{7}
        -\frac{50847840 s_{9 c}}{7}
        -\frac{6168576 s_{8 d}}{7}
        +\frac{5289216 s_{9 d}}{7}
        \nonumber \\ &
        +\frac{21034752 s_{9 e}}{7}
        +465408 s_ {9 f}
  \biggr)
  \Biggr\}
  +\mathcal{O}(\epsilon^6)~.
\end{align}

%- }}}

%\clearpage

%- {{{ Pole cancellation plots:

\section{\label{app::poles}Pole cancellation plots}

In this Appendix we present those pole cancellation plots that we did not
show in Section~\ref{sub::acc}.

\begin{figure}[t]
  \begin{center}
    \begin{tabular}{cc}
      \includegraphics[width=0.47\textwidth]{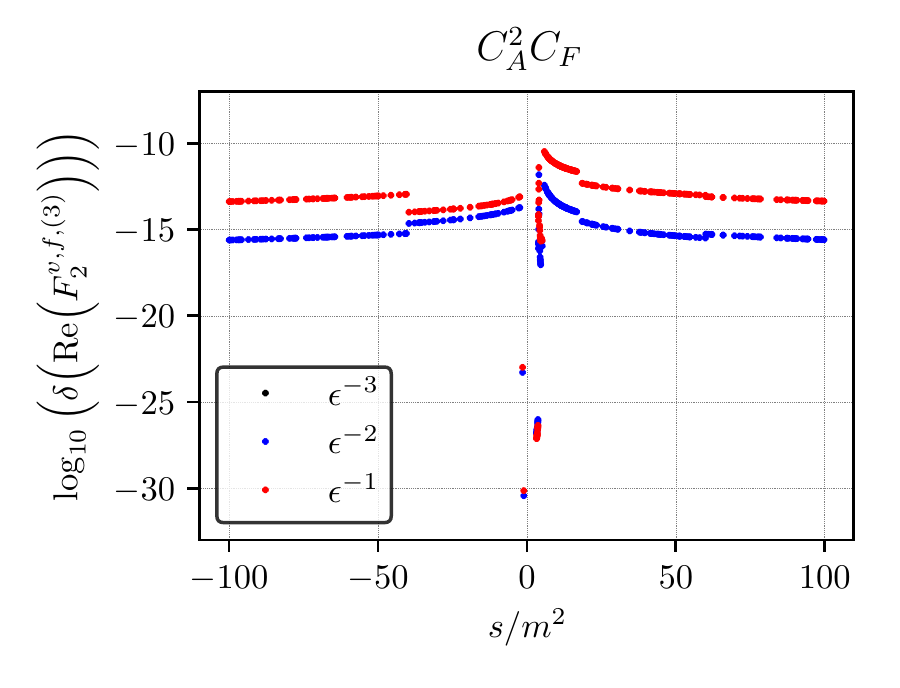}
      &
      \includegraphics[width=0.47\textwidth]{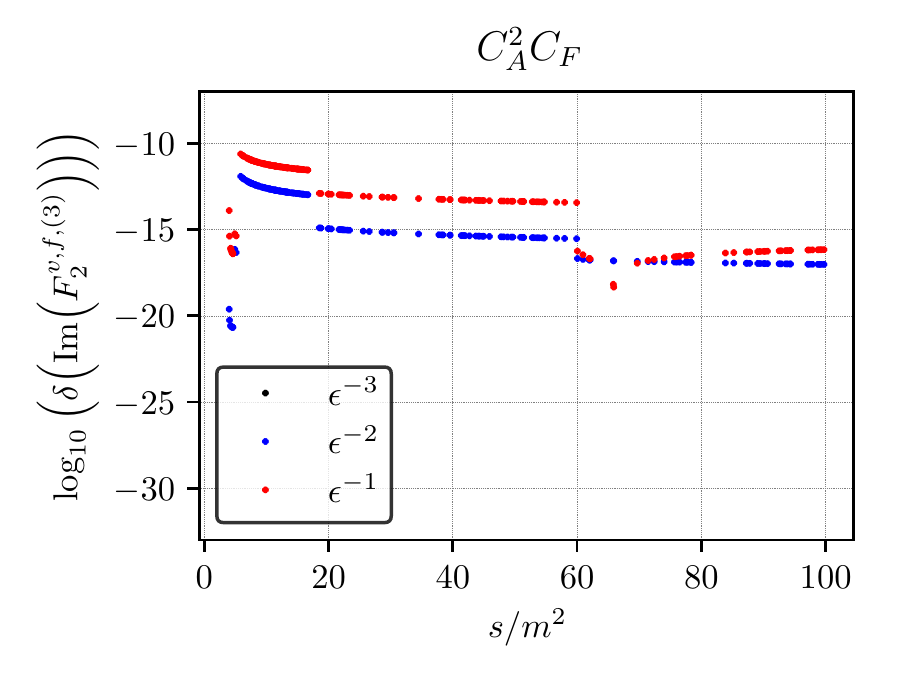}
      \\
      (a) & (b) \\
      \includegraphics[width=0.47\textwidth]{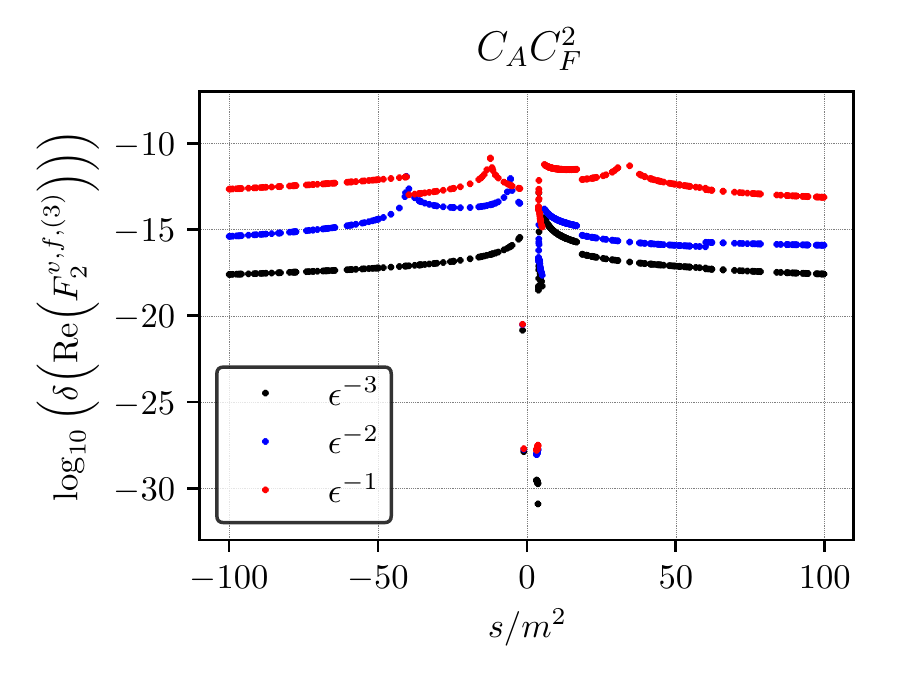}
      &
      \includegraphics[width=0.47\textwidth]{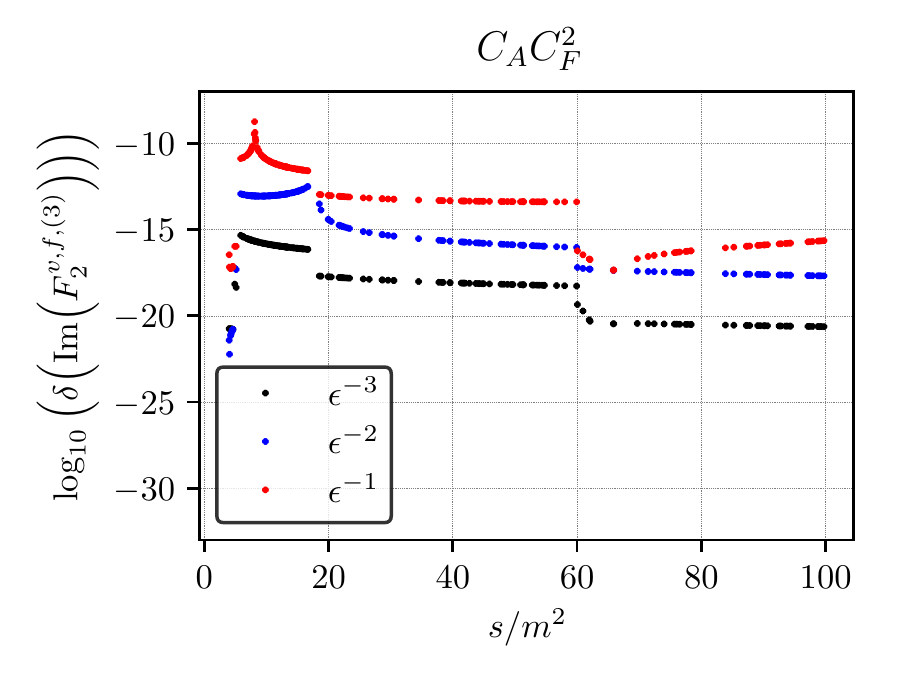}
      \\
      (c) & (d) \\
      \includegraphics[width=0.47\textwidth]{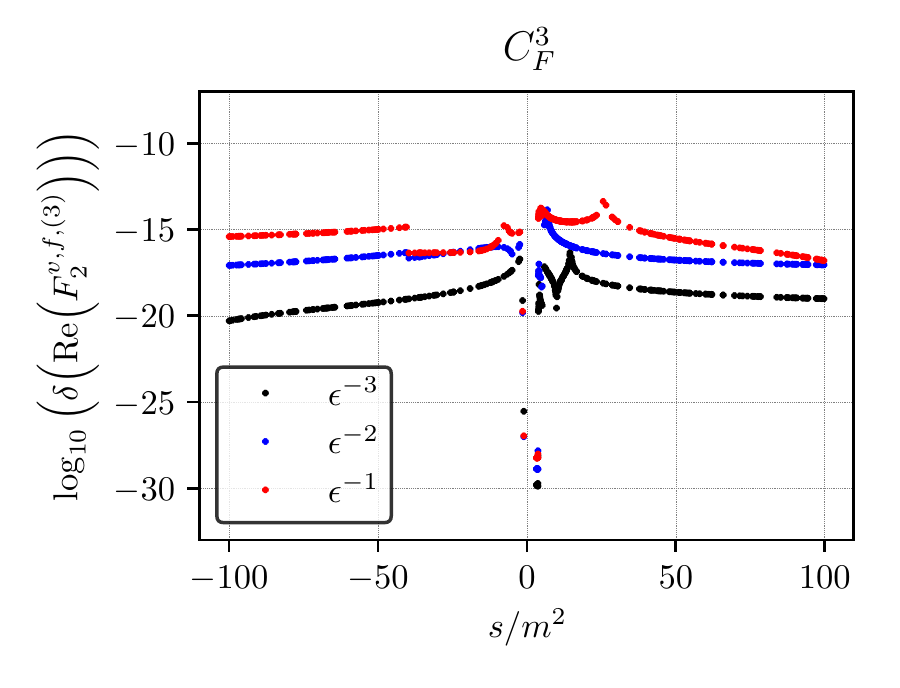}
      &
      \includegraphics[width=0.47\textwidth]{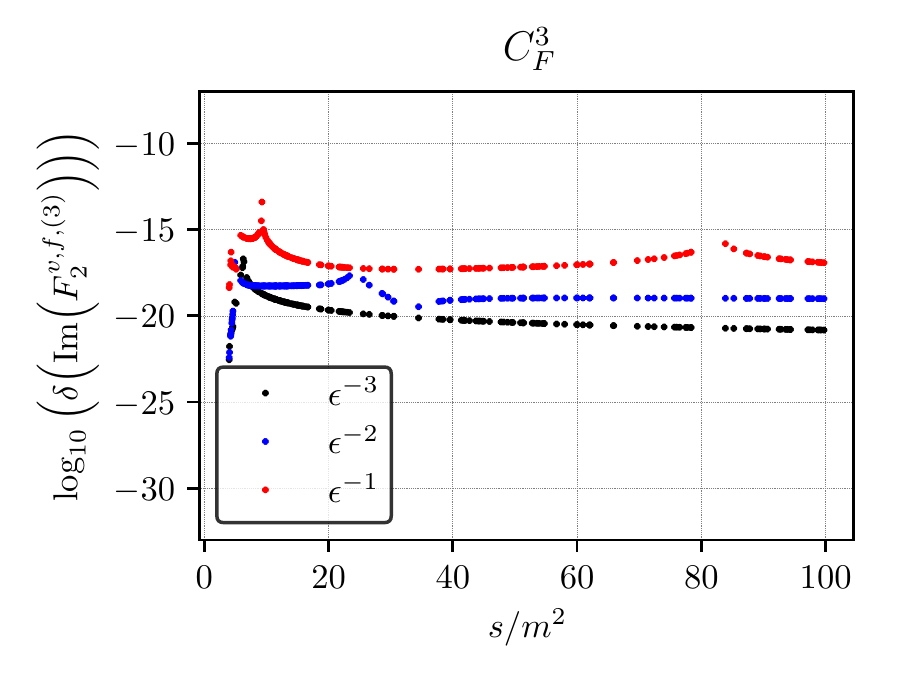}
      \\
      (e) & (f) \\
    \end{tabular}
    \caption{\label{fig::pole-cancellation-veF2}
      Relative cancellation of the real, (a), (c), (e), and imaginary parts, (b), (d), (f), of the poles for the non-fermionic colour structures of $F^{v,f,(3)}_2$, c.f.\ Eq.~(\ref{eq::delta-def}).
      Note that the $1/\epsilon^3$ pole of the colour factor $C_A^2 C_F$
        is zero.
    }
  \end{center}
\end{figure}

\begin{figure}[t]
  \begin{center}
    \begin{tabular}{cc}
      \includegraphics[width=0.47\textwidth]{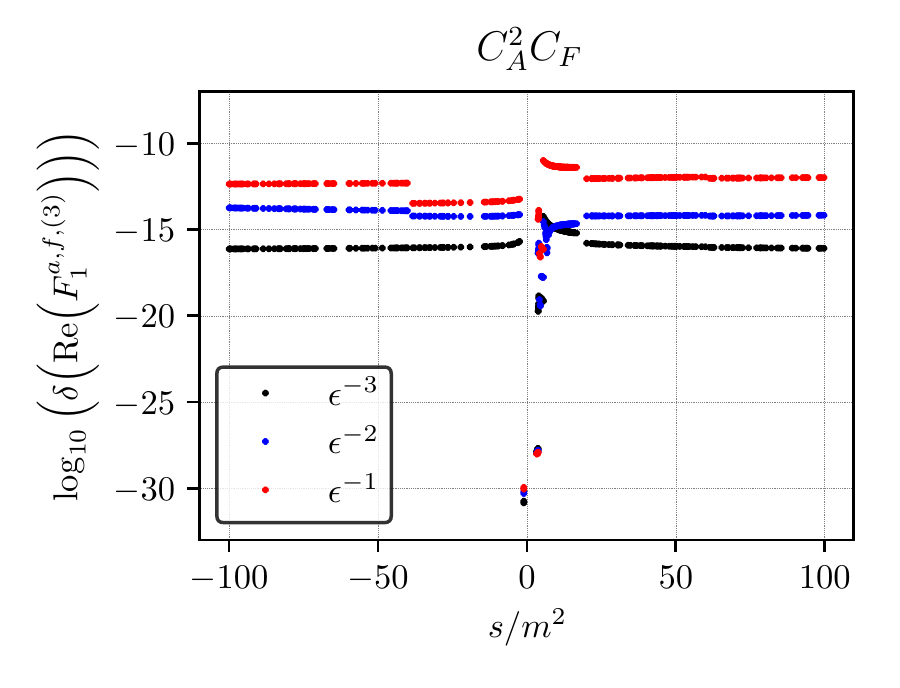}
      &
      \includegraphics[width=0.47\textwidth]{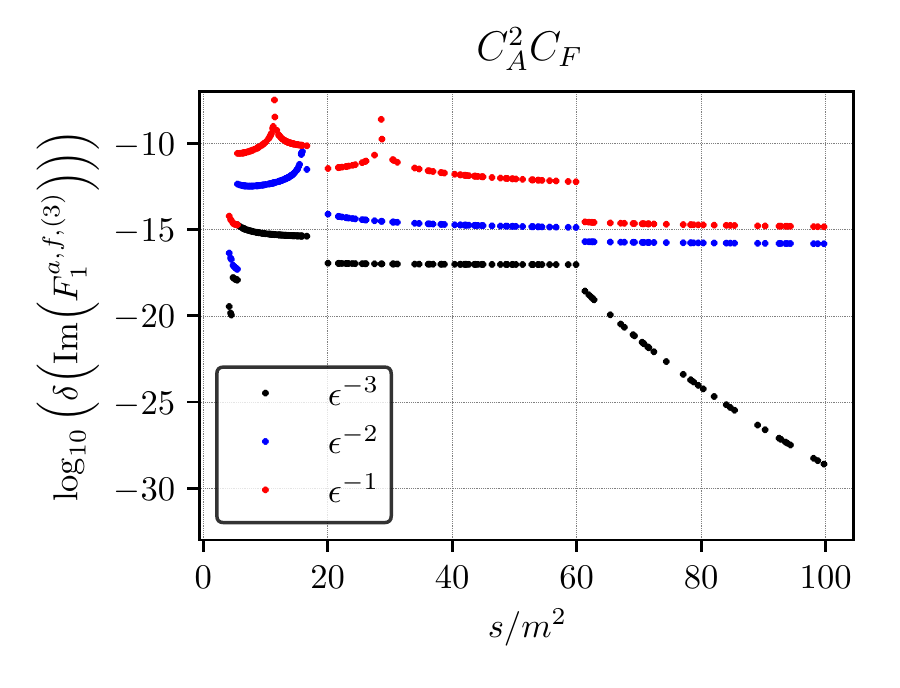}
      \\
      (a) & (b) \\
      \includegraphics[width=0.47\textwidth]{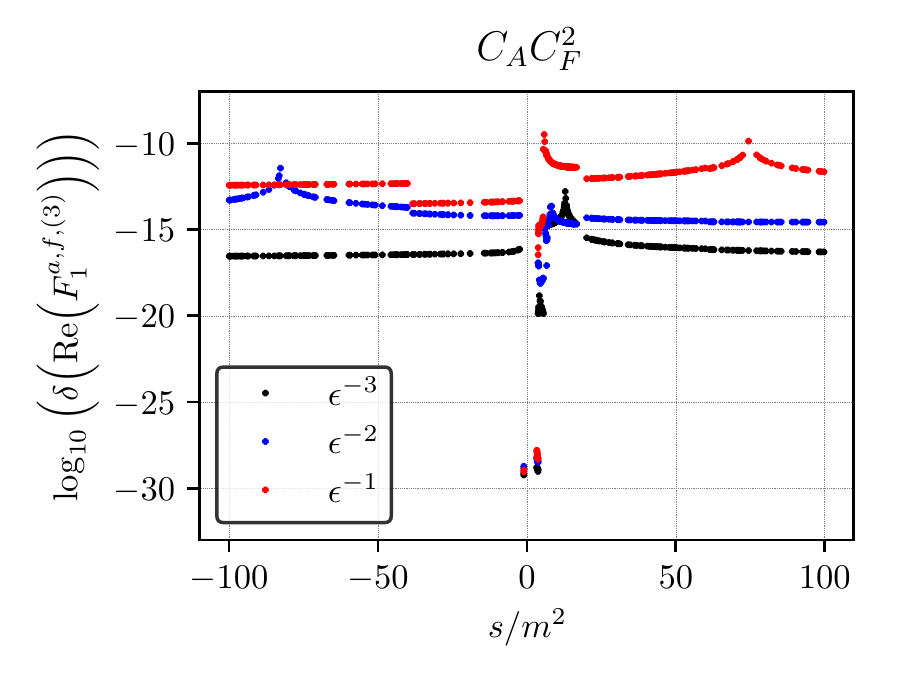}
      &
      \includegraphics[width=0.47\textwidth]{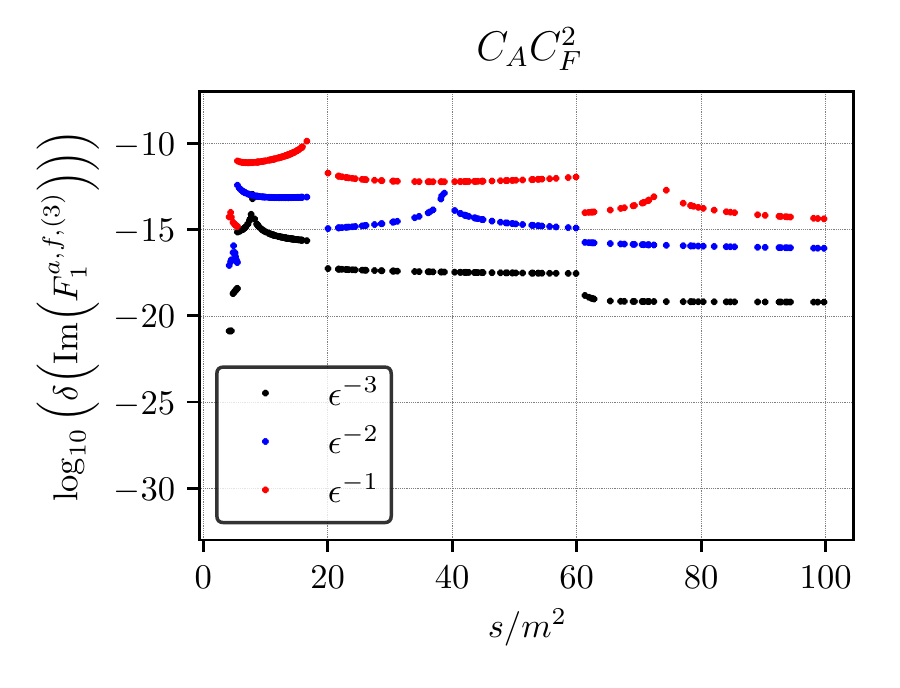}
      \\
      (c) & (d) \\
      \includegraphics[width=0.47\textwidth]{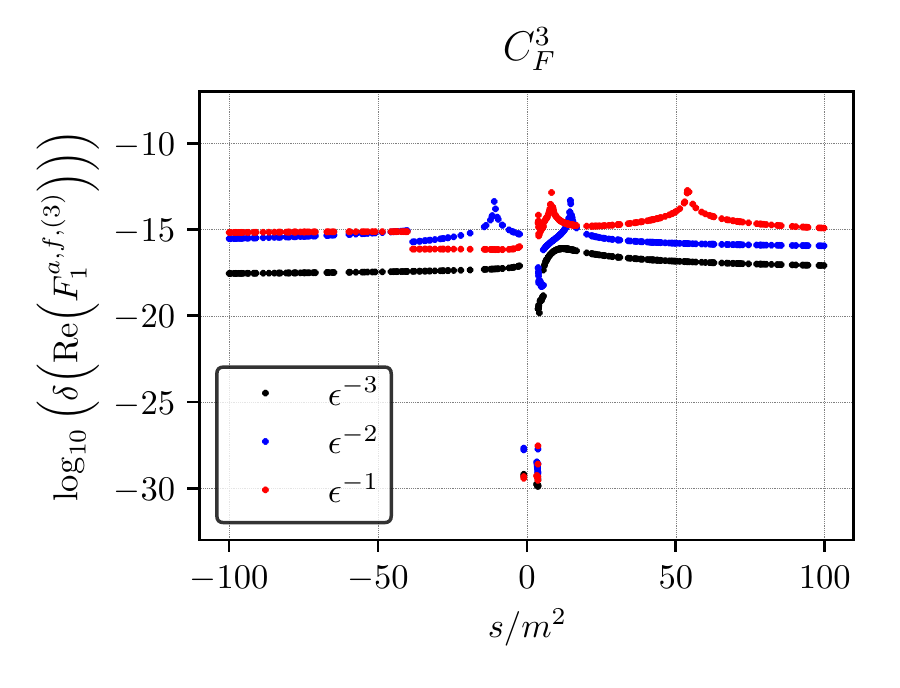}
      &
      \includegraphics[width=0.47\textwidth]{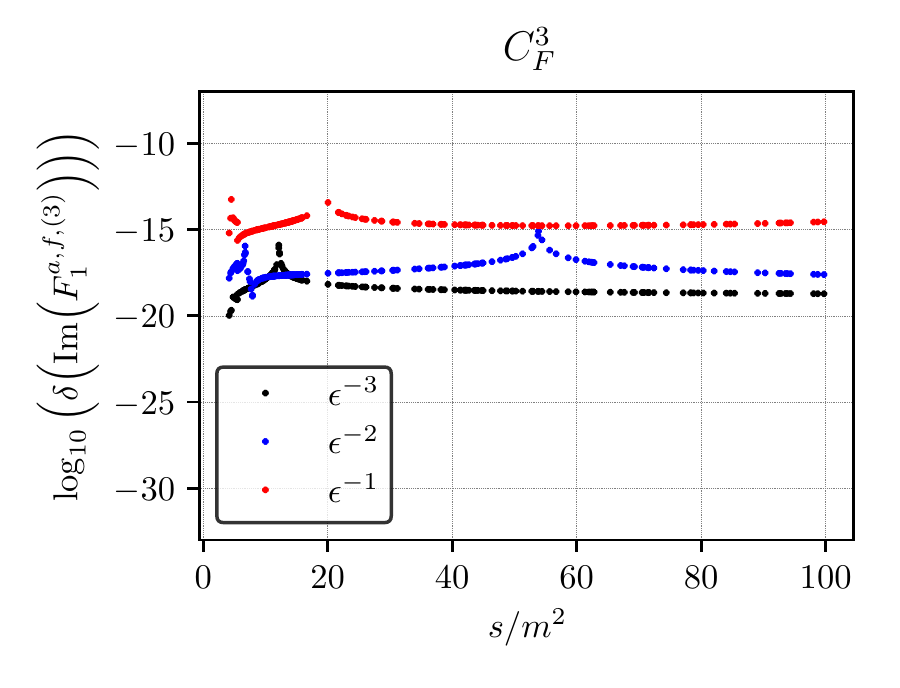}
      \\
      (e) & (f) \\
    \end{tabular}
    \caption{\label{fig::pole-cancellation-axF1}
      Relative cancellation of the real, (a), (c), (e), and imaginary parts, (b), (d), (f), of the poles for the non-fermionic colour structures of $F^{a,f,(3)}_1$, c.f.\ Eq.~(\ref{eq::delta-def}).
    }
  \end{center}
\end{figure}

\begin{figure}[t]
  \begin{center}
    \begin{tabular}{cc}
      \includegraphics[width=0.47\textwidth]{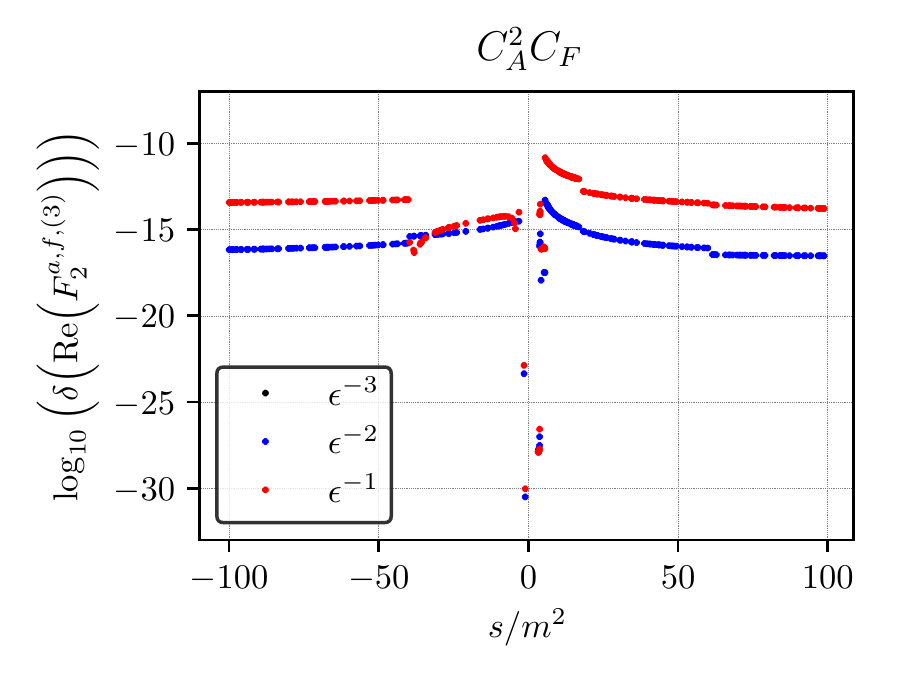}
      &
      \includegraphics[width=0.47\textwidth]{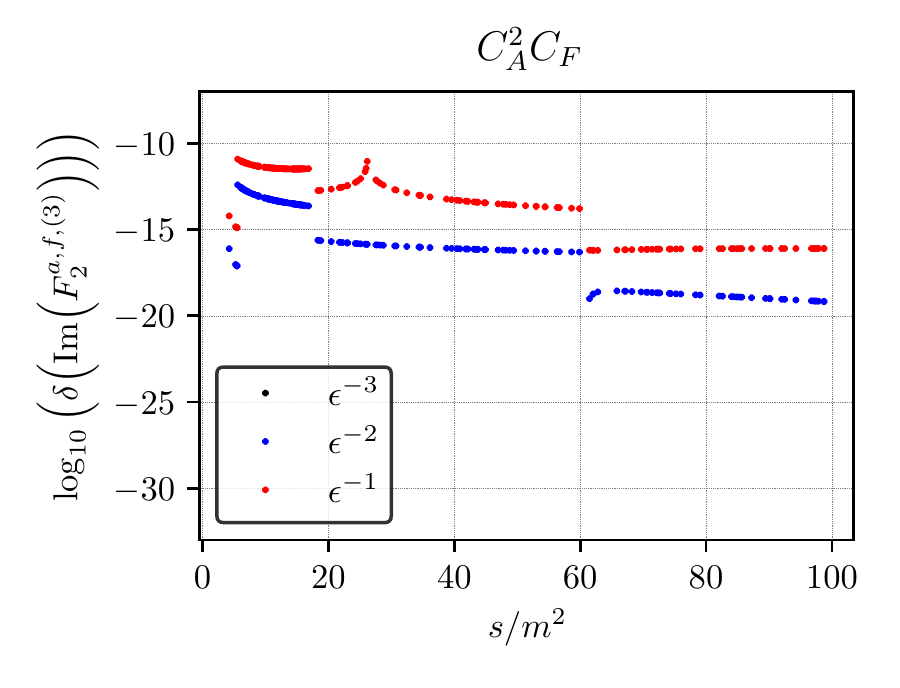}
      \\
      (a) & (b) \\
      \includegraphics[width=0.47\textwidth]{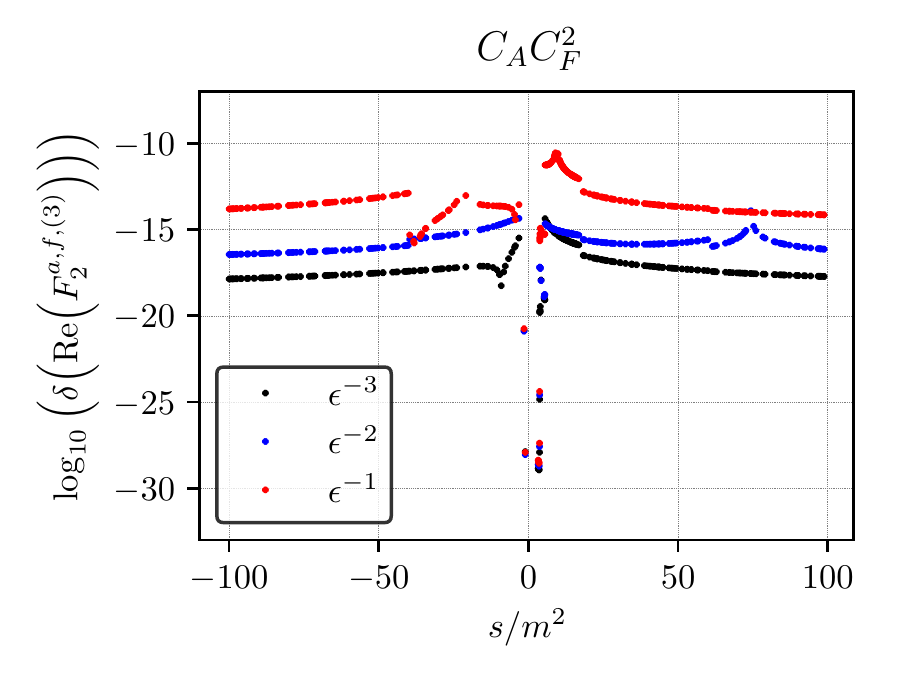}
      &
      \includegraphics[width=0.47\textwidth]{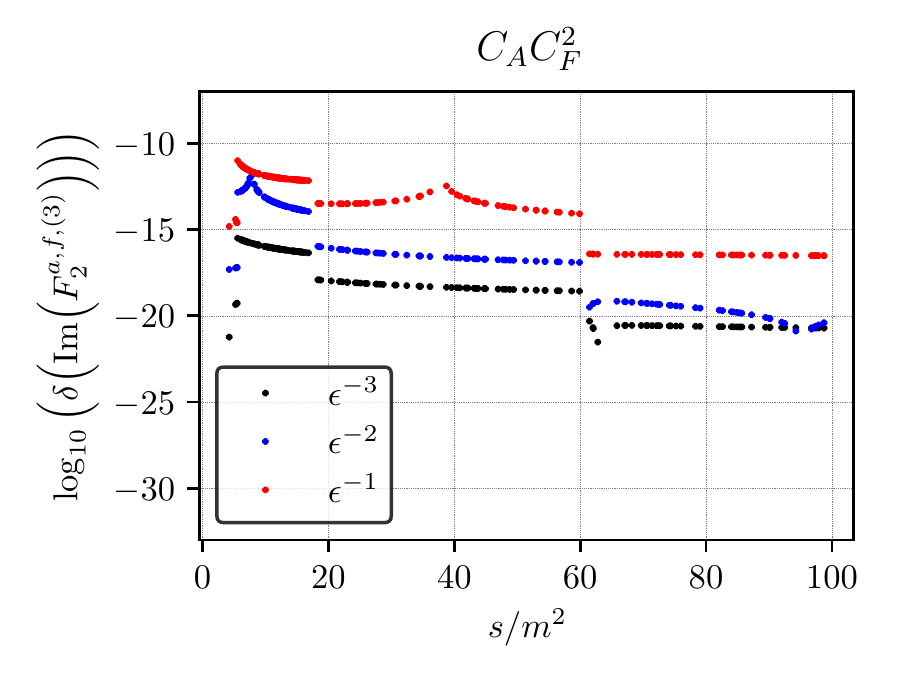}
      \\
      (c) & (d) \\
      \includegraphics[width=0.47\textwidth]{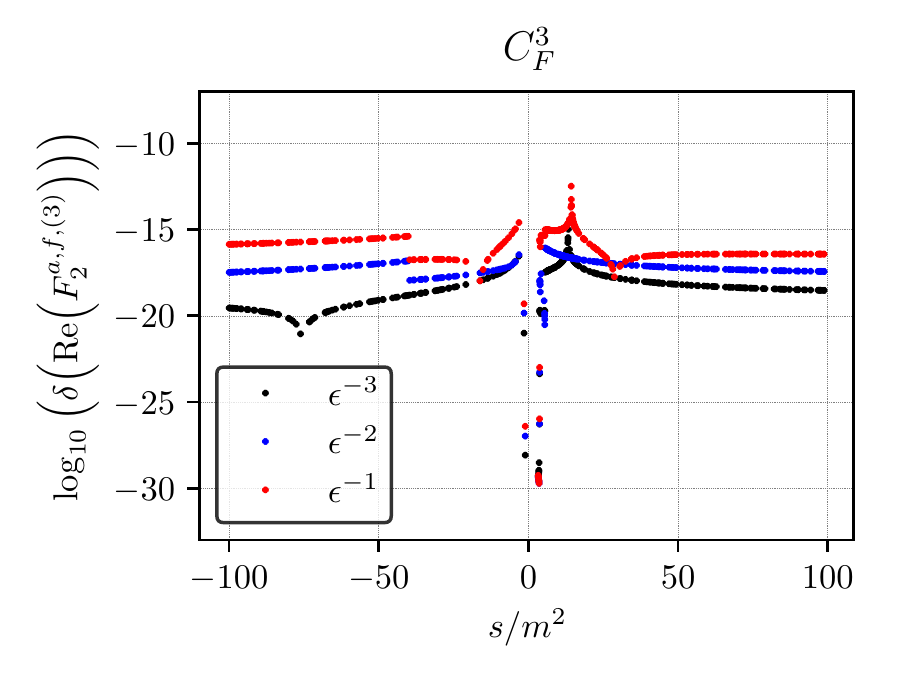}
      &
      \includegraphics[width=0.47\textwidth]{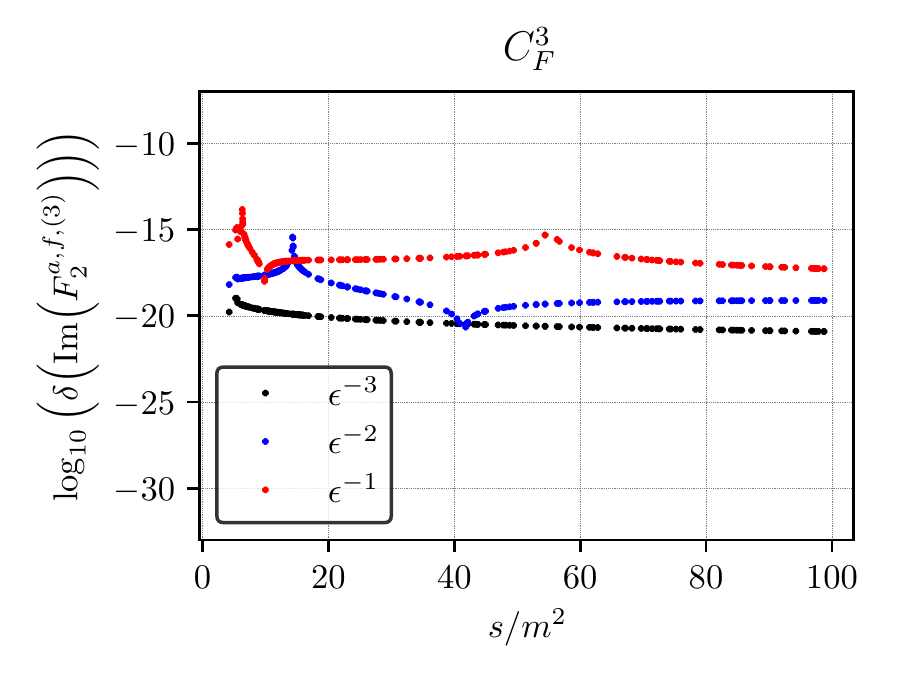}
      \\
      (e) & (f) \\
    \end{tabular}
    \caption{\label{fig::pole-cancellation-axF2}
      Relative cancellation of the real, (a), (c), (e), and imaginary parts, (b), (d), (f), of the poles for the non-fermionic colour structures of $F^{a,f,(3)}_2$, c.f.\ Eq.~(\ref{eq::delta-def}).
      Note that the $1/\epsilon^3$ pole of the colour factor $C_A^2 C_F$
        is zero.
    }
  \end{center}
\end{figure}

\begin{figure}[t]
  \begin{center}
    \begin{tabular}{cc}
      \includegraphics[width=0.47\textwidth]{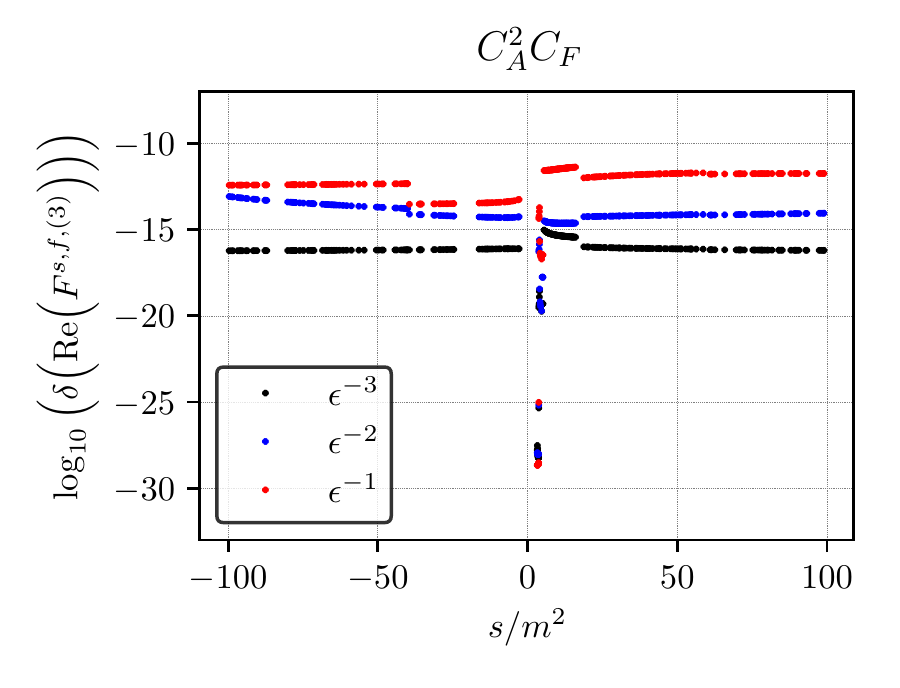}
      &
      \includegraphics[width=0.47\textwidth]{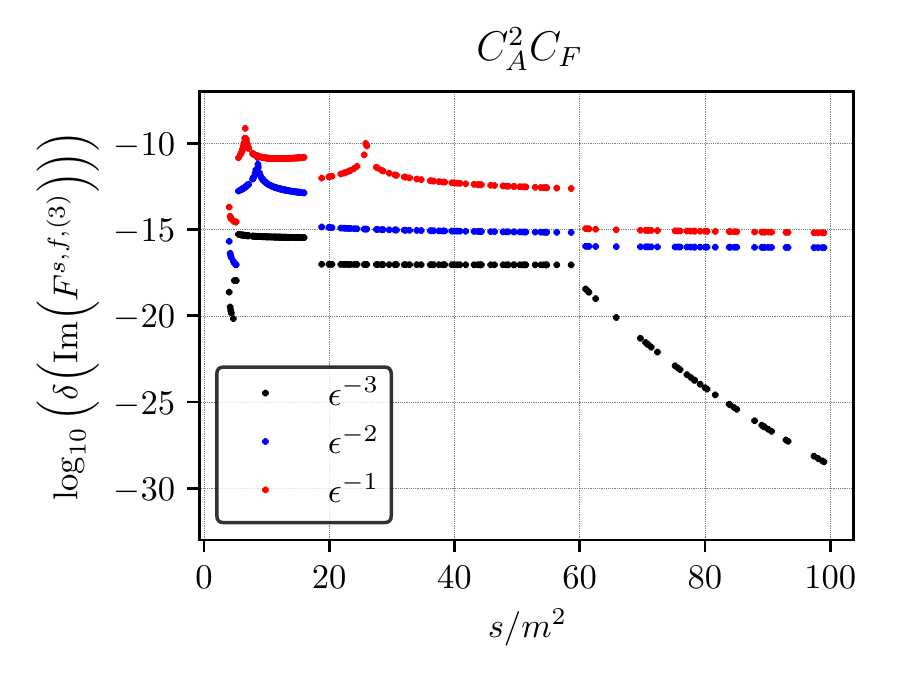}
      \\
      (a) & (b) \\
      \includegraphics[width=0.47\textwidth]{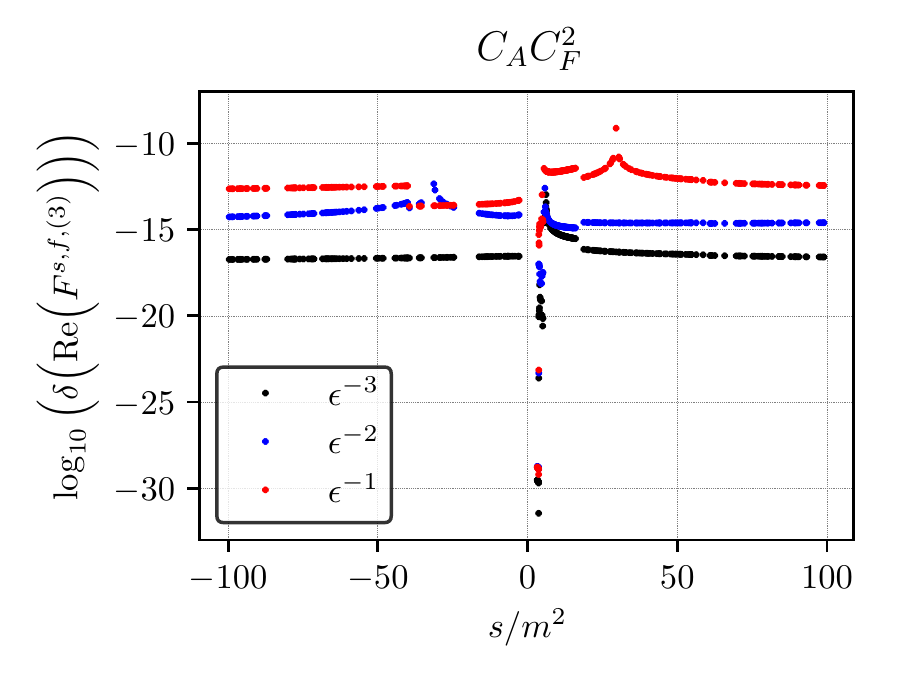}
      &
      \includegraphics[width=0.47\textwidth]{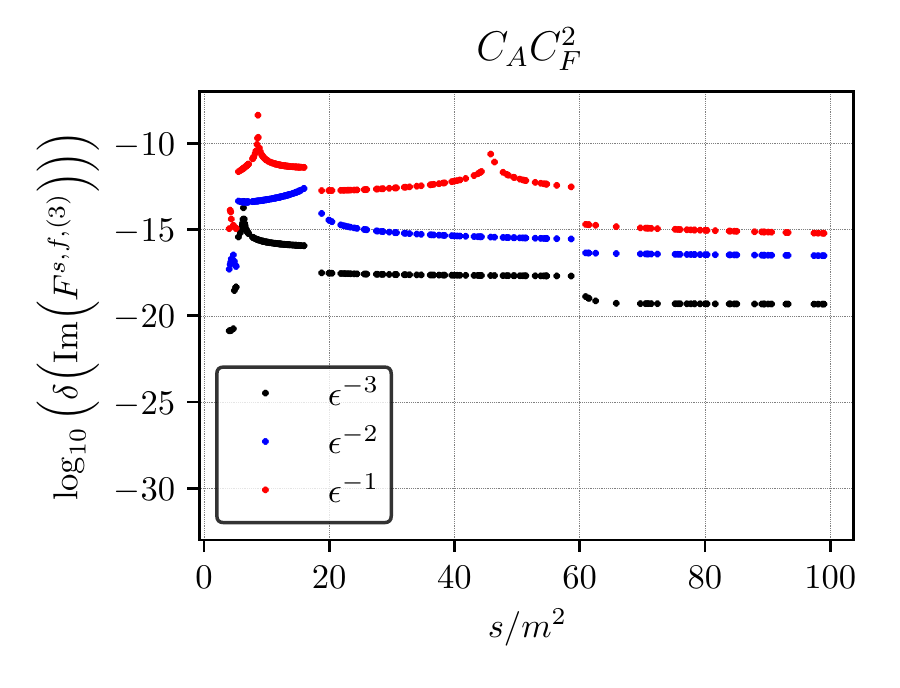}
      \\
      (c) & (d) \\
      \includegraphics[width=0.47\textwidth]{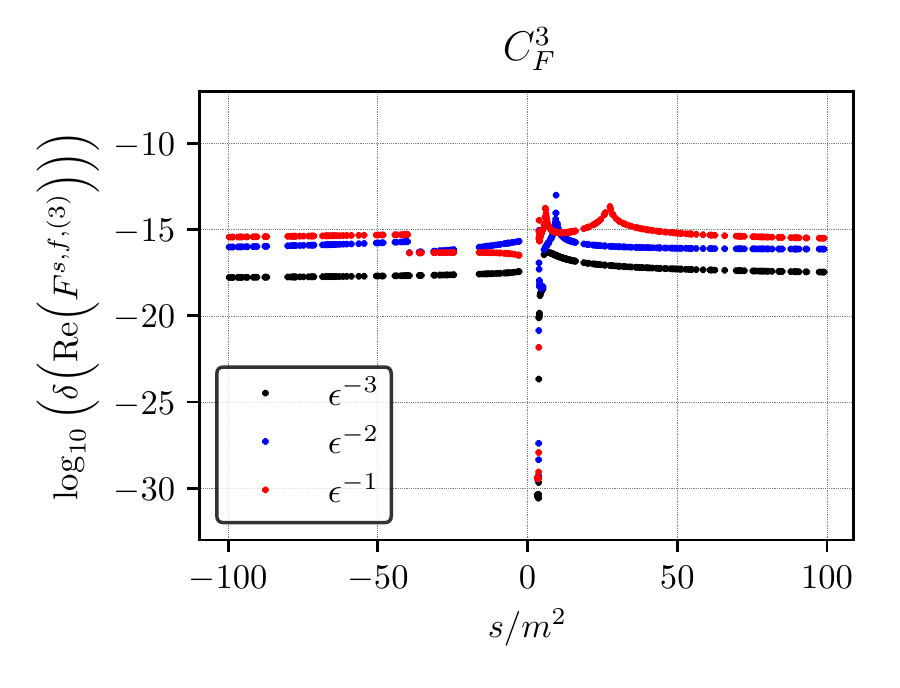}
      &
      \includegraphics[width=0.47\textwidth]{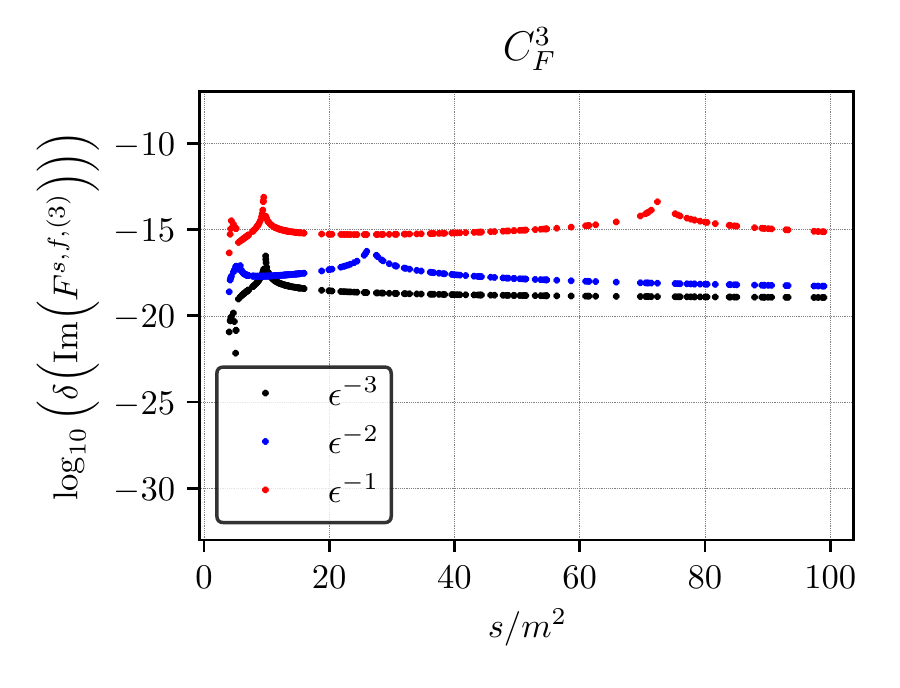}
      \\
      (e) & (f) \\
    \end{tabular}
    \caption{\label{fig::pole-cancellation-sc}
      Relative cancellation of the real, (a), (c), (e), and imaginary parts, (b), (d), (f), of the poles for the non-fermionic colour structures of $F^{s,f,(3)}$, c.f.\ Eq.~(\ref{eq::delta-def}).
    }
  \end{center}
\end{figure}

\begin{figure}[t]
  \begin{center}
    \begin{tabular}{cc}
      \includegraphics[width=0.47\textwidth]{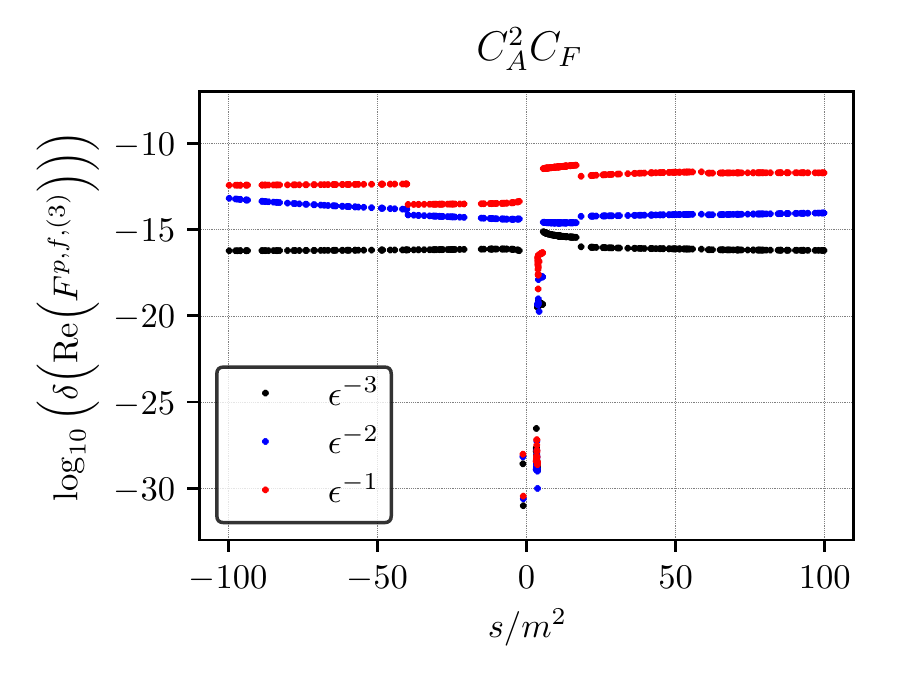}
      &
      \includegraphics[width=0.47\textwidth]{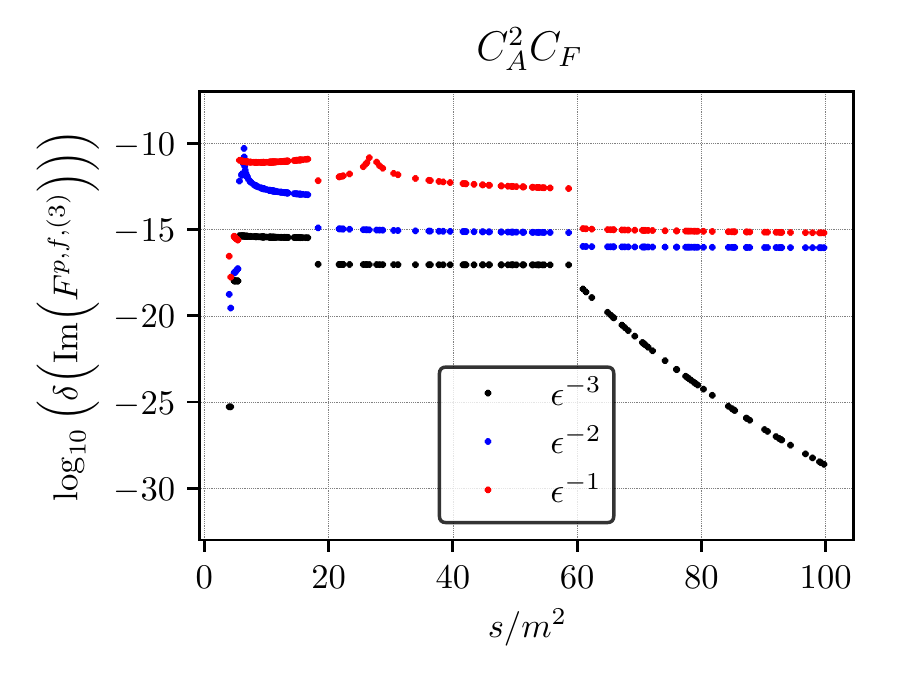}
      \\
      (a) & (b) \\
      \includegraphics[width=0.47\textwidth]{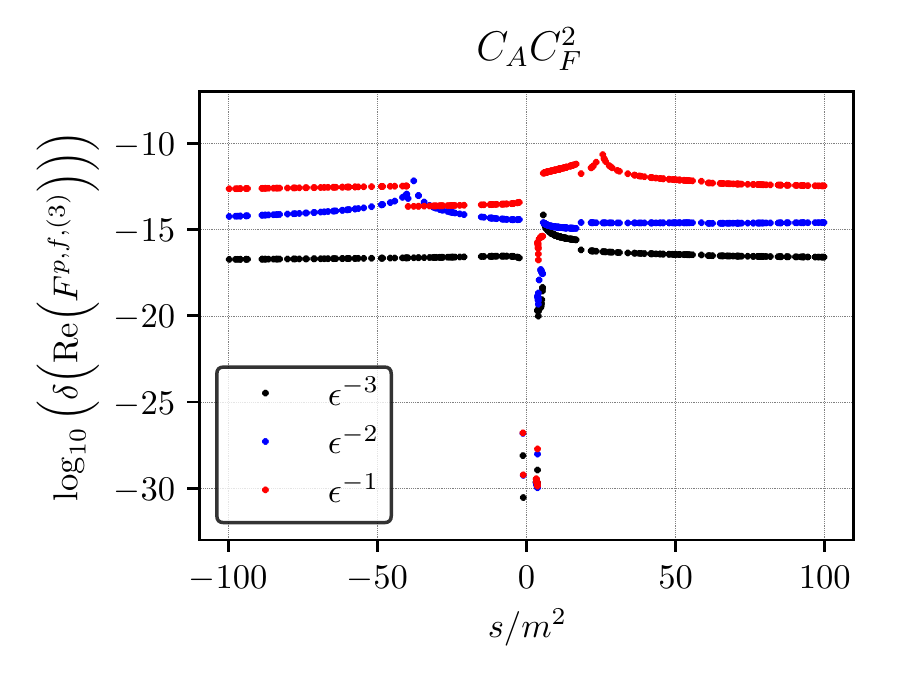}
      &
      \includegraphics[width=0.47\textwidth]{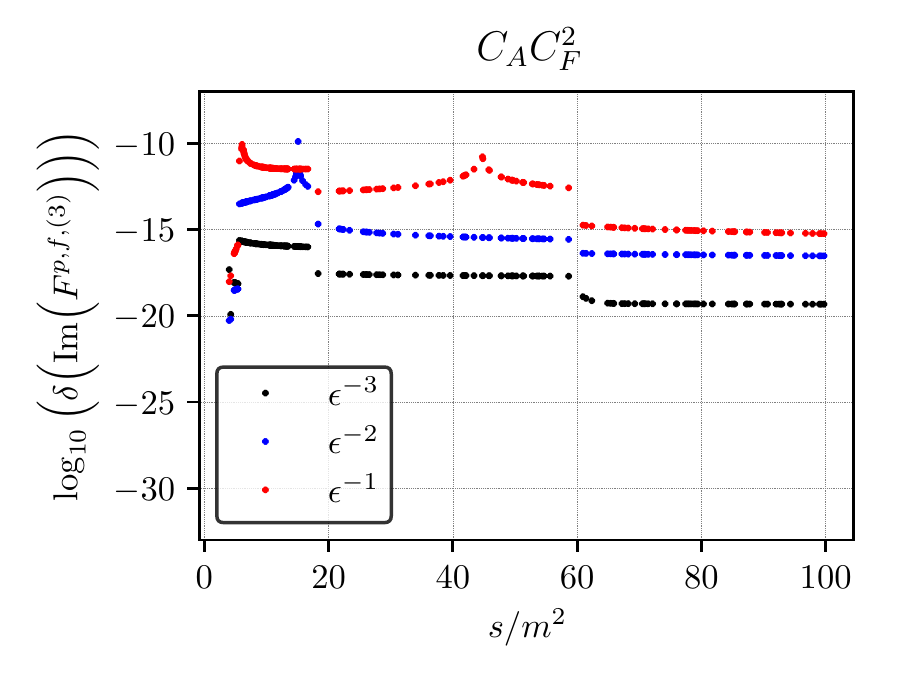}
      \\
      (c) & (d) \\
      \includegraphics[width=0.47\textwidth]{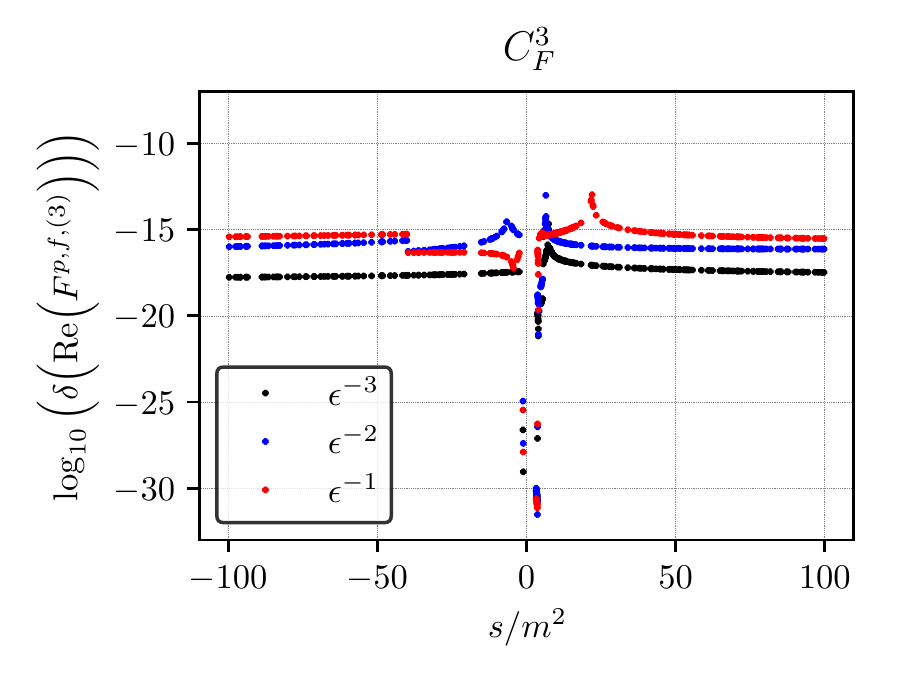}
      &
      \includegraphics[width=0.47\textwidth]{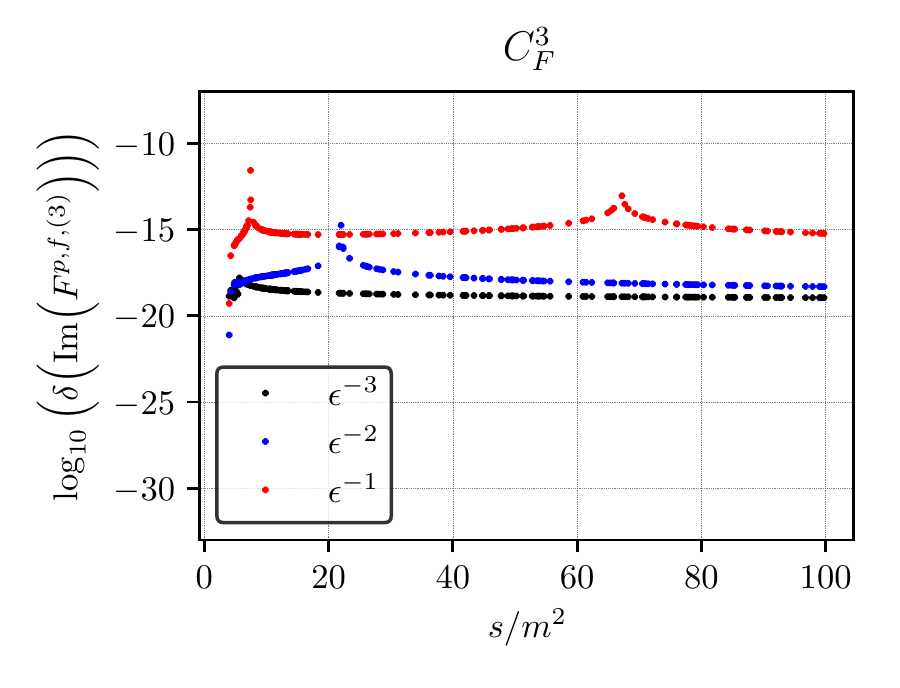}
      \\
      (e) & (f) \\
    \end{tabular}
    \caption{\label{fig::pole-cancellation-ps}
      Relative cancellation of the real, (a), (c), (e), and imaginary parts, (b), (d), (f), of the poles for the non-fermionic colour structures of $F^{p,f,(3)}$, c.f.\ Eq.~(\ref{eq::delta-def}).
    }
  \end{center}
\end{figure}

%- }}}

\clearpage

%- {{{ Analytic results for $s\to0$:

\section{\label{app::s0}Analytic results for \boldmath{$s\to0$}}

In the following we present analytic expansion $s\to0$  of three-loop term for
the non-singlet form factors
$F_2^{v}$, $F_1^{a}$, $F_2^{a}$ and $F^{p}$. The results for $F_1^{v}$
and $F^{s}$ can be found in Section~\ref{subsub::s0}. Our results read
\begin{align}
  &F_2^{v,f,(3)}\Big|_{s\to0} =
  C_A C_F^2 \Bigg[
    -\frac{20 a_4}{3}-\frac{113 \zeta_3}{48}-\frac{5 \pi ^2 \zeta_3}{12}+\frac{185 \zeta_5}{24}-\frac{955}{72}+\frac{1505 \pi ^2}{432}-\frac{35 \pi ^4}{432}-\frac{5 l_2^4}{18} \nonumber\\
    &\quad +\frac{49}{36} \pi ^2 l_2^2-\frac{31}{12} \pi ^2 l_2
  \Bigg]
  + C_A^2 C_F \Bigg[
    \frac{5 a_4}{3}-\frac{\zeta_3}{2}+\frac{29 \pi ^2 \zeta_3}{288}-\frac{65 \zeta_5}{32}+\frac{31231}{2592}-\frac{463 \pi ^2}{216}+\frac{103 \pi ^4}{2880} \nonumber\\
    &\quad +\frac{5 l_2^4}{72}-\frac{11}{18} \pi ^2 l_2^2+\frac{25}{8} \pi ^2 l_2
  \Bigg]
  + C_F^3 \Bigg[
    \frac{20 a_4}{3}+\frac{241 \zeta_3}{24}+\frac{103 \pi ^2 \zeta_3}{72}-\frac{235 \zeta_5}{24}-\frac{101}{64}+\frac{23 \pi ^2}{6} \nonumber\\
    &\quad -\frac{139 \pi ^4}{2160}+\frac{5 l_2^4}{18}-\frac{5}{18} \pi ^2 l_2^2-\frac{22}{3} \pi ^2 l_2
  \Bigg]
  + C_F^2 T_F n_h \Bigg[
    \frac{32 a_4}{3}-\frac{263 \zeta_3}{72}+\frac{2027}{216}+\frac{11 \pi ^2}{162}+\frac{4 \pi ^4}{135} \nonumber\\
    &\quad +\frac{4 l_2^4}{9}-\frac{4}{9} \pi ^2 l_2^2-\frac{16}{9} \pi ^2 l_2
  \Bigg]
  + C_F C_A T_F n_h \Bigg[
    -\frac{20 a_4}{3}-\frac{241 \zeta_3}{36}+\frac{\pi ^2 \zeta_3}{8}-\frac{25 \zeta_5}{24}+\frac{2099}{162} \nonumber\\
    &\quad -\frac{1375 \pi ^2}{648}-\frac{143 \pi ^4}{2160}-\frac{5 l_2^4}{18}+\frac{5}{18} \pi ^2 l_2^2+\frac{32}{9} \pi ^2 l_2
  \Bigg] \nonumber\\
  &\quad + \frac{s}{m^2} \Bigg\{
  C_A C_F^2 \Bigg[
    -\frac{671 a_4}{45}-\frac{589907 \zeta_3}{138240}-\frac{67 \pi ^2 \zeta_3}{360}+\frac{53 \zeta_5}{24}-\frac{317}{60}+\frac{967027 \pi ^2}{1088640}+\frac{4867 \pi ^4}{129600} \nonumber\\
    &\quad -\frac{671 l_2^4}{1080}+\frac{589}{540} \pi ^2 l_2^2+\frac{49 \pi ^2 l_2}{2160}
  \Bigg]
  + C_A^2 C_F \Bigg[
    \frac{28 a_4}{9}-\frac{65 \zeta_3}{864}-\frac{187 \pi ^2 \zeta_3}{2880}-\frac{71 \zeta_5}{192}+\frac{941909}{311040} \nonumber\\
    &\quad -\frac{9189077 \pi ^2}{10886400}+\frac{\pi ^4}{1080}+\frac{7 l_2^4}{54}-\frac{823 \pi ^2 l_2^2}{2160}+\frac{649}{480} \pi ^2 l_2
  \Bigg]
  + C_F^3 \Bigg[
    \frac{782 a_4}{45}+\frac{243809 \zeta_3}{23040}+\frac{671 \pi ^2 \zeta_3}{720} \nonumber\\
    &\quad -\frac{59 \zeta_5}{16}-\frac{16541}{15360}+\frac{8647501 \pi ^2}{3628800}-\frac{7177 \pi ^4}{64800}+\frac{391 l_2^4}{540}-\frac{71}{108} \pi ^2 l_2^2-\frac{589}{108} \pi ^2 l_2
  \Bigg] \nonumber\\
  &\quad + C_F^2 T_F n_h \Bigg[
    \frac{16 a_4}{3}-\frac{2269 \zeta_3}{6912}+\frac{811}{405}+\frac{4403 \pi ^2}{77760}+\frac{8 \pi ^4}{675}+\frac{2 l_2^4}{9}-\frac{2}{9} \pi ^2 l_2^2-\frac{20}{27} \pi ^2 l_2
  \Bigg] \nonumber\\
  &\quad + C_F C_A T_F n_h \Bigg[
    -\frac{37 a_4}{9}-\frac{1493 \zeta_3}{720}+\frac{\pi ^2 \zeta_3}{15}-\frac{\zeta_5}{2}+\frac{56369}{9720}-\frac{14623 \pi ^2}{12960}-\frac{3041 \pi ^4}{129600}-\frac{37 l_2^4}{216} \nonumber\\
    &\quad +\frac{37}{216} \pi ^2 l_2^2+\frac{221}{135} \pi ^2 l_2
  \Bigg]
  \Bigg\} + {\cal O}\left(\frac{s^2}{m^4}\right) + \mbox{$n_l$, $n_l^2$ and $n_h^2$ terms} ,
\end{align}

\begin{align}
  &F_1^{a,f,(3)}\Big|_{s\to0} =
  C_A C_F^2 \Bigg[
    \frac{52 a_4}{9}-\frac{79 \zeta_3}{24}+\frac{\pi ^2 \zeta_3}{18}+\frac{65 \zeta_5}{12}-\frac{2723}{864}+\frac{467 \pi ^2}{288}-\frac{359 \pi ^4}{6480}+\frac{13 l_2^4}{54} \nonumber\\
    &\quad +\frac{10}{27} \pi ^2 l_2^2-\frac{215}{108} \pi ^2 l_2
  \Bigg]
  + C_A^2 C_F \Bigg[
    -\frac{16 a_4}{9}+\frac{215 \zeta_3}{144}+\frac{11 \pi ^2 \zeta_3}{144}-\frac{5 \zeta_5}{3}-\frac{16241}{5184}-\frac{1423 \pi ^2}{1728} \nonumber\\
    &\quad +\frac{97 \pi ^4}{6480}-\frac{2 l_2^4}{27}-\frac{25}{108} \pi ^2 l_2^2+\frac{139}{108} \pi ^2 l_2
  \Bigg]
  + C_F^3 \Bigg[
    -\frac{40 a_4}{9}+\frac{40 \zeta_3}{9}+\frac{7 \pi ^2 \zeta_3}{12}-\frac{20 \zeta_5}{3}-\frac{1141}{576} \nonumber\\
    &\quad +\frac{155 \pi ^2}{216}-\frac{7 \pi ^4}{216}-\frac{5 l_2^4}{27}+\frac{5}{27} \pi ^2 l_2^2-\frac{7}{6} \pi ^2 l_2
  \Bigg]
  + C_F^2 T_F n_h \Bigg[
    \frac{32 a_4}{3}-\frac{355 \zeta_3}{108}+\frac{202}{81}-\frac{433 \pi ^2}{486} \nonumber\\
    &\quad +\frac{\pi ^4}{30}+\frac{4 l_2^4}{9}-\frac{4}{9} \pi ^2 l_2^2+\frac{16}{27} \pi ^2 l_2
  \Bigg]
  + C_F C_A T_F n_h \Bigg[
    -8 a_4-\frac{121 \zeta_3}{27}+\frac{7 \pi ^2 \zeta_3}{36}-\frac{5 \zeta_5}{4}+\frac{449}{24} \nonumber\\
    &\quad -\frac{1019 \pi ^2}{243}-\frac{53 \pi ^4}{1080}-\frac{l_2^4}{3}+\frac{1}{3} \pi ^2 l_2^2+\frac{136}{27} \pi ^2 l_2
  \Bigg] \nonumber\\
  &\quad + \frac{s}{m^2} \Bigg\{
  C_A C_F^2 \Bigg[
    \frac{221 a_4}{50}+\frac{50821 \zeta_3}{38400}-\frac{41 \pi ^2 \zeta_3}{180}+\frac{3 \zeta_5}{32}-\frac{535643}{1036800}-\frac{236713 \pi ^2}{7776000}+\frac{3341 \pi ^4}{259200} \nonumber\\
    &\quad +\frac{221 l_2^4}{1200}+\frac{41}{300} \pi ^2 l_2^2+\frac{413 \pi ^2 l_2}{1800}
  \Bigg]
  + C_A^2 C_F \Bigg[
    -\frac{a_4}{18}+\frac{705841 \zeta_3}{518400}+\frac{7 \pi ^2 \zeta_3}{120}-\frac{91 \zeta_5}{192} \nonumber\\
    &\quad -\frac{13815151}{3499200}+\frac{1480493 \pi ^2}{2332800}+\frac{7 \pi ^4}{259200}-\frac{l_2^4}{432}+\frac{529 \pi ^2 l_2^2}{4320}-\frac{57931 \pi ^2 l_2}{43200}
  \Bigg]
  + C_F^3 \Bigg[
    -\frac{1939 a_4}{225} \nonumber\\
    &\quad -\frac{736801 \zeta_3}{172800}-\frac{7 \pi ^2 \zeta_3}{60}+\frac{\zeta_5}{3}-\frac{1297}{1620}-\frac{386843 \pi ^2}{155520}+\frac{1529 \pi ^4}{25920}-\frac{1939 l_2^4}{5400}-\frac{4121 \pi ^2 l_2^2}{5400} \nonumber\\
    &\quad +\frac{2119}{432} \pi ^2 l_2
  \Bigg]
  + C_F^2 T_F n_h \Bigg[
    \frac{16 a_4}{9}+\frac{413 \zeta_3}{640}-\frac{1693}{960}+\frac{30919 \pi ^2}{388800}+\frac{29 \pi ^4}{4050}+\frac{2 l_2^4}{27}-\frac{2}{27} \pi ^2 l_2^2 \nonumber\\
    &\quad -\frac{4}{27} \pi ^2 l_2
  \Bigg]
  + C_F C_A T_F n_h \Bigg[
    -\frac{163 a_4}{90}+\frac{13291 \zeta_3}{17280}+\frac{23 \pi ^2 \zeta_3}{720}-\frac{7 \zeta_5}{48}+\frac{89623}{38880}-\frac{52861 \pi ^2}{77760} \nonumber\\
    &\quad -\frac{257 \pi ^4}{86400}-\frac{163 l_2^4}{2160}+\frac{163 \pi ^2 l_2^2}{2160}+\frac{323}{540} \pi ^2 l_2
  \Bigg]
  \Bigg\}
  + {\cal O}\left(\frac{s^2}{m^4}\right) + \mbox{$n_l$, $n_l^2$ and $n_h^2$ terms} ,
\end{align}

\begin{align}
  &F_2^{a,f,(3)}\Big|_{s\to0} =
  C_A C_F^2 \Bigg[
    -\frac{1198 a_4}{75}+\frac{52189 \zeta_3}{43200}+\frac{37 \pi ^2 \zeta_3}{180}-\frac{9 \zeta_5}{8}-\frac{304243}{129600}+\frac{3289037 \pi ^2}{972000} \nonumber\\
    &\quad -\frac{1451 \pi ^4}{21600}-\frac{599 l_2^4}{900}+\frac{181}{100} \pi ^2 l_2^2-\frac{18433 \pi ^2 l_2}{5400}
  \Bigg]
  + C_A^2 C_F \Bigg[
    \frac{11 a_4}{3}-\frac{124681 \zeta_3}{43200}-\frac{161 \pi ^2 \zeta_3}{480} \nonumber\\
    &\quad +\frac{27 \zeta_5}{32}+\frac{2720569}{194400}-\frac{113491 \pi ^2}{64800}+\frac{4753 \pi ^4}{129600}+\frac{11 l_2^4}{72}-\frac{19}{24} \pi ^2 l_2^2+\frac{8767 \pi ^2 l_2}{2700}
  \Bigg]
  + C_F^3 \Bigg[
    \frac{432 a_4}{25} \nonumber\\
    &\quad +\frac{211549 \zeta_3}{21600}+\frac{119 \pi ^2 \zeta_3}{360}+\frac{3 \zeta_5}{8}+\frac{61289}{25920}+\frac{43541 \pi ^2}{19440}-\frac{7 \pi ^4}{324}+\frac{18 l_2^4}{25}-\frac{34}{75} \pi ^2 l_2^2-\frac{1109}{180} \pi ^2 l_2
  \Bigg] \nonumber\\
  &\quad + C_F^2 T_F n_h \Bigg[
    -\frac{32 a_4}{9}+\frac{3563 \zeta_3}{1080}+\frac{12199}{3240}+\frac{13421 \pi ^2}{12150}-\frac{44 \pi ^4}{2025}-\frac{4 l_2^4}{27}+\frac{4}{27} \pi ^2 l_2^2-\frac{56}{27} \pi ^2 l_2
  \Bigg] \nonumber\\
  &\quad + C_F C_A T_F n_h \Bigg[
    \frac{76 a_4}{15}+\frac{221 \zeta_3}{2160}-\frac{53 \pi ^2 \zeta_3}{360}+\frac{17 \zeta_5}{24}-\frac{14653}{1215}+\frac{34243 \pi ^2}{9720}-\frac{121 \pi ^4}{32400}+\frac{19 l_2^4}{90} \nonumber\\
    &\quad -\frac{19}{90} \pi ^2 l_2^2-\frac{464}{135} \pi ^2 l_2
  \Bigg] \nonumber\\
  &\quad + \frac{s}{m^2} \Bigg\{
  C_A C_F^2 \Bigg[
    -\frac{131288 a_4}{11025}-\frac{99580907 \zeta_3}{84672000}+\frac{3 \pi ^2 \zeta_3}{40}-\frac{55 \zeta_5}{168}-\frac{106581949}{217728000} \nonumber\\
    &\quad +\frac{33541132409 \pi ^2}{13335840000}-\frac{101909 \pi ^4}{5292000}-\frac{16411 l_2^4}{33075}+\frac{375539 \pi ^2 l_2^2}{264600}-\frac{8183503 \pi ^2 l_2}{2646000}
  \Bigg] \nonumber\\
  &\quad + C_A^2 C_F \Bigg[
    \frac{3874 a_4}{1575}-\frac{32235173 \zeta_3}{67737600}-\frac{2011 \pi ^2 \zeta_3}{20160}+\frac{313 \zeta_5}{1344}+\frac{785687831}{304819200}-\frac{17234629 \pi ^2}{12700800} \nonumber\\
    &\quad +\frac{11279 \pi ^4}{453600}+\frac{1937 l_2^4}{18900}-\frac{7163 \pi ^2 l_2^2}{10800}+\frac{640589 \pi ^2 l_2}{294000}
  \Bigg]
  + C_F^3 \Bigg[
    \frac{51368 a_4}{3675}+\frac{90221321 \zeta_3}{12096000} \nonumber\\
    &\quad +\frac{607 \pi ^2 \zeta_3}{5040}+\frac{5 \zeta_5}{112}-\frac{791347117}{1524096000}+\frac{12650994259 \pi ^2}{13335840000}-\frac{35153 \pi ^4}{496125}+\frac{6421 l_2^4}{11025}-\frac{682 \pi ^2 l_2^2}{3675} \nonumber\\
    &\quad -\frac{478157 \pi ^2 l_2}{189000}
  \Bigg]
  + C_F^2 T_F n_h \Bigg[
    -\frac{16 a_4}{15}+\frac{310831 \zeta_3}{1209600}+\frac{1405037}{680400}+\frac{5910619 \pi ^2}{19051200}-\frac{11 \pi ^4}{1575} \nonumber\\
    &\quad -\frac{2 l_2^4}{45}+\frac{2}{45} \pi ^2 l_2^2-\frac{28}{45} \pi ^2 l_2
  \Bigg]
  + C_F C_A T_F n_h \Bigg[
    \frac{899 a_4}{525}-\frac{17770021 \zeta_3}{24192000}-\frac{\pi ^2 \zeta_3}{24}+\frac{11 \zeta_5}{56} \nonumber\\
    &\quad -\frac{30832783}{15552000}+\frac{23421151 \pi ^2}{34020000}-\frac{3821 \pi ^4}{1512000}+\frac{899 l_2^4}{12600}-\frac{899 \pi ^2 l_2^2}{12600}-\frac{2746 \pi ^2 l_2}{4725}
  \Bigg]
  \Bigg\} \nonumber\\
  &\quad + {\cal O}\left(\frac{s^2}{m^4}\right) + \mbox{$n_l$, $n_l^2$ and $n_h^2$ terms} ,
\end{align}

\begin{align}
  &F^{p,f,(3)}\Big|_{s\to0} =
  C_A C_F^2 \Bigg[
    \frac{64 a_4}{9}+\frac{379 \zeta_3}{96}+\frac{179 \pi ^2 \zeta_3}{144}+\frac{125 \zeta_5}{48}-\frac{36869}{13824}+\frac{305 \pi ^2}{576}-\frac{667 \pi ^4}{3240}+\frac{8 l_2^4}{27} \nonumber\\
    &\quad +\frac{133}{108} \pi ^2 l_2^2-\frac{55}{54} \pi ^2 l_2
  \Bigg]
  + C_A^2 C_F \Bigg[
    -\frac{49 a_4}{9}-\frac{913 \zeta_3}{288}-\frac{415 \pi ^2 \zeta_3}{576}+\frac{35 \zeta_5}{96}+\frac{932761}{124416}-\frac{33 \pi ^2}{128} \nonumber\\
    &\quad +\frac{3461 \pi ^4}{51840}-\frac{49 l_2^4}{216}-\frac{29}{54} \pi ^2 l_2^2+\frac{623}{216} \pi ^2 l_2
  \Bigg]
  + C_F^3 \Bigg[
    \frac{68 a_4}{9}+\frac{1603 \zeta_3}{144}+\frac{31 \pi ^2 \zeta_3}{48}-\frac{175 \zeta_5}{24} \nonumber\\
    &\quad +\frac{1957}{1152}+\frac{8143 \pi ^2}{1728}-\frac{5 \pi ^4}{432}+\frac{17 l_2^4}{54}-\frac{17}{54} \pi ^2 l_2^2-\frac{19}{2} \pi ^2 l_2
  \Bigg]
  + C_F^2 T_F n_h \Bigg[
    8 a_4-\frac{233 \zeta_3}{216}+\frac{6581}{2592} \nonumber\\
    &\quad -\frac{91 \pi ^2}{486}-\frac{19 \pi ^4}{2160}+\frac{l_2^4}{3}-\frac{1}{3} \pi ^2 l_2^2-\frac{8}{27} \pi ^2 l_2
  \Bigg]
  + C_F C_A T_F n_h \Bigg[
    -\frac{20 a_4}{3}-\frac{1609 \zeta_3}{432}+\frac{5 \pi ^2 \zeta_3}{72} \nonumber\\
    &\quad -\frac{5 \zeta_5}{8}+\frac{142537}{15552}-\frac{4181 \pi ^2}{3888}-\frac{\pi ^4}{108}-\frac{5 l_2^4}{18}+\frac{5}{18} \pi ^2 l_2^2+\frac{40}{27} \pi ^2 l_2
  \Bigg] \nonumber\\
  &\quad + \frac{s}{m^2} \Bigg\{
  C_A C_F^2 \Bigg[
    \frac{32 a_4}{75}+\frac{653927 \zeta_3}{345600}-\frac{127 \pi ^2 \zeta_3}{720}-\frac{3 \zeta_5}{16}-\frac{383557}{518400}+\frac{2078087 \pi ^2}{2592000}-\frac{253 \pi ^4}{64800} \nonumber\\
    &\quad +\frac{4 l_2^4}{225}+\frac{707 \pi ^2 l_2^2}{1200}-\frac{14917 \pi ^2 l_2}{21600}
  \Bigg]
  + C_A^2 C_F \Bigg[
    \frac{31 a_4}{36}+\frac{165899 \zeta_3}{259200}-\frac{49 \pi ^2 \zeta_3}{1920}-\frac{101 \zeta_5}{384} \nonumber\\
    &\quad -\frac{3145181}{6998400}+\frac{229537 \pi ^2}{1166400}+\frac{1589 \pi ^4}{172800}+\frac{31 l_2^4}{864}-\frac{163 \pi ^2 l_2^2}{2160}-\frac{7621 \pi ^2 l_2}{14400}
  \Bigg]
  + C_F^3 \Bigg[
    -\frac{967 a_4}{225} \nonumber\\
    &\quad -\frac{348263 \zeta_3}{172800}-\frac{49 \pi ^2 \zeta_3}{1440}+\frac{41 \zeta_5}{96}-\frac{122719}{207360}-\frac{289267 \pi ^2}{155520}+\frac{463 \pi ^4}{8640}-\frac{967 l_2^4}{5400}-\frac{4733 \pi ^2 l_2^2}{5400} \nonumber\\
    &\quad +\frac{1889}{540} \pi ^2 l_2
  \Bigg]
  + C_F^2 T_F n_h \Bigg[
    \frac{8 a_4}{9}+\frac{25403 \zeta_3}{17280}-\frac{13927}{17280}+\frac{140537 \pi ^2}{388800}+\frac{7 \pi ^4}{4050}+\frac{l_2^4}{27}-\frac{1}{27} \pi ^2 l_2^2 \nonumber\\
    &\quad -\frac{2}{3} \pi ^2 l_2
  \Bigg]
  + C_F C_A T_F n_h \Bigg[
    -\frac{49 a_4}{90}+\frac{13733 \zeta_3}{17280}-\frac{7 \pi ^2 \zeta_3}{1440}+\frac{\zeta_5}{32}-\frac{27601}{38880}+\frac{3125 \pi ^2}{15552} \nonumber\\
    &\quad -\frac{1013 \pi ^4}{259200}-\frac{49 l_2^4}{2160}+\frac{49 \pi ^2 l_2^2}{2160}-\frac{47}{180} \pi ^2 l_2
  \Bigg]
  \Bigg\}
  + {\cal O}\left(\frac{s^2}{m^4}\right) + \mbox{$n_l$, $n_l^2$ and $n_h^2$ terms} ,
\end{align}

For the vector and pseudo-scalar singlet form factor we have

\begin{align}
  &F_{1,\,\text{sing}}^{v,f,(3)}\Big|_{s\to0} =
  { - \frac{s}{m^2}} \Bigg\{
  \frac{(d^{abc})^2 n_h}{N_c} \Bigg[
    \frac{50 a_4}{9}-\frac{143 \zeta_3}{216}+\frac{337}{1296}+\frac{43841 \pi ^2}{4860}-\frac{107 \pi ^4}{2592}+\frac{25 l_2^4}{108}- \nonumber\\
    &\quad \frac{25}{108} \pi ^2 l_2^2-\frac{341}{27} \pi ^2 l_2
  \Bigg]
  \Bigg\}
  + {\cal O}\left(\frac{\sqrt{-s}^4}{m^4}\right)\,,
\end{align}
\begin{align}
  &F_{2,\,\text{sing}}^{v,f,(3)}\Big|_{s\to0} =
  \frac{(d^{abc})^2 n_h}{N_c} \Bigg[
    16 a_4-\frac{4 \zeta_3}{3}-\frac{5 \pi ^2 \zeta_3}{18}+\frac{5 \zeta_5}{6}+\frac{5}{9}+\frac{931 \pi ^2}{54}-\frac{41 \pi ^4}{540}+\frac{2 l_2^4}{3}-\frac{2}{3} \pi ^2 l_2^2 \nonumber\\
    &\quad -24 \pi ^2 l_2
  \Bigg] \nonumber\\
  &\quad { - \frac{s^2}{m^2} } \Bigg\{
  \frac{(d^{abc})^2 n_h}{N_c} \Bigg[
    -\frac{116 a_4}{9}-\frac{887 \zeta_3}{360}+\frac{\pi ^2 \zeta_3}{36}-\frac{\zeta_5}{12}+\frac{86}{135}-\frac{234329 \pi ^2}{18900}+\frac{2707 \pi ^4}{32400} \nonumber\\
   &\quad -\frac{29 l_2^4}{54}+\frac{29}{54} \pi ^2 l_2^2+\frac{793}{45} \pi ^2 l_2
  \Bigg]
  \Bigg\}
  + {\cal O}\left(\frac{\sqrt{-s}^4}{m^4}\right)\,,
\end{align}
\begin{align}
  &F_\text{sing}^{p,f,(3)}\Big|_{s\to0} =
  C_A C_F T_F n_h \Bigg[
    8 a_4+\frac{103 \zeta_3}{24}+\frac{1133}{864}+\frac{3401 \pi ^2}{2592}+\frac{2 \pi ^4}{135}+\frac{l_2^4}{3}-\frac{1}{3} \pi ^2 l_2^2-2 \pi ^2 l_2
  \Bigg] \nonumber\\
  &\quad + C_F^2 T_F n_h \Bigg[
    -8 a_4-\frac{71 \zeta_3}{36}+\frac{37}{216}+\frac{1195 \pi ^2}{648}-\frac{\pi ^4}{40}-\frac{l_2^4}{3}+\frac{1}{3} \pi ^2 l_2^2-\frac{4}{3} \pi ^2 l_2
  \Bigg] \nonumber\\
  &\quad + C_F T_F^2 n_h^2 \Bigg[
    -\frac{16 \zeta_3}{9}-\frac{1}{27}+\frac{26 \pi ^2}{81}
  \Bigg]
  + C_F T_F^2 n_h n_l \Bigg[
    -\frac{11}{27}-\frac{17 \pi ^2}{81}
  \Bigg] \nonumber\\
  &\quad + \frac{\sqrt{-s}}{m} \Bigg\{
  C_A C_F T_F n_h \Bigg[
    \frac{19}{48} \pi ^2 l_{\sqrt{-s}/m}+\frac{\pi ^4}{48}-\frac{565 \pi ^2}{576}
  \Bigg]
  + C_F^2 T_F n_h \Bigg[
    -\frac{5 \pi ^2}{16}
  \Bigg] \nonumber\\
  &\quad + C_F T_F^2 n_h n_l \Bigg[
    \frac{19 \pi ^2}{72}-\frac{1}{6} \pi ^2 l_{\sqrt{-s}/m}
  \Bigg]
  \Bigg\} \nonumber\\
  &\quad { - \frac{s}{m^2} } \Bigg\{
  C_A C_F T_F n_h \Bigg[
    -\frac{67 a_4}{90}+\frac{11 l_{\sqrt{-s}/m}^2}{24}+\frac{1}{15} \pi ^2 l_{\sqrt{-s}/m}-\frac{287 l_{\sqrt{-s}/m}}{144}-\frac{20977 \zeta_3}{43200} \nonumber\\
    &\quad -\frac{1139 \pi ^4}{259200}-\frac{10358353 \pi ^2}{27216000}+\frac{1843381}{1036800}-\frac{67 l_2^4}{2160}+\frac{67 \pi ^2 l_2^2}{2160}+\frac{547 \pi ^2 l_2}{1350}
  \Bigg]
  + C_F^2 T_F n_h \Bigg[
    \frac{6 a_4}{5} \nonumber\\
    &\quad -\frac{l_{\sqrt{-s}/m}}{2}+\frac{1463 \zeta_3}{57600}+\frac{17 \pi ^4}{4320}+\frac{56269 \pi ^2}{362880}+\frac{296969}{345600}+\frac{l_2^4}{20}-\frac{1}{20} \pi ^2 l_2^2-\frac{52}{135} \pi ^2 l_2
  \Bigg] \nonumber\\
  &\quad + C_F T_F^2 n_h^2 \Bigg[
    \frac{97 \zeta_3}{216}+\frac{13}{648}-\frac{394 \pi ^2}{6075}
  \Bigg]
  + C_F T_F^2 n_h n_l \Bigg[
    -\frac{1}{6} l_{\sqrt{-s}/m}^2+\frac{19 l_{\sqrt{-s}/m}}{36}+\frac{263 \pi ^2}{4860} \nonumber\\
    &\quad -\frac{353}{1296}
  \Bigg]
  \Bigg\}
  + {\cal O}\left(\frac{\sqrt{-s}^3}{m^3}\right)\,
\end{align}
The result for $F_{\rm sing}^{s,f,(3)}$ is given in Eq.~(\ref{eq::Fssing_s0}).

%- }}}

\end{appendix}

\end{document}